\documentclass[rmp,aps,twocolumn,nofootinbib,amsmath,amssymb]{revtex4}

\usepackage{graphics}
\usepackage{epsfig}

\begin{document}

%%%%%%%%%%%%%%%%%%%%%%%%%%%%%%%%%%%%%%%%%%%%%%%%%%
%\include{newnames}

\newcommand{\dmdt}{\partial M/\partial T}

%-------------------

\newcommand{\URuSi}{URu$_2$Si$_2$}
\newcommand{\urusi}{URu$_2$Si$_2$}
\newcommand{\upt}{UPt$_3$}
\newcommand{\UPt}{UPt$_3$}
\newcommand{\updal}{UPd$_2$Al$_3$}
\newcommand{\unial}{UNi$_2$Al$_3$}

\newcommand{\ube}{UBe$_{13}$}
\newcommand{\uthbe}{U$_{1-x}$Th$_x$Be$_{13}$}

\newcommand{\uge}{UGe$_{2}$}
\newcommand{\urhge}{URhGe}
\newcommand{\uir}{UIr}

\newcommand{\au}{$\alpha$-U}
\newcommand{\ual}{UAl$_{2}$}

\newcommand{\thcrsi}{ThCr$_{2}$Si$_{2}$}
\newcommand{\cemt}{CeM$_{2}$X$_{2}$}
\newcommand{\cerusi}{CeRu$_{2}$Si$_{2}$}
\newcommand{\ceruge}{CeRu$_{2}$Ge$_{2}$}
\newcommand{\cecusi}{CeCu$_{2}$Si$_{2}$}
\newcommand{\cecusige}{CeCu$_{2}$Si$_{2-x}$Ge$_{x}$}
\newcommand{\cecuge}{CeCu$_{2}$Ge$_{2}$}
\newcommand{\cepdsi}{CePd$_{2}$Si$_{2}$}
\newcommand{\cerhsi}{CeRh$_{2}$Si$_{2}$}

\newcommand{\cepdge}{CePd$_{2}$Ge$_{2}$}
\newcommand{\cenige}{CeNi$_{2}$Ge$_{2}$}

%------------------------------------------------------
\newcommand{\hocoga}{HoCoGa$_{5}$}
\newcommand{\hoga}{HoGa$_{3}$}
\newcommand{\mga}{MGa$_{2}$}

\newcommand{\cemin}{Ce$_n$M$_m$In$_{3n+2m}$}
\newcommand{\mint}{MIn$_{2}$}

\newcommand{\cein}{CeIn$_{3}$}
\newcommand{\cesn}{CeSn$_{3}$}

\newcommand{\ceminof}{CeMIn$_{5}$}
\newcommand{\cecoin}{CeCoIn$_{5}$}
\newcommand{\cerhin}{CeRhIn$_{5}$}
\newcommand{\ceirin}{CeIrIn$_{5}$}

\newcommand{\celacoin}{Ce$_{1-x}$La$_x$CoIn$_{5}$}
\newcommand{\cecoinsn}{CeCoIn$_{5-x}$Sn$_x$}
\newcommand{\cecorhin}{CeCo$_{1-x}$Rh$_x$In$_{5}$}
\newcommand{\cecoirin}{CeCo$_{1-x}$Ir$_x$In$_{5}$}
\newcommand{\cerhirin}{CeRh$_{1-x}$Ir$_x$In$_{5}$}
\newcommand{\cerhinsn}{CeRhIn$_{5-x}$Sn$_x$}

\newcommand{\ceminte}{Ce$_{2}$MIn$_{8}$}
\newcommand{\cecointe}{Ce$_{2}$CoIn$_{8}$}
\newcommand{\cerhinte}{Ce$_{2}$RhIn$_{8}$}
\newcommand{\ceirinte}{Ce$_{2}$IrIn$_{8}$}

\newcommand{\larhin}{LaRhIn$_{5}$}

\newcommand{\pucoga}{PuCoGa$_{5}$}
\newcommand{\purhga}{PuRhGa$_{5}$}
\newcommand{\puirga}{PuIrGa$_{5}$}
\newcommand{\npcoga}{NpCoGa$_{5}$}
\newcommand{\nppdal}{NpPd$_5$Al$_{2}$}

\newcommand{\cepdal}{CePd$_5$Al$_{2}$}

%------------------------------------------------------

\newcommand{\ceptsi}{CePt$_{3}$Si}
\newcommand{\celaptsi}{La$_x$Ce$_{1-x}$Pt$_{3}$Si}
\newcommand{\ceptge}{CePt$_{3}$Ge}
\newcommand{\cenigeot}{CeNiGe$_{3}$}
\newcommand{\cenigettf}{Ce$_2$Ni$_3$Ge$_{5}$}
\newcommand{\cenisittf}{Ce$_2$Ni$_3$Si$_{5}$}

\newcommand{\ceptsige}{CePt$_{3}$Si$_{1-x}$Ge$_x$}
\newcommand{\cerhsiot}{CeRhSi$_{3}$}
\newcommand{\larhsiot}{LaRhSi$_{3}$}
\newcommand{\ceirsiot}{CeIrSi$_{3}$}
\newcommand{\ceircosi}{CeIr$_{1-x}$Co$_{x}$Si$_3$}
\newcommand{\ceptsiot}{CePtSi$_{3}$}
\newcommand{\cecogeot}{CeCoGe$_{3}$}

\newcommand{\laptsi}{LaPt$_{3}$Si}
\newcommand{\larhsi}{LaRh$_{3}$Si}
\newcommand{\lairsi}{LaIr$_{3}$Si}
\newcommand{\lapdsi}{LaPd$_{3}$Si}

\newcommand{\rc}{R$_{2}$C$_{3-y}$}
\newcommand{\yc}{Y$_{2}$C$_{3-y}$}
\newcommand{\lac}{La$_{2}$C$_{3-y}$}

\newcommand{\cdreo}{CdRe$_{2}$O$_7$}
\newcommand{\lipdptb}{Li$_2$(Pd,Pt)$_{3}$B}
\newcommand{\lipdb}{Li$_2$Pd$_{3}$B}
\newcommand{\liptb}{Li$_2$Pt$_{3}$B}

%------------------------------------------------------

\newcommand{\prossb}{PrOs$_{4}$Sb$_{12}$}
\newcommand{\laossb}{LaOs$_{4}$Sb$_{12}$}

\newcommand{\prrusb}{PrRu$_{4}$Sb$_{12}$}
\newcommand{\prfesb}{PrFe$_{4}$Sb$_{12}$}
\newcommand{\prrup}{PrRu$_{4}$P$_{12}$}
\newcommand{\prfep}{PrFe$_{4}$P$_{12}$}

%------------------------------------------------------

\newcommand{\ceni}{Ce$_{7}$Ni$_{3}$}
\newcommand{\ceru}{CeRu$_{2}$}

\newcommand{\ccaux}{CeCu$_{6-x}$Au$_{x}$}
\newcommand{\ccautwo}{CeCu$_{5.8}$Au$_{0.2}$}
\newcommand{\ccau}{CeCu$_{5}$Au}
\newcommand{\ceal}{CeAl$_{3}$}
\newcommand{\cealt}{CeAl$_{2}$}

\newcommand{\ccagx}{CeCu$_{6-x}$Ag$_{x}$}
\newcommand{\ccag}{CeCu$_{5}$Ag}

\newcommand{\rrhb}{RRh$_{4}$B$_{4}$}
\newcommand{\errhb}{ErRh$_{4}$B$_{4}$}
\newcommand{\horhb}{HoRh$_{4}$B$_{4}$}
\newcommand{\erhorhb}{Er$_{1-x}$Ho$_x$Rh$_{4}$B$_{4}$}

\newcommand{\dymos}{DyMo$_{6}$S$_{8}$}

\newcommand{\ybrhsi}{YbRh$_{2}$Si$_{2}$}
\newcommand{\ybcusi}{YbCu$_{2}$Si$_{2}$}
\newcommand{\ybnial}{Yb$_{2}$Ni$_{2}$Al}

\newcommand{\efe}{$\epsilon$-Fe}

\newcommand{\zrzn}{ZrZn$_{2}$}
\newcommand{\zrv}{ZrV$_{2}$}
\newcommand{\hfzn}{HfZn$_{2}$}
\newcommand{\zrhfzn}{Zr$_{1-x}$Hf$x$Zn$_{2}$}
\newcommand{\nial}{Ni$_{3}$Al}
\newcommand{\scin}{ScIn$_{3}$}
\newcommand{\crbe}{CrBe$_{12}$}
\newcommand{\cosul}{CoS$_{2}$}

\newcommand{\yni}{YNi$_{3}$}
\newcommand{\ynits}{Y$_{2}$Ni$_{7}$}
\newcommand{\ynitst}{Y$_{2}$Ni$_{17}$}

\newcommand{\rco}{RECo$_{2}$}
\newcommand{\ycoot}{YCo$_{2}$}
\newcommand{\ycons}{Y$_{9}$Co$_{7}$}
\newcommand{\luco}{LuCo$_{2}$}
\newcommand{\scco}{ScCo$_{2}$}
\newcommand{\zrco}{ZrCo$_{2}$}
\newcommand{\hfco}{HfCo$_{2}$}

\newcommand{\auv}{Au$_{4}$V}

\newcommand{\tibe}{TiBe$_{2}$}
\newcommand{\tibecu}{TiBe$_{2-x}$Cu$_{x}$}
\newcommand{\niga}{Ni$_{3}$Ga}

\newcommand{\ymn}{YMn$_{2}$}
\newcommand{\yscmn}{Y$_{1-x}$Sc$_{x}$Mn$_{2}$}
\newcommand{\nis}{NiS$_{2}$}
\newcommand{\mnsi}{Mn$_{3}$Si}

\newcommand{\srsingle}{Sr$_{2}$RuO$_{4}$}
\newcommand{\srdouble}{Sr$_{3}$Ru$_{2}$O$_{7}$}
\newcommand{\srinf}{SrRuO$_{3}$}

\newcommand{\het}{$^{3}$He}
\newcommand{\hef}{$^{4}$He}

\newcommand{\lto}{L$_{2}$1}
\newcommand{\lot}{L1$_{2}$}
\newcommand{\cob}{C1$_{b}$}

\newcommand{\ybco}{YBa$_{2}$Cu$_{3}$O$_{7-\delta}$}

\newcommand{\mgb}{MgB$_{2}$}
\newcommand{\vsi}{V$_{3}$Si}
\newcommand{\nbsn}{Nb$_{3}$Sn}

%%%%%%%%%%%%%%%%%%%%%%%%%%%%%%%%%%%%%%%%%%%%%%%%%%
\title{Superconducting phases of f-electron compounds}

\author{Christian Pfleiderer}
\email{christian.pfleiderer@frm2.tum.de}
\affiliation{Physik Department E21, Technische Universit\"at M\"unchen, D-85748 Garching, Germany}

\begin{abstract}  
Intermetallic compounds containing f-electron elements display a wealth of superconducting phases, that are prime candidates for unconventional pairing with complex order parameter symmetries. For instance, superconductivity has been found at the border of magnetic order as well as deep within ferro- and antiferromagnetically ordered states, suggesting that magnetism may promote rather than destroy superconductivity. Superconductivity near valence transitions, or in the vicinity of magneto-polar order are candidates for new superconductive pairing interactions such as fluctuations of the conduction electron density or the crystal electric field, respectively. The experimental status of the study of the  superconducting phases of f-electron compounds is reviewed.
\end{abstract}                                                                 

\maketitle
\tableofcontents

%%%%%%%%%%%%%%%%%%%%%%%%%%%%%%%%%%%%%%%%%%%%%%%%%
\section{INTRODUCTION
\label{intro}}

Superconductivity was discovered almost a century ago. Yet, unexpected and fascinating new variants of this same old theme are being found at an increasing pace. This is due to great technical advances in materials preparation and an increasingly more systematic screening of new compounds. Prior to the late 1970s all known superconductors could be accounted for in terms of a condensate of Cooper pairs, where the Cooper pairs form due to electron-phonon interactions. With the discovery of the superfluid phases of {\het} this understanding began to change in two ways \cite{oshe72,voll90}. First, {\het} provided an example of non-electron-phonon mediated pairing. Second, it provided an example of a superfluid condensate that breaks additional symmetries. The discovery of heavy-fermion superconductivity as a prime candidate for complex order parameter symmetries and non-electron-phonon mediated pairing in f-electron compounds nearly three decades ago was long recognized as an important turning point in the history of superconductivity. However, progress in heavy fermion superconductivity until not long ago seemed to have been slow.

\begin{figure}
\includegraphics[width=.35\textwidth,clip=]{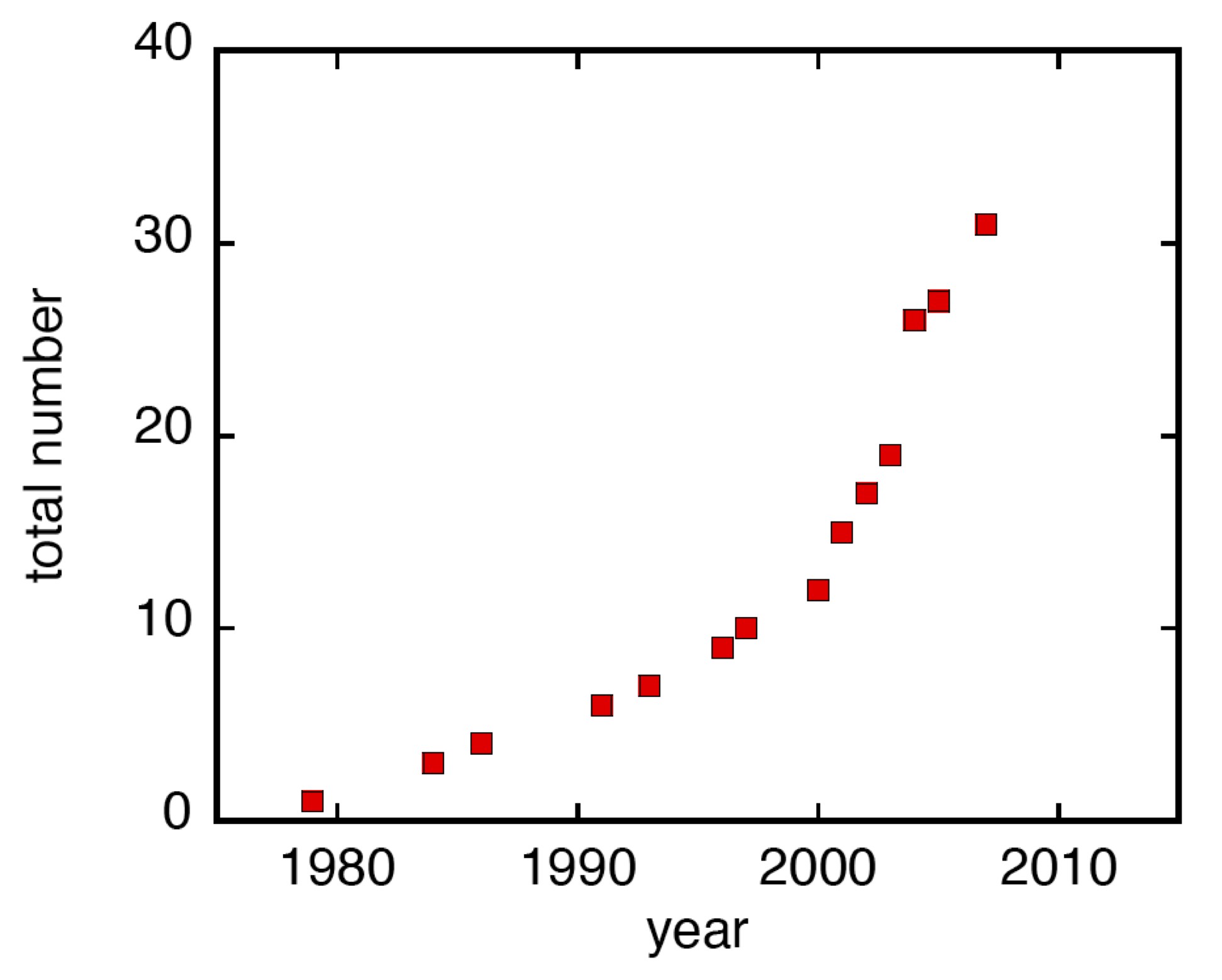}
\caption{Evolution of the total number of f-electron heavy-fermion superconductors. Systems included in this plot and covered in this review: 1979 {\cecusi}; 1984 {\ube}, {\upt}; 1986 {\urusi}; 1991 {\updal}, {\unial}; 1993 {\cecuge}; 1996 {\cepdsi}, {\cenige}; 1997 {\cein}; 2000 {\cerhin}, {\uge}; 2001 {\ceirin}, {\cecoin}, URhGe; 2002 {\pucoga}, {\prossb}; 2003 {\cerhinte}, {\purhga}; 
2004	 {\cenigeot}, {\cenigettf}, UIr, {\prrup}, {\ceptsi}, {\ceirsiot}, {\cerhsiot}; 2005 {\prrusb};
2007 UCoGe, NpPd$_5$Al$_2$, CeCoGe$_3$, CePd$_5$Al$_2$.}
\label{HFS(year)} 
\end{figure}

In recent years especially the superconductivity in the cuprates, ruthenates, cobaltates, pyrochlores and iron-pnictides received great attention. However, a spectacular series of discoveries and developments in f-electron superconductors took place at the same time. While in the first twelve years following the discovery of heavy-fermion superconductivity in {\cecusi} only five more heavy fermion superconductors could be identified, over twenty five additional systems have been found in the past fifteen years (see Fig.\,\ref{HFS(year)}). By now over thirty systems are known, about half of which were discovered in the past five years alone. This illustrates the speed of development the field of f-electron superconductivity has picked up despite its long tradition.

As a result there is growing appreciation that superconducting phases of f-electron compounds frequently exist at the border of competing and coexisting forms of electronic order. For the majority of systems, including the original heavy-fermion superconductors, an interplay with antiferromagnetism is observed. However, there are also several examples of superconductivity that coexists with ferromagnetism. Further examples include superconductivity at the border of polar order and near electron localization transitions. Finally, several  heavy-fermion superconductors have even been discovered with non-centrosymmetric crystal structures and coexistent antiferromagnetic order. The large variety of systems found so far establishes unconventional f-electron superconductivity as a rather general phenomenon. It also suggests the existence of further unimagined forms of superconductivity.

The objective of this review is it to give a status report of the experimental properties of the candidates for unconventional f-electron superconductivity. For a long time the search for a unified microscopic theory of f-electron superconductivity has been hampered by the large differences of the small number of known systems. Even though the increasing number of systems has allowed great progress in understanding, a critical discussion of the theoretical scenarios is well beyond the length constraints of the present review.  For reviews of selected compounds and theoretical scenarios we refer to \cite{grew91,sigr91,saul94,mine99,joyn02,thal05a,thal05b,flou05,flou06,sigr05,mapl08}.

The outline of this paper is as follows. The introduction is continued in section \ref{intro}, with a short account of conventional superconductivity and its interplay with magnetism, Fermi liquid quasiparticle interactions and advances in materials preparation. In section \ref{inte-anti-supe} we address the interplay of antiferromagnetism and superconductivity. Section \ref{inte-ferr-supe} is concerned with ferromagnetism and superconductivity, while we review the properties of emergent classes of new superconductors, discovered very recently, in section \ref{emer-clas-inte-supe}. Finally, in section \ref{mult-supe-phas}, we summarize evidence for multiple superconducting phases in {\upt} and tentative indications for such behavior in other systems as well as for the formation of textures. The paper closes with a short section on the general perspectives of this field. 

\subsection{Superconductivity versus Magnetism}
\label{supe-vers-magn}

Superconductors derive their name from being perfect electrical conductors. However, in contrast to ideal conductors superconductors display, as their second defining property, perfect diamagnetism, i.e., in the superconducting state sufficiently low applied magnetic fields are spontaneously expelled. Flux expulsion identifies superconductivity as a thermodynamic phase. 

Following the discovery of the two defining properties of superconductors, notably perfect conductivity and perfect diamagnetism by Onnes in 1911 \cite{kame11a,kame11b,kame11c} and Meissner and Ochsenfeld in 1933 \cite{meis33}, respectively, it took until 1957 when Bardeen, Cooper and Schrieffer (BCS) proposed a remarkably successful theoretical framework \cite{bard57}. There is a large number of excellent introductory and advanced level textbooks and review papers, e.g., \cite{park69,dege89,tink96,wald96,sigr05}. BCS theory identifies superconductivity as the quantum-statistical condensation of so-called Cooper pairs, which are bound pairs of quasiparticle excitations in a Fermi liquid. For a simple Hamiltonian describing attractively interacting quasiparticles in a conduction band, it is possible to show the formation of an excitation gap $\Delta$ in the quasiparticle spectrum at the Fermi level $E_F$. 

A superconducting transition exists for quasiparticle systems with both attractive and repulsive components of the quasiparticle interactions \cite{ande62}. For instance, in the presence of electron-phonon interactions the Coulomb repulsion of conduction electrons is screened and exhibits a retarded attractive interaction component below the Debye frequency. Physically speaking, the electrons avoid the bare Coulomb repulsion and attract each other in terms of a polarization trace that decays slowly as compared with the speed of travel of the electrons. The mathematical form of the $T_s$ is essentially the same as for purely attractive interactions, but the Coulomb interaction enters in a renormalized form. The same is also true when keeping track of the full retarded solution in the Eliashberg strong-coupling formalism \cite{elia60}, which leads to the MacMillan form of  $T_s$ \cite{macm68,alle75}.  We return to more complex quasiparticle interactions of strongly correlated electron systems in section \ref{road-supe-phas}.

The experimental characteristics of conventional superconductors derive from the formation of an isotropic gap at the Fermi surface. This implies, that bulk properties such as the specific heat show an exponential temperature dependence below $T_s$ and, in the weak coupling limit, an anomaly $\Delta C/\gamma T_s=1.43$. At the heart of the understanding of the superconducting state is the formation of quantum mechanical phase coherence as seen in several microscopic probes. For instance, the NMR spin-lattice relaxation rate  shows coherence effects like the Hebel-Slichter peak and an exponential freezing out below $T_s$ (for a pedagogical discussion with examples see \cite{wald96,tink96}). The rigidity of the superconductivity against external perturbations is expressed by the phase stiffness of superconducting condensate, as measured by the coherence length $\xi$. The length scale of the variations of the superconducting order parameter is expressed by the Pippard or Ginzburg-Landau coherence length.

Taking into account electron-phonon coupling the resulting screened, retarded quasiparticle interactions are short-ranged, representing essentially contact interactions. For the corresponding Cooper pair wave-function, which is composed of the product of an orbital and spin-contribution, this implies that the orbital contribution has to be in the $l=0$ channel (no angular momentum) and the spin part has to have spin-singlet character ($s=0$, opposing spin directions). Otherwise the range of the attractive interaction component is shorter than the average distance of the electrons. 

Characteristic length scales that determine the way applied magnetic fields suppress superconductivity are the coherence length $\xi$, on the one hand, and the penetration depth $\lambda$, on the other hand. If the ratio $\kappa=\lambda/\xi$ exceeds $1/\sqrt{2}$ the energy density of the surface separating the normal and superconducting state turns negative, and the superconducting state is referred to as being type II. Here magnetic field penetrates in flux lines carrying the flux quantum $\Phi_0=h/2e$ \cite{abri52}. The flux lines are organized in a lattice with a geometry that minimizes the ground state energy. All the compounds addressed in this review are strong type II superconductors and the morphology of the flux line lattice yields key information on the nature of the superconductivity (for recent work in Nb see \cite{lave06,mueh09b}). 

Microscopically, applied magnetic fields suppress superconductivity by interacting either with the orbital or spin momentum. For pure orbital limiting the upper critical field, $H^{orb}_{c2}(T\to0)= \Phi_0/(2\pi\xi^2)$ is connected with the initial slope of $H^{orb}_{c2}$ near $T_s$ as $H^{orb}_{c2}(T\to0)=-0.7\,dH_{c2}/dT\vert_{Ts}$ \cite{sain69}. This is contrasted by pure Pauli limiting of the upper critical field, which is related to $T_s$ as $H^{Pauli}_{c2}(T\to0)=1.84 T_s$, where $H$ is in T and $T_s$ in K \cite{chan62,clog62}. The ratio of orbital to Pauli limiting is expressed by the Maki parameter $\alpha=\sqrt{2}H^{Pauli}_{c2}/H^{orb}_{c2}$ \cite{sain69}. It was also noticed that the transition at $H_{c2}$  for pure Pauli limiting becomes first order below $T^{\dagger}=0.56\,T_s$ \cite{kett99,sain69}. 

Since the early days of research in superconductivity, the effect of internal magnetic fields (cf. exchange fields) on the superconductivity was of great interest. Theoretical work suggested that static or dynamic internal magnetic fields would prevent superconductivity \cite{ginz57,berk66}. In the limit of extreme purity and pure Pauli limiting, i.e., large values of $\alpha$, a novel state was predicted to be possible, that consists in real-space modulations of superconductivity with a weakly spin-polarized normal state  \cite{fuld64,lark65}. We return to the experimental status of this so-called FFLO phase in f-electron systems in section \ref{FFLO-phas}.

Experimentally the question for internal magnetic fields in superconducting materials was at first followed up in studies of binary and pseudo-binary systems with rare earth impurities (R) such as La$_{1-x}$R$_x$ and (Y$_{1-x}$R$_{-x}$Os$_2$) \cite{matt58b}. Early studies suffered from metallurgical complexities due to clustering and glassy types of magnetic order and were somewhat inconclusive. They motivated, however, more detailed studies which led to a fairly advanced understanding of paramagnetic impurities in superconductors. Reviews have been given in, e.g.,  \cite{whit79,mapl76,mapl95,mapl05}. Overall it was accepted that magnetic impurities are detrimental to superconductivity, while it was also appreciated that conventional superconductivity is fairly insensitive to nonmagnetic defects \cite{ande59}.

The upshot of these studies has been, that the rate of suppression of $T_s$ is the highest in the middle of the rare-earth series \cite{matt58b,mapl70}, consistent with the strongest pair breaking due to magnetic exchange interactions \cite{herr58,suhl59,abri61}. An exception is Ce, which causes an anomalously large depression of $T_s$ due to the strong hybridization of the f-electrons with the conduction electrons. A more detailed understanding of the effect of magnetic impurities on superconductors requires an understanding of the properties of magnetic moments dissolved in a non-magnetic host. In the Kondo effect, the conduction electrons hybridize with the magnetic moment, eventually forming a screening cloud below a characteristic temperature $T_{K}$, the Kondo temperature (for an introduction see e.g. \cite{hews93}). Alternatively, the moment may by quenched by low lying crystal fields. While the former leads to strong Cooper pair breaking, the latter reduces the effects of pair breaking.

Pioneering studies of Ce$_{1-x}$La$_x$Al$_2$ revealed the presence of Kondo screening with a Kondo temperature $T_{K}\sim0.2$\,K \cite{mapl68,fels73,andr82}. Ce$_{1-x}$La$_x$Al$_2$ displays reentrant superconductivity \cite{ribl71,mapl72}, i.e., the superconducting transition at $T_{s1}$ is followed by a second characteristic temperature $T_{s2}<T_{s1}$ below which superconductivity vanishes again. The reentrance may be understood as resulting from an increasing strength of the pair breaking of the paramagnetic impurities with decreasing temperature, because $T_K<T_s$ \cite{muel71}. The strength of pair breaking due to Kondo screening was also studied in high pressure experiments on La$_{1-x}$Ce$_x$ alloys, where the superconductivity vanishes in a finite pressure interval for $x=0.02$ as the Kondo temperature increases under pressure \cite{mapl72}. As a side-effect of detailed studies in Ce$_{1-x}$La$_x$Al$_2$ it was finally also recognized, that even pure {\cealt} displays a Kondo-effect, thus qualifying as the perhaps first example of a Kondo-lattice \cite{mapl69,vand69,busc70}. The effect of crystal electric fields (CEF) in removing the magnetic moment was studied, e.g., in the series La$_{1-x}$Pr$_x$Tl$_3$, where superconductivity vanishes only slowly, because the crystal fields reduce the pair breaking strength with increasing $x$ \cite{buch72}.

In contrast to a purely competitive form of superconductivity and magnetism doping studies in the series Ce$_{1-x}$Gd$_x$Ru$_2$ also suggested the possibility of a coexistence of superconductivity and magnetism in small parameter regimes \cite{hein59,matt58c,phil61}. By the late 1970s two series of compounds had been discovered, which display such an extremely delicate balance of superconductivity and magnetism intrinsically, notably the series {\rrhb} where R is a rare earth and the Chevrel phases such as {\dymos} \cite{fert77,ishi77,monc77,bula85}. These compounds are frequently referred to as magnetic superconductors. As a key feature $T_s$ in these systems is always larger than the magnetic ordering temperature. 

The interplay of magnetism and superconductivity is exemplified by the series {\erhorhb}, in which the onset of ferromagnetism destroys superconductivity. In a tiny temperature interval for small $x$ magnetic order succeeds in coexisting with superconductivity by forming a modulated state. This firmly suggests  that superconductivity and magnetism are antagonistic forms of order. However, for selected antiferromagnetic members of this series even a constructive interplay of magnetism and superconductivity could be inferred from an increase of the $H_{c2}$ below $T_N$.  In contrast to the systems reviewed here superconductivity and magnetism may be viewed as residing in separate microscopic subsystems. Comprehensive reviews of this field may be found in \cite{fisc82,mapl82,fisc90}.

As a remark on the side, these compounds also provided first hints of the Jaccarino-Peter effect \cite{jacc62}, notably an enhancement of the superconductivity when an applied field cancels any internal magnetic fields. In recent years further compounds have been discovered with a coexistence in separate subsystems, notably the Ruthenocuprates \cite{klam01,fraz01,otzs99} and the Borocarbides RNi$_2$B$_2$C (R=Gd-Lu, Y) \cite{budk06,mazu05}. 

The possibility of unconventional superconducting order parameter symmetries had been anticipated theoretically, when the superfluid phases of {\het} were discovered; excellent reviews may be found in \cite{legg75,whea75,voll90}. In particular {\het} provided a first example of a constructive interplay of superconductivity and the magnetic properties of the system. Theoretically it had been suggested that ferromagnetic fluctuations may mediate superconductive pairing \cite{layz71,fay77} and that superconductivity may even exist in itinerant ferromagnets \cite{fay80}. However, for a long time there was no evidence supporting this suggestion in real materials.

During the 1970s great advances were also made in the understanding of intermediate valence compounds, see e.g. \cite{busc79,whit79}. As a key feature nonmagnetic members of this group of materials  exhibit enhanced Fermi liquid coefficients such as the linear specific heat $\gamma=C/T$ or quadratic temperature dependence of the resistivity $A=\Delta\rho/T^2$. A number of compounds even displayed particularly strong renormalization effects of the Fermi liquid coefficients, like CeAl$_3$ \cite{andr75}. They are known as heavy-fermion systems. Amongst the heavy fermion systems superconductivity was for the first time observed in, {\cecusi} \cite{steg79}. Due to the large specific heat anomaly of {\cecusi} at the superconducting transition it was immediately appreciated, that the strongly renormalized quasiparticle excitations take part in the pairing. Moreover, under tiny changes of stoichiometry the ground state of {\cecusi} was found to become magnetically ordered. This vicinity to magnetic order suggested an important role of magnetic correlations in the superconductive pairing. 

The discovery of heavy fermion superconductivity created intense experimental and theoretical efforts. For early reviews we refer to \cite{grew91,stew84}. However, in the first twelve years following the discovery of superconductivity in {\cecusi} \cite{steg79} only 5 more heavy fermion superconductors were discovered ({\ube} \cite{buch75,ott83}, {\upt} \cite{stew84}, {\urusi} \cite{schl84,mapl86,pals85,schl86}, {\updal} and {\unial} \cite{geib91a,geib91b}). Because the microscopic details of these systems proved to be remarkably different, a unified theoretical understanding turned out to be a great challenge.

\subsection{Roadmap to superconducting phases
\label{road-supe-phas}}

In recent years several ingredients have come to light that prove to be almost universally important in the search for further examples of superconducting phases of f-electron compounds. First, an improved appreciation of the quasiparticle interactions in Fermi liquids.  Second, the experimental ability to tune these interactions in pure metallic systems in a controlled manner by means of a non-thermal control parameter such as pressure, stress or magnetic field. Third, and most important, great advances in materials preparation. In the following we briefly discuss these developments.

A simple plausibility argument shows, that the superconductive pairing in heavy fermion systems is probably not driven by electron-phonon interactions and that the order parameter is most likely unconventional, i.e., the order parameter breaks additional symmetries. Going back to the importance of retardation for electron-phonon mediated pairing and the local character of the interaction, it is helpful to keep in mind that the speed of travel of a quasiparticle excitation in heavy fermion systems typically is reduced by nearly three orders of magnitude. In turn the effects of repulsive quasiparticle interaction components for a conventional pairing symmetry ($l=0$ and $s=0$) can no longer be avoided. However, the repulsive components of the interactions may be avoided in higher angular momentum and spin states of the Cooper pairs. 

A systematic search for novel forms of superconductive pairing interactions and pairing symmetries hence requires a systematic quantitative determination of the quasiparticle interactions in the presence of strong electronic correlations. Second, it requires very clean samples, since unconventional pairing tends to be extremely sensitive to non-magnetic defects. As a rule of thumb, the charge carrier mean free path needs to be substantially larger than the coherence length for superconductivity to occur.

%\subsection{Advances in materials preparation}
%\label{chal-mate-prep}

Quite generally the quasiparticle interactions may be expressed in terms of the generalized, dynamical response function of the system. For instance, in systems at the border of magnetic order this is expressed in terms of the wave vector and frequency dependent magnetic susceptibility $\chi(\vec{q},\omega)$; for a pedagogical introduction see \cite{lonz97}. Experimentally quasiparticle excitation spectra and the related interaction potentials may be explored in quantum oscillatory studies. Careful comparison of the experimentally observed quasiparticle enhancements with the response function determined in, e.g., neutron scattering allows the development of a simple description of the generalized quasiparticle interactions.

A program of this kind was first systematically carried out in the 1980s for weakly and nearly magnetic transition metal compounds and selected f-electron systems. For reviews of this work we refer to \cite{lonz80,lonz87,lonz88}. More recent reviews of quantum oscillatory studies may be found in \cite{onuk95,onuk93,sett07}. As an important aspect of the early work, it became at the same time possible to calculate quantitatively the magnetic ordering temperature of weakly magnetic itinerant-electron systems \cite{lonz85,mori85}. This paved the way for a quantitative analysis of superconducting pairing interactions in weakly ferromagnetic and antiferromagnetic compounds, see e.g. \cite{dung90}, and eventually allowed an educated guess of which systems to study (see also \cite{mont07}).

The quasiparticle interactions were finally tuned by means of high hydrostatic pressures in pure samples. The experiments served to clarify two questions. First, to identify possible examples of magnetically mediated superconductivity (for early attempts see e.g. \cite{cord81}). Second, to investigate the nature of the metallic state in the vicinity of a quantum critical point. Here we briefly note that quantum phase transitions, quite generally, are defined as phase transitions that are driven by quantum fluctuations. In practice this means that quantum phase transitions are zero temperature second order phase transitions. In recent years this definition has been relaxed somewhat and zero temperature phase transition in general are referred to as quantum phase transitions \cite{pfle05b}. It transpires that quantum phase transitions represent an extremely rich field of condensed matter physics. For reviews we refer to \cite{hert76,sach99,vojt03,beli05a,stew01,stew06,vloe07,mont07}. 

%From a metallurgical point of view a 'perfect' crystal implies the absence of impurities and perfect structural order (no vacancies, no dislocations, no site interchanges etc). 
%Suitable methods for growing single crystals, quite generally, must ensure that no impurities are introduced, that stoichiometry is observed and that the individual atoms are arranged ideally with respect to each other. 

Besides the advances in understanding the metallic state in the presence of strong electronic correlations, great advances have also been achieved in the experimental techniques (for a recent review on the 5f states in actinides see \cite{moor09}).  Studies under extreme conditions such as very low temperatures, high pressures and high magnetic fields are now routinely available in numerous laboratories. 

Probably most important are, however, major improvements in materials preparation. For instance, major improvements have been achieved by means of the purification of the starting elements. Electro-transport of, e.g. uranium, under ultra-high vacuum was found to be extremely efficient in removing impurities such as Fe and Cu \cite{haga98,fort87}. Electro-transport in combination with annealing under ultra-high vacuum has also been used to promote the formation of large single-crystal grains and improve the sample quality \cite{schm76,haga07,mats08}. For the growth of high vapor pressure compounds a closed crucible annealing technique was developed \cite{assm84}. In many materials the combination of ultra-high vacuum with an inert gas atmosphere is sufficient to obtain large high-quality single crystals \cite{mcdo96}. This cannot be underestimated given that both the rare earth and actinide elements readily react, especially with oxygen, hydrogen and nitrogen. Advances in the understanding of the phase diagrams of binary and ternary compounds has motivated the improvement and extensive use of techniques like traveling-solvent float-zoning or the controlled use of flux methods, e.g., in the skutterudites or the series of {\cemin} compounds. Finally in recent years an increasing number of groups explores the use of optical floating-zone furnaces for the growth of intermetallic compounds, see e.g. \cite{soup07}. For example large single crystals have been grown of {\unial} \cite{miha97} and of {\urusi} (see e.g. \cite{pfle06}). It is expected that this technique will play a very important role in the future.

\begin{table*}
\centering
\caption{
Key properties of superconductors in the series {\cemt} (M: Cu, Pd, Rh, Ni; X: Si, Ge) and various miscellaneous Ce-based systems. Missing table entries may reflect more complex behavior discussed in the text. References are given in the text. Critical field values represent extrapolated $T=0$ values.
(AF: antiferromagnet, SC: superconductor, ISC: incipient superconductor) 
}
\label{table-cemt}
\begin{tabular}{llllllllll}
\hline\noalign{\smallskip}
     & {\cecusi} &  {\cecuge} & {\cepdsi} & {\cerhsi}  & {\cenige} & {\cenigeot} & {\cenigettf} & {\cepdal} \\
\noalign{\smallskip}\hline\noalign{\smallskip}
structure & tetragonal & tetragonal & tetragonal & tetragonal & tetragonal & orthorh. & orthorh. & tetragonal \\
type & {\thcrsi} & {\thcrsi} & {\thcrsi} & {\thcrsi} & {\thcrsi} & SmNiGe$_3$ & U$_2$Co$_3$Si$_5$ & ZrNi$_2$Al$_5$ \\
space group & I4/mmm & I4/mmm & I4/mmm & I4/mmm & I4/mmm & Cmmm & Ibam & I4/mmm \\
$a$({\AA}) & 4.102 & 4.186 & 4.223 & 4.092 & 4.150 & 21.808 & 9.814 & 4.156 \\
$b$({\AA}) & 9.930 & 10.299 & 9.897 & 10.181 & 9.842 & 4.135 & 11.844 & 4.156 \\
$c$({\AA}) & 9.930 & 10.299 & 9.897 & 10.181 & 9.842 & 4.168 & 5.963 & 14.883 \\
$c/a$ & 2.420 & 2.460 & 2.343 & 2.488 & 2.372 & - & - & -\\
\noalign{\smallskip}\hline
%CEF$^1$ & - & - & $\Gamma_7^{(1)} (0)$ & -\\
 %& - & - & $\Gamma_6 ({\rm 19\,meV})$ & -\\
 %& - & - & $\Gamma_7^{(2)} ({\rm 24\,meV})$ & -\\
%\noalign{\smallskip}\hline
state & AF, SC & AF, SC & AF, SC & AF, SC & ISC & AF, SC & AF, SC & AF, SC \\
$T_{N}$(K) & 0.8 & 4.15 & 10 & 36, 25 & - & 5.5 & 5.1, 4.5 & 3.9, 2.9\\
$\vec{Q}$ & (0.22, 0.22, 0.53) & (0.28, 0.28, 0.53)  & (0.5, 0.5, 0) &  (0.5, 0.5, 0) & - & - & - &  - \\
 & - & -  & - &(0.5, 0.5, 0.5) & - & - & - & - \\
$\mu_{ord}¥$(${\rm \mu_{B}}$) & 0.1 & 1 & 0.62 & 1.42, 1.34 & - & 0.8 & 0.4 & - \\
$\gamma ({\rm J/mol\,K^2}$) & 1 &  & 0.062 & 0.027 & - & - & 0.09 & 0.056 \\
\noalign{\smallskip}\hline
$T_{s}^{\rm max}$(K) & 0.7, 2.5 & 0.64  & 0.4 & 0.42  & 0.3, 0.4 & 0.45 & 0.26 & 0.57 \\
%                                   & 2.5$^*$ & -  & - & -  \\
$p_{s}^{\rm max}$(kbar) & 0, 30 & 70  & 28 & 10 & 0, 18  & 70 & 36 & 108 \\
%					   & 3$^*$ & -  & 2.8 & 1.0  \\
$\Delta÷C/\gamma_{n}T_{s}$ & 1.4 & - & -  & - & - & - & - & -\\
$H^{ab}_{c2}$(T) & 0.45 & 2 & 0.7 & - & -  & 1.55 & 0.7 & 0.25 \\
$\frac{d}{dT}H^{ab}_{c2}$(T/K) & - & -11 & -12.7 & - & -  & -10.8 & - & -1.04 \\
$H^{c}_{c2}$(T) & - & - & 1.3  & 0.28 & - & - & - & - \\
$\frac{d}{dT}H^{c}_{c2}$(T/K) & - & - & -16 & -1 & - & - & - & - \\
\noalign{\smallskip}\hline
$\xi^{ab}_{0}$({\AA}) & - & 90 & 300 & - & - & 100 & 210 & - \\
$\xi^{c}_{0}$({\AA}) & - & - & 230 & 340 & - & - & - & - \\
%$\kappa_{GL,a}$ & - & - & - & - & - \\
%$\kappa_{GL,c}$ & - & - & - & - & - \\
%\noalign{\smallskip}\hline
%$\partial÷T_{s}/\partial÷p_{[100]}$(mK/GPa)& - & - & -  & - & \\
%$\partial÷T_{s}/\partial÷p_{[001]}$(mK/GPa)& - & - & - & - & \\
%$\partial÷T_{s}/\partial÷p_{V}$(mK/GPa)& - &  - & - &  - & \\
%$\partial÷T_{s}/\partial÷p$(mK/GPa)& - & - & - &  - & \\
%\noalign{\smallskip}\hline
%$C/T$ & - & - & - & - & - \\
%$\kappa$ & - & - & - & - & - \\
%$1/T_{1}$ & - & - & - & - & - \\
%$\lambda$ & - & - & - & - & - \\
\noalign{\smallskip}\hline
year of disc. & 1979 & 1993 & 1996 & 1996 & 1996 & 2004 & 2005 & 2007 \\
\noalign{\smallskip}\hline
\end{tabular}
\end{table*}

A frequent objection in materials preparation concerns the relative importance of the various aspects. For instance, it is believed that the accuracy at which perfect stoichiometry can be achieved generally outweighs any efforts put into the purification of the starting elements. Empirically this is contrasted by the impressive list of unusual phenomena such as unconventional superconductivity that have been discovered in ultra-pure compounds. The perhaps most important challenge in materials preparation is the lack of methods for characterization. Generally speaking, high sample quality is proven by the combination of standard characterization (x-ray diffraction, microprobe, etc.) plus the physical properties themselves. This shows that despite all of the technical achievements the growth of high quality single crystals continues to require great physical intuition, systematic work and a fair bit of luck.

%%%%%%%%%%%%%%%%%%%%%%%%%%%%%%%%%%%%%%%%%%%%%%%%%
\section{INTERPLAY OF ANTIFERROMAGNETISM AND SUPERCONDUCTIVITY}
\label{inte-anti-supe}

In this section we review the interplay of antiferromagnetism and f-electron superconductivity. Section [\ref{supe-anti}], reviews systems where superconductivity emerges at the border of itinerant antiferromagnetism. In particular properties of the series {\cemt} and {\cemin} are addressed. Section [\ref{coex-anti}] is concerned with superconductivity in antiferromagnetic compounds. This includes large moment systems like {\updal} and {\unial} as well as small moment systems like {\upt} and {\urusi}.

\subsection{Border of antiferromagnetism}
\label{supe-anti}

\subsubsection{The series {\cemt}
\label{cemt}}
\begin{figure}
%\sidecaption
\includegraphics[width=.45\textwidth,clip=]{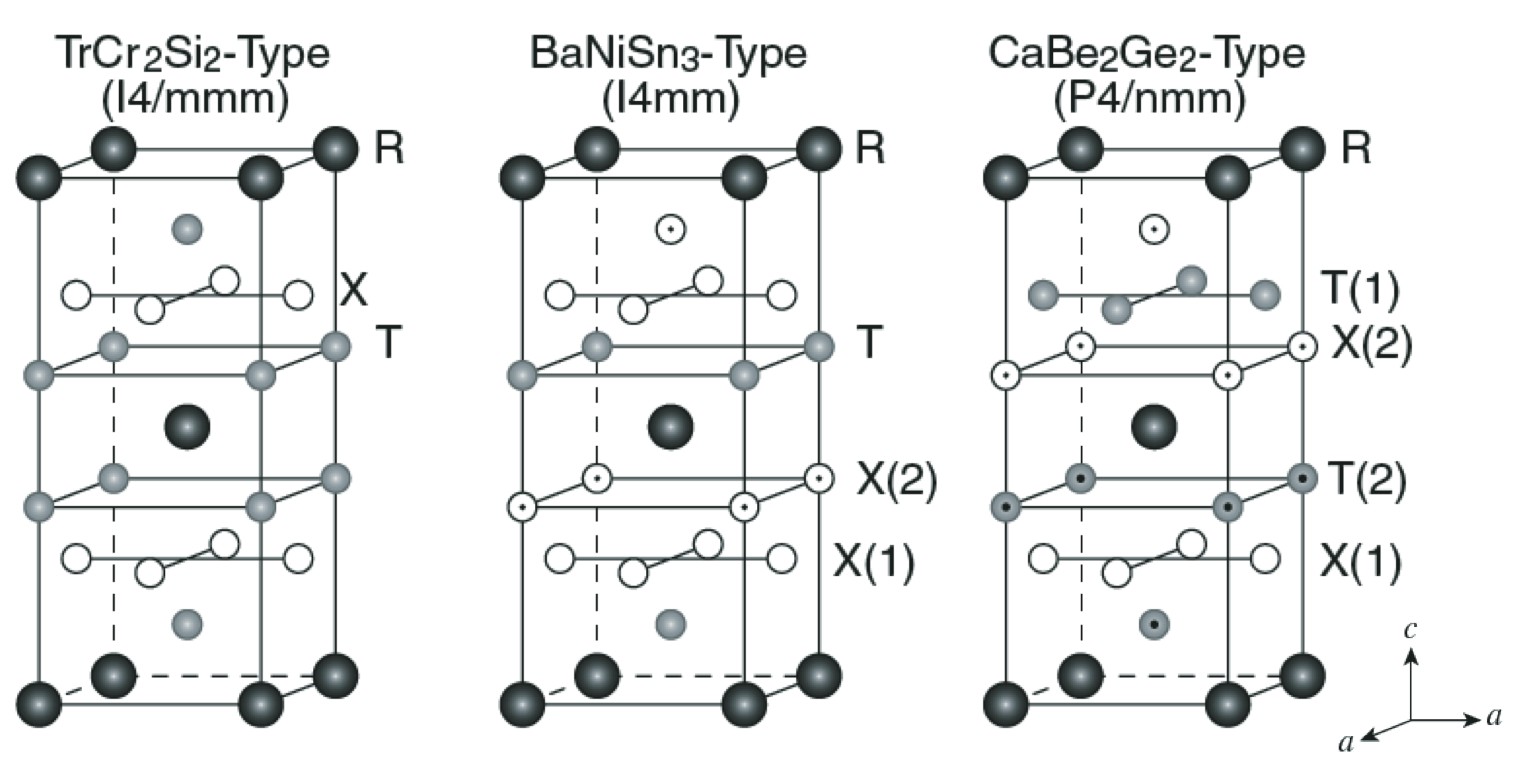}
\caption{Variations of the tetragonal BaAl$_4$ crystal structure. The {\thcrsi} structure is frequently found amongst {\cemt} compounds reviewed in section \ref{cemt}. The BaNiSn$_3$ crystal structure, which lacks inversion symmetry, is typical of the CeMT$_3$ compounds reviewed in section \ref{non-cent-supe}. Amongst f-electron systems with the non-centrosymmetric CaBe$_2$Ge$_2$ structure no compounds are known that exhibit superconductivity. Plot taken from \cite{kimu07a}.}
\label{struc-cerhsi3} 
\end{figure}

The discovery of heavy fermion superconductivity in {\cecusi} \cite{steg79} marked the starting point of unconventional superconductivity \cite{stew84,grew91,spar06,thal05b}. {\cecusi} crystallizes in the tetragonal {\thcrsi} crystal structure as summarized in table \ref{table-cemt}. The heavy-fermion superconductivity in {\cecusi} generated great interest in the isostructural series of {\cemt} compounds, where M is a transition metal (M=Cu, Au, Rh, Pd, Ni) and X=Si or Ge. Because most members of this series exhibit antiferromagnetic order \cite{grie84,thom86}, it represented a major break-through for the entire field, when superconductivity was discovered in {\cecuge}, {\cerhsi}, and in particular {\cepdsi}, as well as incipient superconductivity in {\cenige}. For a summary see also table \ref{table-cemt}. Being a derivative of the BaAl$_4$ parent structure, the {\thcrsi} structure is intimately related to the BaNiSn$_3$ and CaBe$_2$Ge$_2$ types of structures as illustrated in Fig.\,\ref{struc-cerhsi3}. A surprise in recent years was the discovery of superconductivity in several Ce-based compounds with the non-centrosymmetric BaNiSn$_3$ structure, because it was believed that triplet superconductivity cannot exist in crystal structures lacking inversion symmetry. For an account of this work we refer to section \ref{cemx3}. Interestingly no superconductivity has so far been found amongst the CaBe$_2$Ge$_2$ relatives of the {\thcrsi} series, which is also non-centrosymmetric. 

\paragraph{{\cecusi} {\&} {\cecuge}}

The ground state properties of {\cecusi} are extremely sensitive to the precise Cu content, which may be controlled by an annealing procedure under Cu vapor \cite{assm84}. Samples with heavy-fermion superconductivity, antiferromagnetism or a combination thereof may be obtained, which are referred to as (S), (A) or (AS), respectively. For Cu deficient samples the SDW order is stabilized and superconductivity destroyed, while the SDW is destabilized and the superconductivity stabilized for Cu excess. On the level of changes of less than a few \% of composition achieved under Cu annealing, it is believed that the changes of properties originate mostly in changes of unit cell volume \cite{trov97}. This may be inferred also from doping with Ge which, being larger than Si, stabilizes the SDW, while hydrostatic pressure destabilizes the SDW and stabilizes the superconductivity \cite{trov97,krim97}. In the following it proves to be convenient to address S-type samples first. 

Quite generally the normal state of {\cecusi} is characteristic of a heavy Fermi liquid with $C/T=\gamma=1\,{\rm J/mol\,K^2}$ and an equally enhanced Pauli susceptibility. The heavy fermion state develops in a crystal electric field ground state Kramers doublet and a first and second excited doublet at 12.5\,meV and 31\,meV \cite{horn81}. S-type samples of {\cecusi} display $T_s=0.7\,{\rm K}$, accompanied by a distinct specific heat anomaly $\Delta C/\gamma T_s=1.4$. This established for the first time that heavy quasiparticles may undergo superconductive pairing. Under magnetic field the superconductivity exhibits strong Pauli limiting with $H_{c2}\approx0.45\,{\rm T}$ \cite{rauc82}. The leading order temperature dependence of the specific heat $C(T)$, thermal expansion $\alpha$ \cite{lang91} and penetration depth $\lambda$ \cite{gros88} vary as $T^2$ characteristic of line nodes. NMR and NQR show the absence of a Hebel-Slichter peak and a power law dependence of the spin-lattice relaxation rate, also suggesting line nodes \cite{ishi99,kawa04}. 

The magnetic phase diagram of {\cecusi} for magnetic field applied along the $a$-axis in the basal plane is fairly complex \cite{steg01,brul94,brul90}. Ultrasound and thermal expansion measurements early suggested the presence of two spin density wave phases, referred to as A- and B-phase, respectively. Magnetic field suppresses at first the superconductivity above $H_{c2}$, where the A-phase is restored. The $B$-$T$ boundary of the A-phase is reminiscent of $H_{c2}(T)$, as if the A-phase encompasses the superconductivity. Above a critical field $H_c\approx6.4\,{\rm T}$ magnetic field suppresses the A-phase and stabilizes the B-phase. Only little is known about the nature of the B-phase. 

Microscopic evidence for SDW order in {\cecusi} was missing for nearly 25 years. Progress was made only recently by tracking systematically the incommensurate spin-density wave order of {\cecuge} as a function of increasing Si-content. It was found that the ordering wave vector changes little as function Si content, yielding a value of $\vec{Q}=(0.215, 0.215, 0.530)\,{\rm/(r.l.u)}$ in AS samples of {\cecusi} \cite{stoc04}. The neutron scattering studies identified an incommensurate spin-density wave in the A-phase with a small ordered moment $\mu_{ord}\approx0.1\,{\rm \mu_B}$ per Ce site in {\cecusi}, which evolves from the antiferromagnetic order in {\cecuge} continuously with increasing Si-content. The ordering wave vector agrees thereby with the nesting wave vector found in Fermi surface calculations \cite{zwic93}. For the (AS) samples antiferromagnetism and superconductivity are mutually exclusive and separated by a first order phase transition \cite{spar06}. 

\begin{figure}
%\sidecaption
\includegraphics[width=.35\textwidth,clip=]{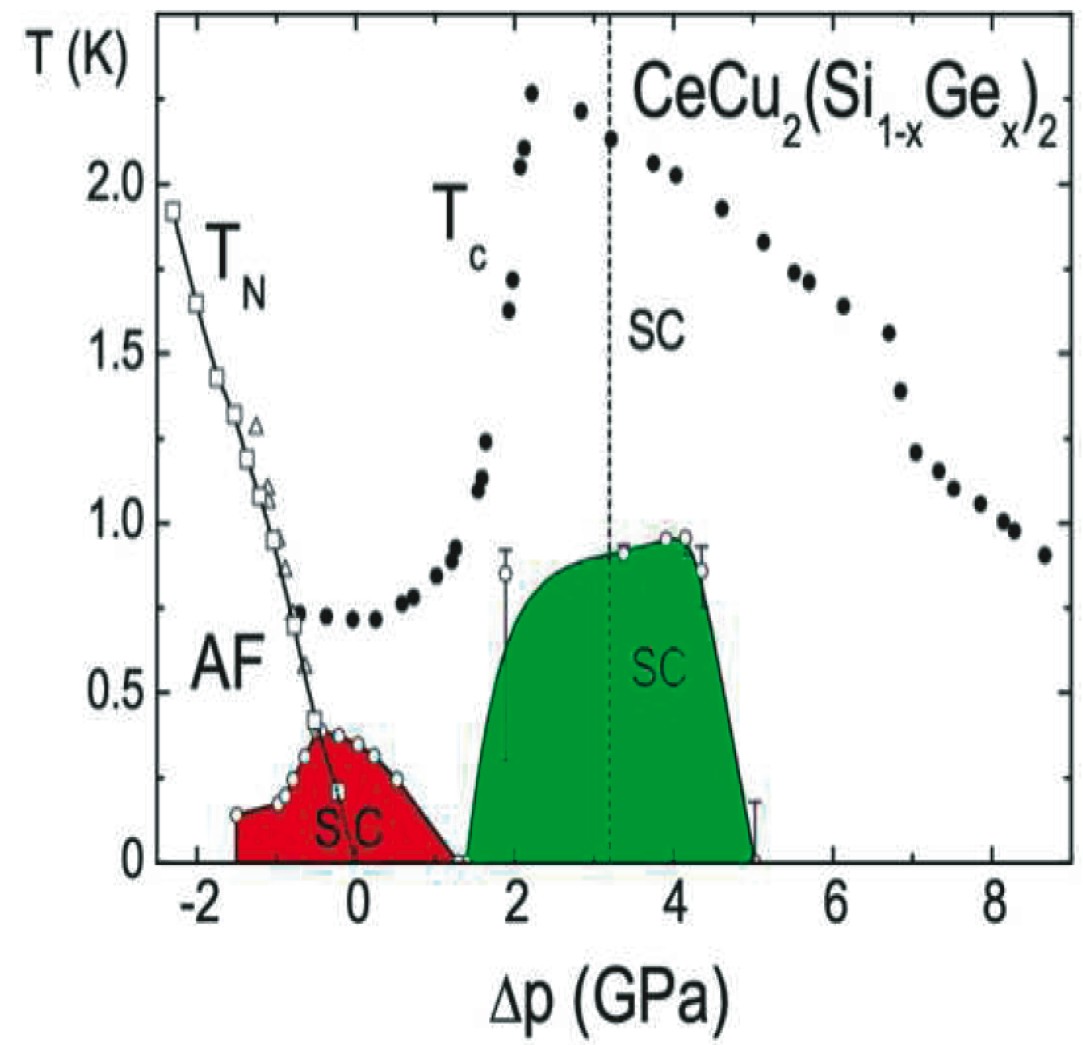}
\caption{Temperature versus pressure phase diagram of {\cecusige}. The pressure dependence of the superconducting transition temperature, here denoted $T_c$, in S-type {\cecusi} exhibits a plateau between 20 and 70\,kbar (black dots). Weak impurity scattering in moderately Ge doped {\cecusi} decomposes the superconductivity into two domes, one (red) at the border of antiferromagnetism and the other (green) at a presumed valence transition. Plot taken from \cite{thal05b}, representing a compilation of several studies as described in the text.}
\label{PD-CeCuSiGe} 
\end{figure}

As a function of pressure $T_s$ in S-type {\cecusi}  increases around 2\,GPa and enters a plateau of 2.25\,K above 2.5\,GPa, followed by a moderate decreases with a small shoulder around 7\,GPa as shown in Fig.\,\ref{PD-CeCuSiGe} \cite{bell84,thom93}. The unusual pressure dependence early on suggested that $T_s(p)$ may be explained in terms of two or more pairing interactions. The discovery of superconductivity in {\cecuge} \cite{jacc92} and {\cerhsi} \cite{movs96}, but in particular in {\cepdsi} \cite{math98}, underscored the idea that superconductivity in {\cecusi} is somehow related to the magnetic properties. At the border of the A-phase NFL properties of the normal metallic state were observed characteristic of quantum critical spin fluctuations, where $\Delta\rho(T)\propto T^{3/2}$ and $C/T=\gamma=\gamma_0-\alpha\sqrt{T}$ \cite{gege98}. 

Under moderate doping with Ge, which introduces pair breaking defects, the superconducting phase disintegrates into two domes as shown in Fig.\,\ref{PD-CeCuSiGe} \cite{holm04,yuan04}. Based on these studies two pairing interactions were proposed: antiferromagnetically mediated pairing in the vicinity of the SDW and pairing by charge density fluctuations in the vicinity of a valence transition at high pressures. The latter consists in fluctuations that originate in a Ce$^{3+}$ to Ce$^{+4}$ change of valence, where the 4f electron is delocalized in the high pressure Ce$^{4+}$ state. We also address the question of charge fluctuation mediated pairing in section \ref{supe-vale-inst}.

The superconducting pairing symmetry has been revisited with the knowledge of the incommensurate SDW and band structure calculations based on the renormalized LDA \cite{thal05b}. A superconducting d-wave singlet state, $d_{x^2-y^2}$, and the SDW order are here treated as two competing ordering phenomena. The model accounts for the change from the incommensurate SDW order in the A-phase to superconductivity, where both order parameters have the $\Gamma_3$ symmetry imposed by the crystal electric fields.

Analogies of the thermopower in superconducting samples of {\cecusi} with {\cecuge} finally motivated high pressure experiments in {\cecuge} \cite{jacc92}. At ambient pressure {\cecuge} orders antiferromagnetically below $T_{N}=4.15\,{\rm K}$ into an incommensurate sinusoidally modulated structure with $\vec{Q}=(0.284,0.284,0.543)\pm0.001$ and an ordered moment $\mu_{ord}=0.74\,{\rm \mu_B}$ \cite{knop89,krim97b}. The low temperature properties develop in a crystal electric field environment of a ground state doublet and a first excited quartet at 19.1\,meV. The metallic state of {\cecuge} is moderately enhanced. 

Under pressure the N{\'e}el temperature of {\cecuge} decreases and vanishes at $p_N\approx70\,{\rm kbar}$ \cite{jacc92}. Superconductivity appears above the critical pressure, where $T_s\approx0.64\,{\rm K}$ is only weakly pressure dependent and extends over a wide range, where $H_{c2}\approx 2\,{\rm T}$ with a large initial slope $dH_{c2}/dT=-11\,{\rm T/K}$. This suggests a coherence length of order $\xi=90\,{\rm \AA}$. The structural similarity and lack of pressure dependence of $T_s$ suggested an intimate similarity with {\cecusi}. In fact evidence for a valence transition at $\sim$150\,kbar, where $T_s$ is largest, has been inferred from x-ray diffraction \cite{onod02}.

\paragraph{{\cepdsi} \& {\cenige}}

The intense studies of the quasiparticle interactions in weakly ferromagnetic transition metal compounds and selected f-electron systems mentioned above \cite{lonz80,lonz87,lonz88} resulted in quantitative estimates of magnetically mediated superconductivity at the border of weak ferromagnetism. This motivated detailed high pressure studies in MnSi \cite{pfle93,pfle97b,pfle01b,pfle04} and related compounds. It inspired also studies of the suppression of antiferromagnetism under pressure of the isostructural and isoelectronic siblings {\cepdsi} and {\cenige} reviewed in the following.

At ambient pressure {\cepdsi} may be described as intermediate valence system. At high temperatures the resistivity varies weakly with temperature, followed by a rapid decrease below $\sim50\,{\rm K}$ and a sharp drop at the onset of antiferromagnetic order at $T_N\approx10\,{\rm K}$. The antiferromagnetic order in {\cepdsi} consists of alternating ferromagnetic sheets with $\vec{Q}=(1/2, 1/2, 0)$  where the moments are oriented along $[110]$, i.e., they reside in the tetragonal basal plane \cite{grie84}. The magnetism has been interpreted as local moment like, with a reduced ordered moment $\mu_{ord}=0.62\,{\rm \mu_B}$ in the crystal field environment \cite{dijk00}. The CEF level scheme of the localized Ce$^{3+}$ 4f$^1$ electrons have been determined from the susceptibility and inelastic neutron scattering as a sequence of three Kramers doublets: $\Gamma_7^{(1)}(0)$,  $\Gamma_6({\rm 19\,meV})$, $\Gamma_7^{(2)}({\rm 24\,meV})$. The metallic state may be described as a Fermi liquid with a moderately enhanced value of $\gamma=0.062\,{\rm J/mol\,K^2}$ \cite{stee88}. 

\begin{figure}
%\sidecaption
\includegraphics[width=.35\textwidth,clip=]{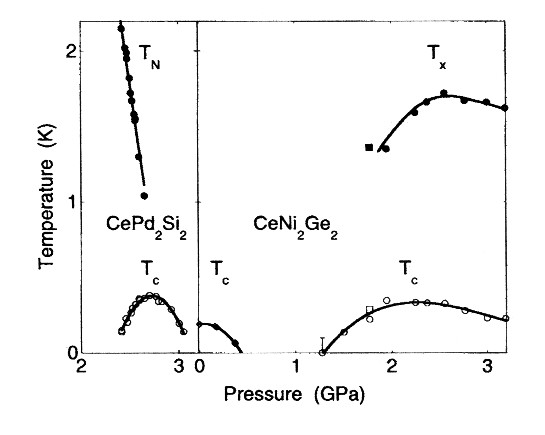}
\caption{Combined temperature versus pressure phase diagram of the isostructural, isoelectronic pair of systems {\cepdsi} and {\cenige}, where superconductivity is observed at the border of antiferromagnetism and at low and high pressures, respectively. Plot taken from \cite{gros00}.}
\label{PD-CePdeSi+CeNiGe}
\end{figure}

Under pressure $T_N$ in {\cepdsi} decreases and vanishes linearly at $p_c\approx28\,{\rm kbar}$ \cite{thom86,gros96,math98}. Superconductivity has been observed in the immediate vicinity of $p_c$ with a maximum of $T_s\approx0.4\,{\rm K}$, as shown in Fig.\,\ref{PD-CePdeSi+CeNiGe}. The observation of a wide superconducting dome in some experiments could be traced to pressure inhomogeneities \cite{raym00,shei01}. The gradual decrease of $T_N$ with pressure suggested that the antiferromagnetic order vanishes continuously at $p_c$. This has been confirmed more recently in neutron scattering experiments of the staggered magnetization \cite{kern05}. The expected abundance of quantum critical spin fluctuations near $p_c$ is consistent with the temperature dependence of the electrical resistivity, which displays a power law dependence $\Delta\rho\sim T^{1.2}$ over a wider temperature range \cite{gros96,math98}. In the context of these fluctuations it has been suggested that the fluctuations in {\cepdsi} exhibit a reduced dimensionality.

Based on its vicinity to a quantum critical point the superconducting pairing interaction was attributed to the exchange of antiferromagnetic spin fluctuations \cite{gros96,math98}. Several pieces of evidence suggest unconventional pairing, where a d-wave state appears to be the most promising candidate. The upper critical field and its initial variation are large and anisotropic, where $H^c_{c2}=1.3\,{\rm T}$ with $dH^c_{c2}/dT=-16\,{\rm T/K}$ and $H^{ab}_{c2}=0.7\,{\rm T}$ with $dH^{ab}_{c2}/dT=-12.7\,{\rm T/K}$ \cite{shei01}. These values suggest anisotropic Pauli limiting, where weak or strong coupling behavior cannot be distinguished unambiguously. The corresponding coherence lengths are quite short with $\xi^{ab}=300\,{\rm \AA}$ and $\xi^c=230\,{\rm \AA}$. 

The search for a compound in the {\cemt} series with lattice parameters and electronic structure at ambient pressure that are akin to {\cepdsi} near $p_c$ motivated further detailed studies of {\cenige} (publication of these studies was delayed for a long time and have been reviewed in \cite{gros00}). The temperature versus pressure phase diagram of single-crystal {\cenige} as determined in resistivity measurements is shown on the right hand side of Fig.\,\ref{PD-CePdeSi+CeNiGe}. The phase diagram is dominated by a non-Fermi liquid form of the resistivity at ambient pressure and indications of incipient superconductivity below $T_s\sim0.3\,{\rm K}$. Further studies of the specific heat and susceptibility of polycrystalline samples even suggest that {\cenige} at ambient pressure displays a genuine non-Fermi liquid ground state \cite{gege99}.  In particular the specific heat shows a logarithmic divergence $C/T\sim\ln T_0/T$ and the susceptibility a square root divergence $\chi\sim\sqrt{T}$ for $T\to0$

Neutron scattering in {\cenige} established high-energy spin fluctuations with a characteristic energy of 4\,meV at an incommensurate wave-vector $\vec{q}=(0.23,0.23,0.5)$ \cite{fak00}. The wave-vector is in remarkable agreement with the ordering wave-vector of the spin density wave in {\cecusi} and {\cecuge}. The spin fluctuations are quasi-two dimensional, characteristic of a sine-modulated structure with the magnetic moments in the $[110]$ plane. With decreasing temperature no critical slowing down of the high-energy spin fluctuations in {\cenige} is observed. 

Another surprise in {\cenige} was the observation of additional hints of superconductivity at high pressure \cite{gros97,gros00}. The origin of this superconductivity could not be related to a particular instability in the spirit of a quantum phase transition, where an anomaly of unknown origin in the normal state resistivity was denoted $T_x$.  One possibility is a valence transition like that considered in {\cecusi}, but this has not been explored further. The evidence for superconductivity in {\cenige} is purely based on the resistivity, while no evidence for bulk superconductivity has been found in the samples studied to date. 

\paragraph{\cerhsi}

A pressure induced transition from an antiferromagnetic ground state to superconductivity exists also in {\cerhsi} \cite{movs96}. The observation of superconductivity in this system is remarkable, because it occurs at a fairly pronounced first order quantum phase transition that may be related to the delocalization of the 4f electron. The properties of {\cerhsi} have been reviewed in \cite{sett07}. 

At ambient pressure {\cerhsi} orders antiferromagnetically below $T_{N1}\approx36{\rm K}$ \cite{thom86}. Neutron scattering establishes an ordering wave vector $\vec{Q}_1=(1/2, 1/2, 0)$ with the moments aligned along $[1,0,0]$ \cite{kawa00}.  The single-$\vec{Q}$ structure changes into a four-$\vec{Q}$ structure below $T_{N2}\approx24\,{\rm K}$ described by two ordering wave vectors, $\vec{Q}_1=(1/2, 1/2, 0)$ and $\vec{Q}_2=(1/2, 1/2, 1/2)$. The ordered moment is quite large and differs slightly for the different Ce sites, namely $1.42\,{\rm \mu_B}$ at the corner site of the tetragonal structure and $1.34\,{\rm \mu_B}$ at the body-centered Ce site. The size of the ordered moment is consistent with CEF-split localized 4f$^1$ state of the Ce-atom, which when taken together with the entropy released at $T_N$,  $\Delta S(T_N)\approx {\rm R \ln 2}$ suggests a CEF Kramers doublet. As a function of magnetic field the antiferromagnetism is suppressed above $\sim 26\,{\rm T}$ \cite{sett97}. The metallic state is described by a weakly enhanced linear term in the specific heat $\gamma=0.027\,{\rm J/mol\,K^2}$ \cite{graf97} and a quadratic temperature dependence of the resistivity \cite{arak02,gros97b,ohas02}.

As a function of pressure both $T_{N1}$ and $T_{N2}$ decrease and vanish at $p_{N1}=10\,{\rm kbar}$ and $p_{N2}=6\,{\rm kbar}$, respectively \cite{kawa00}. A narrow dome of superconductivity emerges precisely at $p_{N1}$, where $T_s^{\rm max}\approx0.42\,{\rm K}$ \cite{movs96} with $H_{c2}=0.28\,{\rm T}$ and $dH_{c2}/dT=-1\,{\rm T/K}$ for the c-axis, corresponding to a coherence length of the order $\xi^{c}\approx340{\rm \AA}$.

Under pressure the ordered moment tracks $T_{N1}$, where both drop fairly abruptly at $p_{N1}$. The specific heat coefficient $\gamma$ increases to $0.08\,{\rm J/mol\,K^2}$ at $p_{N1}$ and gradually decreases at higher pressures \cite{graf97}, while the resistivity exhibits a $T^2$ resistivity at all pressures, where the $T^2$ coefficient tracks the pressure dependence of $\gamma$ consistent with the Kadowaki-Woods ratio \cite{arak02,gros97b,ohas02}. These features early on suggested a first order transition at $p_{N1}$. Unambiguous evidence for a first order suppression of antiferromagnetism was obtained in quantum oscillatory studies as a function of pressure \cite{arak01,arak02b}. More specifically, from the Fermi surface sheets observed it was concluded that the 4f electron changes discontinuously from a local to an itinerant state at $p_{N1}$. This scenario has received further support in recent studies of the thermal expansion under pressure \cite{vill07}. 

%%%%%%%%%%%%%%%%%%%%%%%%%%%%%%%%%%%%%%%%%%%%%%%

\subsubsection{ The series {\cemin}
\label{cemin-system}}

The series \cemin with M=Co, Ir, Rh displays heavy fermion superconductivity with very high transition temperatures, as compared to other Ce-based systems. The systems of interest are summarized in table \ref{table-cemin}. This suggests that a reduction from 3 to 2 dimensions is favorable to superconducting pairing. The superconductivity in these systems appears to be tied to the antiferromagnetic order, where similarities with the cuprates have been pointed out. The interplay of magnetism with superconductivity includes thereby several tentative quantum critical points under pressure and magnetic field, which are all of general interest. There is finally strong evidence for the formation of a FFLO state in {\cecoin}. Status reports on the series of {\cemin} compounds have been given in \cite{sarr07,sett07}.

\begin{figure}
%\sidecaption
\includegraphics[width=.35\textwidth,clip=]{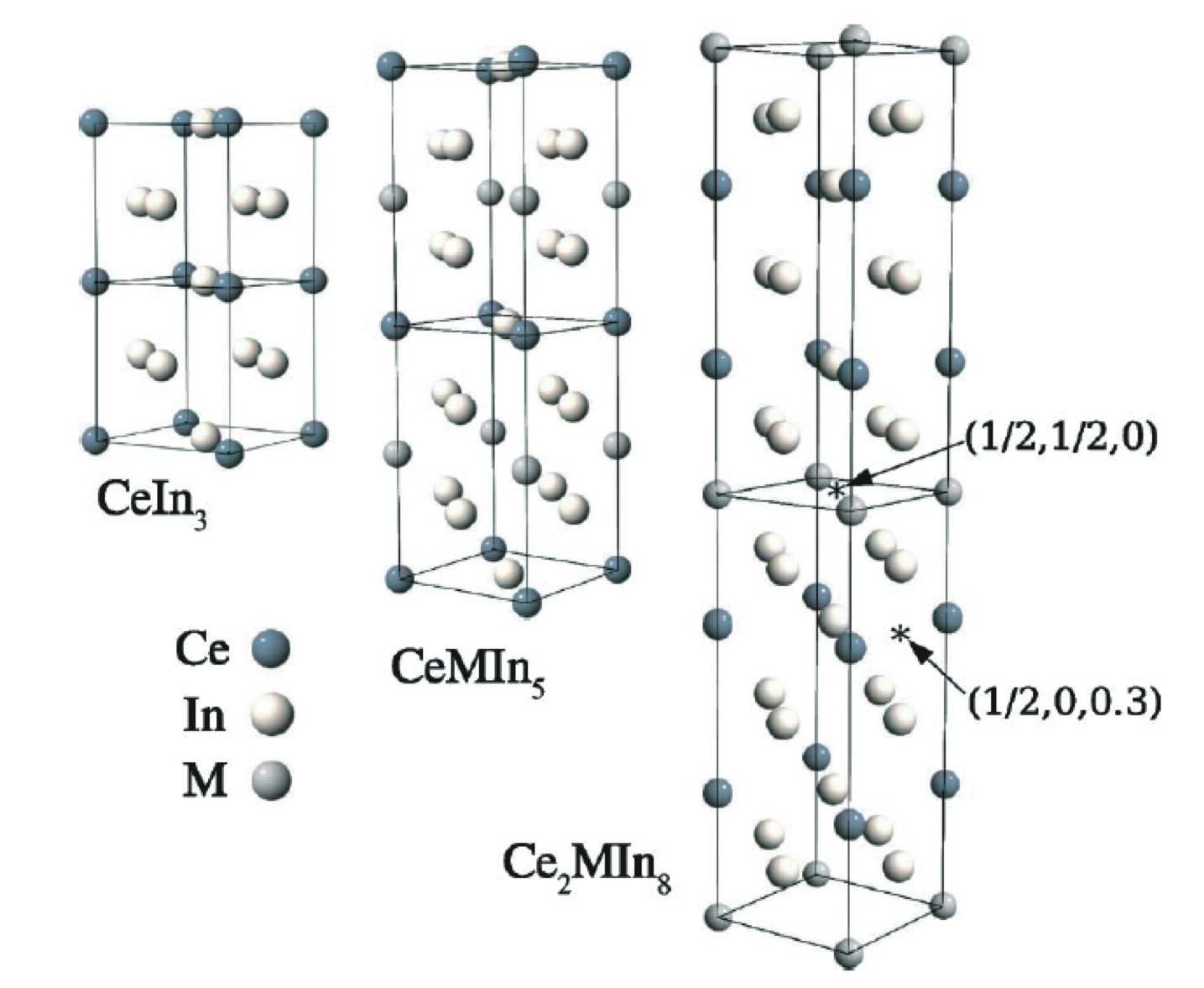}
\caption{Depiction of the structural series {\cemin}. The infinite layer system {\cein} is shown on the left, the single layer systems are shown in the middle, the double layer systems, which are intermediate to the infinite and the single layer systems are shown on the right. Also indicated are typical muon stopping sites in the double layer system. Plot taken from \cite{morr04}}
\label{cmint-structure} 
\end{figure}

\begin{table*}
\centering
\caption{
Key properties of the series {\cemin} and Pu- and Np-based heavy-fermion superconductors.
%Values at high pressure are marked as follows:
%$^{\heartsuit}=1.63$\,GPa; 
%$^{\diamondsuit}=2.1$\,GPa; 
%$^{\spadesuit}=2.5$\,GPa;
%$^{\clubsuit}=2.6$\,GPa.
Missing table entries may reflect more complex behavior discussed in the text. 
References are given in the text.
Values of $H_{c2}$ are extrapolated for $T\to0$.
}
\label{table-cemin}
\begin{tabular}{llllllllll}
\hline\noalign{\smallskip}
{\cemin} & {\cein} &  {\cecoin} & {\cerhin} & {\ceirin} & {\cerhinte} & {\pucoga} & {\purhga} & {\nppdal} \\
\noalign{\smallskip}\hline
structure & cubic & tetragonal & tetragonal & tetragonal & tetragonal & tetragonal & tetragonal & tetragonal \\
space group & Pm 3m &P4/mmm & P4/mmm & P4/mmm & P4/mmm & P4/mmm & P4/mmm & I4/mmm \\
$a$({\AA}) & 4.690 & 4.614 & 4.652 & 4.668 & 4.665(1) & 4.2354 & 4.3012 & 4.148 \\
$c$({\AA}) & - & 7.552 & 7.542 & 7.515 & 12.244(5) & 6.7939 & 6.8569 & 14.716\\
$c/a$ & 1 & 1.63676 & 1.62124 & 1.6099 & 2.624 & 1.604 & 1.594 & 3.547 \\
%$V_0 ({\rm \AA^3})$ & 103.10 & 160.96 & 163.03 & 163.67 & 266.48 & - & - & -  \\
\noalign{\smallskip}\hline
$B_0 {\rm (GPa)}$ & $67.0\pm3.0$ & $78.2\pm1.8$ & $78.4\pm2.0$ & $87.6\pm2.0$ & $71.4\pm1.1$  & - & - & - \\
$dB_0/dp$ & $2.5\pm0.5$ & $3.94\pm0.41$ & $5.60\pm0.62$ & $5.04\pm0.58$ & $3.85\pm0.31$  & - & - & - \\
$\kappa_a {\rm (10^{-3} GPa^{-1})}$ & $4.98\pm0.13$ & $4.35\pm0.08$ & $3.96\pm0.08$ & $3.44\pm0.06$ & $4.20\pm0.04$  & - & - & - \\
$\kappa_c {\rm (10^{-3} GPa^{-1})}$ & $4.98\pm0.13$ & $3.43\pm0.16$ & $4.22\pm0.1 $ & $3.48\pm0.08$ & $4.85\pm0.11$  & - & - & - \\
\noalign{\smallskip}\hline
CEF scheme & ($\Gamma_7$, $\Gamma_8$) & ($\Gamma^1_7$, $\Gamma^2_7$, $\Gamma_6$) & ($\Gamma^1_7$, $\Gamma^2_7$, $\Gamma_6$) & ($\Gamma^1_7$, $\Gamma^2_7$, $\Gamma_6$) & -  & - & - & - \\
$\Delta_1$, $\Delta_2$ (meV) & 12 & 8.6, 25 & 6.7, 29 & 6.9, 24 & -  & - & - & - \\
\noalign{\smallskip}\hline\noalign{\smallskip}
state & AF, SC & SC & AF, SC & SC & AF, SC  & SC & SC & SC \\
$T_{N}$(K) & 10.2 & - & 3.8 & -  & 2.8, 1.65  & - & - & - \\
$\vec{Q}$ & ($\frac{1}{2},\frac{1}{2},\frac{1}{2}$) & - & ($\frac{1}{2},\frac{1}{2},0297$) & -  & ($\frac{1}{2},\frac{1}{2},0)$  & - & - & - \\
$\mu_{ord}¥$($\mu_{B}$) & 0.48 & - & 0.37 & - & 0.55  & - & - & - \\
$\mu_{eff}^{a}¥$($\mu_{B}$) & - & - & - & - & - & 0.75 & 0.8 & 3.22 \\
$\Theta_{CW}^{a}$(K) & - & - & - & - & - & - & - & -42 \\
$\mu_{eff}^{c}¥$($\mu_{B}$) & - & - & - & - & - & 0.75 & 0.8 & 3.06 \\
$\Theta_{CW}^{a}$(K) & - & - & - & - & - & - & - & -139 \\
$\gamma {\rm (J/mol K^2)}$ & 0.14 & - & 0.4 & 0.72 & 0.4 & 0.077 & 0.07 & 0.2 \\
%$A {\rm (\mu\Omega cm/K^2)}$ & 0.15 & - & -  \\
$p_N {\rm (kbar)}$ & 25 & - & 17 & - & $\sim 25$  & - & - & - \\
\noalign{\smallskip}\hline
$T_{s}$(K) & 0.19 ($p_N$) & 2.3 & 2.12 ($p_N$) & 0.4 & 1.1 & 18.5 & 8.7 & 4.9 \\
$\Delta÷C/\gamma_{n}T_{s}$ & - & 4.5 & 0.36 & 0.76 & - & 1.4 & 0.5 & 2.33 \\
$H^{ab}_{c2}$(T) & 0.45 & 11.6-11.9 & - & 1.0 & 5.4 & - & 27 & 3.7 \\
$\frac{d}{dT}H^{ab}_{c2}/$(T/K) & -3.2 & -24  & - & -4.8 & -9.2 & -10 & -3.5 & -6.4\\
$H^{c}_{c2}$(T) & 0.45 & 4.95 & 10.2 & 0.49 & - & - & 15 & 14.3\\
$\frac{d}{dT}H^{c}_{c2}/÷T$(T/K) & -2.5 & -8.2 & -15 & -2.54 & - & -8 & -2 & -3.1 \\
%\noalign{\smallskip}\hline
$dT_{s}/dt\,{\rm (K/month)}$ & - & - & - & - & - & -0.24 & -0.39 & - \\
%$\alpha_{\parallel}$ & - & - & - & - & - &  4.22 & 1.06 & \\
%$\alpha_{\perp}$ & - & - & - & - & - &  5.28 & 2.11 & \\
\noalign{\smallskip}\hline
$\xi^{ab}_{0}$({\AA}) & 300 & 82 & 57 & 260 & - & - & 35 & - \\
$\xi^{c}_{0}$({\AA}) & - & 53 & - & 180 & 77 & - & 45 & - \\
$\kappa_{GL,a}$({\AA}) & - & 108 & - & - & - & - & - & -\\
$\kappa_{GL,c}$({\AA}) & - & 50 & - & - & - & - & - & -\\
$\kappa_{GL}$({\AA}) & - & - & - & - & - & 32 & - & 28\\
%\noalign{\smallskip}\hline
%$\partial÷T_{s}/\partial÷p_{[100]}$(mK/GPa)& - & $+290\pm30$ & -  & $+540\pm40$ & \\
%$\partial÷T_{s}/\partial÷p_{[001]}$(mK/GPa)& - & $+75\pm10$ & - & $-890\pm40$ & \\
%$\partial÷T_{s}/\partial÷p_{V}$(mK/GPa)& - &  $+655\pm50$ & - &  $+190\pm60$ & \\
%$\partial÷T_{s}/\partial÷p$(mK/GPa)& - &  $+400$ & - &  $+250$ & \\
%\noalign{\smallskip}\hline
%$C/T$ & - & $\propto÷T$ & $\propto÷T$ & $\propto÷T$ & - \\
%$\kappa$ & - & $\propto÷T^{3.37}$ & -  & $\propto÷T^{3}$ & - \\
%$1/T_{1}$ & - & $\propto÷T^{3+\epsilon}$ & $\propto÷T^3$ & $\propto÷T^{3.37}$ & - \\
%$\lambda$ & - & $T^{1.5\pm0.2}$ & - & $T-T^{1.65\pm0.2}$ & - \\
\noalign{\smallskip}\hline
discovery of SC & 1997 & 2001 & 2000 & 2001 & 2003 & 2002 & 2003 & 2007 \\
\noalign{\smallskip}\hline
\end{tabular}
\end{table*}
 
For a more detailed review it is helpful to begin with the crystal structure of the series {\cemin}. {\cein} crystallizes in the cubic Cu$_3$Au structure, space group Pm3m, with a lattice constant $a=4.690\,{\rm \AA}$. The tetragonal crystal structure of the series {\cemin} may be derived from the cubic parent compound {\cein} in terms of $n$-fold layers of {\cein} separated by $m$-fold layers of {\mint}. For the single-layer compounds $n=m=1$ ({\ceminof}) one layer of {\mint} is added while in the double-layer compounds  $n=2$, $m=1$ ({\ceminte})  a single layer of {\mint} is added for every two layers of {\cein}. Within this general scheme {\cein} may therefore be referred to as $\infty$-layer system ($n=\infty$, $m=0$).

The low temperature properties of {\cemin} develop in a crystal electric field scheme that is intimately related for all members of the series. For {\cein} the CEFs split the $J=5/2$ manifold into a $\Gamma_7$ ground state doublet and a $\Gamma_8$ quartet at around 12\.meV \cite{lawr80,beno80,gros80,mura93,chri04}. For the series {\ceminof} the quartet is further split into two $\Gamma_7$ and $\Gamma_6$ Kramers doublets, where values of the first and second energy levels are given in table\,\ref{table-cemin} \cite{chri04}.

The series {\cemin} exhibits metallic behavior with a fairly weak temperature dependence of the resistivity at high temperatures. With decreasing temperature the resistivity decreases monotonically with a shoulder around 50 to 100\,K before decreasing drastically to a very low residual value of a few ${\rm\mu\Omega cm}$. The normal state resistivity and magnetic anisotropy for the single- and double-layer series are weakly anisotropic by a factor of two. The susceptibility displays a strong Curie-Weiss temperature dependence, where the effective fluctuating moment for the easy axis corresponds to the free Ce$^{3+}$ ion.  The specific heat is characteristic of strong electronic correlations with a strongly enhanced electronic contribution. However, closer inspection shows that the temperature dependence of these electronic contributions are more complex and typical of non-Fermi liquid behavior, as discussed below.

\paragraph{\cein}

\begin{figure}
%\sidecaption
\includegraphics[width=.4\textwidth,clip=]{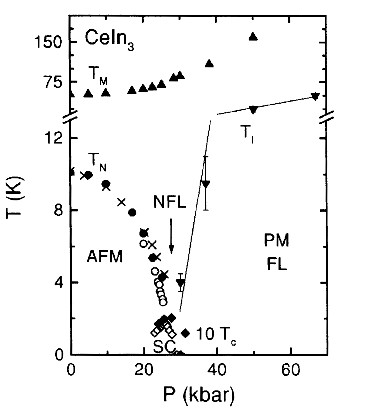}
\caption{Temperature versus pressure phase diagram of {\cein}. $T_M$ denotes the coherence maximum in the resistivity, $T_N$ the N{\'e}el temperature, $T_c$ the superconducting transition temperature and $T_1$ the upper boundary of the regime with Fermi liquid resistivity. Plot taken from \cite{kneb01}.}
\label{pd-cein3} 
\end{figure}

We begin with the cubic system {\cein}, which displays a strikingly simple temperature versus pressure phase diagram shown in Fig.\,\ref{pd-cein3}. Here the superconductivity forms a well-defined dome around an antiferromagnetic QCP.  This makes {\cein} an important point of reference for those systems in the series that are more two-dimensional. 

The properties of {\cein} are typical of a valence fluctuating compound, i.e., by comparison to traditional heavy fermion systems they are moderately enhanced with $\gamma=0.14\,{\rm J/mol\,K^2}$. The characteristic spin fluctuation temperature is fairly high $T_{SF}= 50 - 100\,{\rm K}$ \cite{lawr79,mori88}. At ambient pressure {\cein} orders antiferromagnetically below a N{\'e}el temperature $T_N=10.2\,{\rm K}$ into a type 2 state with ordering wave vector $\vec{Q}=(1/2, 1/2, 1/2)$, i.e., ferromagnetic planes of alternating direction stacked along the $(111)$ cubic space diagonal \cite{lawr80,beno80}. The zero temperature ordered moment $\mu_{ord}\approx0.47\,{\rm \mu_B}$ is reduced as compared to the value of $\sim0.71\,{\rm \mu_B}$, expected in the CEF ground state given by a $\Gamma_7$ doublet. It is also reduced as compared to the Curie-Weiss moment. This is typical of weak itinerant magnetism, where inelastic neutron scattering shows antiferromagnetic magnons as well as quasielastic and crystal field excitations \cite{knaf03}.

Under hydrostatic pressure the N{\'e}el temperature in {\cein} decreases and vanishes continuously at $p_N\approx25\,{\rm kbar}$ \cite{mori88} consistent with a QCP (Fig.\,\ref{pd-cein3}).  The temperature dependence of the electrical resistivity changes from a quadratic temperature dependence at ambient pressure to $\Delta\rho\sim T^{1.5}$ in a narrow interval near $p_N$  \cite{walk97,kneb01}. This suggests scattering of the charge carriers by antiferromagnetic quantum critical fluctuations. In fact, {\cein} is one of the very few systems for which the pressure and magnetic field dependence of the resistivity is in excellent agreement with the predictions of an antiferromagnetic QCP \cite{hert76,mill93}. The existence of a QCP is contrasted by $^{115}$In-NQR measurements, which show that the spin lattice relaxation rate $1/T_1T\sim{\rm constant}$ near $p_N$ as expected of a Fermi liquid \cite{kawa01}. Quantum oscillatory studies through $p_N$ further establish a reconstruction of the topology of the Fermi surface, interpreted as localized to delocalized transition of the 4f-electrons \cite{sett05}. As $p_c$ is approached the cyclotron effective mass becomes strongly enhanced for at least one major Fermi surface sheet reaching $m^*=60m_0$.

In pure samples of {\cein} with residual resistivities below $1\,{\rm \mu\Omega cm}$ the QCP is surrounded by a narrow dome of superconductivity, which exhibits a maximum $T_s\approx0.22\,{\rm K}$ \cite{walk97}. A detailed study up to 100\,kbar with a different set of samples and pressure cells showed that the phase diagram is rather robust and highly reproducible \cite{kneb01}. Under magnetic field $T_s$ initially decreases with $dH_{c2}/dT=-3.2\,{\rm T/K}$, where $H_{c2}(T\to0)=0.45\,{\rm T}$, both characteristic of heavy fermion superconductivity \cite{kneb01,onuk04}. The upper critical field may be accounted for in a strong-coupling framework in the clean limit, where the coherence length $\xi_0=300\,{\rm \AA}$ and the charge carrier mean free path $l=2000\,{\rm \AA}$.

The location of the superconducting dome at the border of antiferromagnetic order, the evidence for quantum critical fluctuations in the resistivity and the sensitivity of the superconductivity to sample purity \cite{walk97,kneb01} provide circumstantial evidence of unconventional pairing. Microscopically this question has been explored in NMR and NQR studies. The spin-lattice relaxation rate $1/T_1$ lacks a Hebel-Slichter peak but the low value of $T_s$ did not permit to determine the temperature dependence below $T_s$ \cite{kawa02}.  From a theoretical point of view it has been argued that the antiferromagnetic quantum critical spin fluctuations may provide a pairing interaction consistent with the size of $T_s$ \cite{math98}. A more detailed theoretical analysis suggests that the gap symmetry due to pairing by antiferromagnetic fluctuations near $\vec{Q}=(111)$ is either $d_{x^2-y^2}$ or $d_{3z^2-r^2}$ \cite{fuka03}.

\paragraph{Introduction to CeMIn$_5$}

We next turn to the single-layer systems in the series of {\cemin} ($n=m=1$). Key properties are summarized in  table\,\ref{table-cemin}, where references to the original publications may be found in the text. Much of the appeal about this series is based on the sequence Co$\to$Rh$\to$Ir representing isovalent substitutions.  In this order the unit cell volume increases, while the $c/a$ ratio of the lattice constants decreases. 

Electronic structure calculations show that the Fermi surface in all three systems is highly two-dimensional with several cylindrical sheets, even though the electrical resistivity and magnetic susceptibility are not strongly anisotropic (see e.g. \cite{sett01}). Band structure calculations suggest, as an important aspect for understanding the evolution of the physical properties within this series, that the transition metal element affects the electronic properties only indirectly \cite{sarr07}. This may be related to the Ce atoms and the transition metal atoms residing in different crystallographic planes, which may also explain why substitutional doping provides a comparatively controlled approach to tuning the ground state properties without metallurgical segregation and excessive effects of disorder \cite{pagl01,pagl02,pagl02c,pagl02b,zapf01}.

The presentation proceeds as follows. We begin with the general phase diagrams of {\cecoin}, {\cerhin} and {\ceirin} and discuss the tentative evidence for QCPs. This is followed by a discussion of the interplay of antiferromagnetism and superconductivity and the evidence for unconventional superconductivity. The section concludes with a brief discussion of possible analogies with the cuprates.

\paragraph{\cecoin}

{\cecoin} is a superconductor with a record high value $T_s=2.3\,{\rm K}$ amongst the Ce-based systems \cite{petr01a}. For the c-axis $H^c_{c2}=4.95\,{\rm T}$ and for the ab-plane $H^{ab}_{c2}=11.6\,{\rm T}$. The anisotropy of $H_{c2}$ may be accounted for by the effective mass model \cite{petr01a,iked01}. Before reviewing the superconducting state of {\cecoin} it is helpful to consider the normal state properties, which are in many ways anomalous. The electrical resistivity of {\cecoin} varies as $\rho(T)=\rho_0+a'T$ \cite{sido02} up to $\sim$4\,K above $T_s$. Taking into account CEF contributions, the normal state electronic specific heat varies as $C/T\propto -\ln T$, and the c-axis susceptibility diverges as $\chi\propto T^{-0.4}$, while the basal-plane susceptibility is essentially constant, $\chi^{-1}\propto a +b T^{0.1}$ \cite{kim01,petr01a}. These normal state non-Fermi liquid temperature dependences differ distinctly from a heavy Fermi liquid state and suggest the vicinity to an antiferromagnetic quantum critical point. 

In applied magnetic fields the normal state retains certain NFL characteristics regardless of field direction, before Fermi liquid behavior is recovered well beyond $H_{c2}$ \cite{bian03c,pagl03,mali05,ronn05}. This is surprising since the NFL characteristics due to a QCP are normally rapidly suppressed in a magnetic field. For instance, at $H^c_{c2}$ the specific heat $C/T$ diverges logarithmically reaching $C/T=1.1\,{\rm J/mol\,K^2}$ at the lowest temperatures studied \cite{petr01a}, while Fermi liquid behavior is only observed above 8\,T. Likewise the {d.c.} susceptibility at $H_{c2}$ diverges as $\chi(T)=\chi_0+C/(T^{\alpha}+a_0)$ with $\alpha=0.8-1$ \cite{taya02}. 

The electronic structure of {\cecoin} has been studied microscopically by angle-resolved photoemission (ARPES). The dispersion and the peak width of the prominent quasi-two-dimensional Fermi surface sheet displays an anomalous broadening near the Fermi level \cite{koit08}. Using resonant ARPES a flat f-band is observed  with a distinct temperature dependence. These observations are consistent with a two-level mixing model. 

Direct microscopic evidence of a NFL normal state is supported by de Haas--van Alphen oscillations for magnetic field along the c-axis. Here strongly spin dependent mass enhancements are observed  in the immediate vicinity of $H^c_{c2}$, that are inconsistent with the Lifshitz-Kosevich expression and thus Fermi liquid theory \cite{mcco05}.  This is supplemented by the spin-lattice relaxation rate $T_1$ in $^{115}$In nuclear quadrupole resonance, which displays a temperature dependence $1/T^4$ characteristic of  antiferromagnetic spin fluctuations \cite{kawa03,koho01}. 

As discussed below, the electronic correlations at the heart of the NFL behavior are likely to be responsible for the superconductivity in {\cecoin}. This raises the question for their origin and the possible nature and location of the QCP. The $T^2$ coefficient of the resistivity for $H$ along the c-axis diverges at an extrapolated field value below $H_{c2}$ suggesting a QCP within the superconducting regime, but the precise location has not been settled \cite{bian03c,pagl03,mali05}. More recently even a dimensional cross-over from three-dimensional to two-dimensional quantum criticality near $H_{c2}$ was inferred from the thermal expansion \cite{dona08}. Entirely unexplained is the observation of a giant Nernst effect in the normal state \cite{bel04,izaw07}. In fact, one scenario offered to explain the giant Nernst effect and scaling of the normal state resistivity as a function of field direction in the basal plane is the formation of a d-density wave \cite{hu06}. 

Further support of unconventional superconductivity with a d-wave gap has been observed in inelastic neutron scattering studies \cite{stoc08}. In the normal state slow commensurate fluctuations ($\hbar\Gamma=0.3\pm0.15\,{\rm meV}$ at $\vec{Q}_0=(1/2,1/2,1/2)$) with nearly isotropic spin correlations are observed. In the superconducting state a sharp spin resonance at $\hbar\omega=0.60\pm0.03\,{\rm meV}$ develops with $\hbar\Gamma<0.07\,{\rm meV}$. The spin resonance is indicative of strong coupling between f-electron magnetism and superconductivity. The similarity of this spin resonance with the properties of {\updal} and the cuprates suggest that it may be understood in a common framework.

The specific heat anomaly of the superconducting transition is exceptionally large, $\Delta C/\gamma T_s=4.5$ when taking the value of $\gamma$ at $T_s$. This would suggest an extreme case of strong coupling superconductivity. However, when considering $\Delta C/\gamma$ normally the extrapolated zero temperature value of $\gamma$ is used, which due to the NFL behavior here is ill-defined. The initial variation of $H_{c2}$ near $T_s$ is large and characteristic of heavy fermion superconductivity, $dH^c_{c2}/dT=-11\,{\rm T/K}$ and $dH^{ab}_{c2}/dT=-24\,{\rm T/K}$ \cite{iked01}. The short coherence length $\xi_a=82\,{\rm \AA}$ and $\xi_c=35\,{\rm \AA}$ and large penetration depth as inferred from microwave measurements, $\lambda(T\to0)=1900\,{\rm \AA}$ \cite{orme02}, along with the low Fermi energy and large charge carrier mean free paths of several 1000\,{\AA} identify {\cecoin} as a type II superconductor ($\kappa_a=108 $ and $\kappa_c=50$) in the super-clean limit \cite{kasa05}. 

A large number of properties suggest an unconventional form of superconductivity in {\cecoin}. For instance the depression of $T_s$ with rare earth substitution correlates with the mean free path \cite{pagl07}. The following experimental evidence suggests line nodes of a $d_{x^2-y2}$ state, notably: (i) the power law temperature dependence of the specific heat, $C\propto T^3$ \cite{movs01}, (ii) the variation of the specific heat with fourfold symmetry for magnetic field in the basal plane \cite{aoki04} (maxima along $[110]$), (ii) the power-law dependence of the thermal conductivity, $\kappa \propto T^3$ \cite{movs01}, (iii) the variation of the thermal conductivity with a fourfold symmetry for magnetic field in the basal plane (maxima along $[110]$) \cite{izaw01}, (vi) the variation of the $H_{c2}$ of 1.2\% with a four-fold symmetry in the basal plane (maxima along $[100]$) \cite{weic06}, and (vii) the differential conductance spectra as interpreted in the extended Blonder-Tinkham-Klapwijk model \cite{park08}.

In contrast, a $d_{xy}$ pairing symmetry has been inferred from the symmetry and the field and temperature dependence of the in-plane torque magnetization \cite{xiao08}. Moreover, the magnetic field and temperature dependence of the thermal conductivity, was found to be inconsistent with unpaired electrons \cite{seyf08}. The latter study points at multi-band superconductivity and a related complex multi-gap state.

Microscopic information on the pairing symmetry may be inferred from the Knight shift, which decreases for both field directions. This shows that the spin susceptibility decreases for all directions, consistent with even parity superconductivity \cite{koho01}. The NMR/NQR spin-lattice relaxation rate shows no Hebel-Slichter peak and a power-law temperature dependence $1/T_1\propto T^3$ \cite{koho01} further suggesting a non s-wave state. 

Small angle neutron scattering (SANS) shows a six-fold symmetry of the flux line lattice at low fields and low temperatures. As a function of magnetic field the flux lattice symmetry undergoes a sequence of transitions from hexagonal to orthorhombic, to square, back to orthorhombic and finally hexagonal symmetry near $H_{c2}$ \cite{eski03,debe06,bian08}. Most remarkably, the form factor of the FLL as traced all the way to $H_{c2}$ \textit{increases} with increasing field, in stark contrast with the predictions of Abrikosov-Ginzburg-Landau theory \cite{bian08}. This behavior has been attributed to a combination of Pauli paramagnetic effects around the vortex cores and the vicinity of the system to a quantum critical point.  The temperature dependence of the penetration depth may be described as $\lambda_{\perp} \propto T^{1.5}$. This has been explained in terms of a temperature dependent coherence length related to the vicinity to quantum criticality  \cite{oezc03}. Alternatively, the penetration depth has been described as $\lambda_{\perp} \propto aT + bT^2$ and $\lambda_{\parallel} \propto T$, where a cross-over from weak to strong coupling superconductivity was proposed \cite{chia03}. 

\begin{figure}
\includegraphics[width=.4\textwidth,clip=]{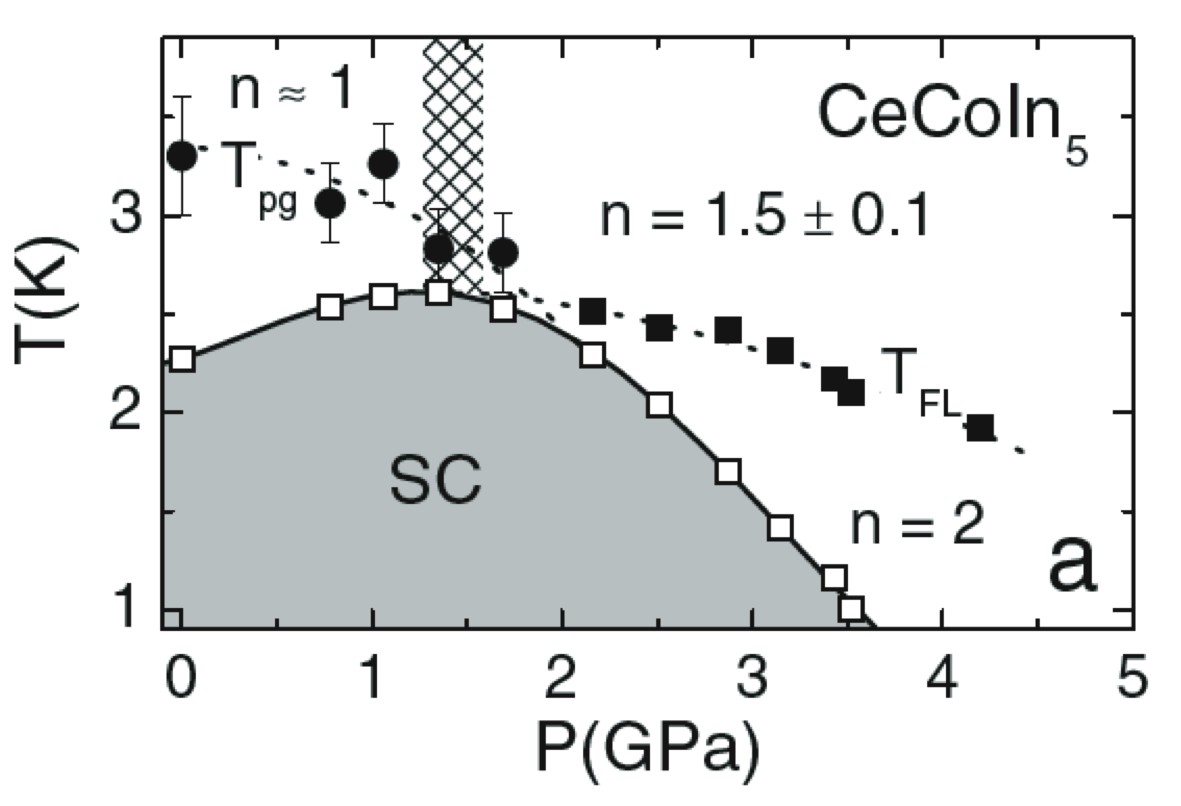}
\caption{Temperature versus pressure phase diagram of {\cecoin}. From the resistivity a 'pseudo-gap' at $T_{pg}$ is inferred that merges with the maximum in the onset of the superconductivity (SC). At high pressures the superconductivity condenses out of a Fermi liquid temperature dependence of the resistivity below $T_{FL}$. Plot taken from \cite{sido02}.}
\label{pd-cecoin5} 
\end{figure}

The properties of {\cecoin} respond sensitively to hydrostatic pressure as shown in Fig.\,\ref{pd-cecoin5} \cite{nick01,spar02,shis03,kneb04,yash04,taya05,micl06,sing07}. Up to 30\,kbar $T_s$ traces out part of a dome; an initial increase is followed by a decrease for $p>16\,{\rm kbar}$. The specific heat anomaly $\Delta C/\gamma T_s$ decreases under pressure monotonically by nearly 80\% up to 30\,kbar \cite{kneb04,spar02}.  $H_{c2}$ increases for the ab-plane while it decreases for the c-axis from 4.95 to 2\,T at 30\,kbar \cite{shis03}, so that the anisotropy of $H_{c2}$ increases from 2.34 at $p=0$ to 3.78 at 30\,kbar \cite{taya05}. 

Despite these rather drastic effects, $^{115}$In NQR shows that the spin-lattice relaxation rate $T_1$ below $T_s$ remains qualitatively unchanged $T_1\propto T^{-3}$ up to 20\,kbar. This suggests that the nature of the superconductivity remains unchanged \cite{yash04}. An increase of the spin fluctuation temperature $T_{SF}$ may be consistently inferred from (i) the decrease of the normal state value of  $\gamma$, (ii) an increase of the coherence maximum in the resistivity from 50 to nearly 100\,K at 15\,kbar and (iii) a change of the normal state spin-lattice relaxation rate. All of these properties suggest, that pressure moves {\cecoin} away from quantum criticality.

We finally mention that {\cecoin} combines a unique set of properties: it shows strong Pauli limiting, the electronic structure is quasi-two dimensional and samples may be grown at ultra-high purity. These are the preconditions for the formation of a FFLO phase. Indeed, striking evidence exists that {\cecoin} stabilizes the first example of such a state as discussed in section \ref{FFLO-phas}.

\paragraph{\cerhin}

In comparison to {\cecoin}, the unit cell volume of {\cerhin} is larger and the c/a ratio smaller as shown in table \ref{table-cemin}. Taking into account the anisotropic compressibility for the a- and c-axis the properties of {\cerhin} at high pressure may be expected to resemble those of {\cecoin}. Considering the bulk modulus of {\cein}, the {\cein} units may be viewed as experiencing an effective pressure of 14\,kbar \cite{hegg00}. The electronic properties of {\cerhin} emerge in a CEF $\Gamma_7$ ground state and $\Gamma_7$ and $\Gamma_6$ first and second excited state at 6 and 29.1\,meV, respectively \cite{chri04}. 

At ambient pressure {\cerhin} orders antiferromagnetically below $T_N=3.8\,{\rm K}$ \cite{hegg00}, with a temperature independent antiferromagnetic ordering wave vector $\vec{Q}=(1/2, 1/2, 0.297)$.  The ordered moment at 1.4\,K of  $\mu_{ord}=0.264(4)\,{\rm \mu_B}$ is strongly reduced as compared to the moment expected in the CEFs of $0.8\,{\rm \mu_B}$ \cite{bao00,bao00err}. It spirals transversely along the c-axis, while the nearest-neighbor moments on the tetragonal basal plane are aligned antiferromagnetically. Based on $\mu$-SR it has been suggested that a small ordered moment also exists at the Rh site \cite{sche02}. The antiferromagnetic transition as seen in neutron scattering and the bulk properties is second order, where the specific heat is characteristic of an anisotropic spin-density wave that gaps nearly 90\% of the Fermi surface \cite{corn01}. The entropy released at $T_N$ corresponds to the small ordered moment \cite{hegg00}.

The normal state specific heat of {\cerhin} is characteristic of a heavy fermion state with $\gamma=0.42\,{\rm J/mol\,K^2}$ \cite{corn00}. In contrast, the thermal expansion shows a non-Fermi liquid divergence of $\alpha/T$ for $[001]$ above $T_N$ while the basal plane is well behaved with $\alpha/T\approx{\rm constant}$ \cite{take01}. Moreover, while the susceptibility displays the Curie-Weiss behavior of nearly free Ce$^{3+}$ moments at high temperatures, $\chi$ keeps increasing even at low temperatures below a shoulder around 30\,K, where $\chi_{ab}^{-1}\propto a +b T^{0.9}$ and $\chi_{c}^{-1}\propto a + T^{1.35}$ \cite{kim01}. Similar anomalous behavior is also seen in the temperature dependence of the normal state electrical resistivity \cite{hegg00,mura01}. 

Microscopic evidence of an abundance of critical antiferromagnetic fluctuations up to 3\,$T_N$ has been seen in inelastic neutron scattering \cite{bao02} and the temperature dependence of the $^{115}$In NQR spin-lattice relaxation \cite{mito01}. The magnetic phase diagram of {\cerhin} as a function of an applied magnetic field has been studied up to 50\,T for the $[110]$ direction. A spin-flop transition is observed at 2\,T and a metamagnetic transition (spin flip) around 45\,T (for 3\,K) \cite{corn01,take01,sett07}. The c-axis is the easy magnetic axis. 

Under pressure the N{\'e}el temperature decreases. Superconductivity was first observed in {\cerhin} above  $15\,{\rm kbar}$, where an abrupt, first order change from antiferromagnetism to superconductivity was reported \cite{hegg00}. Recent studies suggest, that high quality single crystals display superconductivity even in the antiferromagnetic state at ambient pressure below $T_s\approx 0.09-0.11\,{\rm K}$ \cite{chen06,pagl08}. The bulk properties of the superconductivity at ambient pressure by comparison with other systems are characteristic of being far from quantum criticality.

As function of pressure $T_s$ increases, while $T_N$ decreases until $T_N=T_s\approx 2.0\,{\rm K}$ at $p_1\sim 17.7\,{\rm kbar}$. Specific heat and susceptibility suggest a competitive phase coexistence of AFM and superconductivity for pressure below $p_1$ \cite{kneb06}. Neutron scattering shows that the ordering wave vector and ordered moment initially change weakly \cite{maju02,llob04} and a second magnetic modulation emerges \cite{chri05}. For high pressures of 15 and 17\,kbar the incommensurate propagation vector is $\vec{Q}_{hp}=(1/2,1/2,0.4)$, which differs from the ambient pressure propagation vector $\vec{Q}=(1/2,1/2,0.297)$ \cite{raym08}. A competitive coexistence of AF and superconductivity up to $p_1$ is supported by $^{115}$In-NQR \cite{mito03}. Homogenous volume superconductivity is observed above $p_1$ with a maximum value of $T_s\approx2.1\,{\rm K}$ around 20\,kbar \cite{chen06,kneb06}. Resistivity measurements in {\cerhin} extending up to 85\,kbar initially indicated the presence of a second superconducting dome as shown in Fig.\,\ref{pd-cerhin5-b} \cite{mura01}. This finding could not be confirmed later as reviewed by \cite{kneb08}. 

Electronic structure calculations suggest that the 4f electron is localized in {\cerhin} \cite{elga04}. The mass enhancement seen in the specific heat has therefore been attributed to spin fluctuations above frozen magnetic states, which become itinerant and add to the spectrum of fluctuations when going to {\cecoin}. De Haas--van Alphen studies show that the electronic structure of {\cerhin} is highly two-dimensional  \cite{corn00,hall01b}. Under hydrostatic pressure a new branch emerges around 24\,kbar, the extrapolated pressure where $T_N$ vanishes. The similarity with {\cecoin} indeed suggests a delocalization of the 4f electron at this pressure \cite{shis05}. 

\begin{figure}
\includegraphics[width=.35\textwidth,clip=]{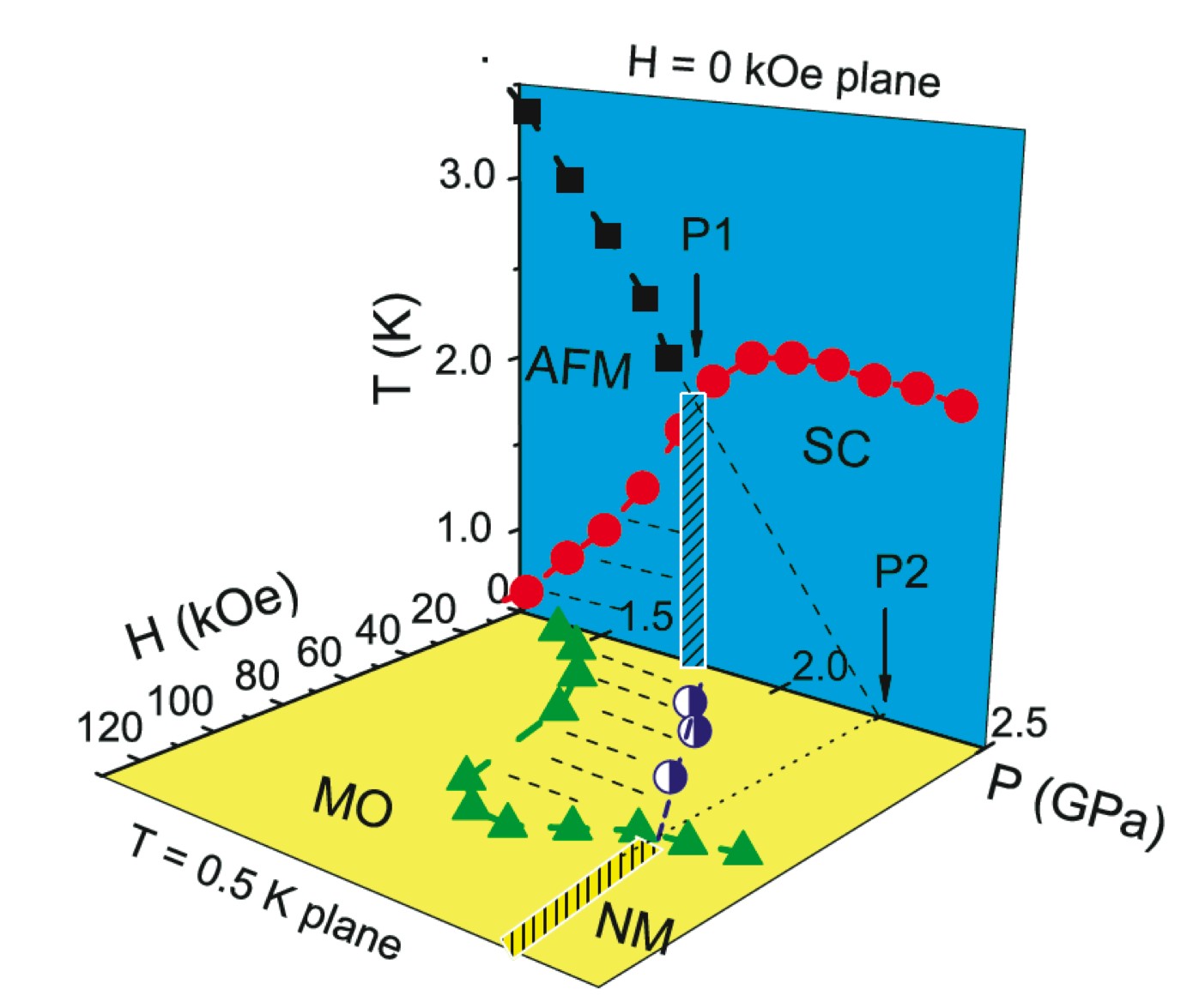}
\caption{Temperature versus magnetic field and pressure phase diagram of {\cerhin} as reported by \cite{park06}. Detailed studies of the specific heat suggests a well defined line of quantum criticality, separating a regime where superconductivity and antiferromagnetism coexist with a regime of homogenous bulk superconductivity. Plot from \cite{park06} as shown in \cite{sarr07}.}
\label{pd-cerhin5} 
\end{figure}

The NFL normal state properties and immediate vicinity of the superconductivity to antiferromagnetism in {\cerhin} are circumstantial evidence suggesting unconventional pairing. The superconductivity is, nevertheless, rather unexplored. The most direct evidence for unconventional pairing may be the spin-lattice relaxation rate of $^{115}$In-NQR in the superconducting state which does not show a Hebel-Slichter peak and varies as $1/T_1\propto T^3$ \cite{mito01}. 

The structural similarity of {\cerhin} with {\cecoin} raises the question for an analogy of the superconducting phase diagram. In {\cecoin} the normal state properties hint at a quantum critical point that is masked by the superconducting dome. Under pressure $H_{c2}$ for $B\perp [001]$ initially tracks the increase of $T_s$ and displays a maximum just above $p_1$. Specific heat measurements under pressure and magnetic field in {\cerhin} reveal a phase boundary separating homogenous volume superconductivity and a phase coexistence of antiferromagnetic order and superconductivity as reported by  \cite{park06} and shown in Fig.\,\ref{pd-cerhin5}.  In the magnetic field versus pressure plane the phase separation line increases from zero at $p_1$ and reaches $H_{c2}$ at $p_2$. The normal state properties in the $B$ versus $p$ plane are consistent with at a quantum critical point for $B\to0$ and $p_2$. Taken together with the de Haas--van Alphen studies this provides evidence of a quantum critical point at $p_2$ that may be related to a delocalization transition of the 4f electrons. 

\paragraph{\ceirin}

The heavy-fermion superconductor {\ceirin}, finally, has the largest unit cell volume and smallest c/a ratio as shown in table\,\ref{table-cemin}. At ambient pressure the normal state properties are characteristic of strong electronic correlations that develop in crystal electric fields related to those of {\cein} \cite{chri04}. The specific heat exhibits a large enhancement with $\gamma=0.72\,{\rm J/mol\,K^2}$ \cite{petr01b}. The susceptibility exhibits a broad shoulder around 7\,K \cite{take01}, but continues to diverge slowly \cite{kim01}. This and the resistivity, which varies as $\Delta \rho\propto T^n$ with $n\approx 1.3$ indicate non-Fermi liquid characteristics of the normal state \cite{petr01b}. 

The bulk properties are consistent with the spin-lattice relaxation rate inferred from $^{115}$In-NQR measurements, which suggests that {\ceirin} is an anisotropic, incipient antiferromagnet \cite{koho01,zhen01}.  More detailed information on the normal state has been inferred from the Hall effect and the magnetoresistance, which also show non-Fermi liquid behavior. Notably, there is a breakdown of Kohler's rule and the Hall angle varies as ${\cot}\Theta_H\propto T^2$ \cite{nair08,naka08}. When taken together, in the $T$ versus $B$ phase diagram the magneto-transport properties suggest a precursor regime of the normal metallic state that shares some similarities with the pseudo-gap regime in the cuprates \cite{nair08}. 

Various properties of {\ceirin} suggest two superconducting transitions. At $T_{s1}=0.75\,{\rm K}$ the resistivity vanishes and there are strong indications of an intrinsic form of filamentary superconductivity. At $T_{s2}=0.4\,{\rm K}$ superconductivity is observed in the specific heat \cite{petr01b} and in $^{115}$In-NQR \cite{kawa05}. Specific heat and In-NQR under pressure show that $T_{s2}$ increases to 0.8\,K at a pressure of 16\,kbar \cite{kawa05,bort02}. The increase of $T_{s2}$ is consistent with the observed decrease of $\gamma$, which may be interpreted as an increase of the characteristic spin fluctuation temperature. $H_{c2}$ is anisotropic where the resistivity and susceptibility show $H^{a}_{c2}=6.8\,{\rm T}$, $H^{c}_{c2}=3.5\,{\rm T}$ and $H^{a}_{c2}=1.0\,{\rm T}$, $H^{c}_{c2}=0.49\,{\rm T}$ for S1 and S2, respectively \cite{petr01b}. The temperature dependence and anisotropy of $H_{c2}$ of the incipient superconducting state and the bulk superconducting state track each other qualitatively \cite{petr01b}, where the anisotropy may be accounted for by the anisotropic mass model \cite{haga01}. 

The specific heat, which varies as $C\propto T^3$ \cite{movs01} and the thermal conductivity for heat current along the a-axis, which varies as $\kappa \propto T^2$ with a finite residual $T=0$ value of $\kappa/T=0.46\,{\rm W/K^2\,m}$ are consistent with an unconventional superconducting state and line nodes. This is supported microscopically by NMR and NQR, which shows (i) no Hebel-Slichter peak, (ii) a temperature dependence of the spin-lattice relaxation rate $1/T_1 \propto T^3$ at pressures up to 21\,kbar, and (iii) a decrease of the Knight shift in the superconducting state with decreasing temperature for all field directions \cite{koho01}. However, the thermal conductivity with heat current along the c-axis does not chow a residual term at low temperatures \cite{shak07}, ruling out line nodes running along the c-axis. Instead the formation of a hybrid gap structure with $E_g$ symmetry has been proposed.

\paragraph{{\cerhinte}}

The properties of the double layer compound {\cerhinte} are intermediate between {\cein} and the single-layer compound {\cerhin} as may be expected from the larger fraction of {\cein} building blocks in the crystal structure. At ambient pressure {\cerhinte} develops antiferromagnetic order below a second order phase transition at $T_{N1}=2.8\,{\rm K}$ with an ordering wave vector $\vec{Q}=(1/2,1/2,0)$ and an ordered moment $\mu_{ord}\approx 0.55\mu_{\rm B}/{\rm Ce}$ \cite{bao01}. The magnetic structure is more akin to that of {\cein}, where the specific heat shows that only 8\,\% of the Fermi surface are gapped in comparison to over 90\,\% in {\cerhin} \cite{corn01}. A second antiferromagnetic transition is observed in the resistivity at $T_{N2}=1.65\,{\rm K}$, which does not appear to be accompanied by an anomaly in the specific heat \cite{nick03}. The magnetic phase diagram at ambient pressure is reminiscent of that of {\cerhin} \cite{corn01}. 

Hydrostatic pressure suppresses both $T_{N1}$ and $T_{N2}$, where $T_{N2}$ vanishes below 1\,kbar and $T_{N1}$ extrapolates to zero around $p_{N1}\approx32{\rm kbar}$, suggesting a quantum critical point as in {\cerhin} \cite{nick03}. Specific heat under pressure shows a broadening of the antiferromagnetic transition and an decrease of $\gamma$ consistent with an increase of the spin fluctuation temperature \cite{leng04}. A superconducting dome surrounds $p_{N1}$ with a maximum $T_s=2.1\,{\rm K}$ \cite{nick03}. At 16.3\,kbar $H_{c2}=5.36\,{\rm T}$ and the initial temperature dependence $dH_{c2}/dT=-9.18\,{\rm T/K}$ are large and comparable with other compounds in this series.

\paragraph{Substitutional doping in {\cemin}}

Particularly appealing in the series CeMIn$_5$ is the relative metallurgical ease with which substitutional doping studies may be carried out. Three different aspects have been at the center of interest: (i) the sensitivity to doping of the f-electron element, (ii) the stability of the ground state under replacement of the transition metal element, and (iii) the sensitivity to disorder on the In site. 

Substitutional doping of Ce in CeMIn$_5$ has been carried out with La, U, Pu and Nd. In {\cecoin} La-doping results in a two-fluid state, notably a combination of single-impurity Kondo and dense Kondo lattice behavior \cite{naka02,naka04}. It is surprising, that La doping does not yield additional complexities.  Further, unconventional superconductivity in principle is very sensitive to disorder. However the superconductivity is remarkably insensitive to La-doping and vanishes only for $x\geq0.15$. In the superconducting state the residual electronic thermal conductivity decreases while the residual electronic specific heat increases with $x$, i.e., the thermal conductivity does not track the electronic degrees of freedom that become available under doping. This has been taken as evidence of extreme multi-band superconductivity in {\cecoin} \cite{tana05}.  Finally, Nd doping allows to study the evolution from local moment magnetism to heavy-fermion superconductivity \cite{hu08}.

Across the series Ce$_m$Rh$_n$In$_{3n+2}$ ($n=0,1$ and $m=1,2$) increasing the La substitution of Ce leads to a suppression of $T_N$. For the tetragonal systems ($n=m=1,2$) the critical concentration is $x_c\approx0.4$ and for {\cein} $x_c\approx0.65$ \cite{pagl02}. La-doping of {\cerhin} leaves the antiferromagnetic wave vector essentially unchanged up to $x=0.1$ \cite{bao02b}. The observation of the same value of $x_c$ for the tetragonal systems suggests, that antiferromagnetic order is essentially controlled by the {\cein} building blocks. The pressure dependence of La and Sn doped {\cerhin}, notably Ce$_{0.9}$La$_{0.1}$RhIn$_5$ and CeRhIn$_{4.84}$Sn$_{0.16}$, shows that La doping shifts the phase diagram to higher pressures, while Sn doping shifts it to lower pressures \cite{ferr08}. This implies that the strength of the on-site Kondo coupling represents the dominant energy scale controlling the phase diagram of {\cerhin}.

\begin{figure}
%\sidecaption
\includegraphics[width=.35\textwidth,clip=]{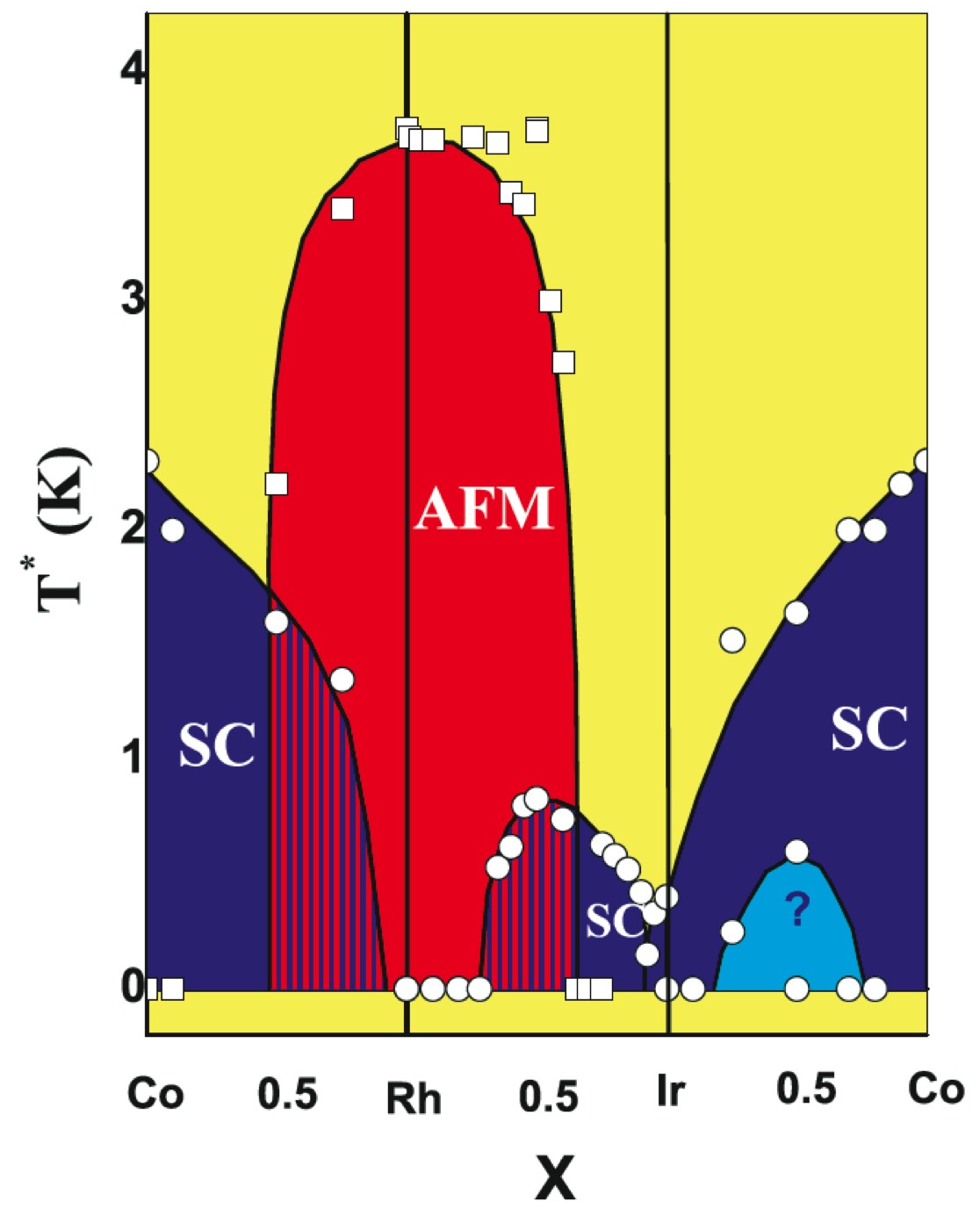}
\caption{Compilation of the evolution of superconductivity and antiferromagnetic order in the series CeMIn$_5$, where M= Co, Rh and Ir. Note the continuous evolution of the superconducting transiton temperature when going from Ir to Co despite the presence of disorder. This continuous evolution is interrupted by an antiferromagnetic dome in the series Co$\to$Rh$\to$Ir. Plot from \cite{pagl02c} as shown in \cite{sarr07}.}
\label{pd-115s} 
\end{figure}

Several studies have explored the evolution of the series CeMIn$_5$ under the isovalent replacement of Co by Rh and Ir, and of Rh by Ir. In the series {\cecorhin} and {\cerhirin} this allows the study of  the evolution between superconductivity and antiferromagnetism, while the series {\cecoirin} allows the study of the evolution between two unconventional superconductors as summarized in Fig.\,\ref{pd-115s}.  In the series {\cecorhin} a coexistence of antiferromagnetism and superconductivity is observed for a large range of $x$ \cite{zapf01}. The total entropy released at the two transitions is thereby constant. This suggests that the two ordering phenomena are intimately related representing two sides of the same coin. NQR studies of the normal state show that Rh doping boosts antiferromagnetic spin fluctuations \cite{kawa06}. The antiferromagnetic structure of {\cerhin} changes from an incommensurate, $\vec{Q}=(1/2,1/2,0.297)$ modulation to a modulated state with two wave vectors $\vec{Q}=(1/2,1/2,1/2)$ and  $\vec{Q}=(1/2,1/2,0.42)$ at intermediate concentrations \cite{yoko08}. Fluctuations with respect to these wave vectors may be relevant for the superconductivity at intermediate concentrations.

Rh doping of {\ceirin} initially suppresses the filamentary transition at $T_{s1}$ so that only the superconducting transition at $T_{s2}$ remains. However, for higher Rh concentrations the bulk $T_s$ increases, until antiferromagnetism emerges for $x>0.3$ \cite{bian01,kawa06,pagl01}. Within the antiferromagnetic state superconductivity in coexistence with antiferromagnetism survives \cite{zhen04,chri05}.  Perhaps most importantly, the superconductivity is insensitive to the disorder associated with the doping. This suggests, that the transition metal element affects  only indirectly those parts of the Fermi surface on which superconductivity is stabilized. 

Finally, substitutional doping on the In site has provided some remarkable hints concerning the nature of the superconductivity. Extensive studies have been carried out in {\cecoinsn}, where the superconductivity vanishes rapidly for $x\geq0.15$  \cite{baue05c}. Note that this represents a much smaller concentration than $x=0.15$ in La-doping. EXAFS studies have thereby established that the Sn atoms preferentially occupy the In(1) site in the {\cein} planes \cite{dani05}, highlighting that the superconductivity is particularly sensitive to disorder in the {\cein} planes. Interestingly the critical Sn concentration, when referred to the {\cein} planes, yields an average distance of the Sn atoms of the size of the superconducting coherence length. The suppression of superconductivity in {\cecoinsn} may be compared with the suppression of antiferromagnetic order in {\cerhinsn} at $x_c\approx0.35$, where a quantum critical point is generated \cite{baue06b}. Assuming that the Sn atoms as for {\cecoinsn} occupy the In(1) site, this reveals that details of the electronic structure within the {\cein} planes control the stability of both antiferromagnetic order and superconductivity.

Rather surprising, substitutional Cd doping of the In site induces a change from superconductivity to long range antiferromagnetic order, where the phase diagram scales with the pressure dependent phase diagram of {\cerhin} \cite{pham06}. Electronically Cd doping, in leading order, acts as the removal of electrons, which in turn compares with the effect of pressure on {\cerhin}. However, NMR studies of the series of Cd doped {\cecoin} establishes a microscopic coexistence of the two forms of order, where the ordered moment of $0.7\,{\rm \mu_B}$ is essentially unchanged and the magnetic order may be attributed to the local environment of the Cd dopant \cite{urba07}. Thus the magnetic order is not the result of a gradual modification of the Fermi surface, but emerges in terms of droplets that coalesce at the onset of long-range antiferromagnetism.

Both the superconductivity and the antiferromagnetism respond sensitively to nonmagnetic disorder within the {\cein} building blocks, while they are relatively insensitive to out-of-plane disorder. As a possible explanation this behavior may be related to the warping of the Fermi surface, which is affected by local distortions created by the replacement of transition metal elements. However, the detailed mechanisms that control the behavior in doping studies have not yet been identified. 

\paragraph{Common features and analogies}
We now discuss the more general features of the entire series of {\cemin} compounds. We begin with material specific aspects and conclude this section with a discussion of possible analogies with other layered superconductors, notably the cuprates.

\begin{figure}
%\sidecaption
\includegraphics[width=.35\textwidth,clip=]{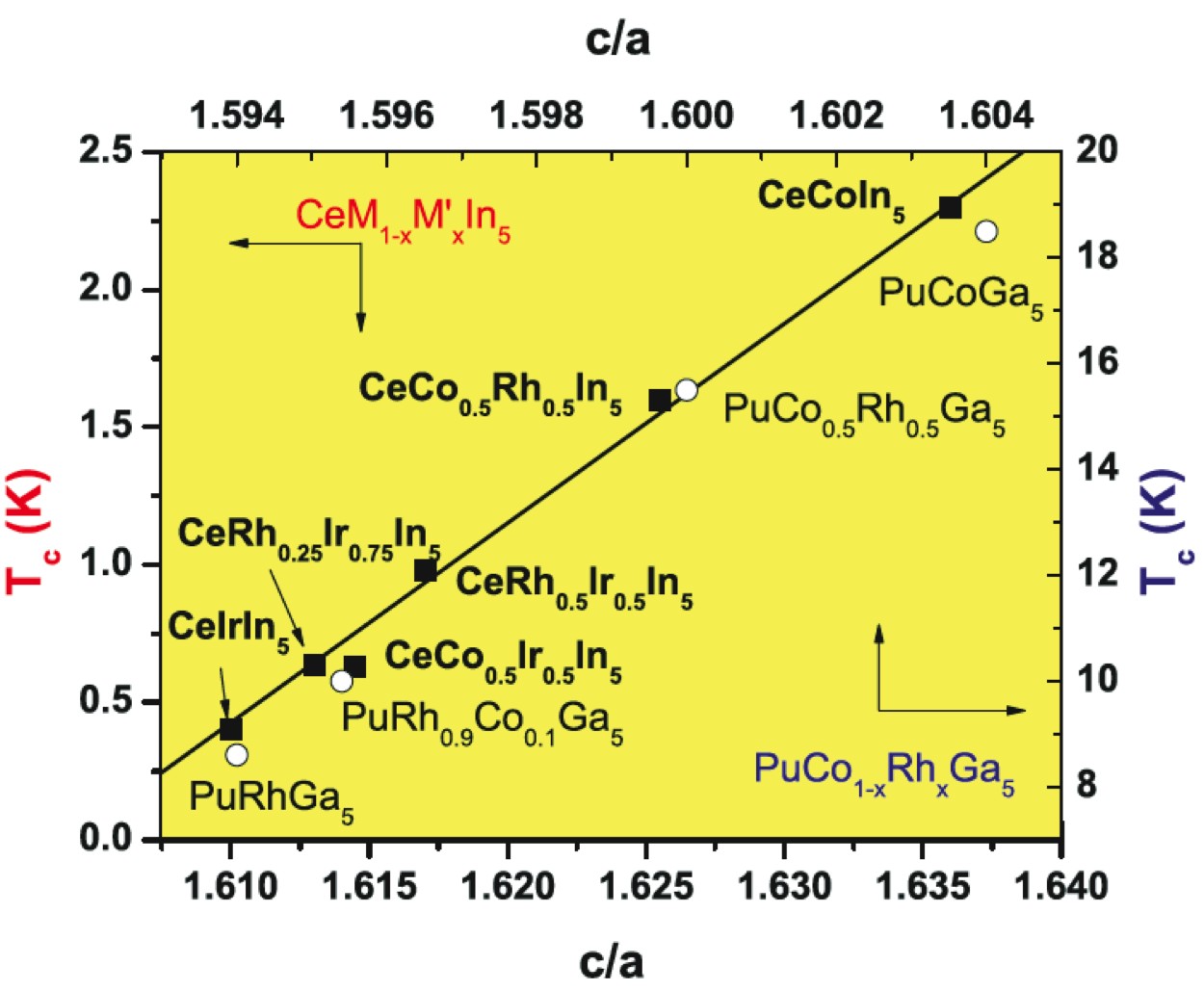}
\caption{Evolution of the superconducting transition temperature as a function of c/a ratio of the lattice parameter in the series CeMIn$_5$ (left-hand and bottom axis) and PuMGa$_5$ (top and right-hand axis). Plot as shown in \cite{sarr07}}
\label{115-covera} 
\end{figure}

A major theme across the literature on the {\cemin} compounds is the tentative role of a quantum critical point in driving superconductivity. This is embodied and was first pointed out with respect to {\cein} \cite{walk97,math98}. A natural question concerns, which mechanisms control the $T_s$. For spin fluctuation mediated pairing it has been pointed out that a reduction of the effective dimension, notably magnetic and/or electronic anisotropy, favour superconductivity \cite{mont01,mont02}. This is consistent with an empirical observation of $T_s$ as a function of the c/a ratio as reported by \cite{baue04b} and shown in Fig.\,\ref{115-covera} .

Similarities of the series {\cemin} with the cuprate superconductors have been taken as evidence that spin fluctuations are responsible for the pairing mechanism in the cuprates (see, e.g., \cite{math98}). It is instructive to summarize these similarities in further detail. The consideration begins with the temperature versus pressure phase diagram, which shows a superconducting dome in the vicinity of antiferromagnetic order. At least in {\cein} a major qualitative difference is, that a proper antiferromagnetic transition vanishes at the putative QCP, while the equivalent feature in the cuprates is a pseudogap of ill-defined nature. Here the phase diagram of {\cecoin} is in better analogy with the cuprates where, however, the temperature ranges of anomalous behavior in {\cecoin} are rather small. 

Likewise, the analogy with the cuprates may also be seen in the sibling pair {\cecoin} and {\cerhin}, where pressure induces superconductivity in {\cerhin} as well as in {\cerhinte}. The similarity of the phase diagrams is also loosely reflected in the doping studies, when keeping in mind that the underlying microscopic processes, notably s-f hybridization in f-electron systems versus pure charge transfer in the cuprates are radically different. Doping with Rh, driving {\cecoin} antiferromagnetic, is akin to hole doping in the cuprates. Likewise Cd doping may be understood as electron doping, stabilizing antiferromagnetic order. Moreover, the complex pressure, magnetic field and temperature phase diagram in {\cerhin} yields another analogy in that magnetic field stabilizes a coexistence of superconductivity and antiferromagnetism \cite{park06}. A related effect of magnetic field has also been found in certain cuprates \cite{lake02}.

The analogy with the cuprates is not just based on qualitative features of the phase diagram, but also on the bulk properties. As for the cuprates the normal metallic state exhibits non Fermi liquid behavior. While the resistivity, susceptibility and specific heat are not in great agreement, there is a remarkable similarity of the Hall effect and regarding the breakdown of Kohlers rule in the magnetoresistance in {\cecoin}. In fact, a quadratic temperature dependence of the Hall angle and a breakdown of Kohler's rule have also been observed in {\ceirin}, where they were interpreted as a precursor phase in the normal metallic state that share similarities with the pseudo-gap in the cuprates \cite{nair08}. Moreover, just as for the cuprates a spin resonance at a frequency $\omega_0$ has now been observed in the spectrum of slow antiferromagnetic fluctuations in {\cecoin}, where the ratio of resonance frequency to gap, $\hbar\omega_0 /2\Delta_0\approx0.74$, is remarkably similar for {\cecoin}, {\updal} and Bi$_2$Sr$_2$CaCu$_2$O$_{8+\delta}$ \cite{stoc08}.

It is at the same time also important to emphasize the differences between {\cemin} and the cuprates. First, {\cerhin} appears to exhibit two superconducting domes. This suggests that at least two pairing mechanisms may exist, where possible candidates are a SDW and a valence instability as in {\cecusi}. The second superconducting dome makes the analogy with {\cecoin} less obvious and it is interesting to ask if there are further superconducting domes in {\cecoin} under pressure. Second, quantum oscillatory studies of the electronic structure show a change of Fermi surface topology through the quantum critical point in {\cein} and {\cerhin} that appears to be related to a delocalization of the f-electron. A stimulating question concerns, whether, quite generally, an instability of the Fermi surface topology drives the superconductivity in the cuprates. 

Finally, {\ceirin} has an even larger unit cell volume than {\cerhin}, but there is no antiferromagnetic order nearby. The phase diagram shown in Fig.\,\ref{pd-115s} may consequently be interpreted differently. Perhaps the entire series of CeMIn$_5$ is superconducting where $T_s$ increases in a linear fashion when going from Co to Rh to Ir. However, a slight change of electronic structure of the Rh system changes the balance from a superconducting to an antiferromagnetic ground state. This does not rule out quantum critical fluctuations as a key ingredient of the superconductivity, though the origin of these fluctuations may differ from the conventional scenario of a simple quantum critical point.

\subsubsection{Miscellaneous Ce-systems}
\label{misc-ce-sys}

We next present several examples of compounds, where evidence for superconductivity has been observed at the border of antiferromagnetic order. These compounds are the first of a given crystal structure, being possibly the first member of a new class of f-electron superconductors. The properties of these 'miscellaneous Ce-systems are summarized in table \ref{table-cmt}. The first two examples, {\cenigeot} and {\cenigettf}, are members of the ternary Ce-Ni-Ge series. The third system, {\cepdal}, is isostructural to {\nppdal} (cf. section \ref{acti-loca}).

For completeness it is worthwhile to also mention briefly {\ccau}, where antiferromagnetic order is suppressed around 40\,kbar and a tiny drop of the resistivity is observed that may be related to superconductivity \cite{wilh01} .

\paragraph{{\cenigeot}}
We begin with the discovery of superconductivity in {\cenigeot}, which crystallizes in the orthorhombic SmNiGe$_3$-type structure (space group Cmmm, see also table \ref{table-cemt}) \cite{naka04,kote06}. At ambient pressure  {\cenigeot} orders antiferromagnetically below a N{\'e}el temperature $T_N=5.5\,{\rm K}$ as determined from the resistivity and susceptibility. The paramagnetic susceptibility in polycrystals at high temperatures follows a Curie-Weiss dependence with $\mu_{eff}=2.54\,{\rm \mu_B}$ as expected of Ce$^{3+}$. The antiferromagnetic transition is accompanied by a distinct anomaly, where the entropy released, $\Delta S=0.65\,{\rm R}\ln 2$, is characteristic of localized moments in a 4f crystal field doublet ground state. The magnetic structure has been explored by powder neutron diffraction, which revealed two transitions at $T_{N1}$ and $T_{N2}$ with ordering wave vectors $\vec{Q}_1=(1,0,0)$ and $\vec{Q}_2=(0,0.409, 1/2)$, respectively, with an ordered moment of $\mu_{ord}=0.8\,{\rm \mu_B}$.

The electrical resistivity of polycrystalline {\cenigeot} is dominated by a maximum around 100\,K and a sharp drop at $T_N$, where no details have been seen of a second transition. The pressure dependence of the polycrystalline samples was investigated with two different pressure techniques: a diamond anvil cell with NaCl \cite{naka04} and Daphne oil  \cite{kote06} as pressure transmitter, respectively. In the following we only address the results obtained with the latter set-up which produces better homogeneity. As a function of pressure $T_N$ initially rises to nearly 8\,K at 40\,kbar followed by a fairly rapid decrease. $T_N$ vanishes at $p_c\approx70\,{\rm kbar}$, where the $T^2$ resistivity crosses over to a temperature dependence $\sim T^{1.5}$ in the range 60 to 70\,kbar. The residual resistivity $\rho_0$ increases and reaches a plateau above $p_c$. 

For pressures in the range 20 to 100\,kbar hints for superconductivity are observed in terms of a zero resistance transition below 40\,kbar and an incomplete resistive transition above 40\,kbar, where $T_s$ is as high 0.45\,K. The transition temperature exhibits two broad maxima separated by a shallow minimum near $p_c$. The ac susceptibility shows diamagnetic screening. $H_{c2}$ increases under pressure from 0.015\,T to 1.55\,T. Correspondingly the coherence length decreases under pressure from 2000\,{\AA} to $\sim$100{\AA}. The initial slope of $H_{c2}$ near $T_s$ increases and reaches $dH_{c2}/dT=-10.8\,{\rm T/K}$ for the maximum $T_s\approx0.45\,{\rm T}$. For low pressures orbital limiting is observed, while there is Pauli limiting for the highest values of $T_s$ around 70\,kbar.

\paragraph{{\cenigettf}}
Another system in the Ce-Ni-Ge series that attracts increasing interest is {\cenigettf} \cite{chev99}. This compound crystallizes in the U$_2$Co$_3$Si$_5$-type structure (space group Ibam; see also table \ref{table-cmt}). A discussion of structural similarities with {\cenige} may be found in \cite{naka05}. The metallic state of  {\cenigettf} is characteristic of a Kondo lattice system with $T_K\approx5\,{\rm K}$,  where a Curie-Weiss susceptibility of free Ce$^{3+}$ moments is observed at high temperatures and antiferromagnetic order at low temperatures \cite{hoss00}. The magnetization shows two transitions at $T_{N1}=5.1\,{\rm K}$ and at $T_{N2}=4.5\,{\rm K}$. The linear term of the specific heat is enhanced and the entropy released $\Delta S = 0.67{\rm R}\ln 2$ at $T_{N1}$ is characteristic of reduced moments. Powder neutron diffraction shows collinear antiferromagnetic order below $T_{N1}$ with the magnetic moments aligned along the a-axis and a small ordered moment of $\mu_{ord}=0.4\,{\rm \mu_B}$ at 1.4\,K \cite{duri02}.

In comparison with {\cenigettf} the unit cell volume in the sibling compound {\cenisittf} is 9.6\% smaller. {\cenisittf} exhibits a nonmagnetic valence fluctuating system \cite{mazu92}. This suggests that hydrostatic pressure suppresses the antiferromagnetic order. Indeed $T_{N1}$ in polycrystalline samples decreases under pressure and vanishes at $p_c=36\,{\rm kbar}$, where a zero resistance transition is observed at $T_s=0.26\,{\rm K}$ \cite{naka05} with $H_{c2}=0.7\,{\rm T}$, corresponding to a coherence length $\xi=210\,{\rm \AA}$.

\paragraph{{\cepdal}}
Another miscellaneous C-based superconductor is {\cepdal} \cite{hond08}, which is isostructural to {\nppdal} reviewed in section \ref{acti-loca}. At ambient pressure {\cepdal} displays two antiferromagnetic transitions at $T_{N1}=3.9\,{\rm K}$ and $T_{N2}=2.9\,{\rm K}$. The metallic state is moderately enhanced with $\gamma=0.056\,{\rm J/mol\,K^2}$. The resistivity and susceptibility as well as the magnetization suggest crystal field levels at $\Delta_1=197\,{\rm K}$ and $\Delta_2=224\,{\rm K}$. Under pressure $T_{N1}$ and $T_{N2}$ at first increase, where $T_{N2}$ can only be tracked as high as $\sim30\,{\rm kbar}$. $T_{N1}$ displays a maximum around 50\,kbar and appears to vanish around 90\,kbar. The resistivity displays a superconducting transition in the pressure range 80 to 120\,kbar, with a maximum $T_s=0.57\,{\rm K}$ at 108\,kbar.

\subsection{Coexistence with antiferromagnetism}
\label{coex-anti}

In a number of f-electron systems superconductivity emerges deep inside an antiferromagnetically ordered regime, i.e., $T_s\ll T_N$. The presentation of these systems may be grouped in two parts, large and small moment systems. We first discuss the large moment antiferromagnets {\updal}, {\unial} and {\ceptsi}. For these compounds the coexistence of antiferromagnetism and superconductivity appears to be homogenous. The second class are antiferromagnets with tiny ordered moments, notably {\upt} and {\urusi}. While the tiny moments in {\upt} appear to be homogenous, there is growing evidence for a small volume fraction of large ordered moments in {\urusi}. 

\begin{table*}
\centering
\caption{Key properties of uranium based heavy-fermion superconductors. 
Missing table entries may reflect more complex behavior discussed in the text. 
$H_{c2}$ represents the extrapolated value for zero temperature.
References are given in the text. 
(AF: antiferromagnet; FM,F: ferromagnetism; HO: hidden order; SC: superconductor)}
\label{table-ubased}
\begin{tabular}{llllllllll}
\hline\noalign{\smallskip}
U-based & {\ube} &  {\upt} & {\urusi} & {\updal}  & {\unial} & {\uge} & {\urhge} & UCoGe & UIr \\
\noalign{\smallskip}\hline\noalign{\smallskip}
structure & cubic & hexagonal  & tetragonal & hexagonal & hexagonal & orthorh. & orthorh. & orthorh. & monoclinic\\
type & NaZn$_{13}$ & -  & ThCr$_2$Si$_2$ & PrNi$_2$Al$_3$ & PrNi$_2$Al$_3$ & - & - & - & - \\
space group & $O_h^6$ Fm3c & P6$_3$/mmc & I4/mmm & P6/mmm & P6/mmm & Cmmm & Pnma & Pnma & P2$_1$ \\
$a$({\AA}) & 10.248 & 5.764 & 4.128 & 4.189 & 5.207 & 3.997 & 6.875 & 6.845 & 5.62 \\
$b$({\AA}) & - & - & - & - & - & 15.039 & 4.331 & 4.206 & 10.59 \\
$c$({\AA}) & - & 4.899 & 9.592 & 5.382 & 4.018 & 4.087 & 7.507 & 7.222 & 5.60 \\
%$c/a$ & - & - & - & - & -  \\
%$V_0 ({\rm \AA^3})$ & - & - & - & - & - \\
%\noalign{\smallskip}\hline
%CEF scheme & - & ($\Gamma^1_7$, $\Gamma^2_7$, $\Gamma_6$) & ($\Gamma^1_7$, $\Gamma^2_7$, $\Gamma_6$) & ($\Gamma^1_7$, $\Gamma^2_7$, $\Gamma_6$) & - \\
%$\Delta_1$, $\Delta_2$ (meV) & - & 8.6, 25 & 6.7, 29 & 6.9, 24 & - \\
\noalign{\smallskip}\hline
state & SC & AF, SC & HO, SC & AF, SC & AF, SC & FM, SC & FM, SC & FM, SC & F1, F2, F3, SC\\
$\gamma\,{\rm (J/mol\,K^2)}$ & - & 0.44 & 0.07 & 0.2 & 0.12 & 0.032 & 0.164 & 0.057 & 0.049 \\
$T_{N}$,$T_C$(K) & - & 5 & 17.5 & 14.2 & 4.6 & 52 & 9.5  & 3 & 46 \\
easy axis & - & - & - & - & - & a & b, c & c & $[10\bar{1}]$\\
hard axis & - & - & - & - & - & b, c & a & a,b & $[010]$ \\
$\vec{Q}$ & - & $(\pm1/2,0,1)$ & (0,0,1) & ($0,0,\frac{1}{2}$) & $(1/2\pm\delta,0,1/2)$ & - & - & - & -\\
 &  &  &  &  & $\delta=0.110\pm0.003$\\
$\mu_{ord}¥$($\mu_{B}$) & - & 0.01 & 0.03 & $0.85\pm0.03$ & $0.24\pm0.10$  & 1.48 & 0.42 & 0.07 & 0.5, 0.05, 0.1 \\
$\mu_{eff}¥$($\mu_{B}$) & - & - & - & - & - & 2.7 & 1.8 & 1.7 & - \\
\noalign{\smallskip}\hline
$T_{s}$(K) & 0.95 & 0.530, 0.480 & 1.53 & 2.0 & 1.1 & 0.8 & 0.25 (S1) & 0.8 & 0.15 (in F3)\\
&  &  &  &  &  &  & 0.4 (S2) &  &  \\
$\Delta÷C/\gamma_{n}T_{s}$ & 2.5 & 0.545, 0.272  & 0.93 & 1.48 & 0.4 & 0.2-0.3 & 0.45 & 1 & - \\
$H^{\parallel}_{c2}$(T) & 14 & 2.1 & 3 & 3.9 & 0.9 & - & - & - & - \\
$\frac{d}{dT}H^{\parallel}_{c2}$(T/K) & -45 & $-7.2\pm0.6$ & -5.3 & -5.45 & -1.14  & - & - & - & - \\
$H^{\perp}_{c2}$(T) & - & 2.8 & 14 & 3.3 & 0.35  & - & - & - & -  \\
$\frac{d}{dT}H^{\perp}_{c2}$(T/K) & - & $-4.4\pm0.3$ & -14.5 & -4.6 & -0.42  & - & - & - & - \\
$H_{c2}$(T) & - & - & - & - & - & - & - & - & 0.0265  \\
$\frac{d}{dT}H_{c2}$(T/K) & - & - & - & - & - & - & - & -10.8 & - \\
\noalign{\smallskip}\hline
$\xi_{\perp}$, $\xi_{\parallel}\,{\rm (\AA)}$ & 50 & $\sim$120 & 100, 25 & 85 & -  & - & - & 150 & 1100 \\
%$\xi^{ab}_{0}$({\AA}) & - & - & - & - & - & - & - & - & 1100 \\
%$\xi_{GL}$({\AA}) &  - & - & - & - & - & - & - &150 & - \\
$\lambda_{\parallel}$,$\lambda_{\perp}\,{\rm (\AA)}$ & 4000 & 4500, 7400 & $\sim$7000 & 4500, 4800 & -  & - & - & - & - \\
$\lambda_{GL}$({\AA}) & - & - & - & - & - & - & 9100 (S1) & - & - \\
%$\lambda_{\perp}\,{\rm (\AA)}$ & - & & - & - & - \\
$\kappa$ & 80 & 44 & 70 & 52 & 11 & - & - & - & - \\
%\noalign{\smallskip}\hline
%$C/T$ & - & $\propto T$ & $\propto T$ & - & - \\
%$\kappa$ & - & - & - & - & - \\
%$1/T_{1}$ & - & $\propto T^3$ & $\propto T^3$ & - & - \\
%$\lambda$ & - & - & - & - & - \\
\noalign{\smallskip}\hline
year of disc. & 1984 & 1984 & 1986 & 1991 & 1991 & 2000 & 2001 & 2007 & 2004 \\
\noalign{\smallskip}\hline
\end{tabular}
\end{table*}
 
\subsubsection{Large moment antiferromagnets}

Superconductivity in the sibling pair of low temperature antiferromagnets {\updal} and {\unial} was discovered in 1991 \cite{geib91a,geib91b}. Both compounds crystallize in the hexagonal PrNi$_2$Al$_3$ structure (space group P6/mmm) as summarized in table\,\ref{table-ubased}. Large single crystals may be grown of {\updal}, while the metallurgy of {\unial} is more complex, i.e., there are fewer single-crystal studies for {\unial}. In turn the body of work on {\updal} is much more complete. In the following we first review the present understanding of {\updal} before turning to the properties of {\unial} at the end of the section. We address only briefly the coexistence of superconductivity and antiferromagnetism in {\ceptsi}, which is reviewed extensively in section \ref{ceptsi3}.

\paragraph{{\updal}
\label{updal}}

The electrical resistivity of {\updal} decreases monotonically as a function of temperature below a broad maximum around $85\,{\rm K}$ \cite{sato92}. In single crystals the resistivity is weakly anisotropic by a factor of two with $\rho_c>\rho_{ab}$. The susceptibility exhibits a broad maximum around 35\,K in the basal plane and an anisotropy of $\sim3.5$ ($\chi_c<\chi_{ab}$) \cite{geib91a}. Above $\sim100\,{\rm K}$ a Curie-Weiss dependence is observed with a fluctuating moment $\mu_{eff}$ that changes from 3.2 to $3.4\,{\rm \mu_B/U}$ around 300\,K \cite{grau92}. To account for the temperature dependence of the susceptibility the following crystal electric field scheme of a tetravalent uranium configuration has been proposed: $\Gamma_4$ singlet ground state, $\Gamma_1$ singlet first excited state at 33\,K, two $\Gamma_6$ doublets at 102\,K, two $\Gamma_5$ doublets at 152\,K; $\Gamma_3$ singlet at 562\,K; $\Gamma_5$ at 1006\,K \cite{grau92}. Crystal field excitations at a temperature around 30\,K have also been inferred from a dip in the elastic constants \cite{modl93}. 

A key characteristic of crystal electric fields in uranium compounds is, that they hybridize very strongly with the conduction electrons. This is also the case in {\updal}, where time-of-flight inelastic neutron scattering fails to detect well defined crystal field excitations. Instead very broad spectra consisting of quasielastic Lorentzians plus additional inelastic scattering are observed \cite{krim96}. The quasi-elastic scattering thereby limits to an intrinsic width of 5\,meV consistent with the 50\,K energy scale seen in the susceptibility and resistivity. When subtracting lattice contributions by means of reference measurements in ThPd$_2$Al$_3$, the remaining inelastic scattering is consistent with the crystal field scheme given above. 

{\updal} develops strong electronic correlations at low temperatures with an enhanced linear temperature dependence of the specific heat in the paramagnetic state $\gamma\sim0.21\,{\rm J/mol\,K^2}$ \cite{geib91b}. Antiferromagnetic order is observed below $T_N=14\,{\rm K}$ \cite{geib91a}. The magnetic entropy released at $T_N$ is a substantial fraction of the local Zeeman entropy, $S_m=0.65\,{\rm R}\ln 2$. The resistivity displays a change of slope at $T_N$ \cite{casp93}. There is no evidence suggesting the formation of a density-wave, e.g., like the small maximum in the resistivity near $T_0$ in {\urusi}. In the ordered state the linear temperature dependence of the specific heat is also enhanced, $\gamma\sim0.15\,{\rm J/mol\,K^2}$ \cite{geib91b}. 

Early single-crystal studies suggested that the magnetic moments in {\updal} are oriented in the basal-plane of the hexagonal crystal structure, i.e., {\updal} has an easy magnetic plane \cite{sato92}. In zero magnetic field neutron scattering shows commensurate antiferromagnetic order with a wave vector $\vec{Q}=(0,0,1/2)$ and an ordered moment $\mu_{ord}=0.85\,{\rm \mu_B/U}$ \cite{krim93}. This corresponds to ferromagnetic planes stacked antiferromagnetically along the c-axis. The ordered moment in {\updal} displays a mean-field temperature dependence that corresponds essentially to the form of a $S=1/2$ Brillouin function. However, the moment is systematically larger than $S=1/2$ consistent with a tetravalent uranium state. 

The onset of antiferromagnetic order in {\updal} may be seen in a large number of properties. For instance, (i) the thermal expansion, which shows a large sensitivity to uniaxial stress \cite{link95}, (ii) the longitudinal and transverse elastic constants \cite{modl93,luet93}, (iii) a kink in the $^{27}$Al spin-lattice relaxation rate and a gradual increase of the $^{27}$Al NMR line width \cite{kyog93}, (iv) the emergence of a gap in tunneling spectroscopy \cite{aart94}, and (v) an increase of the thermal conductivity \cite{hiro97}.

The magnetic phase diagram of {\updal}, which yields key information of the nature of the magnetic and superconducting order, has been studied in considerable detail. For magnetic fields applied in the basal-plane three transitions at $H_1=0.6\,{\rm T}$, $H_2=4.2\,{\rm T}$ and $H_m=18\,{\rm T}$ may be distinguished \cite{devi92b,sugi93,oda94,sugi94}. In contrast, for the c-axis no field induced transition may be observed up to 50\,T, the highest field studied.  At $H_1$ the ordered state changes from commensurate antiferromagnetism to a canted state \cite{grau92,kita94}.  The metamagnetic transition at $H_m$ has attracted considerable interest. At $H_m$ the magnetization increases from $\sim0.5\,{\rm \mu_B/U}$ to $\sim1.5\,{\rm \mu_B/U}$ \cite{devi92b}. Below 4.2\,K the transition becomes hysteretic \cite{sako01b}. The magnetoresistance displays a peak at $H_m$ for $H$ and $i\parallel\langle100\rangle$, while there is a discontinuous step in the magnetoresistance for $H\parallel\langle010\rangle$ and  $i\parallel\langle001\rangle$ \cite{devi93}. 

The critical field $H_m$ increases when tilting the field direction towards the c-axis. It exceeds 50\,T, the highest field measured, for an angle larger than 60$^{\circ}$ \cite{oda94,sugi94}. The angular dependence is consistent with XY-type of order. Torque magnetization measurements show that the basal plane anisotropy persists up to 60\,K \cite{suel96}. The metamagnetic transition field changes only weakly as a function of temperature, terminating in a tricritical point around 12\,K \cite{kim01b}. For temperatures well above the tricritical point a cross-over survives at $H_m$, reminiscent of the metamagnetic transition \cite{oda99}.  When taken together, the latter properties suggest that crystal electric fields and the electronic structure at the Fermi level play an important role in controlling the metamagnetic transition, possibly related to a change of 5f localization.

Superconductivity in {\updal} is observed below $T_s=2\,{\rm K}$. Even though $T_s$ is amongst the highest of all heavy fermion systems, it is nearly an order of magnitude smaller than $T_N$.  This distinguishes {\updal} and {\unial} from the systems reviewed above. The superconducting transition is accompanied by a distinct anomaly in the specific heat, with $\Delta C/\gamma T_s\approx 1.48$. Below $T_s$ the specific heat varies as $C(T)=\gamma T+AT^3$, suggesting the presence of line nodes \cite{casp93}. Also consistent with lines nodes is the cubic temperature dependence of the thermal expansion, $\alpha\propto T^3$ \cite{modl93}. The ratio of the thermal conductivity divided by the temperature, $\kappa/T$, shows a finite contribution for $T\to0$ of the order of 10\% of the normal state value \cite{chia97}. Near $T_s$ a cross-over is observed rather than a sharp kink, followed by a dependence $\kappa/T\propto T$ providing further evidence for line nodes \cite{hiro97}. In magnetic field $\kappa/T$ increases, a kink appears at $T_s(H)$ and the temperature dependence changes slightly. Recently angle-resolved magneto-thermal transport measurements showed the absence of an orientation dependence in the basal plane, while a two-fold symmetry exists in the plane perpendicular to the basal plane \cite{wata04b}. From this it was concluded that the gap has a single line node orthogonal to the c-axis, while the gap is isotropic in the basal plane and may be given as $\vec{\Delta}(\vec{k})=\Delta_0\cos(k_z c)$.

The upper critical fields  $H^a_{c2}=3.3\,{\rm T}$ and $H^c_{c2}=3.9\,{\rm T}$ and the initial slopes $\partial H^a_{c2}/\partial T\vert_{T_s}\approx-4.6\,{\rm T/K}$ and $\partial H^c_{c2}/\partial T\vert_{T_s}\approx-5.45\,{\rm T/K}$ are remarkably isotropic \cite{ishi95,sato96}. They corresponds to a coherence length $\xi_{\rm GL}\approx85\,{\rm \AA}$, where the penetration depths $\lambda_{\perp}(0)=4800\pm500\,{\rm \AA}$ and $\lambda_{\parallel}(0)=4500\pm500\,{\rm \AA}$ (with respect to the c-axis) inferred from magnetization and $\mu$-SR measurements \cite{geib91b,feye94}. It establishes {\updal} as strong type 2 superconductor with $\kappa_{\rm GL}\approx52$. The anisotropy of $H_{c2}$ for $T\to0$ may be accounted for by an anisotropic mass model \cite{sato97c}. It is instructive to compare the observed field values with the conventional weak-coupling orbital and paramagnetic limiting fields, where $H_{p0}=3.7\,{\rm T}$, $H^*_{a}=6.4\,{\rm T}$ and $H^*_{c}=7.6\,{\rm T}$. Thus the upper critical fields are smaller than orbital limiting and close to the paramagnetic limit. 

The charge carrier mean free path inferred from the residual resistivity or the Dingle temperature in quantum oscillatory studies clearly shows large values of the order $10^3\,{\rm \AA}$. {\updal} exhibits hence type 2 superconductivity in the clean limit, that is dominated by paramagnetic limiting. This motivated an interpretation of an anomalous dip in the AC susceptibility and magnetization near $H_{c2}$ in terms of a FFLO state \cite{gloo93,norm93}. However, further studies suggest that the anomalous dip exists at all temperatures below $T_s$ in contrast to the finite temperature range predicted theoretically. Taken together the anomalous behavior near $H_{c2}$ is more characteristic of the peak effect \cite{haga96}. For a further discussion of FFLO states we refer to section \ref{FFLO-phas}.

For the further discussion of the interplay of superconductivity and antiferromagnetism it is instructive to consider the nature of the 5f electrons. The U-U spacing in {\updal} and {\unial}, given by $d_{\rm U-U}=4.186\,{\rm \AA}$ and $d_{\rm U-U}=4.018\,{\rm \AA}$, respectively, are well above the Hill limit of 3.4\,{\AA} \cite{hill70}. This implies that any itineracy of the f-electrons must be related to a hybridization with other electrons. The larger spacing in {\updal} is thereby consistent with the evidence of stronger localization of the 5f electrons. Several properties of {\updal} are characteristic of local uranium moments. For instance, the susceptibility at high temperatures shows a Curie-Weiss dependence with a fluctuating moment $\mu_{eff}=3.2\,{\rm \mu_B/U}$. Polarized neutron scattering of the magnetic form factors in an applied field of 4.6\,T shows the lack of magnetic polarization at the Pd site, i.e., the magnetic polarization is well localized at the uranium site \cite{paol93}. It was however not possible to infer unambiguously from the magnetic form factor, whether uranium is tetravalent. The observed ratio of the orbital to the spin moment $R=\mu_L/\mu_S\approx -2.01$ is closer to the value of U$^{3+}$ ($R=-2.56$) than for  U$^{4+}$ ($R=-3.29$). Uranium 5f x-ray circular dichroism is characteristic of strong interactions between the 5f states and their environment \cite{yaou98}. 

The evidence for local moment magnetism is contrasted by optical conductivity and quantum oscillatory studies of the Fermi surface, which clearly show strongly renormalized quasiparticle conduction bands  \cite{tera97,inad99}. In the optical conductivity Drude behavior is observed with ultra-slow relaxation rates \cite{sche05b}. At the metamagnetic transition at $H_m$ a reconstruction of the Fermi surface topology is observed without substantial variation of the renormalization. This may be related to a magnetic field induced transition from an antiferromagnetic to a ferromagnetic exchange splitting, but does not appear to be driven by a localization of the f-electrons. 

Experimentally several properties of {\updal} suggest a dual state, where part of the 5f electrons are localized and the other part are itinerant, i.e., a combination of both characteristics may be seen in the same physical quantity. This was first noticed in measurements of the specific heat under pressure up to 10.8\,kbar, where amongst other things the size of the anomaly at the antiferromagnetic transition is strongly suppressed, while the superconducting transition is not \cite{casp93}. Also, the magnetic properties are anisotropic as opposed to the superconducting properties which are isotropic \cite{ishi95,sato96,feye94}.  Neutron scattering \cite{krim93} and NMR/NQR studies \cite{koho94} further show that the antiferromagnetic order survives essentially unchanged in the superconducting state. This suggests that both forms of order may be carried by different subsystems. Finally, the spectrum of excitations exhibits different contributions. Resonant 5d - 5f photoemission shows a sharp peak near $E_F$ and a broad hump at a binding energy $\sim 1\,{\rm eV}$ characteristic of the features expected of itinerant and localized 5f electrons, respectively \cite{ejim94,taka95}. As a function of temperature photoemission establishes that the electronic properties change from itinerant to localized \cite{sato99a,fuji07}.  Inelastic neutron scattering shows a weakly dispersive mode at an energy of $\sim$1.5\,meV that softens at $T_N$, consistent with early studies \cite{pete94}, and a quasi-elastic signal at the antiferromagnetic ordering wave vector \cite{sato97b}. {\updal}-Pb tunnel junctions show a superconducting gap around 0.235\,meV and antiferromagnetic spin wave mode around 1.5\,meV, consistent with the neutron scattering studies \cite{jour99}.

At first sight the dispersive and the quasi-elastic excitations in {\updal} seen in neutron scattering may appear to be disconnected. However, polarized neutron scattering shows that the dispersive mode and the quasi-elastic signal are both transversely polarized. This suggests a common origin \cite{bern98}. As part of this study it was further shown that the spectrum of antiferromagnetic spin fluctuations in the framework of conventional paramagnon theory \cite{mori85,lonz85} is quantitatively consistent with $T_N$ and the enhancement of the normal state specific heat. As the temperature decreases below $T_s$ the quasi-elastic spectrum changes and a steep maximum emerges at very small $\omega$. The maximum is also referred to as resonance mode. When plotting the maximum as a function of temperature, a remarkable agreement with the temperature dependence of a BCS gap is found, where $2\Delta=3.86k_{\rm B}T_s$ \cite{meto98,bern99}. Under magnetic field the resonance vanishes at $H_{c2}$ \cite{blac06a}. In a spin-echo neutron scattering study the vanishing of spectral weight in the superconducting state was investigated at ${\rm \mu eV}$ resolution \cite{blac06b}. The experiments establish that the intensity vanishes completely, placing a strong constraints on the pairing symmetries.

Self-consistent LDA band structure calculations treating the 5f states in {\updal} as being itinerant reproduce the ordered magnetic moment, magneto-crystalline anisotropy and de Haas-van Alphen spectra \cite{sand94}. These studies also showed that an antiferromagnetic and ferromagnetic ground state are nearly degenerate, consistent with the metamagnetic transition at $H_m=18\,{\rm T}$. In these calculations the two largest Fermi surface sheets have markedly different 5f contributions, where one is almost purely 5f and the other yields 30\,\% 5f character, respectively \cite{knoe96}.  These differences may provide a tentative explanation for the dual behavior. 

In recent years a controversy has developed concerning the interplay of antiferromagnetic order and superconductivity in {\updal}. In the traditional view of heavy-fermion systems the f-electron orbitals are screened by a singlet coupling with the conduction electrons and then condense into a heavy Fermi liquid at low temperatures. In this scenario the f-electrons are itinerant and the superconductivity is due to an abundance of soft magnetic fluctuations. The effects of spin-orbit coupling may then be treated by a two-component susceptibility \cite{bern95,bern98}. Here the observation that the correlation length associated with the resonance peak matches the superconducting coherence length inspired an interpretation of the resonance peak as a key feature of the Copper pairs themselves. The main objection against the traditional scenario is its lack of material specific aspects.

In an alternative scenario it has been proposed that only one of the three 5f uranium electrons is itinerant, whereas the other two are localized \cite{sato01,zwic02b}. The microscopic underpinning of this so-called duality-model are strong intra-atomic correlations that are subject to Hund's rules and weak anisotropic hopping \cite{efre04}. In the duality-model the exchange interaction between the itinerant- and localized-electron subsystems drives the superconductivity in terms of a magnetic exciton. The main objection against the duality-model and a pairing mediated by crystal field excitations is, that the crystal field levels cannot be distinguished experimentally. The model nevertheless proves to be quite powerful.  In a first analysis an $A_{1g}$ order parameter symmetry was predicted \cite{miya01}. Further implications have been worked out in a strong coupling approach which were found to be compatible with experiment \cite{mcha04}. The theoretical analysis established that the emergence of unconventional superconductivity results in a resonance peak in the spectrum of magnetic excitations, consistent with neutron scattering \cite{chan07}.

We conclude this section with a brief review of the properties of {\updal} at high pressure. The electrical resistivity under pressure shows that $T_N$ decreases from 14\,K to about 8\,K at a pressure of 65\,kbar, while the normal state maximum in the resistivity increases \cite{link95}. At low pressures elastic neutron scattering shows an initial increase of $\mu_{ord}$, followed by a decrease above 5\,kbar with a rate $d\mu_{ord}/dp=-016\,{\rm \mu_B/kbar}$. This is tracked by $T_N$ which decreases at a rate $dT_N/dp\approx-0.05\,{\rm K/kbar}$ at high pressures \cite{honm99}. Up to 11\,kbar the lattice constants decrease at a rate $c_0^{-1}dc/dp=7.5\times10^{-4}\,{\rm kbar^{-1}}$ and $a_0^{-1}da/dp=4.728\times10^{-4}\,{\rm kbar^{-1}}$. 
 
High-pressure x-ray diffraction in {\updal} and {\unial} up to 400\,kbar shows that both compounds have essentially the same bulk modulus $B_0=159(6)$\,GPa \cite{krim00}. In {\updal} these studies further revealed a structural phase transition at $p_c=$250\,kbar from a high-symmetry hexagonal to low-symmetry orthorhombic state with space group Pmmm. Up to 230\,kbar the c/a ratio remains essentially constant.  In the high-pressure phase the compressibility is a factor of two larger. The structure above $p_c$ belongs to space group Pmmm, which is a subgroup of Cmmm, which in turn is a non-hexagonal non-isomorphic subgroup of P6/mmm. The shortest metal-metal spacing in {\updal} is the U-Pd distance, which reaches 1.51\,{\AA} at $p_c$. Interestingly this corresponds to the sum of ionic radii of U$^{4+}$ and Pd$^{4+}$, suggesting  a U$^{4+}$ valence fluctuating state below $p_c$ and U$^{4+}$  to U$^{5+}$ transition at $p_c$, where the ionic radius of U$^{5+}$ is reduced by 15\%. A combination of resonant inelastic x-ray scattering with first principles structure calculations is consistent with a delocalization from U$^{+4-\delta}$ to U$^{+4+\delta}$ \cite{ruef07}. Finally, the extrapolated pressure, where the superconductivity in {\updal} vanishes corresponds to the critical pressure of the structural transition \cite{link95}. While this may be completely fortuitous, it might alternatively identify the tetravalent U configuration as a precondition for superconductivity.

\paragraph{{\unial}}

In comparison to {\updal} the magnetism and superconductivity in {\unial} are much more typical of itinerant 5f electrons. The antiferromagnetic order is an incommensurate spin density wave, and the superconductivity is a candidate for spin-triplet pairing.  Further, at the antiferromagnetic transition at $T_N=4.6\,{\rm K}$ the anomalies in the physical properties, such as the specific heat, are fairly weak. The corresponding magnetic entropy released at $T_N$ is small, $S_m=0.12{\rm R}\ln 2$ \cite{tate98}. Likewise the resistivity only shows a faint feature at $T_N$ \cite{dali92}. As compared with {\updal} the smaller U-U distance, $d_{\rm U-U}=4.018\,{\rm \AA}$ in {\unial}, is also compatible with the more itinerant character of the 5f electrons. As mentioned above, because the U-U distance in both compounds is above the Hill limit ($3.4\,{\rm \AA}$) the itineracy must be due to hybridization with other electrons. 

The normal state susceptibility displays a broad maximum at $T^*\sim100\,{\rm K}$, characteristic of a dominant energy scale, but the coherence temperature may be as high as 300\,K \cite{sato96}. The normal state properties of {\unial} at low temperatures show the presence of strong electronic correlations. This is best seen in the specific heat, which shows an enhanced Sommerfeld coefficient $\gamma=0.12\,{\rm J/mol\,K^2}$ and an enhanced $T^2$ resistivity \cite{geib91b}. 

Selected microscopic probes nevertheless suggest a certain degree of 5f localization. Photoemission exhibits a combination of a sharp peak near $E_F$, a smaller feature around 0.6\,eV and broad hump at 2\,eV \cite{yang96}. The features near $E_F$ and at 0.6\,eV have been attributed to itinerant and localized 5f electrons, respectively, while the hump at 2\,eV is related to the Ni 3d states. The photoemission studies compare with polarized neutron scattering and circular dichroism measurements, which show a nearly spherical magnetization distribution at the uranium sites of the order 86\% in both {\updal} and {\unial}. In {\updal} the remaining 14\% are due to diffuse background, while in {\unial} the remaining 14\% can be attributed to the Ni site (7\%) and diffuse background (7\%) \cite{kern00}. The 5f orbital contribution observed in circular dichroism is consistent with that inferred from the polarized neutron scattering study. A local character of the 5f electrons has finally also been inferred from $\mu$-SR measurements \cite{amat00,sche00}. A peculiarity of the $\mu$-SR studies in {\unial} are extended muon stopping sites, where the muon may tunnel along a ring of six m-sites that surrounds the b-site $(0,0,1/2)$.

The bulk magnetic anisotropy of {\unial} is comparable to {\updal} and of the order 3 to 5 depending on the temperature \cite{sato96,suel97}. The proposed crystal electric field scheme to account for the susceptibility is the same for {\updal}, however, with larger values. Specifically, a $\Gamma_4$ singlet ground state is followed by a $\Gamma_1$ first excited singlet at 100\,K, two $\Gamma_6$ doublets at 340\,K, two $\Gamma_5$ doublets at 450\,K, one $\Gamma_3$ singlet at 1300\,K and a $\Gamma_5$ doublet at 1800\,K \cite{suel97}. It is interesting to note, that the ratio of ordered moment to $T_N$ in both compounds is consistent with the crystal field scheme. As for {\updal} the experimental evidence hence also supports a tetravalent uranium configuration. 

Neutron scattering experiments in {\unial} at first failed to detect the antiferromagnetic order \cite{krim92}, while $\mu$SR and NMR showed numerous features hinting at incommensurate antiferromagnetism with a small ordered moment \cite{amat92,kyog93}.  Moreover, $^{27}$Al NMR shows an enhancement of the spin lattice relaxation rate near $T_N$ characteristic of an abundance of spin fluctuations \cite{kyog93}. Single-crystal elastic neutron scattering eventually revealed a second order phase transition of incommensurate antiferromagnetic order at $T_N=4.6\,{\rm K}$ with a wave vector $\vec{Q}=(1/2\pm\delta,0,1/2)$, where $\delta=0.110\pm0.003$, and a magnetic correlation length, $\xi_m\approx400\,{\rm \AA}$, are typical of heavy fermion systems. The ordered moment $\mu_{ord}=(0.24\pm0.10)\,{\rm \mu_B}$ is indeed small \cite{schr94,luss97} with a critical exponent $\beta=0.34\pm0.03$ characteristic of three-dimensional order. In particular the latter feature contrasts the small moment antiferromagnetism in {\upt} and {\urusi}. Spherical neutron polarimetry finally established that the magnetic structure may indeed be viewed as a spin-density wave, where the moments point in the $\vec{a}^*$ direction and the amplitude is modulated \cite{hies01a}. The antiferromagnetic planes are stacked along the c-axis. The magnetic phase diagram of {\unial} as inferred from the bulk properties is fairly isotropic \cite{suel97}. An exception is the crystallographic b-axis, where an additional transition has been taken as evidence of an incommensurate to commensurate phase transition, i.e., magnetic field allows to tune the commensurability. Taken together, the magnetic order in {\unial} and {\updal} differ considerably.

{\unial} superconducts below a temperature $T_s=1.06\,{\rm K}$. In polycrystalline samples the specific heat anomaly is distinct but small, with $\Delta C/\gamma T_s\approx0.4$. Magnetization measurements of $H^a_{c1}\approx0.002\,{\rm T}$ and $H^a_{c2}\approx0.52\,{\rm T}$ imply type II superconductivity with a Ginzburg-Landau $\kappa\approx11$ \cite{sato96}. In contrast to {\updal}, which shows a fairly isotropic $H_{c2}$ and initial slope near $T_s$ and paramagnetic limiting {\unial} displays marked anisotropies where $H^c_{c2}\approx0.9\,{\rm T}$, $dH^c_{c2}/dT=-1.14\,{\rm T/K}$ and $H^a_{c2}\approx0.35\,{\rm T}$, $dH^a_{c2}/dT=-0.42\,{\rm T/K}$, respectively \cite{sato96}.  As for {\updal} these values may be compared with the expected paramagnetic limit $H_{p0}=0.18\,{\rm T}$ and orbital limits $H^{*a}_{c2}=0.79\,{\rm T}$ and $H^{*c}_{c2}=0.29\,{\rm T}$. Thus $H_{c2}$ exhibits orbital limiting $H_{c2}\approx H^*_{c2}$ and $H_{c2}<H_{p0}$ in stark contrast to {\updal}. At first sight this comparison suggests pure orbital limiting consistent with triplet pairing \cite{ishi02}. However, it may also be reconciled with the coexistence of superconductivity and antiferromagnetic order \cite{sato96}. In any case, the superconductivity clearly shows numerous hints for unconventional pairing. For instance, NMR measurements show the absence of a Hebel-Slichter peak at $T_s$ \cite{kyog93}, where the decrease of $1/T_1$ in the superconducting state is consistent with line nodes \cite{tou97}. The Knight shift remains, moreover, unchanged in the superconducting state characteristic of spin-triplet pairing \cite{ishi02}. This contrasts the behavior observed in {\updal}, where the decrease of the Knight shift indicates spin-singlet pairing. Spin triplet pairing in bulk samples of {\unial} is also contrasted by preliminary studies of thin epitaxial films of {\unial}. These studies suggest that $T_s$ depends on the current direction, where $H_{c2}$ implies spin-singlet pairing \cite{jour04}.

Early $\mu$-SR measurements suggested a genuine coexistence of superconductivity and antiferromagnetism \cite{amat92}. Elastic neutron scattering shows an effective increase of the ordered magnetic moment in the superconducting state \cite{luss97}. Inelastic neutron scattering shows quasi-elastic scattering around $\vec{Q}=(0.39,0,0.5)$ similar to what is observed in {\updal}, but with a reduced intensity of about 10\%. However, there is neither a build-up of additional intensity nor a gap developing, nor a gapped spin wave excitation \cite{aso00}. Further studies established quasi-elastic scattering along $(H,0,n/2)$, where $n$ is an odd integer, and the width is $\sim6\,{\rm meV}$. This scattering shifts with increasing temperature from an incommensurate to a commensurate position \cite{gaul02}.

As for {\updal} only a small pressure dependence of $T_N$ and $T_s$ is observed in {\unial}, given by $dT_N/dp\approx-0.12\,{\rm K/kbar}$ and $dT_s/dp=-(0.024\pm0.003)\,{\rm K/kbar}$ \cite{wass94b}. In fact, substitutional doping of Ni by Pd appears to act dominantly like pressure. Likewise the bulk modulus determined by x-ray diffraction up to 385\,kbar is similar and given by $B_0=150(5)$GPa without evidence for a structural phase transition up to 385\,kbar \cite{krim00}. In {\unial} the pressure where an U-Pd spacing is reached that is equivalent to {\updal} at $p_c$ may be extrapolated as 725\,kbar.

In summary both {\updal} and {\unial} do not seem to be located in the immediate vicinity of a zero temperature instability, that may be reached with hydrostatic pressure. This may provide an important hint, that crystal electric fields indeed provide a key ingredient for the superconductivity to occur in both compounds. 

\paragraph{{\ceptsi}}

The discovery of heavy-fermion superconductivity in the antiferromagnetic state of {\ceptsi} has attracted great interest, not so much because it coexists with antiferromagnetic order, but because the crystal structure of {\ceptsi} lacks inversion symmetry \cite{baue04}. The low temperature properties of this compound are characterized by the onset of commensurate antiferromagnetic order at $T_N=2.2\,{\rm K}$ with an ordering wave vector $\vec{Q}=(0,0,1/2)$. Even though band structure calculations show dominant effects of the Rashba spin-orbit coupling on the electronic structure \cite{samo04a,samo04b}, chiral components or a canting of the magnetic order have so far not been observed. 

The value of $T_s=0.75\,{\rm K}$ first reported for {\ceptsi} is fairly high. In contrast, more recent work suggest a lower $T_s=0.45\,{\rm K}$ in combination with sharper magnetic and superconducting transitions \cite{take07}. Due to the lack of inversion symmetry {\ceptsi} may be viewed as the first representative of a new class of heavy-fermion superconductors. Further members of this class discovered so far are {\cerhsiot}, {\ceirsiot} and {\cecogeot}. The properties of the non-centrosymmetric superconductors including {\ceptsi} is reviewed in section \ref{non-cent-supe}.

\subsubsection{Small moment antiferromagnets}

\paragraph{{\upt}
\label{upt}}

The heavy fermion compound {\upt} exhibits two forms of order at low temperatures. At $T_N\approx5\,{\rm K}$ {\upt} orders antiferromagnetically. This is followed by a superconducting transition at $T_s=0.54\,{\rm K}$. Because {\upt} so far is the only intermetallic compound, which unambiguously displays multiple superconducting phases with different order parameter symmetries, it has been studied in great detail. In the following we briefly review key features of the magnetic order and superconductivity to put them in perspective with the antiferromagnetic compounds addressed so far. The evidence for multiple superconducting phases is addressed in section \ref{mult-supe-phas}. For an extensive review of the properties of {\upt} we refer to \cite{joyn02}. 

{\upt} crystallizes in a hexagonal structure, space group P6$_3$/mmc, point group $D_{6h}$. The lattice parameters are $a=5.764\,{\rm \AA}$ and $\tilde{c}=4.899\,{\rm \AA}$, where $\tilde{c}$ is the distance between neighboring planes. It is convenient to define the b-axis perpendicular to the a-axis (and thus parallel to the a$^*$ axis). The molar volume is $V_m=42.43\times10^{-6}\,{\rm m^3/mol\,U}$ and the nearest U-U distance with $d_{\rm U-U}=4.132\,{\rm \AA}$ quite large. The compressibilities have been inferred from measurements of the sound velocity. They are given by $\kappa_a=-a^{-1}da/dp=0.164\,{\rm Mbar^{-1}}$, $\kappa_c=-c^{-1}dc/dp=0.151\,{\rm Mbar^{-1}}$ and for the volume $\kappa_V=2\kappa_a+\kappa_c=0.479\,{\rm Mbar^{-1}}$ \cite{devi87}. Several transmission electron microscopy studies have reported a possible incommensurate structural modulation. However, it is now generally believed that this modulation results from ion milling and is not present in bulk samples \cite{ellm95,ellm97}.

The normal state properties of {\upt} at low temperatures are well described as a heavy Fermi liquid. The normal state specific heat in {\upt} up to 1.5\,K is linear in temperature with $C/T\approx0.44\pm0.02\,{\rm J/K^2\,mol}$ and a weak cubic term $T^3\ln (T/T^*)$ as discussed in \cite{devi87}. At higher temperatures an additional $T^3$ contribution emerges consistent with a Debye temperature $\Theta_D\approx210\,{\rm K}$. For $H>H_{c2}$ an unexplained additional strong upturn in $C/T$ emerges below $\sim0.1\,{\rm K}$ \cite{bris94}. 

As a function of temperature the resistivity of {\upt} decreases monotonically from a room temperature value $\rho_{ab}\approx230\,{\rm \mu\Omega cm}$ and $\rho_{c}\approx130\,{\rm \mu\Omega cm}$ \cite{devi87,kimu95}. At low temperatures a quadratic temperature dependence of the resistivity is observed $\rho(T)=\rho_0+AT^2$ where $A_{ab}\approx1.55\pm0.1\,{\rm \mu\Omega\,cm\,K^{-2}}$ and $A_{c}\approx0.55\pm0.05\,{\rm \mu\Omega\,cm\,K^{-2}}$ (e.g., \cite{luss94,kimu95,sude97b}. At low temperatures the anisotropy of the resistivity is essentially temperature independent with $\rho_{b}/\rho_c\approx 2.6$.  The anisotropy is attributed to differences of Fermi velocities. The charge carrier mean free path inferred from the residual resistivity and quantum oscillatory studies is of the order 5000\,{\AA}. Under pressure the $A$ coefficient of the resistivity decreases at a rate $d \ln A/dp\approx-40\,{\rm Mbar^{-1}}$ \cite{will85,ponc86}. A comparison of the $T^2$ resistivity with the linear temperature dependence of the specific heat establishes consistency of the ratio $\gamma/\sqrt{A}$ with other heavy fermion systems \cite{kado86}. The observation that {\upt} forms a slightly anisotropic three-dimensional Fermi liquid with strong electronic correlations is underscored by temperature dependence observed in thermal conductivity measurements \cite{luss94,sude97b}.

The normal state magnetic properties of {\upt} are strongly enhanced. The uniform susceptibility in the basal plane exhibits a strong Curie-Weiss dependence at high temperature and a broad maximum around 20\,K \cite{frin83}.  The susceptibility is anisotropic with $\chi_c<\chi_{ab}$. The behavior seen in the uniform susceptibility is tracked in $^{195}Pt$ NMR \cite{tou96}. Inelastic neutron scattering establishes a complex spectrum of antiferromagnetic fluctuations \cite{aepp88a,aepp88b}. At moderate temperatures a fluctuation spectrum characteristic of large uranium moments ($\sim2\,{\rm \mu_B}$) is observed with a characteristic energy of 10\,meV. Below $\sim20\,{\rm K}$ antiferromagnetic correlations develop at $\vec{Q}=(0,0,1)$ that peak around 5\,meV. These fluctuations correspond to correlations between adjacent nearest-neighbor uranium sites. When decreasing the temperature well below 20\,K additional antiferromagnetic correlations develop around $\vec{Q}=(\pm 1/2,0,1)$ with a characteristic energy $\sim0.3\,{\rm meV}$ and an effective moment $\sim0.1\,{\rm \mu_B}$. These fluctuations correspond to intersite correlations within each hexagonal plane. Finally, slow magnetic fluctuations with a dispersive relaxation rate exist at low temperatures \cite{bern95}. Thus the excitation spectrum yields a duality of slow and fast excitations somewhat similar to {\updal}. In what way these fluctuations affect the unconventional superconductivity in {\upt} is an open issue.

The magnetic properties of {\upt} finally include also an elastic component of the magnetic correlations at $\vec{Q}=(\pm1/2,0,1)$ with a tiny ordered moment around 0.01 to $0.03\,{\rm \mu_B/U}$. The antiferromagnetic order was first noticed in $\mu$-SR and later confirmed by neutron scattering \cite{aepp88a}. The magnetic order is collinear and commensurate with fairly short correlation lengths $\sim300\,{\rm \AA}$. It appears to be insensitive to sample quality. Perhaps most remarkably, the only experimental probes that are sensitive to the antiferromagnetic order are neutron scattering and $\mu$-SR. Notably, the antiferromagnetism is not seen in NMR \cite{tou96}, specific heat \cite{fish91} and magnetization. It has therefore been suggested that the magnetic order is essentially dynamic in nature. 

Microscopic evidence that {\upt} forms a heavy-fermion ground state par excellence was obtained in quantum oscillatory studies \cite{tail87,tail88}. The studies revealed a wide range of mass enhancements up to 120 times of the free electron mass. Despite these strong mass enhancements the spectra were found to be in remarkable agreement with density functional theory taking the 5f electrons to be itinerant (see \cite{joyn02} and references therein). Most of the frequencies, especially those corresponding to large portions of the Brillouin zone could be identified satisfactorily. In summary the Fermi surface consists of six sheets of uniformly high effective masses. In fact, the Fermi velocities on the observed sheets are extremely slow $\langle v_F \rangle_{bc}\approx 5000\,{\rm m/s}$ and do not differ by more than 15\%. In contrast to the topology of the Fermi surface, functional density theory fails to account for these large mass renormalizations. 

It has been proposed that the mass enhancement in {\upt} is due to a duality of the 5f electrons in the spirit of that discussed for {\updal} and {\unial} \cite{zwic02b}. In this scenario one f electron is itinerant while the other two are localized. The mass enhancement in {\upt} can be accounted for, when assuming a crystal field level scheme similar to {\updal} with a $\Gamma_4$ ground state and $\Gamma_3$ first excited state. A potential weakness of this assumption is that the crystal field levels hybridize so strongly with the conduction electrons, that inelastic neutron scattering fails to detect them. The relationship of the duality model as applied to {\upt} and the experimentally observed tiny ordered moments is thereby also an unresolved issue. The recent thorough analysis of quantum oscillatory studies of the Fermi surface are, finally, in much better agreement with fully itinerant f-electrons \cite{mcmu08}. 

Measurements of the resistivity, specific heat and AC susceptibility establish {\upt} as a bulk superconductor \cite{stew84}. Early studies of the ultrasound attenuation in magnetic field \cite{muel87,qian87,sche89} and of $H_{c2}$ \cite{tail88} suggested the possibility of two superconducting phase transitions. This was eventually confirmed in high resolution specific heat measurements \cite{fish89,hass89}. Further studies establish that there are three superconducting phases, denoted A, B and C. The antiferromagnetic order can be shown to introduce an additional symmetry breaking that stabilizes these phases. In summary three pieces of evidence identify {\upt} as unconventional superconductor. First, several transport quantities display marked anisotropies, most notably the ultrasound velocity and the thermal conductivity. Second, there is evidence for phase transitions within the superconducting state as seen in the specific heat and ultrasound attenuation. Third, several properties show activated temperature dependences instead of the exponential freezing out of excitations. The superconducting phases of {\upt} will be described in further detail in section \ref{mult-supe-phas}.

\paragraph{{\urusi}}

The body-centered tetragonal uranium compound {\urusi}, space group I4/mmm, crystallizes with lattice constants $a=4.128\,{\rm \AA}$ and $c=9.592\,{\rm \AA}$. At low temperatures it undergoes two phase transitions \cite{schl84}: a transition to an hitherto unknown form of order at $T_0\approx 17.5$\,K, and a second transition at $T_s\approx1.4$\,K to unconventional superconductivity \cite{pals85,schl86,mapl86}. The entropy released at $T_0$ is given by $\Delta S\approx0.2 \rm R \ln 2$. Despite intense experimental and theoretical efforts the ordering phenomenon accounting for this entropy reduction has still not been identified. The phase below $T_0$ in {\urusi} has in turn become known as "hidden order" (HO). The hidden order exhibits many characteristics of an electronic condensation: (i) the specific heat is consistent with a BCS gap \cite{mapl86}, (ii) the resistivity at $T_0$ is strongly reminiscent of the density-wave system chromium \cite{fawc88}, (iii) slight doping suppresses the resistivity anomaly rapidly \cite{kim04}, (iv)  the magnetization at $T_0$ suggests the formation of a spin gap \cite{park97}, while optical conductivity indicates a charge gap \cite{bonn88}. Recent thermal conductivity measurements also point towards a gap formation \cite{shar05}. The Hall effect and magnetoresistance suggest near compensation of particle- and hole-carriers and a strong interplay between the stability of the hidden order under Rh-doping and the degree of polarization of the Fermi liquid and the Fermi surface topology \cite{jo07,oh07}.

\begin{figure}
\includegraphics[width=.4\textwidth,clip=]{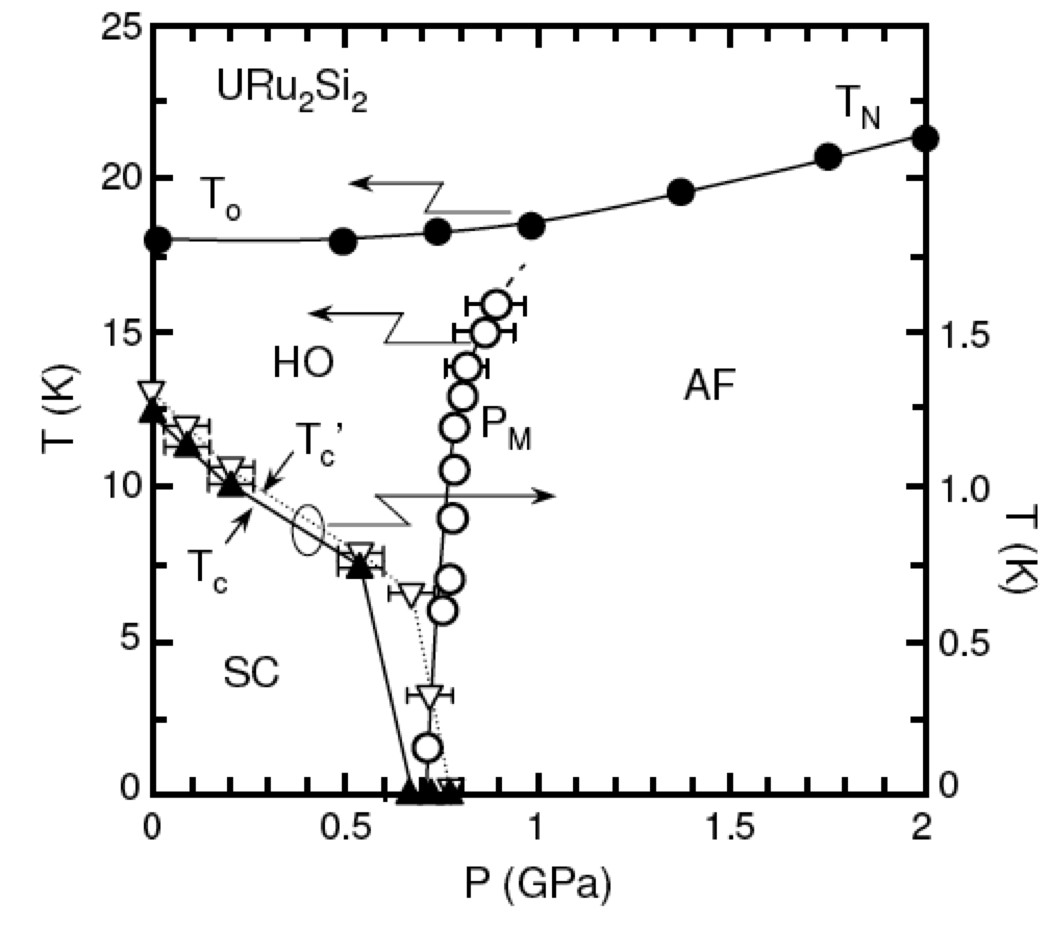}
\caption{Temperature versus pressure phase diagram of {\urusi} inferred from various experimental probes. The onset of the hidden order $T_0$ is weakly pressure dependent. The hidden order changes to large moment Ising antiferromagnetism above $7\,{\rm kbar}$ without pronounced effect on the evolution of $T_0$/$T_N$. Superconductivity vanishes with the appearance of the antiferromagnetism. HO: hidden order; AF: large moment antiferromagnet; SC: superconductivity. Plot taken from \cite{amit06}.}
\label{pd-uru2si2}
\end{figure}

Neutron diffraction in {\urusi} shows antiferromagnetic order below $T_0$ with a $[001]$ modulation of tiny moments, $(0.03\pm0.01)\mu_{\rm B/U}$, and the spins aligned along the c-axis \cite{broh87}. The magnetic order is three-dimensional with strong Ising-type spin anisotropy. Within a local-moment scenario the antiferromagnetism does not account for $\Delta S$. This contrasts antiferromagnetism with a large moment of $0.4\,\rm \mu_{\rm B}$/U and the same Ising anisotropy, which emerges under large hydrostatic pressure \cite{amit99}. A recent phase diagram is shown in Fig.\,\ref{pd-uru2si2} \cite{amit06}. NMR \cite{mats01,mats03} and $\mu$-SR \cite{amit03b} measurements suggest that the tiny-moment antiferromagnetism at ambient pressure represents a tiny volume fraction of large moment antiferromagnetism. As function of pressure $T_0$ increases, where $dT_0/dp$ increases $p^*\approx14\,{\rm kbar}$. In fact, the increase of $dT_0/dp$ at $p^*$ even persists under Re-doping \cite{jeff07}.  There is currently growing consensus, that the small antiferromagnetic moment is not an intrinsic property of the hidden order. However, a spin-density-wave close to perfect nesting may exhibit the combination of a small moment with a large reduction of entropy \cite{chan03,mine05}.

The hidden order in {\urusi} is bounded by more conventional behavior at high excitation energies, high pressure and high magnetic fields. Inelastic neutron scattering shows a gap $\Delta(T\!\to\!0)\approx1.8$\,meV in the excitation spectrum on top of the anisotropy gap  \cite{broh91}. At low energies and temperatures, dispersive crystal-field singlet--singlet excitations at the antiferromagnetic ordering wave vector are observed. These propagating excitations merge above 35\,meV or for $T>T_0$, respectively, into a continuum of quasi-elastic antiferromagnetic spin fluctuations, as normally observed in heavy-fermion systems. The excitations exhibit the Ising anisotropy up to the highest energies investigated experimentally. A rough integration of the fluctuation spectra suggests that the size of the fluctuating moments would be consistent with $\Delta S$, provided that these moments are involved in the ordering process \cite{broh91,wieb07}. Under large applied magnetic fields parallel to the c-axis the antiferromagnetic moment and $T_0$ decrease, where $T_0$ collapses to zero at $B_m=38$\,T \cite{maso95,sant00,bour03,bour05}. At $B_M$ a cascade of metamagnetic transitions is observed, in which a large uniform magnetization is recovered \cite{harr03,kim03b}. Up to $B_m$ the entropy reduction at $T_0$ stays approximately constant \cite{kim03a}, while the gap $\Delta$, as seen in neutron scattering, increases at least up to 17\,T \cite{bour03}. For a recent review see, e.g., \cite{harr04}.

The antiferromagnetic order in {\urusi} is stabilized under uniaxial stress along certain crystallographic directions and hydrostatic pressure. NMR \cite{mats01}, $\mu$SR \cite{amit03b} and neutron scattering \cite{amit99} measurements suggest, that the AF volume fraction increases and reaches 100\% above $p_c\sim 14$\,kbar. An analogous increase of the AF signal is also seen in neutron scattering under uniaxial stress of a few kbar along the [100] and [110] directions \cite{yoko02,yoko05}, but not under uniaxial stress along the $c$-axis [001]. Inelastic neutron scattering under pressure shows that the dispersive crystal-field singlet excitations at low energies vanish at high pressures \cite{amit00}, consistent with them being a property of the HO volume fraction. 

A major challenge are measurements of the Fermi surface. For instance, de Haas--van Alphen (dHvA) studies under hydrostatic pressure \cite{naka03} do not resolve abrupt changes of the dHvA frequencies and cyclotron masses at $p_c$. This contrasts naive expectation of a distinct phase separation at $p_c$. In these studies the most important observation is a considerable increase of the cyclotron mass with increasing pressure. New insights may be achieved with ultra-pure samples, that have recently become available \cite{mats08,kasa08}.

A large number of microscopic scenarios have been proposed to explain the hidden order. These include various versions of spin- and charge-density wave order \cite{maki02,mine05}, forms of crystal electric field polar order \cite{sant94,ohka99,kiss05}, unconventional density waves \cite{iked98} and orbital antiferromagnetism \cite{chan02}, Pomeranchuk instabilities \cite{varm06} or nematic electronic phases \cite{barz93}, combinations of local with itinerant magnetism \cite{okun98} and dynamical forms of order \cite{fak99,bern03}. None of the models was able to satisfactorily explain all of the available experimental data; some models are purely phenomenological yet lack material-specific predictions that can be readily verified by experiment, while others focus only on selected microscopic features. This leaves considerable space for fresh theoretical input.

The nature of the superconductivity in the hidden order of {\urusi} is still comparatively little explored. $T_s$ depends sensitively on sample quality. It is as high as $T_s=1.53\,{\rm K}$ in the purest samples, which have residual resistivities as low as several ${\rm \mu\Omega cm}$ and charge carrier mean free paths $l\sim1000\,{\rm \AA}$ as inferred from quantum oscillations \cite{bris95}. In the specific heat the onset of superconductivity is accompanied by a pronounced anomaly, where $\Delta C/\gamma T_s\approx0.8$ to 0.93 \cite{fish90,bris94}. However, this value is reduced by comparison to the weak coupling BCS value of 1.43. Between $T_s$ and $0.2\,T_s$ the specific heat varies approximately as $C\propto T^2$ akin that seen in {\upt}. This is consistent with line nodes of either a $E_u(1,1)$ or $B_g$ state \cite{hass93}. Line nodes and unconventional superconductivity has also been inferred from $^{29}$Si NMR and $^{101}$Ru NQR, where $1/T_1$ is found to show no coherence peak and decreases as $1/T_1\propto T^3$ below $T_s$, while the Knight shift is unchanged \cite{koho96,mats96}. However, it has been pointed out, that the specific heat data are equally well explained in terms of s-wave pairing in the presence of antiferromagnetism, where nodes are generated by the magnetic order \cite{bris94}. 

Further information of the possible location and nature of the nodal structure has been inferred from the angular field dependence of the specific heat, where the absence of an agular dependence in the tetragonal basal plane and marked anisotropy between a- and c-axis suggests that the gap nodes are rather localized near the c-axis \cite{saka07}. An anisotropic gap has also been inferred from point contact spectroscopy, consistent with d-wave pairing \cite{hass92,dewi94,naid96}.   If the experimental evidence for nodes is indeed due to the antiferromagnetism as suggested above, this requires, that the small antiferromagnetic moments are an intrinsic property of the hidden order, or that the hidden order interacts with the superconductivity in the same way antiferromagnetism would do. 

A different scenario of the superconductivity has recently been proposed based on the electrical and thermal transport properties in ultra-pure {\urusi} \cite{mats08,kasa08}. Here the Hall effect and magnetoresistance suggest multiband superconductivity in a compensated electronic environment. Most remarkably, in the low temperature limit the thermal conductivity divided by temperature, $\kappa/T$ displays a rapid increase at low fields followed by a plateau up to some intermediate field $H_s<0.2\,H_{c2}$. Above $H_s$ evolves differently for field parallel and perpendicular to the c-axis, but $\kappa/T$ drops abruptly just below $H_{c2}$ characteristic of $H_{c2}$ being first order (the first order behavior occurs below $\sim0.5\,K$. Based on their observations \cite{kasa08} suggest a two-component order parameter, with two distinct gaps: line nodes perpendicular to the c-axis on a spherical light hole band and point nodes along the c-axis on the elliptical heavy electron band. This scenario, notably the first order behavior and point nodes are consistent with the magnetic field dependence of the specific heat in the superconducting state \cite{yano08}. Interestingly, the thermal conductivity in the same ultra-pure samples also suggest a melting transition of the flux line lattice and the formation of a coherent quasiparticle Bloch state \cite{okaz08}.

The lower critical field of the superconductivity in {\urusi} of $H_{c1}(T\to0)\approx3.3\times10^{-3}\,{\rm T}$, is essentially isotropic and displays a weak temperature dependence \cite{wuec93}. $H_{c2}$ is in contrast strongly anisotropic with $H^a_{c2}=14\,{\rm T}$ and $H^c_{c2}=3\,{\rm T}$. This implies strong type 2 behavior and short coherence lengths $\xi_a\approx100\,{\rm \AA}$ and $\xi_c\approx25\,{\rm \AA}$. The anisotropy of $H_{c2}$ may be accounted for reasonably well by an anisotropic mass model \cite{bris94}. For the c-axis $H_{c2}$ can be explained by Pauli limiting, while it can be described by a combination of Pauli and orbital limiting for the a-axis with strongly anisotropic Pauli limiting between the a- and c-axis \cite{bris95}. 

An additional weak increase of $H_{c2}$  for the c-axis at low temperatures that exceeds Pauli limiting has been considered as tentative evidence for an FFLO phase. Also unusual is the temperature dependence of the anisotropy $H^a_{c2}/H^c_{c2}$, which initially increase below $T_s$ and becomes constant below $\sim0.6T_s$. In fact, the Ginzburg-Landau parameter inferred from the magnetization exhibits a gradual decrease well below $T_s$, somewhat slower than the behavior anticipated from $H_{c2}$ but consistent with paramagnetic limiting \cite{teny00}. Finally, a small positive curvature in the temperature dependence of $H_{c2}$ near $T_s$ has been considered as possible evidence of a multicomponent order parameter that couples to an antiferromagnetic moment \cite{kwok90,thal91}. Taken together, it is presently accepted that {\urusi} does not display multiple superconducting phases in terms of real-space or momentum-space modulations (cf sections \ref{urusi-mult-phase} and \ref{FFLO-phas}).

The thermal expansion displays pronounced anomalies at $T_s$ with $\Delta\alpha_a=-0.68\times10^{-6}\,{\rm K^{-1}}$ and $\Delta\alpha_c=0.47\times10^{-6}\,{\rm K^{-1}}$ \cite{vand95}. Thus, the superconductivity varies sensitively with uniaxial pressure, notably $dT_s/dp_a=-0.062\,{\rm K/kbar}$ and  $dT_s/dp_c=+0.043\,{\rm K/kbar}$, consistent with experiment \cite{bakk92}. The qualitative temperature dependence of $H_{c2}$ for uniaxial pressure applied along the a-axis remains thereby unchanged \cite{pfle97}. For comprehensive information on the elastic constants we refer to \cite{luet95}.

The interplay of hidden order, small moment antiferromagnetism and superconductivity in {\urusi} is largely unresolved. Early neutron scattering studies suggested, that the small antiferromagnetic moments remain either unchanged in the superconducting state \cite{broh87,maso90,wei92} or may be decreasing by 1 to 2\% \cite{honm99}. This may be consistent with a microscopic coexistence of hidden order and superconductivity. Under hydrostatic pressure $T_s$ decreases and vanishes between 5 and 14\,kbar \cite{mcel87,bris94,jeff07}. The magnetization and specific heat thereby shows, that the superconducting volume fraction decreases or, alternatively, that the superconducting gap vanishes \cite{fish90,teny05,uemu05}.  Since the suppression of superconductivity is accompanied by an increase of volume fraction of large antiferromagnetic moments, the large moment antiferromagnetism and superconductivity must represent competing forms of order. In contrast, the HO may even represent a precondition for the superconductivity in {\urusi} to occur, which points at an unknown superconducting pairing interaction.

\subsection{The puzzling properties of {\ube}}

In the following we briefly review the properties of {\ube}. Being the second system in which heavy-fermion superconductivity was identified this compound remains one of the most puzzling materials amongst the systems known to date. For a long time {\ube} seemed to be outside any of the patterns observed in the other systems. Recent work suggests the possible vicinity to an antiferromagnetic quantum critical point under magnetic field \cite{gege04}. It is not unlikely, however, that incipient antiferromagnetism is only part of the story. 

{\ube} crystallizes in the cubic NaZn$_{13}$ structure, space group $O_h^6$ or Fm3c with lattice constant a=10.248\,{\AA} \cite{pear58}. There are 8 formula units per unit cell, with two Be sites; the uranium atoms are surrounded by cages of 24 Be atoms \cite{gold85}. The U-atoms form a simple cubic sublattice, with a large U-U spacing $d_{\rm U-U}=5.13\,{\rm \AA}$, well above the hill limit of 3.4\,{\AA}, suggesting that any broadening of the uranium f-states into bands is due the hybridization with the conduction bands and not the result of direct overlap of the f-orbitals. By comparison with other heavy-fermion superconductors the properties of {\ube} are fairly insensitive to sample quality. In the normal metallic state of {\ube} the specific heat exhibits a shallow maximum around 2\,K, with a large linear term $C/T=\gamma \approx 1.1\,{\rm J/mol\,K^2}$ \cite{ott83,ott84c}. The susceptibility displays a strong Curie Weiss dependence with $\mu_{\rm eff}\approx3\,\mu_{\rm B}$ and a Curie-Weiss temperature $\Theta\approx-70\,{\rm K}$. The electrical resistivity increases with decreasing temperature and reaches a large value of order $240\,{\rm \mu\Omega\,cm}$ before it decreases around 2\,K and reaches value of $130\,{\rm \mu\Omega cm}$ at the onset of superconductivity. The extrapolated zero temperature residual resistivity is $\rho_0=60\,{\rm \mu\Omega cm}$ \cite{mapl85}.

Superconductivity was first observed in the resistivity of {\ube} in 1975 \cite{buch75} - four years prior to the discovery of superconductivity in {\cecusi}. However, the zero-resistance transition at $T_s=0.9\,{\rm K}$ was erroneously attributed to a filamentary uranium segregation. The superconducting transition was eventually identified as onset of heavy-fermion superconductivity in 1983 by means of specific heat measurements \cite{ott83}. The specific heat anomaly is characteristic of strong coupling superconductivity with $\Delta C/\gamma T_s\approx 2.5$. The initial variation of $H_{c2}$ near $T_s$ is exceptionally large $dH_{c2}/dT=-45\,{\rm T/K}$ \cite{mapl85,thom95}. In the zero temperature limit $H_{c2}(T\to0)= 14\,{\rm T}$.  $H_{c2}$ exhibits strong Pauli limiting and as an additional feature a change of curvature at $T/T_s\sim0.5\,{\rm K}$.  The unusual temperature dependence of $H_{c2}$ has been attributed to a combination of very strong coupling superconductivity and the tendency to form a FFLO state (see also section \ref{FFLO-phas}). While the coupling constant $\lambda=15$ in these calculations is suspiciously large and exceeds coupling constants in comparable systems by an order of magnitude, this scenario finds further support in the pressure dependence of $\lambda$, which tracks the mass enhancement inferred from $dH_{c2}/dT\vert_{T_s}$ and the specific heat \cite{glem99}.

\begin{figure}
\includegraphics[width=.35\textwidth,clip=]{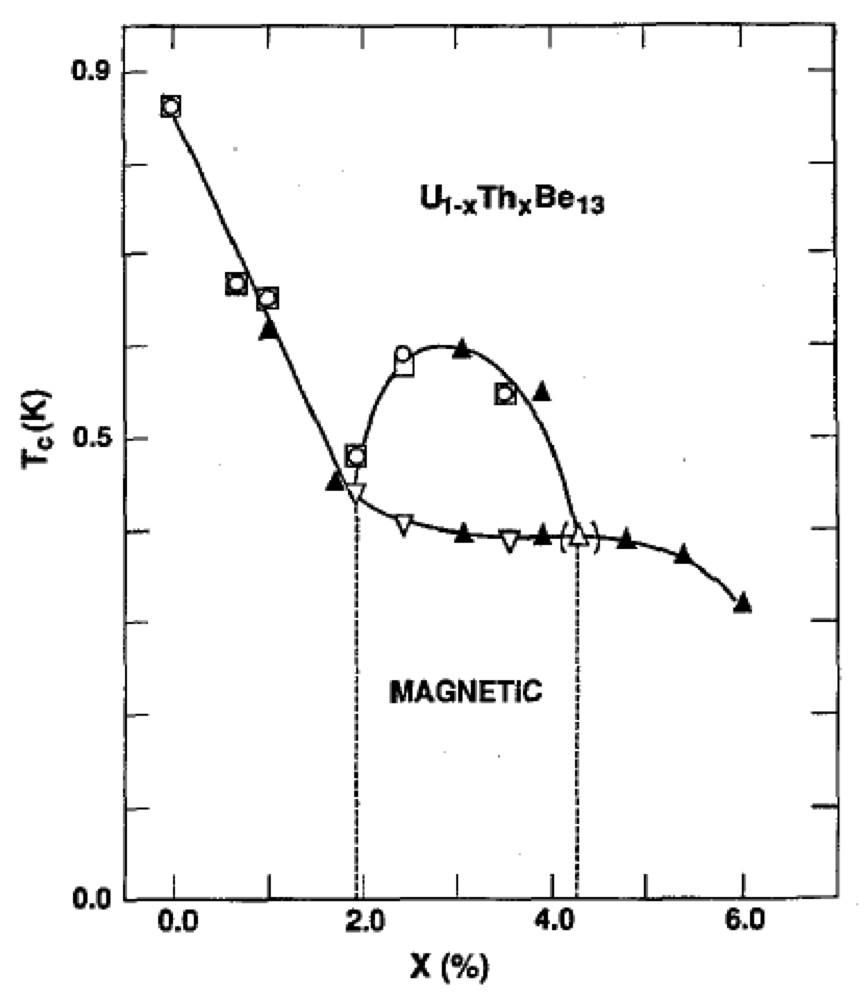}
\caption{Temperature versus Th concentration in {\uthbe}. The upper curve corresponds to the onset of superconductivity in the resistivity. In the range $2\%<x<4\%$ a second transition is observed in the specific heat, which may be related to magnetic order and/or another superconducting phase. Plot taken from \cite{mapl95}.}
\label{pd-uthbe13} 
\end{figure}

Several properties suggest the presence of zeros of the superconducting gap. The power law dependence of the specific heat $C\sim T^3$ \cite{ott84,maye86,ott87c} and penetration depth $\lambda\sim T^2$ \cite{einz86,gros86} suggest point nodes, whereas the NMR spin lattice relaxation rate suggests lines nodes \cite{macl87}. This identifies {\ube} as unconventional superconductor, a conjecture that is supported by the behavior under substitutional Th doping \cite{lamb86}. {\uthbe} displays a complex phase diagram as show in Fig.\,\ref{pd-uthbe13} with multiple superconducting phases \cite{ott86}. Thermal expansion and specific heat measurements identify a precursor of this effect in pure {\ube} \cite{krom98,krom00}.  We refer to section \ref{uthbe-phases} for a brief discussion of the details of this phase diagram.

The calculated electronic structure of {\ube} is relatively simple for itinerant f-electrons \cite{norm87,take00}. The nature of the heavy fermion state in {\ube} has nevertheless provided a major puzzle. By comparison to other heavy fermion systems the susceptibility and specific heat vary only weakly under magnetic field. This is contrasted by a strong negative magnetoresistance \cite{reme86,rauc85} providing tentative evidence that {\ube} is a low density carrier system \cite{norm87,take86}. Under hydrostatic pressure the normal metallic state assumes the more conventional form of a coherent Kondo lattice with a broad maximum at several 10\,K and a decreasing resistivity at low temperatures \cite{aron89,mcel90}. The superconductivity is suppressed under pressure and $T_s$ extrapolates to zero around 40\,kbar. Interestingly the residual resistivity decreases strongly around 40\,kbar, suggesting that the scattering mechanism causing the residual resistivity may be involved in the superconducting pairing {\ube}.

First neutron scattering studies revealed a broad quasi-elastic Lorentzian spectrum of magnetic  fluctuations with a half-width of 13.2\,meV \cite{gold86}. They failed to observe evidence for a narrow f-resonance of antiferromagnetic correlations \cite{gold86,land92}. Recent studies however, reveal short-range antiferromagnetic correlations below $\sim20\,{\rm K}$ for $\vec{Q}=(1/2,1/2,0)$ with a characteristic energy width of 1 to 2\,meV \cite{coad00,hies02}. 

New studies of the normal metallic state as a function of magnetic field establish non-Fermi liquid behavior with $\rho\sim T^{3/2}$ and a related logarithmic divergence of the specific heat  \cite{gege04}. For field above $H_{c2}$ a regime with $T^2$ resistivity emerges. This Fermi liquid behavior has been linked with the suppression of a feature in the thermal expansion that has been interpreted as a freezing of three-dimensional antiferromagnetic fluctuations. When taken together this has motivated speculations on a field-tuned antiferromagnetic quantum critical point $\sim5\,{\rm T}$ in {\ube}, at least as a facet of the complex cobination of properties of {\ube}.

%%%%%%%%%%%%%%%%%%%%%%%%%%%%%%%%%%%%%%%%%%%%%%%%%
\section{INTERPLAY OF FERROMAGNETISM AND SUPERCONDUCTIVITY}
\label{inte-ferr-supe}

Several f-electron ferromagnets have been found in recent years, that exhibit superconductivity with $T_s\ll T_C$ (cf table\,\ref{table-ubased}). These systems contrast the reentrant superconductivity observed in ErRh$_4$B$_4$ and related compounds, where ferromagnetic order appears well below the superconducting transition temperature and both forms of order originate in separate microscopic subsystems. We begin this section with a review of systems that exhibit superconductivity in the ferromagnetic state, notably {\uge} and URhGe. We next address superconductivity at the border of ferromagnetism in UCoGe and UIr.

\subsection{Superconducting Ferromagnets}

\subsubsection{\uge
\label{uge}}

The superconducting ferromagnet {\uge} crystallizes in the orthorhombic ThGe$_2$ crystal structure, space group Cmmm (no. 65), with lattice constants  $a=3.997(3)$\,{\AA}, $b=15.039(7)$\,{\AA} and $c=4.087(2)$\,{\AA} \cite{boul97,oika96}. The crystal structure of {\uge} is dominated by zig-zag chains of the U atoms along the a-axis, where the U-spacing, $d_{\rm U-U}=3.85\,{\rm \AA}$. As for {\updal} and {\unial} the U-U distance is above the Hill limit and without hybridization with other electrons the f-electrons would be localized. The U-chains are stacked with Ge atoms at interstitial positions to form corrugated sheets. These sheets are separated by further Ge atoms along the b-axis, giving the crystal structure a certain two-dimensional appearance perpendicular to the b-axis. As discussed below the two-dimensional crystallographic appearance manifests itself in the electronic structure, which is dominated by a large cylindrical Fermi surface sheet along the b-axis \cite{shic01,shic04}. 

At ambient pressure {\uge} develops ferromagnetic order below $T_C=52\,{\rm K}$ with a zero temperature ordered moment $\mu_{s}=1.48\,{\rm \mu_B/U}$ aligned along the a-axis. By comparison with the a-axis, the b- and c-axis exhibit large magnetic anisotropy fields ($\sim100\,{\rm T}$ for the c-axis) \cite{onuk92}. The magnetic anisotropy imposes a strong Ising character on the magnetic properies. In turn the temperature dependence of the ordered moment varies as $M(T) \propto (T-T_C)^{\beta}$ between $0.9\,T_C$ and $T_C$, where $\beta=0.33$ is close to calculated value $\beta\approx0.36$ of a 3D Ising ferromagnet \cite{huxl01,kern01}. 

Neutron depolarization measurements down to 4.2\,K establish, that the magnetic moments are strictly aligned along the a-axis, with a typical domain size in the bc-plane of the order $4.4\,{\rm \mu m}$  \cite{saka05a}. This contrasts earlier reports of macroscopic quantum tunneling of the magnetization below 1\,K, where the inferred domain size was only $\sim40\,{\rm \AA}$ \cite{nish02,llho03}. 

The susceptibility of the paramagnetic state is anisotropic exhibiting a Curie-Weiss dependence for the a-axis with a corresponding fluctuating moment of $\mu_{CW}=2.7\,{\rm \mu_B}$, that exceeds the ordered moment considerably. Taken by itself, the reduced ordered moment as compared with the free uranium ion value does not proof itinerant magnetism, but may be reconciled with the presence of strongly hybridized crystal electric fields. We note that inelastic neutron scattering fails to detect distinct evidence for crystal electric fields, as common for uranium based compounds. However, the reduction of the ordered moment as compared with the Curie-Weiss moment provides clear evidence of 5f itineracy.

The degree of delocalization of the 5f electrons has been explored by a variety of experimental techniques. The perhaps most direct probe is a combination of quantum oscillatory studies with band structure calculations, showing dominant f-electron contributions at $E_F$ \cite{tera01,shic01,shic04}. We will discuss these studies in further detail below. Polarized neutron scattering shows that the magnetic order is strictly ferromagnetic without additional modulations \cite{kern01}. The magnetic form factor of the uranium atoms is equally well accounted for by a U$^{3+}$ or U$^{4+}$ configuration \cite{huxl01,kern01}, where a magnetic field of 4.6\,T does not induce any magnetic polarization at the Ge sites. However, the ratio of the orbital to spin moment, $R$, does not vary substantially as a function of temperature between the paramagnetic and ferromagnetic states.  As compared with the free ion value it is systematically reduced, suggesting a delocalization of the 5f electrons. Also, the value of $R$ for U$^{3+}$ is in better agreement with circular dichroism measurements \cite{okan06} and LDA+U band structure calculations \cite{shic01}, which support a trivalent uranium state. 

Evidence for some delocalization of the f-electrons in {\uge} may also be seen in the specific heat and the spectrum of low lying magnetic excitations. At the Curie temperature the specific heat displays a pronounced anomaly, where $\Delta C/T\approx0.2\,{\rm J/mol\,K^2}$. This compares with a moderately enhanced Sommerfeld contribution $C/T=\gamma=0.032\,{\rm J/mol\,K^2}$ at low temperatures \cite{huxl01}. The strong uniaxial anisotropy causes a large anisotropy gap for spin wave excitations. In turn inelastic neutron scattering near $T_C$ only shows strongly enhanced spin fluctuations, that are characterized by a finite relaxation rate $\Gamma_q$ for $q\to0$ due to strong spin-orbit coupling \cite{huxl03b}. In a one-band approximation the finite relaxation at $q=0$ would imply that the magnetization is not conserved, which is not true in multi-band systems. The Ising character of the spin fluctuations underscores, that they are intermediate between local moment and itinerant electron fluctuations.

Itinerant ferromagnetism may finally be inferred from the fact, that {\uge} forms a very good metal. High quality single crystals may be grown with residual resistivities well-below $1\,{\rm \mu\Omega cm}$. As a function of decreasing temperature the resistivity decreases monotonically with a broad shoulder around 80\,K. At the ferromagnetic transition the resistivity shows a pronounced decrease characteristic of the freezing out of an important scattering mechanism. As an additional feature the resistivity displays a down-turn around $T_x\approx 25\,{\rm K}$, that is best seen in terms a broad maximum in the derivative $d\rho/dT$ \cite{oomi95}. Further evidence for anomalous behavior at $T_x$ has been observed in terms of a minimum in the a-axis thermal expansion \cite{oomi93}, a drastic decrease of thermal conductivity \cite{misi05}, a pronounced minimum in the normal Hall effect \cite{tran04} and a broad hump in the specific heat \cite{huxl01}. Finally, high resolution photoemission shows the presence of a narrow peak in the density of states below $E_F$ that suggests Stoner-like itinerant ferromagnetism \cite{ito02}.

As explained below, the behavior at $T_x$ yields the key to an understanding of the superconductivity in {\uge}. The available experimental evidence suggests that the density of states near $T_x$ is increased, i.e., thermal fluctuations with respect to the Fermi level are sensitive to fine-structure of the density of states such as local maxima or changes of slope. It is helpful to briefly comment on two specific scenarios that have been proposed to account for the features at $T_x$. 

The first scenario is inspired by the chain-like arrangement of the uranium atoms in {\uge}.  The structural similarity with $\alpha$-U, which develops a charge density wave at low temperatures \cite{land94}, has motivated considerations that the anomaly at $T_x$ may be related to a coupled spin-and charge-density wave instability \cite{wata02}. Electronic structure calculations predict a dominant cylindrical Fermi surface sheet with strong nesting \cite{shic01}. However, because the U-U spacing in {\uge} is larger than for $\alpha$-U, nesting is less important. Moreover, despite great experimental efforts so far no direct microscopic evidence has been observed that would support a density-wave instability \cite{aso06,huxl01,huxl03}. In fact, detailed inelastic neutron scattering studies of the phonons in {\uge} show that the hump in the specific heat near $T_x$ does not hint at soft phonons \cite{raym06}. This contrasts the structural softness expected of an incipient charge density wave. 

The second scenario is also based on the electronic structure calculations in the LDA+U, which account for the ordered moment and the magneto-crystalline anisotropy \cite{shic01}. In these calculations the ordered moment is identified as the sum of large, opposing spin and orbital contributions. Closer inspection of the results shows the presence of two nearly degenerate solutions, that differ in terms of the orbital moment \cite{shic04}. The upshot of these calculations is, that the anomaly at $T_x$ may be related to fluctuations between these two orbital states. 

\begin{figure}
\includegraphics[width=.33\textwidth,clip=]{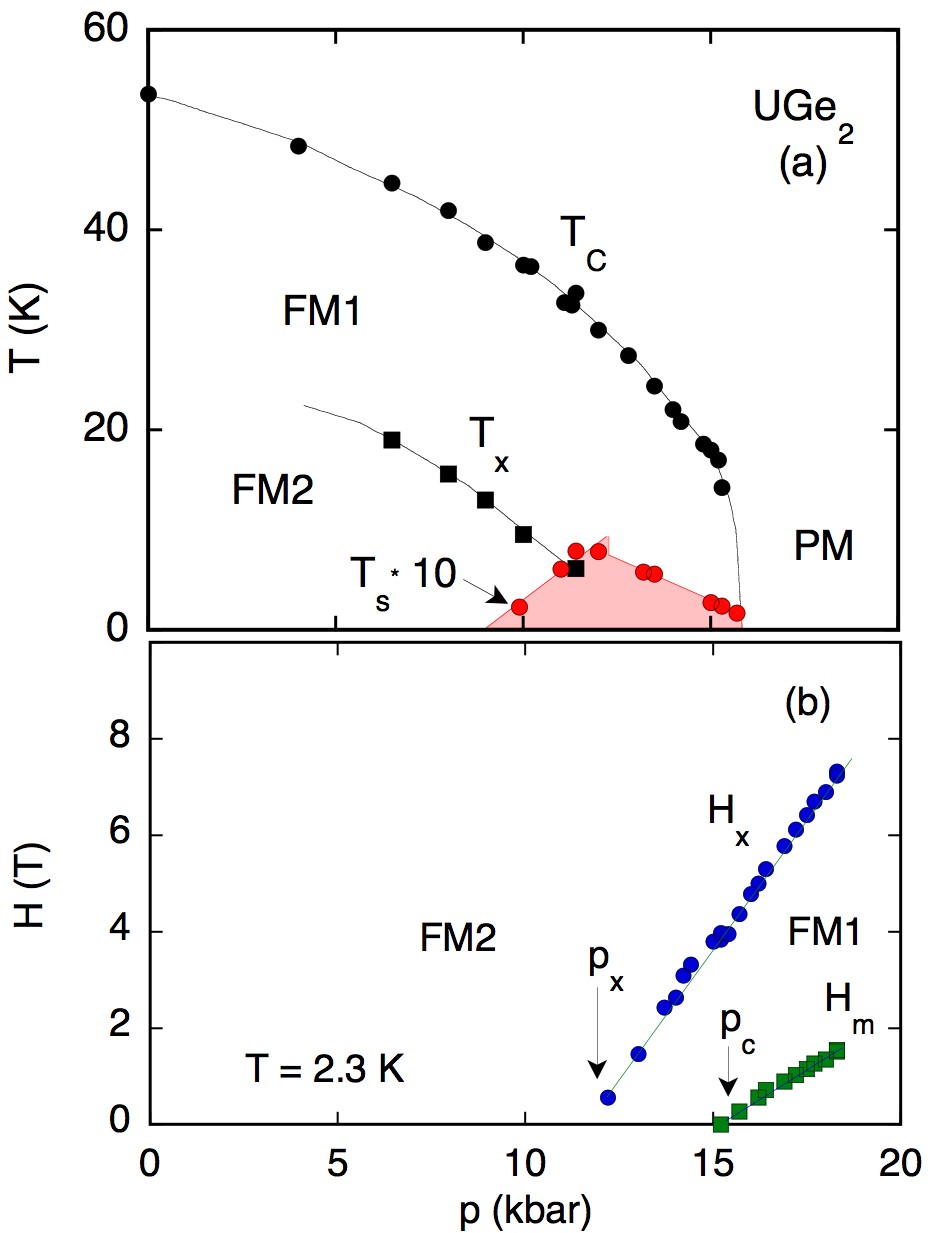}
\caption{(a) Pressure versus temperature phase diagram of {\uge}. Superconductivity is observed well within the ferromagnetic state in the vicinity of transition between a large moment and small moment ferromagnet. (b) Magnetic field versus pressure phase diagram of {\uge}. The ferromagnetic transition at $p_x$ and at $p_c$ are both first order as seen by first order metamagnetic transitions at $H_x$ and $H_m$.  Plot taken from \cite{pfle02}.}
\label{pd-uge2} 
\end{figure}

Under modest hydrostatic pressures {\uge} exhibits a rich phase diagram as shown in Fig.\,\ref{pd-uge2}. The Curie temperature is suppressed monotonically and collapses continuously at $p_c=16\,{\rm kbar}$.  AC susceptibility studies establish that the ferromagnetic transition changes from second to first order for pressures above $\sim12\,{\rm kbar}$ \cite{huxl00}. A first order transition at $p_c$ is confirmed by the DC magnetisation, which drops discontinuously at $p_c$ \cite{pfle02}. Note that the discontinuous change of the ordered moment is perfectly consistent with the continuous variation of $T_C$. Further evidence for a first order transition at $p_c$ is provided by a discontinuous change of the spin-lattice relaxation rate $1/T_1T$ \cite{kote05} and quantum oscillatory studies \cite{tera01}.  

The broad anomaly at $T_x$ is also suppressed under pressure and vanishes at $p_x=12\,{\rm kbar}$. This was first inferred from the derivative of the resistivity \cite{oomi95,huxl01}, but may also be seen in the thermal expansion \cite{ushi03} and the specific heat \cite{tate04}. In the magnetization a broad hump emerges near $T_x$, which turns into a sharp ferromagnetic phase transition near $p_x$ with increasing pressure \cite{tate01b,huxl01,pfle02}. Below $T_x$ the ferromagnetic moment increases. The low temperature, large moment phase is referred to as FM2, while the high temperature low-moment phase is referred to as FM1 (cf. Fig.\,\ref{pd-uge2}). 

Neutron scattering of the magnetic order is comparatively straight forward. Due to the cancellation of nuclear scattering lengths certain Bragg peaks are purely magnetic. Comparison of selected Bragg peaks strongly suggests, that both FM1 and FM2 are strictly ferromagnetic \cite{huxl03}. Moreover, neutron scattering at a pressure just below $p_x$ shows that the intensity of the $(100)$ Bragg spot scales with the square of the bulk magnetization. This shows that the FM2 state does not break up just below $p_x$. 

Finally, within a finite pressure interval ranging from $\sim9\,{\rm kbar}$ to $p_c$ the resistivity and AC susceptibility show a superconducting transition \cite{saxe00,huxl01}. As a function of pressure $T_s$ increases below $p_x$ and decreases above $p_x$ with the possibility of a small discontinuity exactly at $p_x$ \cite{huxl01,naka05}. 

As a function of pressure the zero temperature ferromagnetic moment drops discontinuously by $\sim30$\,\% at $p_x$, followed by a discontinuous drop at $p_c$ \cite{pfle02}. Application of a magnetic field along the a-axis at pressures above $p_x$ restores the full ordered moment at a characteristic transition field $H_x$, that emerges at $p_x$ and increases rapidly under pressure (Fig.\,\ref{pd-uge2} (b)). For pressures above $p_c$ the application of a magnetic field restores initially the ordered moment of the FM1 phase when crossing the transition field $H_m$ that emerges at $p_c$. This is followed by the recovery of the full moment at $H_x$. At low temperatures the transition at $H_x$ and $H_m$ both are discontinuous \cite{pfle02}. The lines of first order transitions at $T=0$ at $H_x(p)$ and $H_m(p)$ are expected to end in a quantum critical point for very high fields. Likewise, as a function of increasing temperature at constant pressure the transition fields $H_x$ and $H_m$ terminate in critical end-points. The importance of this finite temperature criticality to the superconductivity is an open issue.

The Sommerfeld contribution $\gamma$ to the specific heat is essentially unchanged at pressures well below $p_x$.  Just below $p_x$ the value of $\gamma$ increases and settles in a nearly four-fold larger value $\gamma\approx0.11\,{\rm J/mol\,K^2}$ above $p_x$ \cite{tate01a,tate04}. Even though the pressure depencence of $\gamma$ is sometimes described as a maximum at $p_x$, real data rather display the shape of a plateau characteristic of an increased linear specific heat term in the FM1 phase. This is supported by the temperature dependence of the resistivity, which shows a $T^2$ form everywhere. The $T^2$ coefficient $A$ increases as a function of pressure from below to above $p_x$. For magnetic fields above $H_x$ it varies as $A\propto1/\sqrt{H-H_x}$ \cite{tera06}.

To explore the nature of the transitions at $p_x$ and $p_c$ detailed quantum oscillatory studies have been carried out for magnetic fields parallel to the b-axis \cite{tera01,sett01}. This probes the predicted cylindrical Fermi surface sheets, without adding the complexities of the transitions at $H_x$ and $H_m$ \cite{shic01}. In the FM2 phase starting from ambient pressure three fundamental frequencies are observed with $F_{\alpha}=6800\pm30\,{\rm T}$, $F_{\beta}=7710\pm10\,{\rm T}$ and $F_{\gamma}=9130\pm30\,{\rm T}$. These frequencies exhibit considerable mass enhancements of $m_{\alpha}^*/m=23\pm3$, $m_{\beta}^*/m=12\pm1$, $m_{\gamma}^*/m=17\pm2$, that are weakly pressure dependent with $dF_{\alpha}/d\ln p=3.9\pm0.1\times 10^{-3}\,{\rm kbar^{-1}}$, $dF_{\beta}/d\ln p=-2.1\pm0.1\times 10^{-3}\,{\rm kbar^{-1}}$, $dF_{\gamma}/d\ln p\approx0\pm0.1\times 10^{-3}\,{\rm kbar^{-1}}$ \cite{tera01}. The mass enhancement is consistent with the specific heat. 

Between 11.4 and 15.4\,kbar, the regime of the FM1 phase, the de Haas--van Alphen spectra change in the following manner: (i) the $\alpha$ and $\gamma$ branches vanish, (ii) the $\beta$ branch initially decreases followed by a steep rise with a substantial increase of the mass enhancement to $39.5\pm5$ and a reduction of signal size to just 2.5\,\%, (iii) a new $\delta$-branch emerges, which is similar to the $\beta$ branch, where $F_{\delta}=4040\pm40\,{\rm T}$, $m_{\delta}^*/m=22\pm9$ and $dF_{\delta}/d\ln p=15\pm4\times 10^{-3}\,{\rm kbar^{-1}}$.

It is interesting to note that no minority-spin counterpart to the $\beta$-branch is observed, characteristic of a fully spin-polarized state. Under the assumption that the Fermi surface volume remains unchanged through $p_x$, it is not necessary to invoke a complete reconstruction of the Fermi surface to understand the data. When assigning the $\beta$- and $\alpha$-branches to extremal orbits of the majority-spin Fermi surface and the $\gamma$ branch to a Fermi surface sheet with hole character, the $\delta$ branch may be understood as resulting from a shrinking and breaking-up of the $\gamma$ hole surface. Again the mass enhancement is consistent with the specific heat.

For the paramagnetic state above $p_c$ the situation differs. Here the spectra consist of four new branches, that are not connected in any obvious manner with the spectra in the FM1 and FM2 phase. This suggests that the Fermi surface completely reconstructs at $p_c$. Because the change of the frequencies is abrupt, the reconstruction appears to be first order. Preliminary studies have also been carried out for magnetic field along the a-axis \cite{haga02,tera02}. For fields above $H_x$ in the FM2 phase the spectra and mass enhancements vary weakly with pressure. In contrast, very little information could be obtained below $H_x$.

The very weak pressure dependence of the ordered moment in the FM1 and FM2, and the fact that the transition between FM1, FM2 and paramagnetism may be controlled either by pressure and/or magnetic field suggests an important role of maxima in the density of states \cite{huxl01,pfle02,sand03}. However, several properties show that purely spin-based models or the delocalisation of the 5f electrons would be too simple as an explanation. For instance, the derivative of the magnetization $\chi_{\parallel}=dM/dH$ measures the longitudinal susceptibility, i.e., the sensitivity for changes of amplitude of the ordered moment. A comparison of the pressure dependence of $\chi_{\parallel}$ for the a- and c-axis establishes, that the anisotropy of the longitudinal susceptibility increases strongly under pressure, i.e., the magnetic response becomes more anisotropic instead of less \cite{pfle02,huxl03}. 

We further note, that the transition at $p_x$ is probably not controlled by a density wave instability either. Neutron scattering of the crystal structure at high pressure shows that U-U spacing at 14\,kbar reduces to $d_{\rm U-U}\approx3.5\,{\rm \AA}$ and the zig-zag chain straightens \cite{huxl01}.  It is conceivable that the requirements for nesting would be much too sensitive to survive these fairly large structural changes up to $p_x$. Second, the observation of quantum oscillations on large Fermi surface sheets seems inconsistent with a charge-density wave gap in the FM2 phase. Moreover, measurements of the uranium magnetic form factor show, that it may still be accounted for by either a U$^{3+}$ or U$^{4+}$ configuration, but the ratio of orbital to spin moment, $R=\mu_L/\mu_S$ increases across $p_x$ so that $R_{FM1}/R_{FM2}\approx1.10\pm0.05$ \cite{kuwa02,huxl03}. This contrasts a delocalization of the 5f electrons, since the orbital contribution should then decrease. It is interesting to note, that the increase of $R$  through $p_x$ is consistent with the proposed degeneracy of orbital contributions in the FM1 and FM2 phases as calculated in the LDA+U \cite{shic04}. This suggests that the FM2 to FM1 transition at $p_x$ and related properties may be driven by fluctuations between two different orbital moments.

Having reviewed the metallic and magnetic state extensively, we finally turn to the superconductivity in the ferromagnetic state of {\uge}. The initial experiments suggested that the superconductivity in {\uge} is extremely fragile. The critical current density, of order $j_c\approx0.1\,{\rm A/cm^2}$, is between one and two orders of magnitude smaller than for heavy-fermion systems such as {\upt} and even three orders of magnitude smaller than for conventional superconductors \cite{huxl01}. The reduced values of $j_c$ may be reconciled with flux flow resistance, where the flux lattice forms spontaneously even at ambient field due to the internal field (the ordered moment corresponds to 0.19\,T). The expected flux line spacing at this field is of the order 600 to 1000\,{\AA} \cite{huxl01}. Further, the susceptibility depends sensitively on the excitation amplitude, consistent with very low $j_c$, and reaches full diamagnetic screening only for very small amplitudes \cite{saxe00}. The diamagnetic shielding as seen in the AC susceptibility is largest at $p_x$. Note that this does not show the volume fraction of Meissner flux expulsion. Instead it may be the result of changes of sensitivity to the AC excitation amplitude. Interestingly, the diamagnetic screening and the pressure dependence of $T_s$ do not reflect in a simple manner the difference of 30\% of the ordered moment in the FM1 and FM2 phase.

Bulk superconductivity in {\uge} was at first inferred from the magnetic field dependence of the flux flow resistance, which displays the characteristic convex increase up to $H_{c2}$ \cite{huxl01}. Less ambiguous information provided the specific heat, which was found to show a small, yet distinct, anomaly $\Delta C/\gamma T_s\approx0.2$ to 0.3 \cite{tate01a}. The spin lattice relaxation rate in Ge NQR shows a change of slope at $T_s$.  However, in contrast to the resistivity and susceptibility the specific heat suggests that bulk superconductivity exists only in a very narrow interval surrounding $p_x$ \cite{tate04}. Such a narrow interval of bulk superconductivity at $p_x$ is supported by $dH_{c2}/dT\vert_{T_s}$, which in the same narrow interval is ten-fold increased, exceeding $dH_{c2}/dT\vert_{T_s}<-20\,{\rm T/K}$ \cite{naka05}.

The superconductivity in {\uge} is remarkable, because $T_s$ is always at least two orders of magnitude smaller than $T_C$. The superconductivity hence emerges in the presence of a strong ferromagnetic exchange splitting, estimated to be of the order 70\,meV. This suggests an unconventional form of superconductivity pairing. For what is known about the Fermi surface, odd-parity equal-spin triplet pairing is thereby the most promising candidate. This state is equivalent to the A1 phase of $^3$He. 

As a first experimental hint for an unconventional state the superconducitivity in {\uge} is fairly sensitive to the sample purity, i.e., superconductivity vanishes when the charge carrier mean free path becomes shorter than the coherence length \cite{shei01}. Inferring triplet pairing from the mean free path dependence when doping with selected impurities was previously employed in studies of {\upt} \cite{dali95} and Sr$_2$RuO$_4$ \cite{mack98}. As for UGe$_2$ the conclusion of triplet pairing has been questioned on the basis of superconductivity observed in polycrystalline {\uge} samples with $\rho_0\approx3\,{\rm \mu\Omega cm}$ \cite{baue01}.  However, the purity dependence in polycrystals is still within the uncertainty at which the charge carrier mean free path can be inferred from from $\rho_0$. Interestingly the specific heat of the polycrystals only shows a faint superconduting anomaly and thus bulk superconductivity at 14.7\,kbar. This may be caused by the presence of internal strains between the crystal grains \cite{voll02}. 

In single-crystals the maximum specific heat anomaly $\Delta C/\gamma T_s\approx0.2$ to 0.3 and the finite residual specific heat in the zero temperature limit, $\gamma_0/\gamma(T>T_s)\approx0.3$ are characteristic of nodes in the superconducting gap, where the linear $T$ dependence of $C/T$ more specifically suggests line nodes. 

The strongest evidence supporting p-wave superconductivity thus far are comprehensive studies of $H_{c2}$ \cite{shei01}. Absolute values of $H_{c2}$ vary strongly as a function of pressure and crystallographic direction, where typical values are in the range of a few T. Below $p_x$ the coherence lengths inferred from $H_{c2}$ are fairly isotropic and of the order 100\,{\AA}. In contrast, above $p_x$ the coherence lengths display a marked anisotropy, e.g., for 15\,kbar $\xi_a=210\,{\rm \AA}$, $\xi_b=140\,{\rm \AA}$ and $\xi_c=700\,{\rm \AA}$.

It is helpful to address at first two unusual features for the a-axis, that are outside the more general pattern of behavior. At small magnetic fields $H^a_{c2}$ displays negative curvature, that may be attributed to the internal fields associated with the ferromagnetic order. Second, for pressures just above $p_x$, pronounced reentrant behavior is observed in $H_{c2}$, when the magnetic field crosses the transition at $H_x$ \cite{huxl01}. This reentrant behavior in $H_{c2}$ may also provide a possible explanation for the pronounced extremum in $dH_{c2}/dT$ \cite{naka05}. Keeping these two aspects in mind, the more general features of $H_{c2}$ may be summarized as follows: (i) $H_{c2}$ exceeds conventional paramagnetic and orbital limiting for all field directions, except very close to $p_c$, where the a- and b-axis show more conventional limiting, (ii) the anisotropy of $H_{c2}$ in the vicinity of $T_s$ may be described by the effective mass model, (iii) the anisotropy seems to relate to the inverse of the magnetic anisotropy, i.e., $H_{c2}$ for the c-axis is always the largest. 

A remarkable feature of the critical field for the c-axis  is the presence of positive curvature at temperatures as low as $0.1\,T_s$. The general form of $H^c_{c2}$ is reminiscent of that observed in {\ube}. It may be accounted for in a strong-coupling scenario, where the coupling parameter $\lambda$ decreases rapidly with increasing pressure from $\lambda=14$, 7 and 1.7 at $p=12$, 13.2 and 15\,kbar, respectively. We note that for conventional electron-phonon mediated superconductivity these high values of $\lambda$ would imply an incipient lattice instability. 

Neutron scattering shows that the ferromagnetic scattering intensity at $(100)$ remains unchanged to within less than a percent when entering the superconducting state \cite{huxl01,huxl04c,pfle05,aso05}. However, these studies were probably not carried out sufficiently close to $p_x$ to provide information on the narrow regime, where bulk superconductivity is seen in the specific heat. When taken together the available experimental evidence makes it highly unlikely, that the superconductivity is carried by tiny sections of the Fermi surface, where the exchange splitting vanishes. 

The observation that the superconductivity in {\uge} is confined to the ferromagnetic state has created great theoretical interest. We conclude this section with a very brief account of some of the theoretical contributions {\uge} has inspired. The microscopic coexistence of ferromagnetism and superconductivity has been addressed in a number of contributions, e.g., \cite{mach01,kirk03,abri01,suhl01,spal01,sa02}. Possible order parameter symmetries of superconducting ferromagnets for given crystal structures and easy magnetizations axis have been classified in \cite{mine02,mine02b,mine04,mine04b,mine05b,samo02,samo02b,mine05a}. For instance, it has been pointed out that ferromagnetic superconductors with triplet pairing and strong spin-orbit coupling are at least two-band superconductors. Without spin-orbit coupling it is generically expected that separate superconducting transitions take place, for the majority and minority Fermi surface sheet \cite{kirk04,beli04}. The upper critical field in these systems is determined by a novel type of orbital limiting, and the precise order parameter symmetry depends on the orientation of the ordered magnetic moment. The latter property, in principle, allows to switch the superconducting order parameter through changes of orientation of the magnetization. The precise impact of spin-orbit coupling in this scenario, which of course is strong in f-systems, awaits further clarification.

The greatest fascination has generated the absence of superconductivity above $p_c$, because it even suggests ferromagnetism as a precondition for superconductivity. Experimentally the reconstruction of the Fermi surface topology supports a less generic explanation. It is however interesting to note, that a large number of mechanisms could be identified that promote superconductivity as confined to the ferromagnetic state. These include hidden quantum criticality, the enhancement of longitudinal (pair-forming) spin fluctuations in the ferromagnetic state, special features of the density of states and the possible coupling of spin- and charge density wave order \cite{karc03,sand03,kirk01,wata02}. As a new thread several studies have considered the possible interplay of magnetic textures with the superconductivity and spontaneous flux line lattices. We briefly return to this question in section \ref{doma-stru-supe-ferr}.

%\begin{itemize}
%\item discovery of superconductivity in {\uge}: \cite{saxe00,huxl01}
%\item selected papers on the superconductivity in {\uge}: \cite{baue01, tate01a, kern01, shei01, moto01, shic01, tera01, nish02, yaou02, pfle02, tera02}
%\end{itemize}

\subsubsection{URhGe}

The series UTX, where T is a higher transition metal element and X=Si or Ge, crystallize in the orthorhombic TiNiSi crystal structure, space group Pnma \cite{sech98,tran98}. Even though the crystal structure of this series differs from that of {\uge} it also shares certain similarities. In particular, as for {\uge} the uranium atoms form zig-zag chains. For URhGe the U-U spacing $d_{\rm U-U}\approx3.48\,{\rm \AA}$ compares well with the U spacing in {\uge} at a pressure of 13\,kbar. This has motivated detailed studies of high quality crystals, which let to the discovery of superconductivity in the ferromagnetic state of URhGe \cite{aoki01}. Further studies have revealed a metamagnetic transition within the ferromagnetic state, surrounded by superconductivity \cite{levy05}. For clarity we refer in the following to the superconductivity at ambient field as S1 and for that the metamagnetic transition at S2.

At ambient pressure URhGe displays a paramagnetic to ferromagnetic transition with a Curie temperature $T_C=9.6\,{\rm K}$ and an ordered moment $\mu_{ord}=0.42\,{\rm \mu_B/U}$ \cite{aoki01,prok02}. Neutron scattering studies show that superconducting samples (S1) are strictly ferromagnetic. This contrasts earlier studies of polycrystalline samples which displayed a non-colinear magnetic structure \cite{tran98}. Electronic structure calculations in the LSDA  \cite{shic02} and LAPW+ASA \cite{divi02} reproduce the ordered moment and magneto-crystalline anisotropy (LDA+U appears to be not necessary).  These calculations also show the possibility for a canted antiferromagnetic state. In any case, as for {\uge} the ordered moment is the result of strongly opposing spin and orbital contributions. In the following we discuss the properties of ferromagnetic URhGe only.

The ferromagnetic moment in URhGe is aligned with the crystallographic c-axis \cite{huxl03}. In contrast to {\uge} the magnetic anisotropy field is only large for the a-axis. As discussed in further detail below, a magnetic field $H_R=11.7\,{\rm T}$ applied along the b-axis rotates the ordered moment into the field direction. URhGe hence exhibits an quasi-easy magnetic plane, rather than the Ising anisotropy observed in {\uge}. The easy-axis susceptibility in URhGe follows a Curie-Weiss dependence above $T_C$ with a fluctuating moment $\mu_{eff}=1.8\,{\rm \mu_B/U}$ \cite{aoki01}, while the b-axis susceptibility varies with temperature as expected of antiferromagnetic order at low temperatures \cite{huxl03}. This strongly suggests itinerant ferromagnetism with strongly delocalized 5f electrons. 

The ferromagnetic transition shows a $\lambda$-anomaly at $T_C$, where the magnetic entropy released at $T_C$ is small $S_m=0.4\,{\rm R}\ln2$ \cite{hagm00}. At low temperatures the specific heat follows a dependence $C\sim \gamma T+ bT^2$ where $\gamma=0.164\,{\rm J/mol\,K^2}$. The $\lambda$ anomaly is rapidly suppressed for magnetic fields applied parallel to the c- and b-axes, where $\gamma$ in a field of 15\,T decreases by $\sim27\%$ and $\sim19\%$, respectively. This underscores that URhGe has an easy magnetic plane. 

High quality polycrystalline and single crystal specimen of URhGe undergo a superconducting transition with $T_s\approx0.25\,{\rm K}$ (S1) \cite{aoki01}. In polycrystalline samples $H_{c2}=0.71\,{\rm T}$ corresponds to a Ginzburg-Landau coherence length $\xi_{GL}\approx180\,{\rm \AA}$. Measurements of the magnetization show the onset of weak flux expulsion to be consistent with a penetration length $\lambda_l=9100\,{\rm \AA}$. The specific heat shows a clear anomaly at $T_s$ characteristic of bulk superconductivity, where $\Delta C/\gamma T_s\approx 0.45$ is strongly reduced as compared with the weak-coupling BCS value. The superconductivity in URhGe is sensitive to the sample purity. With increasing residual resistivity $T_s$ decreases and vanishes for low sample quality, consistent with unconventional superconductivity. 

$H_{c2}$ of the S1 state is anisotropic, where the anisotropy compares with the inverse of the magnetic anisotropy, i.e., $H_{c2}$ is largest for the a-axis and and smallest for the c-axis \cite{hard05a}. This suggests an intimate connection between superconductivity and ferromagnetism. For all directions $H_{c2}(T\to0)$ exceeds paramagnetic limiting. As a function of sample quality it is found that $H_{c2}(T\to0)$ varies $\propto T_s^2$, showing the intrinsic nature of the large critical field values. The comparatively small anisotropy shows, that large critical field values are not due to a reduced $g$-factor or electronic anisotropies. 

Because the superconductivity (S1 and S2) occurs in the ferromagnetic state, it is expected that the pairing dominantly occurs on the spin-majority Fermi surface akin the odd-parity equal-spin p-wave pairing of the A1 phase of $^3$He. This is consistent with the reduced specific heat anomaly as compared to the BCS value of $\Delta C/\gamma T_s=1.43$ and residual zero temperature specific heat $\gamma(T\to0)=\gamma/2(T>T_s)$ \cite{aoki01}. 

For the crystallographic point group of URhGe, a ferromagnetic moment parallel to the c-axis and strong spin-orbit coupling only two odd-parity states are possible \cite{hard05a}. The temperature dependence of the ratios of the upper critical fields allows to distinguish between these two states. The observed combination of 20\% increase of $H^a_{c2}/H^b_{c2}$ with decreasing temperature while $H^a_{c2}/H^b_{c2}=$constant strongly supports an odd parity p-wave state with gap node parallel to to magnetic moments. Finally, the temperature dependence of $H_{c2}$ is in excellent agreement with strong coupling calculations, when the initial slope $dH_{c2}/dT$ near $T_s$ is taken from experiment. 

It is interesting to note that the ratio of the Curie temperature to the maximal superconducting transition in {\uge}  ($T_C/T_s\approx30/0.8=37.5$) compares well that in URhGe ($T_C/T_s\approx9.6/0.25=38.4$). Together with the structural similarity of the uranium zig-zag chains this raises the question for further similarities notably the pressure dependence. The thermal expansion of the ferromagnetic transition in URhGe shows positive anomalies for all three crystallographic axes:  
$\Delta\alpha^a=3.4(1)\times10^{-6}\,{\rm K^{-1}}$ so that $dT^{a}_C/dp=0.052(3)\,{\rm K/kbar}$, 
$\Delta\alpha^b=1.7(1)\times10^{-6}\,{\rm K^{-1}}$ so that $dT^{b}_C/dp=0.026(2)\,{\rm K/kbar}$ and 
$\Delta\alpha^c=2.7(1)\times10^{-6}\,{\rm K^{-1}}$ so that $dT^{c}_C/dp=0.041(2)\,{\rm K/kbar}$.
This yields a volume thermal expansion and pressure dependence of $T_C$,
$\Delta V^a=7.8(2)\times10^{-6}\,{\rm K^{-1}}$ and 
$dT_C/dp=+0.119(6)\,{\rm K/kbar}$, respectively \cite{saka03}, i.e., $T_C$ increases under pressure. This has been confirmed in experimental studies up to 140\,kbar \cite{hard05b}. In these studies $T_s$ is found to be suppressed for pressures above $\sim30\,{\rm kbar}$. Despite the increase of $T_C$ under pressure the ordered moment decreases with $d\mu_{ord}/dp=-6.3\times10^{-3}\,{\rm \mu_B/kbar}$ \cite{hard04}.

\begin{figure}
%\sidecaption
\includegraphics[width=.38\textwidth,clip=]{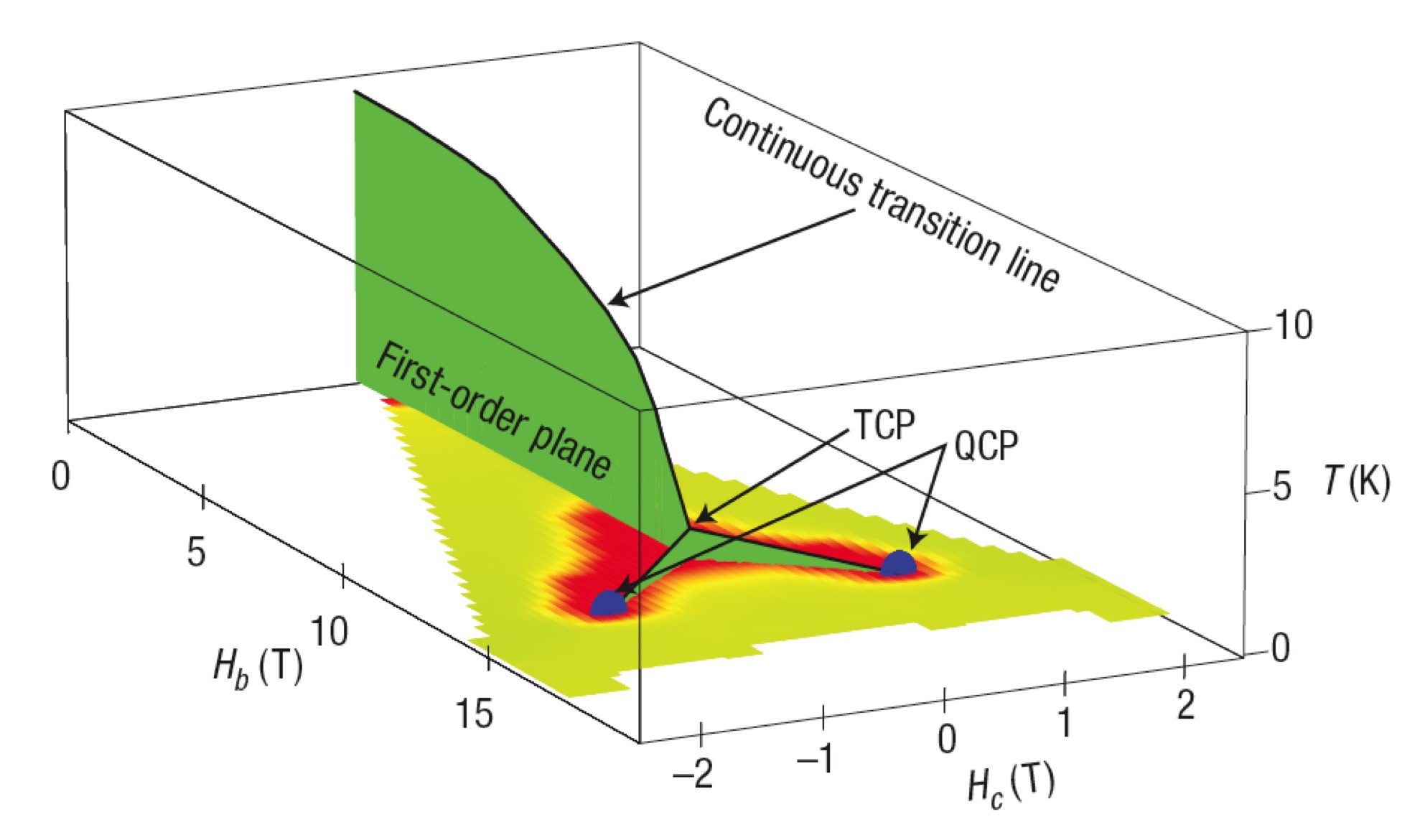}
\caption{Temperature versus magnetic field phase diagram of URhGe, for magnetic fields in the bc-plane. The critical end point of the reorientation transition of the magnetic order is surrounded by a dome of superconductivity (S2). Plot taken from \cite{levy07}.}
\label{pd1-urhge} 
\end{figure}

As depicted in Fig.\,\ref{pd1-urhge} magnetic field applied parallel to the b-axis may be used to tune the ferromagnetic transition (green shading) towards zero. Close to the field value where $T_C$ would vanish the dependence of $T_C$ versus field bifurcates. Because the transition is continuous throughout, the bifurcation represents a tricritical point (TCP).  Application of magnetic field with suitably chosen components along the b-axis and c-axis allows to further reduce the $T_s$, until it vanishes at a field tuned quantum critical point (QCP) for $\vec{H}=(0, H^b=\pm 12\,{\rm T}, H^c=\pm 2\,{\rm T})$. Neutron scattering and the torque magnetization establish that the transition is driven by a change of orientation of the ordered magnetic moment, where the moment in general is not parallel to the applied field. The bifurcation in $T_C(H)$ suggests that the excitation spectrum includes longitudinal fluctuations.

In the vicinity of the TCP and QCPs of URhGe superconductivity (S2) emerges  \cite{levy05}.  For a magic angle in the range 30$^{\circ}$ to 55$^{\circ}$ S2 even stabilizes for field components along the c-axis. The maximum value of $T_s=0.4\,{\rm K}$ exceeds that observed at zero applied field. The wide range of orientations of the ordered moment under which superconductivity is seen shows, that the superconductivity is not related to the Jaccarino-Peter effect (a cancelation of the internal field by the applied field). Instead the phenomenology of the phase diagram suggests that the superconductivity is driven by the field tuned quantum critical point.  The possible connection of the superconductivity at ambient field with that at high field as different manifestations of this quantum critical point and the associated changes of the triplet pairing has been discussed in \cite{mine06}.

The perhaps most spectacular characteristic of the superconducting state is the upper critical field for the hard magnetic a-axis. Here $H_{c2}$ diverges and exceeds 28\,T, the highest field studied \cite{levy07}. The anisotropy of the upper critical field may be accounted for in terms of an anisotropic mass model, where $H_{sc1}=\Phi_0/(2\pi\xi_c)\sqrt{\xi_a^2\cos^2(\gamma)+\xi_b^2\sin^2(\gamma)}$, with $H_{sc1a}=\Phi_0/(2\pi\xi_c\xi_b)=2.53\,{\rm T}$, $H_{sc1b}=\Phi_0/(2\pi\xi_c\xi_a)=2.07\,{\rm T}$ and $H_{sc1c}=\Phi_0/(2\pi\xi_b\xi_a)=0.69\,{\rm T}$. Further, assuming that the anisotropy of the critical fields that is observed at zero applied field remains unchanged for the high field superconductivity, a geometric average of the coherence length, $\xi=\sqrt{\xi_a\xi_b\xi_c}$, can be inferred. Remarkably, the coherence length $\xi$ as a function of applied magnetic field for the b-axis diverges at $H_R$, where the magnetic field dependence of $\xi$ of both superconducting phases fall on the same line. The coherence length thereby decreases from $\xi(H_b=0)=143\,{\rm \AA}$ to $\xi(H_R)<44\,{\rm \AA}$. The common field dependence of the coherence length suggests, that both superconducting phases have the same origin, notably the quantum critical point at high fields.

\subsection{Border of ferromagnetism}
\label{tran-meta-ferr}

Recently two superconducting ferromagnets have been discovered, notably UIr \cite{akaz04a,akaz04b} and UCoGe \cite{huy07}, in which the ordered moment of the ferromagnetic state is small as compared with the compounds introduced so far. The superconductivity in both compounds is observed at the border of ferromagnetism, rather than deep inside the ferromagnetic state.

\paragraph{UCoGe}
UCoGe is orthorhombic and isostructural to URhGe with lattice constants $a=6.645\,{\rm \AA}$, $b=4.206\,{\rm \AA}$ and $c=7.222\,{\rm \AA}$. It was long thought the UCoGe is paramagnetic, but polycrystalline samples were recently found to exhibit ferromagnetic order with a small ordered moment $\mu_{ord}=0.03\,{\rm \mu_B/U}$ below $T_C=3\,{\rm K}$. The ordered moment is much smaller than the fluctuating moment observed in the paramagnetic state $\mu_{eff}=1.7\,{\rm \mu_B/U}$. The specific heat shows a small anomaly at $T_C$, where the magnetic entropy released is tiny, $S_m=0.03\,{\rm R}\ln 2$, and the normal state specific heat is moderately enhanced, $C/T=\gamma=0.057\,{\rm J/mol\,K^2}$. The thermal expansion shows a volume contraction, where the idealized discontinuity in $\alpha$ is estimated to by $\Delta\alpha=-1.1\times 10^{-6}\,{\rm K^{-1}}$. Thus, according to the Ehrenfest relation, the Curie temperature would decrease under pressure at a rate $dT_C/dp=V_mT_C\Delta\alpha/\Delta C=-0.25\,{\rm K/kbar}$ and is expected to vanish around 12\,kbar ($V_m=3.13\times10^{-5}\,{\rm m^3/mol}$ is the molar volume). 

Polycrystalline samples of UCoGe display superconductivity with $T_s\approx0.8\,{\rm K}$, i.e., $T_s$ is much smaller than $T_C$. The superconducting transition is seen in the resistivity, AC susceptibility, specific heat and thermal expansion. In the AC susceptibility the diamagnetic screening is of the order 60 to 70\%. In the specific heat the anomaly corresponds to $\Delta C/\gamma T_s\approx 1$, which is smaller than the weak-coupling BCS value. The thermal expansion displays a positive anomaly, with an idealized change of length at $T_s$ of the order $\Delta L/L\approx-1\times 10^{-7}$. This implies that $T_s$ increases under pressure at a rate $dT_s/dp\approx+0.048\,{\rm K/kbar}$. 

Experimentally it is found, that $T_C$ in polycrystals rapidly drops under pressure and appears to vanish between 8 and 20\,kbar, while $T_s$ is essentially unchanged consistent with the thermal expansion. The width of the superconducting transition and additional features of the normal state resistivity, such as the residual resistivity and the temperature dependence, suggest a quantum critical point already at 7\,kbar \cite{hass08}. In any case, the superconductivity in UCoGe appears to survive in the non-ferromagnetic state at high pressure. However, data available to data do not rule out that a ferromagnetic moment survives at high pressures. The pressure dependence is supplemented by the variation of the superconductivity and ferromagnetism as a function of Si substitution in polycrystals, UCoGe$_{1-x}$Si$_x$, which shows a simultaneous suppression of $T_C$ and $T_s$ at the same critical concentration $x_c\approx0.12$ \cite{nijs08}. 

The upper critical field of polycrystalline UCoGe varies near $T_s$  as $dH_{c2}/dT\approx-5.2\,{\rm T/K}$ for the sample with the largest $T_s$. This implies a fairly short coherence length $\xi\approx150\,{\rm \AA}$ as compared with the charge carrier mean free path $l=500\,{\rm \AA}$ inferred from the residual resistivity $\rho_0=12\,{\rm \mu\Omega cm}$. In other words the samples are in the clean limit, a precondition for unconventional superconductivity. An unconventional superconducting state is also inferred from $H_{c2}$, which exceeds 1.2\,T, the highest field measured, which is thus clearly larger than the Pauli limit. 

NMR and NQR measurements in polycrystals also point at unconventional pairing \cite{ohta08}. In the normal state the spin-lattice relaxation and the Kight shift are characteristic of ferromagnetic quantum critical fluctuations, where $T_C\approx2.5\,{\rm K}$ was observed. This underscores that the system is indeed at the border of ferromagnetism. However, in the superconducting state the spin-lattice relaxation rate appears to yield two contributions. These may be related to a superconducting and a normal volume fraction, which are either due to bad sample quality or the result of the spontaneous formation of flux lines due to the ferromagnetism.

Recently single crystals of UCoGe have become available with $T_C=2\,{\rm K}$ and $T_s=0.6\,{\rm K}$, that are also in the clean limit \cite{huy08}. The ferromagnetic moment $m_s=0.07\,{\rm \mu_B}$ is aligned with the c-axis and the a- and b-axis are magnetically hard. Thus UCoGe is an easy-axis ferromagnet like {\uge}, in contrast with the hard-axis ferromagnetism in URhGe. $H_{c2}$ of single-crystal UCoGe shows a marked anisotropy between the ab-plane and the c-axis, where $B_{c2}$ for field parallel to the a- and b-axis exceeds the Pauli limit with $B^a_{c2}\simeq B^b_{c2}\approx{\rm 5\,T}\gg B^c_{c2}\approx{\rm 0.6\,T}$. The initial slope $dB^{a,b}_{c2}/dT\approx-8\,{\rm T/K}$ is also large. This suggests an equal-spin pairing state with an axial symmetry of the gap function and with point nodes along the c-axis. Moreover, an upward kink of $B^a_{c2}$ may indicated multiband superconductivity.

\paragraph{UIr}
The signatures of the superconductivity in UIr are still rather incomplete as the superconductivity exists at high pressures and very low temperatures. Because the crystal structure of UIr lacks inversion symmetry, the properties of UIr are presented in more detail in section \ref{uir}, which deals with non-centrosymmetric superconductors.

\paragraph{Note on d-electron ferromagnets}
It is worthwhile to comment briefly on two ferromagnetic d-electron systems in which superconductivity has been reported. First, high-purity samples of iron exhibit superconductivity above 140\,kbar \cite{shim01,jacc02}. It turns out that the superconductivity occurs in the hexagonally closed packed $\epsilon$-phase of iron, which is believed to represent an incipient antiferromagnet \cite{saxe01,mazi02}. Nevertheless several hints, such as great sensitivity to sample purity and a non Fermi liquid temperature dependence of the resistivity near the highest value of $T_s$, suggest unconventional pairing. 

The other system is the weak itinerant electron magnet {\zrzn}, where an incomplete resistivity transition has been reported \cite{pfle01}. Here more recent work suggests that the superconductivity is not intrinsic, but due to the Zn depletion of spark eroded sample surfaces \cite{yell05}. 

%%%%%%%%%%%%%%%%%%%%%%%%%%%%%%%%%%%%%%%%%%%%%%%%%
\section{EMERGENT CLASSES OF SUPERCONDUCTORS}
\label{emer-clas-inte-supe}

A growing number of intermetallic compounds exhibit unusual forms of superconductivity that do not fit into the general category of magnetism and superconductivity covered in sections \ref{inte-anti-supe} and \ref{inte-ferr-supe}. These compounds promise to be representatives of new classes of superconductors. The following section is dedicated to a review of these emergent classes of f-electron superconductors. We distinguish non-centrosymmetric systems, materials at the border to a valence transition and systems at the border of polar order.

\subsection{Non-centrosymmetric superconductors
\label{non-cent-supe}}

In general the strong electronic correlations in heavy fermion systems may be viewed as an abundance of magnetic fluctuations, which, being pair-breaking, suppress conventional s-wave superconductivity. This is contrasted by spin-triplet pairing, which may occur as long as time reversal symmetry and inversion symmetry are satisfied \cite{ande84}. In turn it is was long believed that pure spin-triplet heavy-fermion superconductivity cannot exist in non-centrosymmetric systems. This is contrasted by the recent discovery of supercondcuctivity in the antiferromagnets {\ceptsi} \cite{baue04}, {\cerhsiot} \cite{kimu05}, {\ceirsiot} \cite{sugi06} and {\cecogeot} \cite{kawa08,sett07b}. Perhaps most remarkably superconductivity has even been discovered at the border of ferromagnetism in the non-centrosymmetric compound UIr \cite{akaz04a,akaz04b}. In this section we will review the current understanding of these compounds. Because their properties may be explained by a mixed s- plus p-wave pairing state they may be representatives of  a new class of superconductors, outside the traditional scheme of classification.

From a theoretical point of view non-centrosymmetric heavy-fermion superconductors are interesting, because in these materials the Fermi surface exhibits a splitting due to antisymmetric spin-orbit coupling $\alpha(\vec{k}\times\nabla\phi)\cdot\sigma$. In two-dimensional electron gases this splitting is referred to as Rashba- and in bulk compounds as Dresselhaus-effect \cite{rash60,dres55}. As a reminder, spin-orbit coupling is a purely relativistic effect that is due to gradients of the electric potential $\nabla\phi=\vec{E}$ transverse to the motion of the electrons. It can be shown that antisymmetric spin-orbit coupling leads to a splitting of the Fermi surface along $\vec{k}_F\times\nabla\phi$. In magnetic materials the asymmetric spin-orbit coupling also generates a contribution to the exchange interaction that is akin to superexchange, where the role of the nonmagnetic atom is played by an empty orbital \cite{mori63}. This superexchange is also known as Dzyaloshinsky-Moriya interaction. 

\begin{figure}
%\sidecaption
\includegraphics[width=.38\textwidth,clip=]{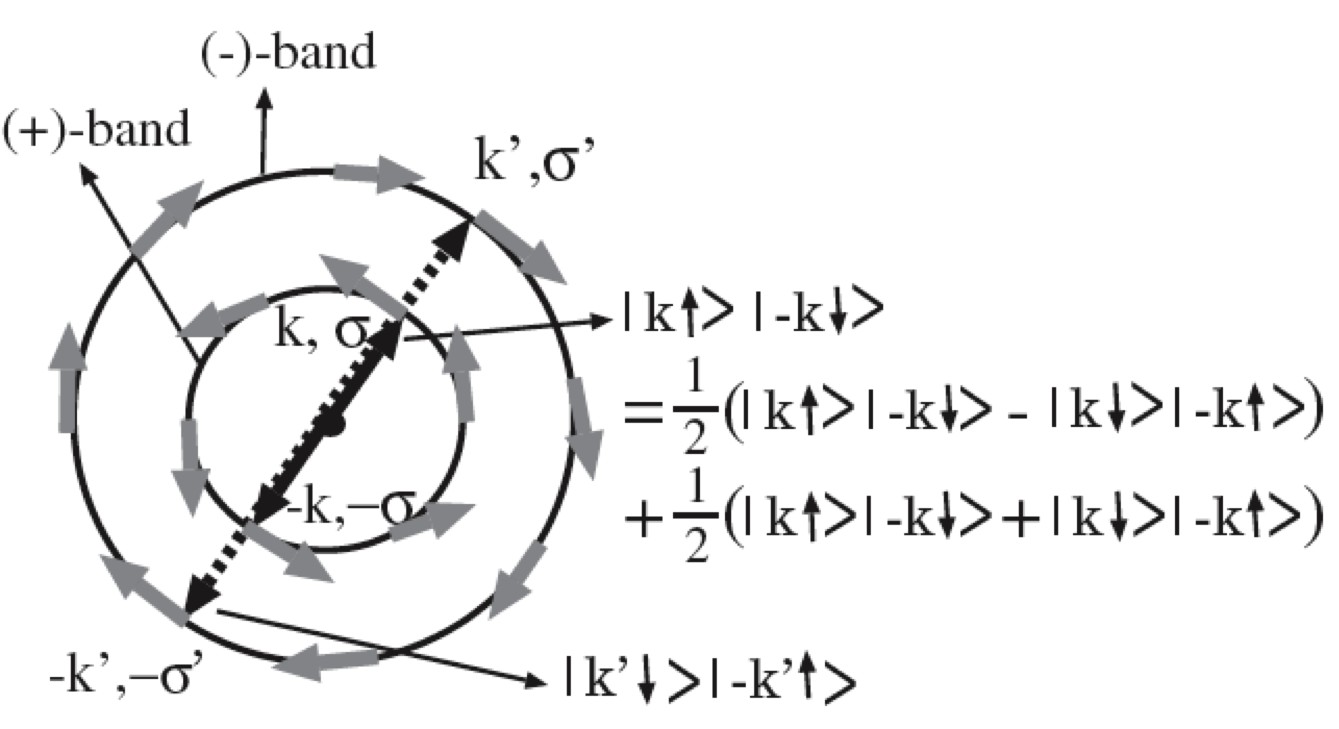}
\caption{Qualitative depiction of chiral exchange splitting of a spherical Fermi surface by Rashba spin-orbit interactions. Also shown are Cooper pairs that may form under such an exchange splitting, notably a mixed singlet with triplet state. Plot taken from \cite{fuji07}.}
\label{fs-rashba} 
\end{figure}

In a simple-minded view the asymmetric spin-orbit coupling leads to a highly unusual chiral exchange splitting of the Fermi surface (see e.g. \cite{fuji07}). A qualitative depiction is shown in Fig.\,\ref{fs-rashba}, where $\nabla\phi$ is along the z-axis (the x- and y-axis are in the plane). The exchange splitting translates into dispersion curves that energetically favor a precessional motion of the electron spin with a particular handedness, where the axis of the precession is denoted by the gray arrows in Fig.\,\ref{fs-rashba}. For a Fermi surface with chiral exchange splitting a Cooper pair forming between electrons with momentum $\vec{k}$ and spin $\uparrow$ and momentum $-\vec{k}$ and $\downarrow$ do not correspond to a spin singlet state, because ${\vert\vec{k}\uparrow}\rangle\vert-\vec{k}\downarrow\rangle$ and $\vert\vec{k}\downarrow\rangle\vert-\vec{k}\uparrow\rangle$ form on different Fermi surface sheets. Instead it has long been predicted \cite{edel89,gork01}, that superconductive pairing requires an admixture of a spin singlet with a spin triplet state. 

%, for instance,
%\begin{equation}
%\begin{split}
%\vert\vec{k}\uparrow\rangle\vert-\vec{k}\downarrow\rangle
%&=\frac{1}{2}(\vert\vec{k}\uparrow\rangle\vert-\vec{k}\downarrow\rangle
%-\vert\vec{k}\uparrow\rangle\vert-\vec{k}\downarrow\rangle)\\
%&+\frac{1}{2}(\vert\vec{k}\uparrow\rangle\vert-\vec{k}\downarrow\rangle
%-\vert\vec{k}\uparrow\rangle\vert-\vec{k}\downarrow\rangle)\\
%\end{split}
%\end{equation}
%where the first term on the right hand side corresponds to a spin singlet and the second term to a spin triplet state. Here the in-plane spin projection $S_{inplane}=0$. Since the spin quantization axis is parallel to the $xy$-plane, the triplet state corresponds to the $S^z=\pm1$ state for the spin quantization axis along the $z$-direction. Thus $\vec{d} $ of the triplet component is parallel to the plane. 

The formation of parity violating Cooper pairs in terms of the singlet-triplet mixing applies also in more general cases with more complicated forms of $\nabla\phi$. To quantify the absence of inversion symmetry it is convenient to introduce a vector $\alpha \vec{g}_{\vec{k}}$, where $\alpha$ is the Rashba parameter and $\langle\vert\vec{g}_{\vec{k}}\vert^2\rangle_0=1$ is the normalization condition in terms of the average over the Fermi surface. The vector $\vec{g}_{\vec{k}}$ may be determined by symmetry arguments. For the analysis of the allowed superconducting pairing symmetry $\vec{g}_{\vec{k}}$ is compared with $\vec{d}(\vec{k})$. For simple cases, like {\ceptsi}, the gap function $\Delta_{\pm}$ may then be expressed as $  \Delta_{\pm}(\vec{k})=(\psi\pm d\vert\vec{g}_{\vec{k}}\vert) $ where $\psi$ corresponds to the wave function of the singlet condensate and $d\vert\vec{g}_{\vec{k}}\vert$ to the triplet state \cite{gork01,agte06}.

The discovery of the non-centrosymmetric heavy-fermion superconductors reviewed in the following has revived the interest in non-centrosymmetric superconductors with weak or modest electronic correlations. For instance, the boride superconductor {\lipdptb} \cite{badi05} displays conventional BCS behavior for {\lipdb} \cite{toga04} and unconventional superconductivity for {\liptb}. Interestingly the evolution from conventional to unconventional superconductivity may be studied as function composition where superconductivity is observed for all compositions; i.e., the disorder introduced by the alloying does not affect the superconductivity. 

The gap symmetry is explained as $s$-wave with spin-singlet and spin-triplet admixtures \cite{yuan06}. Another example is the Pyrochlore oxide system {\cdreo} \cite{hana01,saka01}. For this system a structural phase transition leads to a loss of inversion symmetry at low temperatures, so that strong spin-orbit coupling effects change the electronic structure \cite{hana02}. Old examples from the literature are rare earth sesquicarbides, {\rc}, where R=La or Y. For instance {\yc} displays a high $T_s=18$\,K, but only weak spin-orbit coupling \cite{krup69,gior70,aman04}.  On a more general note we also mention that amorphous superconductors have no centre of inversion. However, they are characterized by a very low diffusivity and associated large Ginzburg-Landau parameter $\kappa\gg1$, i.e., in contrast to the materials listed so far they are in the extreme dirty limit. Likewise thin superconducting films lack inversion symmetry, i.e., they also fall into a different category.

\begin{table}
\centering
\caption{Key properties of on-centrosymmetric intermetallic superconductors. 
Missing table entries may reflect more complex behavior discussed in the text. 
$H_{c2}$ represents the extrapolated value for zero temperature.
References are given in the text. 
*Samples with much sharper antiferromagnetic and superconducting transitions.}
\label{table-noncentro}
\begin{tabular}{lllllllll}
\hline\noalign{\smallskip}
& {\ceptsi} &  {\ceirsiot} & {\cerhsiot} & {\cecogeot}\\
\noalign{\smallskip}\hline\noalign{\smallskip}
structure & tetrag. & tetrag. & tetrag. & tetrag.\\
space group & P4mm & I4mm & I4mm & I4mm \\
$a$({\AA}) & 4.072(1) & 4.252 & 4.244 & -\\
%$b$({\AA}) & - & - & -\\
$c$({\AA}) & 5.442(1) & 9.715 & 9.813 & -\\
$c/a$ & 1.336 & - & -& -\\
\noalign{\smallskip}\hline
CEFs & $\Gamma_6$, $\Gamma_7$, $\Gamma_6$ & $\Gamma_7$, $\Gamma_6$, $\Gamma_7$ & $\Gamma_7$, $\Gamma_6$, $\Gamma_7$\\
& $\Gamma_7$, $\Gamma_6$, $\Gamma_7$ & \\
$\Delta_1\,{\rm(meV)}$ & 13 & 13.7 & 22.4  &  \\
& or 1.4 &\\
$\Delta_2\,{\rm(meV)}$ & 24 & 43 & 23.3\\
\noalign{\smallskip}\hline
state & AF, SC & AF, SC & AF, SC & AF, SC\\
$T_{N}$(K) & 2.2 & 5.0 & 1.6 & 21, 12, 8\\
$\vec{Q}$ & (0,0,$\frac{1}{2}$) & - & $(.215,0,\frac{1}{2})$ & -\\
$\mu_{ord}¥$($\mu_{B}$) & 0.2 & - & - & - \\
$\gamma {\rm (J/mol K^2)}$ & 0.39 & 0.12 & 0.11 & 0.032 \\
& 0.35*\\
$A {\rm (\mu\Omega cm/K^2)}$ & - & 0.33 & 0.2  & 0.011\\
$T_K {\rm (K)}$ & 7--11 & - & 50 & - \\
\noalign{\smallskip}\hline
$p_{c}$(kbar) & 8 & 25 & 20 & 55 \\
$T_{s,max}$(K) & $\sim0.75$ & 1.6 ($p_c$) & 0.72 & 0.69 \\
& $\sim0.45$* \\
$\Delta÷C/\gamma_{n}T_{s}$ & $\sim0.25$ & - & - \\
$H^{ab}_{c2}$(T) & $\sim3.6$ & 9.5 & -\\
& $\sim2.3$*\\
$dH^{ab}_{c2}/dT$ & -8.5 & -13  & - \\
& -7.2*\\
$H^{c}_{c2}$(T) & $\sim4$ & $>30$ & $>30$ & $\sim 45$\\
%& $\sim2.3$*\\
$dH^{c}_{c2}/dT$ & -8.5 & -20 ($p_N$) & -23 & -20\\
%& -7.2*\\
\noalign{\smallskip}\hline
$\xi^{ab}_{0}$({\AA}) & $\sim82$ & - & - \\
$\xi^{c}_{0}$({\AA}) & $\sim90$ & 54 & 70 \\
%$\kappa_{GL,a}$ & - & - & - \\
%$\kappa_{GL,c}$ & - & - & - \\
%$\partial÷T_{s}/\partial÷p_{[001]}$(mK/GPa)& - & - & - & - & \\
%$\partial÷T_{s}/\partial÷p_{V}$(mK/GPa)& - &  - & - &  - & \\
%$\partial÷T_{s}/\partial÷p$(mK/GPa)& - &  - & - &  - & \\
%\noalign{\smallskip}\hline
%$C/T$ & $\propto \gamma_s+\beta_s T$ & - & - \\
%$\kappa$ & $\propto T$ & - & - \\
%$1/T_{1}T$ & $\propto\sim T^3$ & - & - \\
%$\lambda$ & $\propto T$ & - & - \\
\noalign{\smallskip}\hline
year & 2004 & 2006 & 2007 & 2008 \\
\noalign{\smallskip}\hline
\end{tabular}
\end{table}
 
\subsubsection{CePt$_3$Si}
\label{ceptsi3}

The compound {\ceptsi} was the first non-centrosymmetric compound in which heavy-fermion superconductivity was discovered \cite{baue04}. Recent reviews may be found in \cite{baue07,baue05,baue05b,sett07}. In the following we first introduce the structural and physical properties of the normal metallic state. We then proceed to present the magnetic properties. This sets the stage for the superconducting properties presented at the end of this section.

{\ceptsi} crystallizes in the tetragonal  CePt$_3$B type structure with space group $P4mm$ (No. 99). The lattice constants are $a=4.072(1)$\,{\AA} and $c=5.442(1)$\,{\AA}. The perhaps most important feature of the crystal structure is a lack of inversion symmetry, i.e., for the generating point group $C_{4V}$ the mirror plane $z\to -z$ is missing. The crystal structure of {\ceptsi} may be derived from the cubic AuCu$_3$-type crystal structure of CePt$_3$, which is isostructural to {\cein}. In contrast to the series of {\cemin} compounds discussed above, which are related to {\cein} in terms of additional MIn$_2$ layers, the structure of {\ceptsi} evolves from the AuCu$_3$ structure by filling a void with Si.  It is interesting to note that filling a void in cage like structures, as for the skutterudites discussed in this review, plays a key role for their properties in terms of soft rattling modes. Concerning {\ceptsi} this does not seem to be the case. Rather, there is an important indirect role played by the Si atom in generating the lack of inversion symmetry and a considerable tetragonal distortion with $c/a= 1.336$. 

Below $T_N=2.2$\,K {\ceptsi} displays long range antiferromagnetic order followed by the superconducting transition at $T_s$ \cite{baue04}. Recent studies suggest that samples either exhibit $T_s=0.75\,{\rm K}$ or $T_s\approx0.45\,{\rm K}$, where the samples with lower $T_s$ show much sharper magnetic and superconducting transitions \cite{take07,sett07}. Neutron scattering  in samples with $T_s=0.75\,{\rm K}$ revealed subtle metallurgical segregations and a broad distribution of lattice constant \cite{pfle08}. Similar conclusions were drawn from the pressure dependence of $T_N$ and $T_s$, i.e., the larger $T_s$ seems related to a distribution of lattice constant \cite{aoki08}. A systematic study of small specimens that were all cut from the same polycrystalline ingot showed the same $T_N$ for all pieces. However, samples either displayed  $T_s=0.45\,{\rm K}$ or $T_s=0.75\,{\rm K}$ \cite{moto08}, characteristic of two different superconducting states. This may be consistent with the earlier observation of a superconducting double transition \cite{sche05}. To date the majority of studies have been carried out in samples with the larger $T_s$.

The antiferromagnetism and superconductivity in {\ceptsi} emerge from a normal state above $T_N$ that is typical of $f$-electron heavy-fermion systems. The specific heat exhibits an enhanced Sommerfeld contribution $C/T=\gamma\sim0.39\,{\rm J/mol\,K^2}$ as extrapolated from $T>T_N$. The resistivity varies quadratically just above $T_N$ with $A=2.23\,{\rm \mu\Omega cm/K^2}$, where the samples exhibiting superconductivity have low residual resistivities $\rho_0$ of a few ${\rm \mu\Omega cm}$. 

The low-temperature properties of {\ceptsi} may be attributed to an interplay of three energy scales. RKKY interactions as the origin of the magnetic order, Kondo screening as the origin of strong correlations and finally crystal electric fields. The Kondo temperature may be deduced in various different ways, where values are in the range 7 to 11\,K \cite{baue07}. This uncertainty is rather typical for f-electron systems. Nevertheless, $T_N$ clearly falls below $T_K$. The CEFs lift the $2j+1=6$-fold degenerate ground state of the $j=5/2$ total angular momentum of the Ce atom. However, the crystal electric field levels have not been identified conclusively. 

Based on inelastic neutron scattering measurements it has been proposed that the CEF levels consist of three doublets \cite{baue05}, where the $\Gamma_7$ first excited state and  $\Gamma_6$ second excited state are separated from the  $\Gamma_6$ ground state by 13\,meV and 24\,meV, respectively. The proposed CEF level scheme has been compared with the bulk properties, where the sibling La- compound {\laptsi} and substitutional doping with La have been considered additionally. Because {\laptsi} is metallic with a Debye temperature $\Theta_D\approx 160$\,K, the 2nd excited CEF level cannot be identified quantitatively in the bulk properties. 

It is also interesting that the ratio $A/\gamma^2\approx1.0\times10^{-3}\,{\rm m\,mol\,K^2/J^2}$ is consistent with a small degeneracy of the ground state as lifted by the CEFs. This differs from another group of materials in the Kadowaki-Woods plot with large degeneracy and small CEF splitting, where $A/\gamma^2\approx0.4\times10^{-3}\,{\rm m\,mol\,K^2/J^2}$ \cite{baue05b}. The CEF assignment given by \cite{baue07} is contrasted by inelastic neutron scattering measurements providing evidence of excitations at 1.4\,meV and 24\,meV \cite{meto04}.  The associated CEF level scheme is a $\Gamma_7$ ground state and $\Gamma_6$ and $\Gamma_7$ first and second excited states, respectively, where the lower two doublets originate from a $\Gamma_8$ quartet in the cubic point symmetry. The CEF scheme with the low lying $\Gamma_6$ was found to account for the magnetization as measured up to 50\,T \cite{take04}. We return to a discussion of the high-field magnetization below.

The magnetic properties of {\ceptsi} are dominated by a strong Curie-Weiss susceptibility  at high temperatures with an effective fluctuating moment $\mu_{eff}=2.54\,{\rm \mu_B}$ of the free Ce$^{3+}$ ion and a Curie temperature $\Theta_p=-46\,{\rm K}$ characteristic of antiferromagnetic exchange coupling of the moments. The Curie-Weiss susceptibility extends down to $\sim 11\,{\rm K}$, where a broad maximum in $\chi$ signals Kondo-type screening of the fluctuating moments. At $T_N=2.25\,{\rm K}$ a $\lambda$-anomaly in the specific heat shows the onset of long range antiferromagnetic order. The size of the specific heat anomaly implies a release of entropy  through $T_N$ of $\Delta S\approx 0.22\,{\rm R \ln 2}$, characteristic of small ordered moments. Under pulsed magnetic fields up to 50\,T the magnetization increases almost linearly up to $\sim23\,{\rm T}$ for $H\parallel [100]$ and $[110]$, respectively. Above 23\,T the magnetization levels of and settles around $0.8\,{\rm \mu_B/Ce}$. This implies a rather small in-plane magnetic anisotropy.

Microscopic evidence for antiferromagnetic order below $T_N$ has been obtained in neutron diffraction \cite{meto04} and $\mu$-ion spin rotation \cite{amat05} experiments. Neutron diffraction is consistent with a magnetic ordering vector $\vec{k}=(0,0,1/2)$ and small ordered moments $\mu\approx0.2\,{\rm \mu_B}$ at $T=1.8\,{\rm K}$. The antiferromagnetic order consists of ferromagnetic planes stacked along the $c$-axis, where the ordered moments are oriented in-plane. By comparison to the slightly reduced moment expected of a degenerate $\Gamma_8$ CEF quartet, $\mu(\Gamma_8)=1.96\,{\rm \mu_B}$, the ordered moment is strongly reduced. This may by due to the doublet ground state, but suggests also significant Kondo-screening. $\mu$-SR measurements show that the small ordered moments exist throughout the entire sample volume \cite{amat05}. 

Keeping in mind what is presently known about the antiferromagnetic order of {\ceptsi}, it is interesting to consider the spin-orbit splitting of the Fermi surface due to the lack of inversion symmetry. Quantum oscillatory studies in {\ceptsi} have shown a small number of branches \cite{hash04}. Cyclotron masses up to 20 times of the bare electron mass have been observed, where masses of up to 65 times of the bare electron mass are expected. While this is not a definitive identification, it represents nevertheless strong evidence for a  fairly conventional heavy fermion state. The lack of inversion symmetry generates a spin-orbit splitting consistent with the de Haas -- van Alphen data in {\laptsi} \cite{hash04}. In principle this splitting may generate chiral components of the magnetization.

The crystal structure and magnetic properties of {\ceptsi} set a stage where no superconductivity is expected. Yet, superconductivity emerges in {\ceptsi} well below $T_N$. A number of properties suggest unconventional superconductivity. For instance, under substitutional doping of Ce by La superconductivity is suppressed in {\celaptsi} for $x\geq0.02$ \cite{youn05}. A high sensitivity to nonmagnetic impurities is considered a hall mark of unconventional pairing. However, removing a magnetic atom from the structure may not qualify as nonmagnetic impurity, rather than a magnetic defect.

Near $T_s$ the upper critical field varies strongly as $dH_{c2}/dT\vert_{T_s}=-8.5\,{\rm T/K}$ for samples with $T_s=0.75\,{\rm K}$ \cite{baue07} and with $dH_{c2}/dT\vert_{T_s}=-7.2\,{\rm T/K}$ for samples with $T_s=0.45\,{\rm K}$ \cite{take07}. This is characteristic of heavy-fermion superconductivity and consistent with the mass enhancement inferred from the normal state specific heat. The upper critical field reaches $H_{c2}\sim5\,{\rm T}$ for the samples with the highest $T_s\approx0.75\,{\rm K}$ \cite{baue07}. The value of $H_{c2}$ exceeds by a large margin the Pauli-Clogston limit, $H_{PC}\sim1.1\,{\rm T}$, when not taking into account conventional spin-orbit coupling. Most remarkably, $H_{c2}$ does not display a sizable anisotropy, namely $H^c_{c2}/H^{ab}_{c2}\approx1.18$. 

Calculations of the electronic structure of {\ceptsi} show a large Rashba parameter 
$\alpha={\rm 100\,meV}$ \cite{samo04b}. This implies that spin-orbit splitting of the Fermi surface plays an important role for the superconductivity. An analysis of the Rashba splitting in {\ceptsi} in terms of group theory for the space group $P4mm$ and generating point group $C_{4V}$ suggests that $\vec{g}_{\vec{k}}=k^{-1}_F(k_y, -k_x, 0)$ \cite{frig04}. The most stable spin pairing state expected for $\vec{d}(\vec{k})\parallel\vec{g}_{\vec{k}}$ corresponds then to a $p$-state $\vec{d}(\vec{k})=\hat{x}k_y-\hat{y}k_x$. This state is characterized by point nodes.  However, for mixing of this triplet state with a singlet state the experimental properties are expected to display the behavior of line nodes. Alternatively a Balian-Werthamer (BW) state $\vec{d}(\vec{k})=\hat{x}k_x+\hat{y}k_y+\hat{z}k_z$ is possible, which would have no nodes. However, the BW state is expected to be less stable \cite{frig04}. In fact, one may consider the $\hat{x}k_y-\hat{y}k_x$ p-state as being protected by the Rashba exchange splitting. 

A key characteristic of the $\hat{x}k_y-\hat{y}k_x$ state for magnetic field parallel to the $c$-axis is the absence of paramagnetic limiting, while there would be considerable paramagnetic limiting for magnetic field perpendicular to the c-axis. To account for the nearly isotropic behavior of $H_{c2}$ it has been suggested that the superconducting wave function develops a helical phase factor $\exp(i\vec{q}\vec{R})$ in applied magnetic fields \cite{kaur05,agte06}. It is, however, also important to take into account the interplay of antiferromagnetic order with the superconductivity, which accounts in parts for the reduced anisotropy \cite{yana07}.

The superconducting transition in samples with larger $T_s$ is accompanied by a fairly broad anomaly in the specific heat with  $\Delta C/\gamma T_s\approx0.25$. This value is strongly reduced as compared with the isotropic BCS value $\Delta C/\gamma T_s\approx1.43$ \cite{baue04}. It deviates in particular from strong coupling behavior frequently observed in heavy fermion superconductors. The reduced specific heat anomaly may be explained by a vanishing of the superconducting gap on parts of the Fermi surface, e.g., due to an unconventional Cooper pair symmetry. It also raises the question how the antiferromagnetic order and superconductivity coexist microscopically. In single-crystal samples with $T_s\approx0.6\,{\rm K}$ the specific heat varies as $C_e/T=\gamma_s+\beta_s T$ with $\gamma_s=34.1\,{\rm mJ/mol\,K^2}$ and $\beta_s=1290\,{\rm mJ/mol\,K^3}$ below $0.3\,{\rm K}$ \cite{take07}. Moreover, the specific heat displays a nonlinear magnetic field dependence $\gamma_s\propto\sqrt{H}$. When taken together this suggests line-nodes of the superconducting gap.

The presence of line nodes has also been inferred from measurements of the temperature and magnetic field dependence of the thermal conductivity $\kappa(T,H)$ in single crystals \cite{izaw05}. The key results of this study are (i) a residual term in $\kappa(T,0)$ for $T\to0$ in quantitative agreement with the universal conductivity \cite{lee93,graf96,sun95}, (ii) a linear temperature dependence $\kappa(T,0)\propto T$ at low $T$, and (iii) a magnetic field dependence that exhibits one-parameter scaling of the form $T/\sqrt{H}$. This behavior is taken as evidence that the magnetic field dependence is due to a Doppler shift of the quasiparticles (Volovik effect) \cite{vekh99,kueb98,huss02,volo93}.

Measurements of the penetration depth $\lambda(T)$ represent the most direct probe of the superfluid density. They are not connected with any other preponderant interaction process, notably the long range antiferromagnetic order. The experimental data in polycrystals and single crystals displays a broad transition with a point of inflection around 0.5\,K \cite{bona05b,bona07}. Below $\sim 0.165\,T_s$ changes of the penetration depth are linear in temperature, characteristic of line nodes in the gap \cite{haya06}.

While the low temperature specific heat, thermal conductivity and penetration depth only shed light on the behavior well below $T_s$, NMR measurements of the spin-lattice relaxation rate $1/T_1$ and the Knight shift provide insights into the full temperature dependence \cite{yogi04,yogi06}. For the spin-lattice relaxation rate the behavior appears to be a mixture of differing contributions. Near $T_s$ a Hebel-Slichter peak is observed. The variation below $T_s$ clearly deviates from conventional exponential activation, but does not fully settle into a power-law dependence either. The data have been interpreted in two different ways. In the first scenario the temperature dependence is attributed to a mixing of the singlet and triplet states \cite{haya06b}. This accounts for the Hebel-Slichter peak and shows that the low temperature data essentially limits to the $T^3$ dependence expected of line nodes. In the second scenario the interplay of the antiferromagnetic order with the triplet contribution to the superconducting pairing symmetry is considered \cite{fuji06}. In particular it is pointed out that the signatures of line nodes may be found even in fully gaped triplet superconductors in the presence of suitably chosen magnetic order. 

The NMR Knight shift as measured at a field of 2\,T perpendicular and parallel to the field does not change when entering the superconducting state \cite{yogi06}. The Knight shift may be taken as a probe of the spin susceptibility $\chi$. The experimental result are contrasted by the theoretical prediction for a spin-singlet and spin-triplet state in the presence of asymmetric spin-orbit interactions \cite{frig04b}. For the spin-singlet state both  $\chi_{\perp}$ and $\chi_{\parallel}$ are expected to decrease in the superconducting state with some anisotropy, where the decrease gets smaller with increasing $\alpha$. For the spin-triplet state the susceptibility becomes independent of the size of the spin-orbit coupling. Here $\chi_{\parallel}$ shows no decrease, while $\chi_{\perp}$ shows a modest decrease. Thus the experimental absence of any decrease is taken as evidence for the dominant spin-triplet component, where the in-plane behavior awaits further clarification.

In a phase diagram that combines the effects of pressure and substitutional Ge-doping assuming a bulk modulus,  $B_0=1000\,{\rm kbar} $\cite{nick05,take07,tate05,yasu04,baue07} two features may be noticed. First the antiferromagnetism vanishes for pressures in excess of $p_N\approx 6$ to 8\,kbar. Second, the superconductivity decreases with decreasing volume and thus increasing pressure and vanishes around $p_s\approx15$\,kbar. Interestingly the superconducting transition broadens considerable in the range between $p_N$ and $p_s$ in samples with higher $T_s$ \cite{nick05} (note that the crystal structure of {\ceptsige} is not stable for $x>0.06$). An unresolved issue concerns the pressure dependence of samples with the lower $T_s$. In samples with larger $T_s$ the DC susceptibility is rapidly suppressed \cite{moto08b}. This might be the result of the large distribution of lattice constants in these samples, mentioned above.

In any case the phase diagram suggests the vicinity of a quantum critical point.  This is underscored by tentative evidence for critical fluctuations in the specific heat and inelastic neutron scattering of pure {\ceptsi} at ambient pressure. Above $T_N$ electronic contributions to the specific heat decrease as $\Delta C/T\sim\log(T)$. Preliminary inelastic neutron scattering measurements suggest that there is $Q$-independent quasi-elastic scattering at high temperatures typical of conventional heavy-fermion systems. At low temperatures, however, short range correlations are observed for $Q^{-1}\approx0.8\,{\rm \AA}$ at energy transfers of a few meV that may be related to the NFL behavior.

In the light of the possible Rashba splitting of the Fermi surface, the nature of the excitations that mediate the superconductivity is an open issue. In the spirit of the studies of {\cepdsi} and {\cein} the circumstantial evidence suggests that quantum critical spin fluctuations may be a key ingredient. However, due to the lack of inversion symmetry these may include complex textures, like the skyrmion ground states observed recently \cite{mueh09a,neub09}. 

It is finally interesting to point out that several LaMX$_3$ compounds are superconducting. In particular, {\larhsi}, {\lairsi} and {\lapdsi} show superconductivity with $T_s$ of 1.9, 2.7 and 2.6\,K respectively \cite{muro00}. For {\larhsi} the upper critical field is low $H_{c2}=0.03\,{\rm T}$.

\subsubsection{CeMX$_3$
\label{cemx3}}

Following the discovery of superconductivity in the antiferromagnetic state of the non-centrosymmetric heavy-fermion system {\ceptsi} superconductivity was also discovered in the non-centrosymmetric systems {\cerhsiot} \cite{kimu05}, {\ceirsiot} \cite{sugi06} and {\cecogeot} \cite{sett07b,kawa08}. These compounds belong to the class of isostructural CeMX$_3$ systems, where M$=$Co, Ru, Pd, Os, Ir, Pt, Fe, Rh and X$=$Si and Ge. The BaNiSn$_3$ crystal structure, space group $I4mm$ (No. 107), of the CeMX$_3$ series derives from the body-centered tetragonal BaAl$_4$ crystal structure, space group I4/mm. We note that BaAl$_4$ is also the parent structure of the body-centered ThCr$_2$Si$_2$ systems of the heavy fermion superconductors {\cecusi}, {\cecuge}, {\cepdsi}, {\cerhsi} and  {\urusi} (cf Fig.\,\ref{struc-cerhsi3}). It is moreover the parent structure of the series of CaBe$_2$Ge$_2$ body-centered tetragonal ternary systems, none of which have so far been found to be superconducting. As shown in Fig.\,\ref{struc-cerhsi3} the BaSnNi$_3$ structure is composed of a sequence of planes along the c-axis R-M-X(1)-X(2)-R-M-X(1)-X(2)-R, where X(1) and X(2) denote different lattice positions of the X atom. Thus the structure lacks inversions symmetry along the c-axis. The generating point group $C_{4V}$ is identical to that of {\ceptsi}.

\begin{figure}
%\sidecaption
\includegraphics[width=.38\textwidth,clip=]{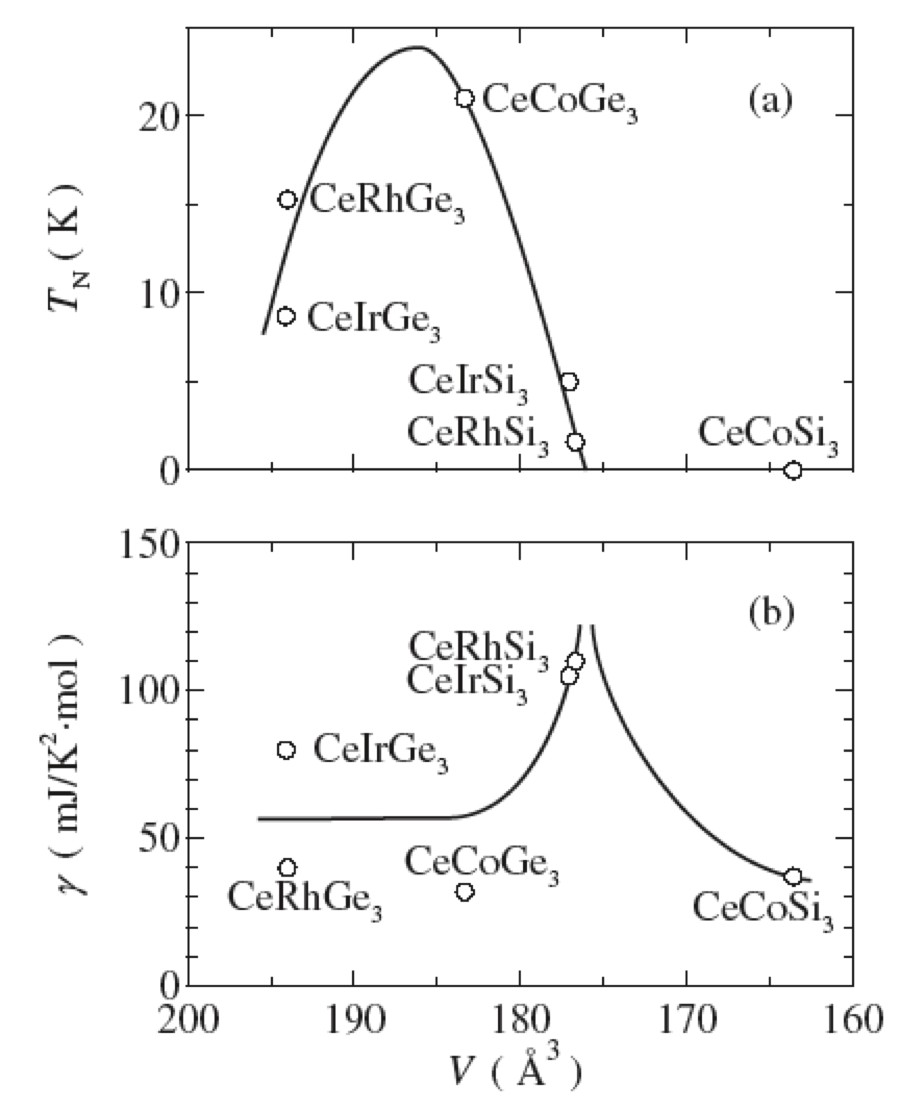}
\caption{(a) Neel temperature $T_N$ versus unit cell volume $V$ in the series CeMX$_3$ (M: Rh, Ir, Co; X: Si, Ge). (b) Sommerfeld coefficient $\gamma$ of the specific heat versus unit cell volume $V$ in the series CeMX$_3$. Plot taken from \cite{kawa08}.}
\label{pd-cemx3} 
\end{figure}

For a brief review of the properties of the series CMX$_3$ we refer to \cite{kawa08}. This review includes considerations on the crystal electric fields, which play a prominent role in the ground state properties. Amongst the systems studied, {\cecogeot} has the highest antiferromagnetic ordering temperature. Interestingly those compounds with low values of $T_N$ have the a-axis as easy magnetic axis, while in {\cecogeot} the c-axis is magnetically soft. It is interesting to note, that the antiferromagnetic transition temperatures, their pressure dependence and the Sommerfeld coefficient of the specific heat as function of decreasing unit cell volume are consistent with a Doniach phase diagram (Fig.\,\ref{pd-cemx3}). So far, superconductivity near a magnetic quantum phase transition has been observed in those systems that are on the right hand border of the phase diagram.

In passing we note that the system CeNiGe$_3$, which also displays superconductivity when antiferromagnetism is suppressed at high pressure \cite{naka04c,kote06}, crystallizes in a centro-symmetric orthorhombic structure (see section \ref{misc-ce-sys}).

\paragraph{{\cerhsiot}}
The first system in this class for which superconductivity was observed is {\cerhsiot}. For a recent review we refer to \cite{kimu07a}. The lattice constants are $a=4.244\,{\rm \AA}$ and $c=9.813\,{\rm \AA}$ and the single crystals studied had very low residual resistivities of a few tenths of a ${\rm \mu\Omega\,cm}$. At ambient pressure {\cerhsiot} orders antiferromagnetically below $T_N=1.6\,{\rm K}$. The antiferromagnetic order is anisotropic, where the basal plane a-axis is the easy axis. Neutron scattering shows that the antiferromagnetic order is incommensurate with $\vec{Q}=(\pm0.215, 0, 0.5)$ \cite{aso07}. The magnetic field dependence of the magnetization suggests an anisotropy of about 2. Only a small magnetization is seen up to 8\,T. At high temperatures an isotropic Curie-Weiss susceptibility is observed with a fluctuating moment $\mu_{eff}=2.65\,{\rm \mu_B}$ as expected of the full Ce$^{3+}$ moment. 

The specific heat is interpreted in terms of a Schottky anomaly around 100\,K and Kondo temperature of order $T_K\approx50\,{\rm K}$. The Kondo screening is affected by the CEF level scheme, where inelastic neutron scattering has been interpreted as three doublets with a $\Gamma_6$ ground state and $\Gamma_7$ and $\Gamma_6$ first and second excited state at 260 and 270\,K, respectively \cite{muro07}. The low temperature specific heat above $T_N$ exhibits a strongly enhanced Sommerfeld contribution $\gamma=0.110\,{\rm J/mol\,K^2}$. This suggests that a heavy fermion state forms despite a large Kondo temperature, in which incommensurate antiferromagnetism stabilizes at very low temperatures. The Fermi surface has been investigated by means of quantum oscillations \cite{kimu07a,kimu01} and compared to {\larhsiot}. Substantial differences are interpreted as evidence for an itinerant f-electron and spin density wave type of antiferromagnetism. Moreover, several branches show a small splitting with similar angular dependences. This is seen as evidence for Rashba splitting.

The pressure dependence of $T_N$ in {\cerhsiot} is quite unique. Up to 9\,kbar $T_N$ increases moderately before decreasing again gradually up to 20\,kbar.  For pressure above 2\,kbar \cite{kimu05,kimu07a} superconductivity emerges in the antiferromagnetic state, where $T_s$ increases up to 30\,kbar, the highest pressure measured. The superconducting dome is exceptionally wide. The AC susceptibility shows susperconducting screening with additional features that require further clarification. Together with the zero resistance state the susceptibility is a strong indication of superconductivity. However, it does not establish spontaneous Meissner flux expulsion and thus volume superconductivity.  The initial slope of $H_{c2}$ is strongly enhanced and becomes anomalously large around 26\,kbar with  $dH_{c2}/dT\vert_{T_s}=-23\,{\rm T/K}$. This suggests that $H_{c2}$ is exceptionally large and may even exceed 30\,T \cite{kimu07b}. 

\paragraph{{\ceirsiot}}
The compound {\ceirsiot} is isostructural to {\cerhsiot} with lattice constants $a=4.252\,{\rm \AA}$ and $c=9.715\,{\rm \AA}$ \cite{muro98}. The ambient pressure properties of {\ceirsiot} are characteristic of a heavy-fermion system with an enhanced Sommerfeld contribution to the normal state specific heat $\gamma=0.12\,{\rm J/mol\,K^2}$ and antiferromagnetic order below $T_N=5.0\,{\rm K}$. The CEF levels have been inferred from magnetization data, where the ground state is a $\Gamma_6$ doublet and the first and second excited states are $\Gamma_7$ and $\Gamma_6$ doublets at 149\,K and 462\,K, respectively \cite{okud07}. The magnetization is anisotropic by a factor of about two, where the a-axis is the easy axis. Quantum oscillatory studies of {\lairsi} suggest show that the Fermi surface is similar to that of LaCoGe$_3$, characteristic of a compensated metal where branches with an exchange splitting of 1000\,K exhibit an angular dependence that track each other rather closely. This suggests the presence of Rashba splitting due to the lack of inversion symmetry. 

\begin{figure}
%\sidecaption
\includegraphics[width=.35\textwidth,clip=]{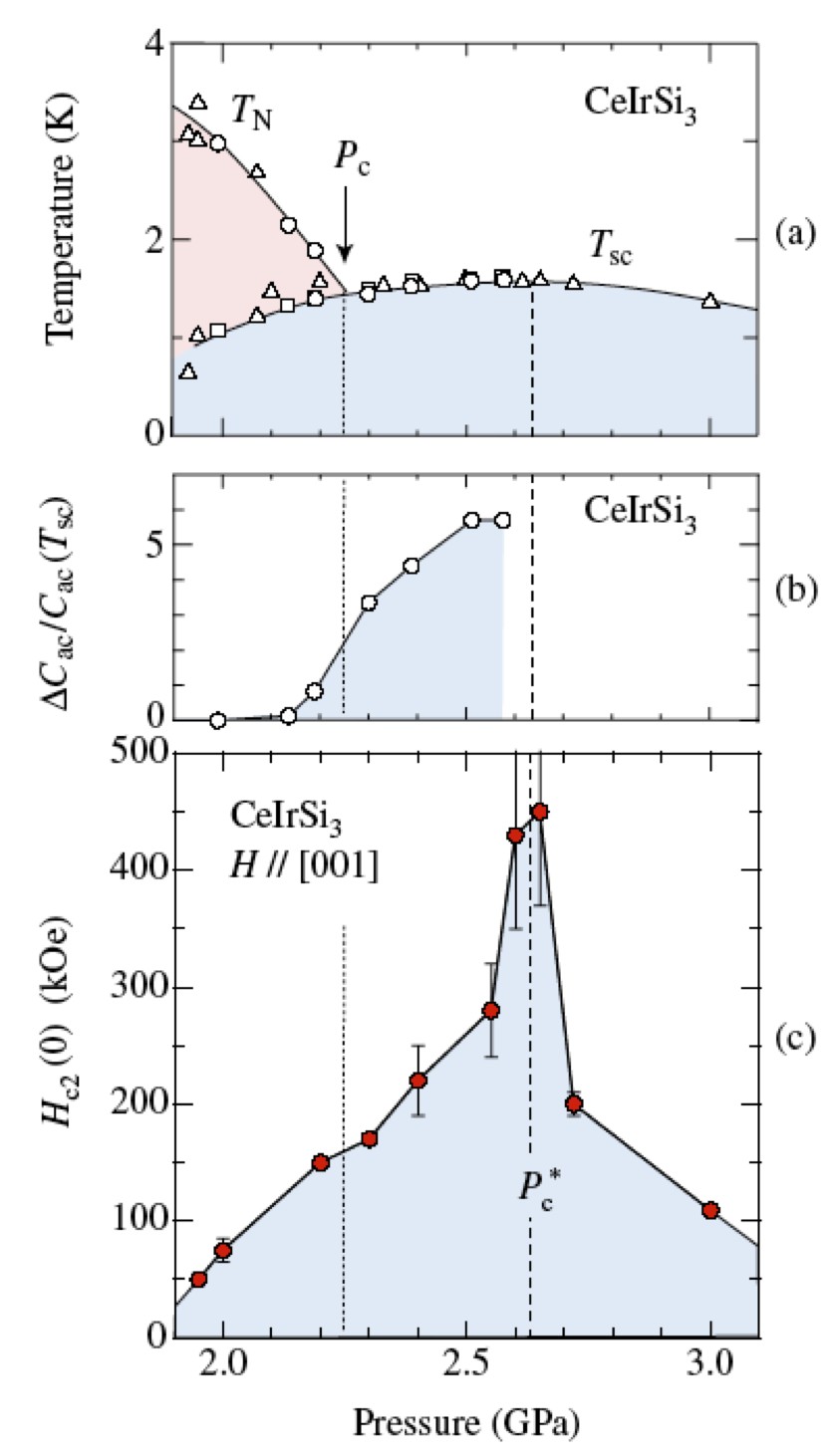}
\caption{(a) Temperature versus pressure phase diagram of {\ceirsiot}. At the border of antiferromagnetic order a wide superconducting dome emerges. Note that the pressure axis begins at 19\,kbar. (b) Superconducting specific heat anomaly as a function of pressure. (c) Extrapolated zero temperature upper critical field. Plot taken from \cite{sett08}.}
\label{pd-ceirsi3} 
\end{figure}

The antiferromagnetism in {\ceirsiot} vanishes for pressures in excess of $p_N=22.5$\,kbar \cite{sugi06,tate07} and a superconducting dome emerges, with $T_s^{max}=1.65\,{\rm K}$ for pressure in excess of $p_N$ as shown in Fig.\,\ref{pd-ceirsi3}. $H_{c2}$ exhibits a strong temperature dependence near $T_s$. For the basal plane $dH_{c2}^{ab}/dp=-13\,{T/K}$ at $p_N$ with $H_{c2}^{ab}(T\to0)=9.5\,{\rm T}$. For the c-axis $dH_{c2}/dp=-20\,{T/K}$ and $H_{c2}^c$ reaches 18\,T just below 1\,K, suggesting a extremely large value in excess of 30\,T \cite{sett08}. This is strikingly similar to {\cerhsiot}. Recent specific heat and AC susceptibility measurements up to 35\,kbar show distinct specific heat anomalies for both the antiferromagnetic and superconducting transitions, i.e. they may be tracked very well as a function of pressure using an AC method, but quantitative information is not available \cite{tate07}. Above $p_N$ the specific heat anomaly is particularly pronounced and suggests strong coupling superconductivity. NMR studies show the absence of a coherence peak in the spin lattice relaxation rate and a cubic temperature dependence characteristic of line nodes \cite{muku08}. The normal state spin-lattice relaxation rate is thereby characteristic of an abundance of antiferromagnetic spin fluctuations, which are likely to be implicated in the superconducting pairing.

Rather remarkable is the behavior observed under substitutional doping in {\ceircosi} \cite{okud07}. Replacing Ir with Co represents to leading order a reduction of unit cell volume equivalent to the application of pressure. For $x=0.2$ and $x=0.35$ the N{\'e}el temperature is reduced and superconductivity is observed. Metallurgical tests suggest that the compound for $x=0.35$ is not single phase with a dominant contribution also of the $x=0.2$ phase. In any case the result suggests that superconductivity is not particularly sensitive to disorder.

\paragraph{{\cecogeot}}
Amongst the CeMX$_3$ compounds {\cecogeot} has the highest magnetic ordering temperature $T_{N1}=21\,{\rm K}$, followed by two more transitions at $T_{N2}=12\,{\rm K}$ and $T_{N3}=8\,{\rm K}$ \cite{sett07b,kawa08}. The metallic state is well described as a moderately enhanced Fermi liquid with a Sommerfeld coefficient of the specific heat $\gamma=0.032\,{\rm J/mol\,K^2}$ and a coefficient of the quadratic temperature dependence of the resistivity $A=0.11\,{\rm \mu\Omega cm/K^2}$. The easy magnetic axis is the c-axis, as opposed to other members of the CeMX$_3$ series, where the a-axis is the easy axis. Under pressure $T_{N1}$ decreases and vanishes around 55\,kbar, where the rate of suppression drops around 30\,kbar. Superconductivity is observed in the range 54 to 75\,kbar with $T_s=0.69\,{\rm K}$ at a pressure around 65\,kbar. For this pressure the $H_{c2}$ along the c-axis, as extrapolated from the very large increase near $T_s$, given by $dH_{c2}=-20\,{\rm K/T}$, is exceptionally large and may reach 45\,T.  

The Fermi surface of the series LaTGe$_3$ (T: Fe, Co, Rh, Ir) has been reported in \cite{kawa08b}. All systems exhibit strong Rashba spin-orbit splitting. It will be interesting to see how the characteristics of these superconductors relate to those of the Ce-systems. For instance, the La-compounds may display the singlet state superconductivity to which the triplet state pairing gets admixed in the Ce-systems.

We finally note that superconductivity has also been reported in CeCoSi$_3$ at 0.5\,K \cite{haen85}. However this observation could not be confirmed down to 50\,mK in a subsequent study \cite{eom98}.

\subsubsection{UIr
\label{uir}}

Superconductivity in non-centrosymmetric heavy-fermion systems also exists at the border of ferromagnetism in UIr \cite{akaz04a,akaz04b}. The structure of UIr is monoclinic of PbBi-type (space group P2$_1$) and lacks inversion symmetry \cite{domm88}. Four different uranium sites may be distinguished. In the paramagnetic state the susceptibility follows a Curie-Weiss dependence with an effective moment ${\rm \mu_{eff}=2.4\mu_B/U}$. Below a Curie temperature $T_{C1}=46\,{\rm K}$ Ising ferromagnetism develops with a reduced ordered moment of $0.5\mu_B/U$, characteristic of itinerant electron magnetism. The easy axis is  $[10\bar{1}]$. The properties are summarized in table \ref{table-ubased}.

\begin{figure}
%\sidecaption
\includegraphics[width=.38\textwidth,clip=]{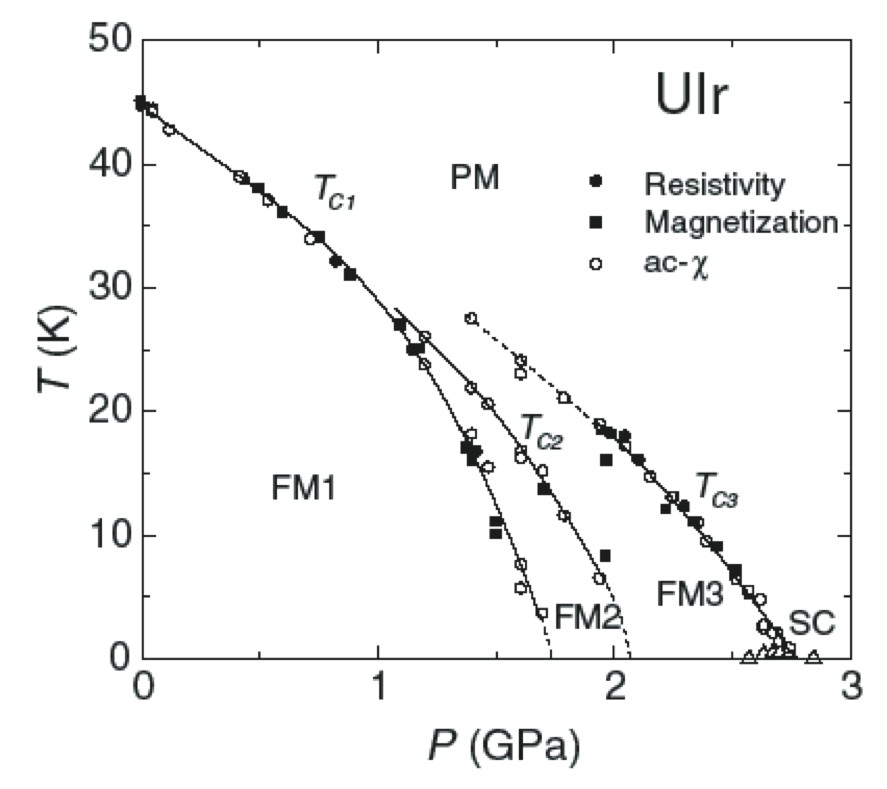}
\caption{Temperature versus pressure phase diagram of UIr. Three ferromagnetic phases have been identified. Superconductivity is only observed at the border of the FM3 phase with very low superconducting transition temperatures. Plot taken from \cite{koba07}.}
\label{pd-uir} 
\end{figure}

A recent review of the temperature-pressure-magnetic field phase diagram of UIr may be found in \cite{koba07}. Several samples of varying quality have been studied so far, where an indenter pressure cell was used. The pressure technique leaves room for uncertainties regarding the possible role of non-hydrostatic conditions. As shown in Fig.\,\ref{pd-uir}, the resistivity, AC susceptibility and magnetization establish, that three magnetic phases may be distinguished under pressure. Data were mostly collected for the $[10\bar{1}]$ easy axis and $[010]$ hard axis. The nature of the magnetic states has not been identified by means of microscopic probes yet. Based on the available bulk data the phases are referred to as ferromagnetic states.

Under pressure the FM1 state vanishes for pressure in excess of $p_{c1}=17\,{\rm kbar}$. The transition at $T_{c1}$ may be readily seen in the resistivity, AC susceptibility and magnetization.  The ordered moment decreases gradually between $0.5\mu_B/$U and $0.27\mu_B/$U before dropping discontinuously at $p_{c1}$. In the limit $T\to0$ the FM2 phase exists between $p_{c1}$ and $p_{c2}=21\,{\rm kbar}$. As a function of temperature the FM2 transition may be seen in the AC susceptibility, but not in the resistivity. The ordered moment in the FM2 state is strongly reduced and not larger than $0.05\mu_B/$U. The FM3 phase exists in the limit $T\to0$ for pressures up to $p_{c3}=27.5\,{\rm kbar}$. The ordered moment in the FM3 phase vanishes continuously at $p_{c3}$, where the behavior between $p_{c1}$ and $p_{c3}$ is complex with the possibility of a metamagnetic transition from the FM2 to the FM3 phase. The magnetic ordering temperature of the FM3 phase at $T_{c3}$ may be seen in the resistivity and AC susceptibility. As a rather peculiar feature of the FM3 phase $T_{c3}(p)$ is not directly connected with either $T_{c1}(p)$ or $T_{c2}(p)$, but begins in the middle of the paramagnetic regime as shown in Fig.\,\ref{pd-uir}. Clearly, based on symmetry considerations there must be another transition line along which the symmetry breaking takes place.

Superconductivity is observed in the FM3 phase of UIr for pressures in the range $26\,{\rm kbar} < p < p_{c3}$ \cite{akaz04a,akaz04b}, reaching $T_s=0.15\,{\rm K}$ and $H_{c2}=0.0258\,{\rm T}$ is quite low. The superconductivity has been seen in the resistivity and AC susceptibility, i.e., bulk superconductivity has not been established yet. No superconductivity is observed in the paramagnetic regime above $p_{c3}$. Also, the superconductivity is only observed for samples with fairly high residual resistivity ratios (${\rm > 170}$). The observation of superconductivity in the ferromagnetic state and for high purity samples suggests unconventional pairing. A possible Ir-based impurity phase has been ruled out on the basis of the pressure dependence of $T_s$ of Ir, which does not match or track the behavior observed experimentally.

We finally return to the question of the nature of the FM1, FM2 and FM3 phases. The FM1 phase appears to be a straight-forward Ising ferromagnet. In contrast, the dominant feature of the FM2 phase is a 25-fold increase of the residual resistivity for the magnetically hard $[010]$ axis \cite{hori06} and a strongly reduced spontaneous moment. Moreover, quantum oscillations vanish outside the FM1 phase \cite{shis06}. This let to the speculation of a multilayer-like phase separation along the $[010]$ axis. It is presently not clear, whether this structure is related to a structural modification, so far not supported by high pressure x-ray diffraction. The easy and hard axis of the magnetization are unchanged in the FM3 phase, which supports the superconducivity at low temperatures \cite{koba07}. Finally, the FM3 phase again appears to be a straight forward Ising ferromagnet with strongly reduced ordered moment. It has been argued that there is no additional modulation in the FM3 state, because the easy and hard axis are unchanged. Finally, it appears unlikely that the crystal structure has recovered the centro-symmetric symmetry under pressure, because this would require major rearrangements of the atomic positions. The ordered magnetic moment in the FM3 phase ($\sim0.05\,{\rm \mu_B/U}$) corresponds to a fairly small internal field, also consistent with conventional superconductivity. Also, the coherence length of $\xi=1100\,{\rm \AA}$ as inferred from $H_{c2}$ is comparable to the charge carrier mean free path of $l=1240\,{\rm \AA}$. The role of the different U-sites has not been addressed at all. Clearly the interplay of magnetism and superconductivity in UIr poses a large number of experimental and theoretical challenges for the future. 

\subsection{Superconductivity near electron localization}
\label{supe-near-elec-loca}

The degree of itineracy of the f-electrons in intermetallic compounds provides a major source of scientific debate. The transition from an itinerant to a localized state creates variations in the charge density that also drive strong correlations in the spin density. Interestingly, heavy fermion superconductivity is found in materials at the border of such a localization transition. This suggests that the nature of the superconductive interactions is related to charge density fluctuations as a new route to superconductivity. The interplay of these fluctuations with spin fluctuations and further degrees of freedom is an open issue.

\subsubsection{Border of valence transitions
\label{supe-vale-inst}}

It has recently been suggested that the superconductivity maximum in {\cecusi} at high pressures is related to a Ce$^{3+}$ to Ce$^{+4}$ valence transition (cf Fig.\ref{PD-CeCuSiGe}), where the 4f electron is delocalized in the high pressure Ce$^{4+}$ state \cite{holm04,yuan04}. This type of QPT transition is non-symmetry breaking in the spirit of itinerant-electron metamagnetism. The suggestion was inspired by the analogy of the temperature versus pressure phase diagrams of {\cecuge} and {\cecusi}. In {\cecuge} x-ray diffraction suggests a valence transition at a pressure $p_{c2}\approx15\,{\rm GPa}$ \cite{onod02}. However, there is no microscopic evidence for a valence transition in {\cecusige} except for faint features seen in the $\rm L_{III}$-x-ray absorption \cite{roeh88} and changes of the metallic state notably the electrical resistivity. 

Studies of the magnetic phase diagram under pressure establish, that the $T^2$ coefficient of the resistivity qualitatively tracks $dH_{c2}/dT$ up to $\sim4.5\,{\rm GPa}$, but drops to a value about two orders of magnitude smaller above $\sim4.5\,{\rm GPa}$ \cite{varg98}. Further studies established, that the $T^2$ coefficient of the resistivity drops abruptly, when the characteristic temperature scale $T_1^{\rm max}$ varies under pressure or Ge-doping reaches a value of $\sim70\,{\rm K}$  \cite{holm04}. Under the same conditions a five-fold enhancement of the residual resistivity is observed and a tiny maximum in the specific heat coefficient.

It is conceivable that the superconductivity in {\cecusi} at high pressure develops with a rather different pairing symmetry. A microscopic pairing mechanism has been proposed in which the pairing is dominantly mediated by the exchange of charge fluctuations between the conduction bands and the f-site \cite{onis04}. In the limit of a spherical Fermi surface and weak coupling this model predicts a d-wave superconducting state, where the value of $T_c$ scales with the slope of the continuous valence transition as a function of the f-level energy.

\subsubsection{Plutonium and neptunium based systems}
\label{acti-loca}

Another surprise in recent years has been the discovery of heavy-fermion superconductivity in the actinide compounds {\pucoga} \cite{sarr02}, {\purhga} \cite{wast03} and {\nppdal} \cite{aoki07}. The properties of these systems are summarized in table\,\ref{table-cemin}. Status reports for {\pucoga} and {\purhga} have been given by \cite{thom06a,thom06b,thom06c,sarr07,haga07}. The striking feature about the superconductivity in  {\pucoga}, {\purhga} and {\nppdal} are values of $T_s$ of 18.5, 8.7 and 4.9\,K, respectively, which are the highest of all f-electron systems. It seems natural to assume that the key ingredients responsible for the high transition temperatures in these systems are related to the special electronic properties of the 5f electrons in the elements. 

First, plutonium is delicately placed at the border between a large and small Wigner-Seitz radius characteristic of the transition between delocalized and localized f-electrons. Second, because Coulomb screening is stronger for 4f than 5f electrons, the typical band width of 5f systems is intermediate between 3d and 4f systems. Moreover, the effects of spin-orbit coupling in 5f systems vary quite strongly along the series and change from weak for U to very strong for Pu, Am and Cm \cite{moor09}. Qualitatively this suggests that certain 5f superconductors are intermediate between the traditional 4f heavy fermion superconductors and 3d high-$T_c$ superconductors. This conjecture is strongly supported by the experimentally observed properties, especially when plotting $T_s$, versus a temperature characteristic of the electronic correlations $T_0$ (cf. the band width). 

\paragraph{{\pucoga} \& {\purhga}}
Both {\pucoga} and {\purhga} crystallize in the tetragonal {\hocoga} structure, space group P4/mmm \cite{sarr02,wast03}. The structure is identical to the series of {\cemin} compounds reviewed in section \ref{cemin-system} and derives from the cubic {\hoga} in terms of {\mga} layers stacked sequentially along the $[100]$ axis (for further information see \cite{wast03}). The normal state of both of {\pucoga} and {\purhga} is characterized by a Curie-Weiss susceptibility with an effective fluctuating moment $\mu_{eff}\sim0.75\,{\rm \mu_B/Pu}$, respectively.  The effective moment is close to the 5f$^5$ (Pu$^{3+}$) configuration of $0.84\,{\rm \mu_B}$.  In {\pucoga} the Curie-Weiss temperature, $\Theta=-2\,{\rm K}$, is remarkably low \cite{sarr02}. Above $\sim100\,{\rm K}$ the effective moment in {\purhga} assumes the free ion value \cite{haga07}. The susceptibility exhibits Curie-Weiss behavior throughout the normal state. The temperature dependence of the electrical resistivity is anomalous, showing a power law dependence $\propto T^{n}$ with $n\sim1.35$ instead of the conventional quadratic temperature dependence of an enhanced Fermi liquid. In both systems the specific heat is well described as that of a heavy Fermi liquid state plus lattice term $C(T)=\gamma T+\beta T^3$, where $\gamma=0.077\,{\rm J/mol\,K^2}$ and $\gamma=0.07\,{\rm J/mol\,K^2}$ for {\pucoga} and {\purhga}, respectively. The value for $\beta$ corresponds to a Debye temperature $\Theta_D\sim240\,{\rm K}$ for {\pucoga} and {\purhga}.

Below $T_s=18.5\,{\rm K}$ {\pucoga} exhibits superconductivity. The initial change of the upper critical field near $T_s$ in polycrystals is unusually large $dH_{c2}/dT=-5.9\,{\rm T/K}$. This implies $H_{c2}=74\,{\rm T}$ \cite{wert66}, which exceeds the estimated Pauli limit ($H_P=34\,{\rm T}$) by a factor of two. The estimated value of $H_{c2}$ corresponds to a Ginzburg-Landau coherence length $\xi_{GL}\approx21\,{\rm \AA}$. The heat capacity confirms bulk supercoductivity, where the size of the anomaly, $\Delta C/\gamma T_s=1.4$, is consistent with conventional BCS superconductivity. Further specific heat studies in single-crystals confirm these conjectures and show a quadratic temperature dependence, consistent with an axial gap symmetry with line nodes \cite{javo07}. This study also establishes the possibility of an FFLO state in {\pucoga}, where a large Maki parameter $\alpha$ is inferred. The magnetization is characteristic of strong type II superconductivity. 

It has been noticed that the anisotropy of the superconductivity in {\pucoga} and {\purhga} to an applied magnetic field qualitatively matches the anisotropy of the antiferromagnetism in NpCoGa$_5$ and NpRhGa$_5$ to an applied magnetic field \cite{wast06,coli06}. This supports the notion that the magnetic interactions arise on the same grounds than the superconductivity. However, using polarized neutron scattering orbital and spin contributions to the Curie-Weiss susceptibility have been discriminated \cite{hies08}. While the microscopic magnetization in {\npcoga} agrees with the bulk susceptibility, there is a large discrepancy in {\pucoga}. In fact, the polarized neutron scattering data imply that orbital contributions to the fluctuating moment are dominant. In turn this suggests that the superconductivity is not straight forwardly due to the more traditional versions of spin fluctuation mediated pairing.

Microscopic evidence for unconventional superconductivity has been inferred from measurements of the $^{69,71}$Ga and $^{59}$Co Knight shift $K_s$ and nuclear spin-lattice relaxation rate $T_1$ in {\pucoga} \cite{curr05,saka06} and {\purhga} \cite{saka05,bang06}.  The Knight shift provides information on the orbital susceptibility $\chi_o$, which is essentially constant, and the spin susceptibility, $\chi_s$, which decreases in the superconducting state. This clearly identifies {\pucoga} and {\purhga} as spin-singlet d-wave superconductors. The spin-lattice relaxation rate in both systems drops abruptly when entering the superconducting state without evidence of a Hebel-Slichter peak. Below $T_s$ the relaxation rate initially varies as $T_1^{-1}\propto T^3$ and settles into a dependence $T_1^{-1}\propto T$ at the lowest temperatures, presumably due to impurity scattering. 

\begin{figure}
%\sidecaption
\includegraphics[width=.38\textwidth,clip=]{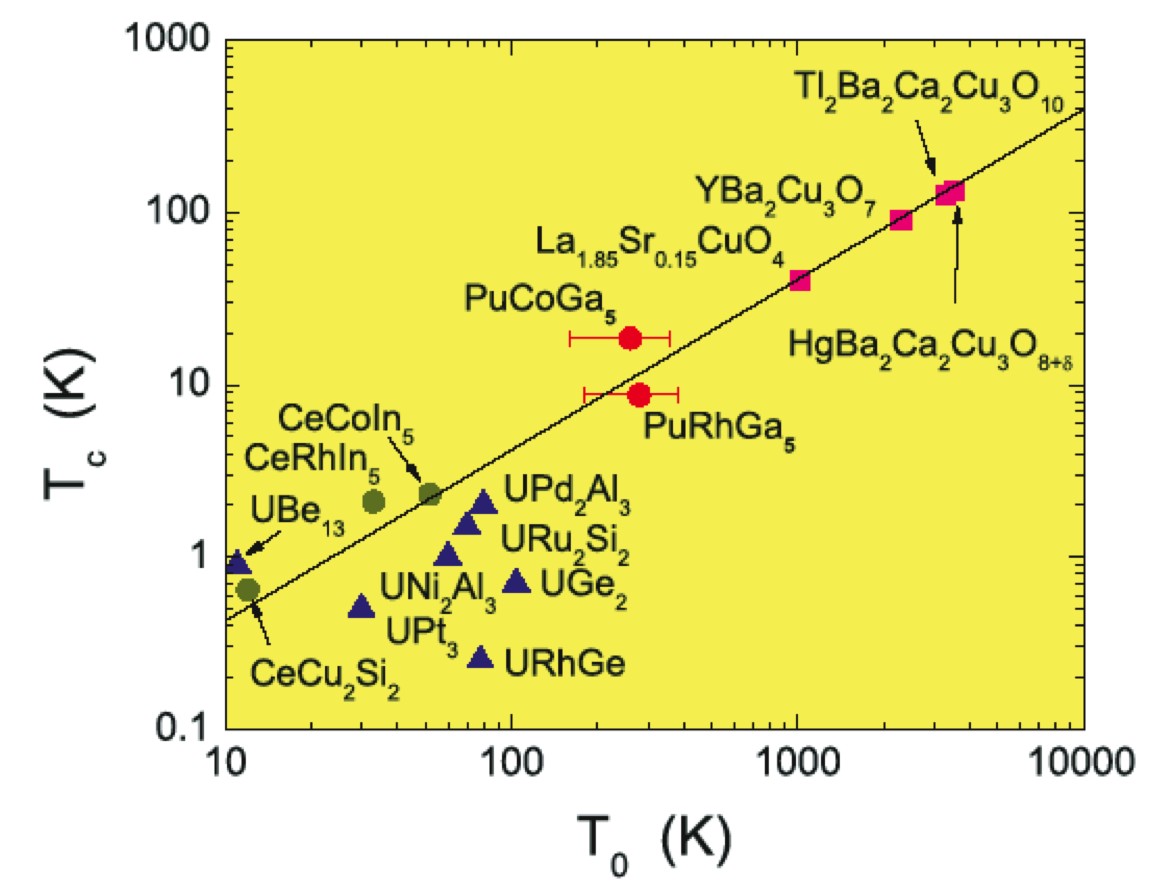}
\caption{Comparison of superconducting transition temperature with the characteristic spin fluctuation temperature. The latter is essentially a band width and may be insensitive to the precise microscopic nature of the correlations. Plot from \cite{curr05} as shown in \cite{sarr07}.}
\label{tc-tsf} 
\end{figure}

The spin-lattice relaxation in {\pucoga} and {\purhga} differs markedly from conventional electron-phonon mediate superconductivity observed in Al or MgB$_2$, corresponding to the predictions of antiferromagnetically mediated superconductive pairing and scales with the behavior observed in {\cecoin} and YBa$_2$Cu$_3$O$_7$. Thus the observed form of $T_1$ suggests common microscopic features of the superconductivity for materials with vastly different values of $T_s$ which, however, are all strong contenders for antiferromagnetic pairing. In fact, when plotting $T_s$ versus spin fluctuation temperature, which measures the effective band width, {\pucoga} and {\purhga} are found to be intermediate to the class of 4f heavy fermion superconductors and the 3d high $T_c$ cuprates as shown in Fig.\,\ref{tc-tsf}. Interestingly the temperature dependence of $T_1$ in the normal state of {\purhga} deviates from that observed in {\pucoga}, {\cecoin} and YBa$_2$Cu$_3$O$_7$. This has been interpreted as  a pseudogap consistent with the canonical phase diagram of a superconducting dome surrounding a quantum phase transition.

The strong radioactivity of Pu imposes several experimental constraints. Self-heating generates a considerable heat load that does not allow to perform experiments at very low temperatures. A typical value is $\sim 0.45\,{\rm \mu W}$ per mg for {\purhga}. More important is the structural damage incurred by the radioactive decay of $^{239}$Pu, which results in a uranium nucleus and a high-energy alpha particle. The uranium nucleus displaces by a mean distance of 120\,{\AA} and creates on average of 2300 Frenkel pairs of vacancies and displaced interstitials distributed over a range of 75\,{\AA} \cite{wolf00}. 

Several studies have addressed the effects of self-irradiation \cite{ohis07,juti06,ohis06,juti05,boot07}, which may be seen as an unique opportunity to study the evolution of a superconducting state as a function of increasing defect concentration. Experimentally it is observed that $T_s$ decreases in both compounds under self-irradiation, where $\Delta T_s/\Delta t \sim -0.39\,{\rm K/month}$ for {\purhga} and $\Delta T_s/\Delta t \sim -0.24\,{\rm K/month}$ for {\pucoga} \cite{juti05}. For doped samples with PuCo$_{0.1}$Rh$_{0.9}$Ga$_5$ and PuCo$_{0.5}$Rh$_{0.5}$Ga$_5$ the rates of decrease are intermediate \cite{juti06}. $H_{c2}$ and the critical current density show more complex behavior. The initial variation of $H_{c2}$ near $T_s$ increases in 553 days for {\pucoga} from $dH_{c2}/dT=-5.5\,{\rm T/K}$ to $dH_{c2}/dT=-13\,{\rm T/K}$, while is decreases strongly for {\purhga} from $dH_{c2}/dT=-3.4\,{\rm T/K}$ to $dH_{c2}/dT=-0.8\,{\rm T/K}$. The same trends are reflected in the critical currents. These studies suggest that self-irradiation generates point defects, where defects of the size of the coherence length are known to represent effective pinning centers \cite{camp72}.

The nature of the damage caused by self-irradiation has been studied microscopically by $\mu$-SR \cite{ohis06,ohis07}. The $\mu$SR line widths are found to narrow dramatically with increasing self-irradiation. This is seen as the result of an abundance of pinning centers that trap flux lines thereby reducing the internal field distribution. The absolute value of the penetration depth as inferred from the $\mu$SR data strongly depends on the defect concentration. Yet, the low temperature variation, $\lambda\propto T$, consistently shows d-wave behavior for the prestine and the irradiated samples. When taken together this suggests that the superconducting state is rather robust against the damages incurred by self-irradiation. 

Finally, when monitoring the consequences of self-irradiation over a period of four years, the degradation of the superconductivity actually deviates from a strictly linear behavior \cite{juti08}. This deviation has been explained in the framework of an Eliashberg theory of a 'dirty' d-wave superconductor, consistent with the NMR measurements. These authors point out that a phononic mechanism reproduces the experimental data, leaving open the role of the spin and orbital fluctuations.

We now turn to the possible nature of the superconductivity in {\pucoga} and {\purhga}. The Curie-Weiss dependence may be taken as evidence of localized 5f electrons. Yet, PuIn$_3$, shows a similar strong Curie-Weiss susceptibility, but quantum oscillatory studies establish that the 5f electrons are in an itinerant state. An itinerant f-electron state in {\pucoga} and {\purhga} is also supported by the temperature dependence of the resistivity. This is supported further by band structure calculations for {\pucoga} in the local density approximation which suggest that the origin of the high value of $T_s$ indeed lies in the 5f electrons \cite{opah03,maeh03,szaj03}. 

Also, a comparison of the resistivity of the series ACoGa$_5$ (A$=$U, Np, Pu and Am) establishes that the resistivity for the systems {\cecoin}, {\pucoga} and UCoGa$_5$ scale with each other characteristic of a single spin fluctuation energy. Moreover, the physical properties of the ACoGa$_5$ systems suggests that {\pucoga} resides near an itinerant to localized crossover of the 5f electrons that is reminiscent of the itinerant to localized crossover that occur near Pu in the actinide series \cite{moor09}. The peculiar emergence of the superconductivity out of a metallic state with strong Curie-Weiss susceptibility has inspired theoretical considerations concerning the symplectic symmetry of the spin in {\pucoga} and {\npcoga} and how a coupling of local spins with the conduction electrons may promote superconductivity \cite{flin08}. 

The specific heat of {\pucoga} suggests a Debye temperature $\Theta_D=240\,{\rm K}$, which, using the McMillan equation with a Coulomb pseudopotential $\mu^*=0.1$ and weak electron-phonon coupling $\lambda=0.5$ and $\lambda=1$, suggests $T_s\approx2.5\,{\rm K}$ and $\sim14\,{\rm K}$, respectively \cite{thom04}. Thus conventional electron-phonon mediated pairing cannot be ruled out. However, it is difficult to reconcile it with the large fluctuating magnetic moments seen in the normal-state susceptibility. Moreover, because the temperature dependence of the resistivity is best explained in terms of scattering by antiferromagnetic spin fluctuations it has been concluded that superconductivity in {\pucoga} is unconventional. In fact, taking into account the presence of defects as measured by the residual resistivity $\rho_0=20\,{\rm \mu\Omega cm}$ transition temperatures as high as $\sim40\,{\rm K}$ may be expected \cite{bang04}. 

The lattice dynamics of {\pucoga} was studied experimentally by room temperature inelastic x-ray scattering \cite{raym06} and compared to first principles calculations using the generalized gradient approximation (GGA) in density functional theory  \cite{piek05}. Excellent quantitative agreement was obtained when the on-site Coulomb repulsion was taken into account with $U=3\,{\rm eV}$ (GGA+U) and Hund's rule exchange. The estimated averaged electron-phonon constant is calculated to be $\lambda=0.7$ \cite{piek05}. In the Allen-Dynes or equivalently McMillan formalism this value of $\lambda$, when taken together with the Debye temperature and a pseudo-Coulomb interaction $\mu*$ below 0.1, implies $T_s$ to be in the range 7 to $14\,K$. In other words electron-phonon coupling alone cannot be responsible for the superconductivity in {\pucoga}. However, the detailed understanding of electron-phonon interactions in {\pucoga} and {\cecoin} requires to resolve also why UCoGa$_5$ is not superconducting even though the phonon spectra are similar.

A dual nature of the 5f electrons was inferred from a photoemission study of {\pucoga} \cite{joyc03}, where excellent agreement with a so-called mixed-level calculation (MLL) in density functional theory was observed. In this calculation one f-electron is in an itinerant state and four f-electrons are localized 1.2\,eV below $E_F$. The data are in stark contrast with the predictions of purely itinerant f electrons in a generalized gradient approximation (GGA). The conclusion of the MLL calculation has been questioned by a first principles calculation of the ground state \cite{soed04}. It transpires that the photoemission spectra can be accounted for by fully itinerant f electrons when the spin and orbital degrees of freedom are allowed to be correlated. 

Using relativistic linear augmented-plane-waves the Fermi surface was found to be dominated by several large cylindrical f-electron sheets in fair agreement with the Fermi surface of CeMIn$_5$ \cite{maeh03}. In particular, the band width of the 5f electrons is intermediate to typical 3d and 4f systems.  While the calculated Fermi surface of {\pucoga} and CeMIn$_5$ (M=Co, Ir, Rh) is similar it differs from the calculated Fermi surface of the pair of actinide systems UCoGa$_5$ and NpCoGa$_5$, which consists of several small sheets plus a single large sheet for the case of NpCoGa$_5$. 

The similarities of the Fermi surface in {\pucoga} and CeMIn$_5$ can be explained in terms of  a tight-binding calculation taking into account $j$-$j$ coupling \cite{maeh03,hott03}. The analogy may be traced to the pseudo-spin representation of the $j$-$j$ coupling, where one electron exists in the $j=5/2$ sextett for Ce$^{3+}$, while there is one hole for the five electrons of Pu$^{3+}$ in the sextett. Thus Pu$^{3+}$ may be viewed as the hole analogue of the one electron state of Ce$^{3+}$. The increased value of $T_s$ may then be attributed to the increased width of the 5f bands, where an additional role of the orbital structure of the Pu systems is likely.

The role of the transition metal element in controlling the nature of the ground state in {\pucoga} and related compounds has been explored experimentally by means substitutional replacement of Pu by U and Np and of Co by Fe, Rh and Ni \cite{boul05b}. Superconductivity is most dramatically suppressed for U and Np substitution, while isoelectronic substitution is the least destructive. These results are theoretically underpinned by DFT calculations in the full-potential linear-muffin-tin-orbital (FP LMTO) approximation, where the transition metal element does not contribute directly to the density of states at the Fermi level \cite{oppe06}. Rather the transition metal effectively hole- or electron-dopes the Pu atom.

Ab initio total energy calculations in the local spin density approximation suggest antiferromagnetic ground states for {\pucoga} and {\purhga} \cite{opah04}. When taking into consideration that LSDA calculations do not treat correlation effects properly these results suggest that {\pucoga} and {\purhga} are at least close to antiferromagnetic order. The effects of Coulomb correlations have been addressed in a study using the relativistic LSDA+U \cite{shic05}. This study unexpectedly shows a considerable reconstruction of the LSDA results suggesting $j$-$j$ like coupling for the Pu 5f manifold similar to what is observed for pure Pu metal. The dynamical mean field theory (DMFT), finally, suggests an important role of van Hove singularities in the $\vec{k}$-resolved spectral density that may provide strong enhancements of the magnetic susceptibility leading to d-wave superconductivity \cite{pour06}.

The analogy of PuMGa$_5$ and CeMGa$_5$ has been explored experimentally in several studies. Besides the evidence for an important role of critical antiferromagnetic fluctuations and the general considerations based on the calculated band structure given above, there is striking similarity concerning the dependence of $T_s$ on the ratio $c/a$ of the lattice parameters as shown in Fig.\,\ref{115-covera} in section \ref{cemin-system} \cite{baue04b}. This trend is consistent with trends predicted for magnetically mediated pairing \cite{mont01,mont02}. However, the experimental investigation of the lattice parameters under high pressure establishes that for none of the  PuMGa$_5$ and CeMGa$_5$ systems the value of $T_s$ scales with $c/a$. This suggests that there are other important aspects besides the $c/a$ ratio \cite{norm05}. On another note it has been suggested, that the normalized pressure dependence of the superconductivity is consistent with a dome, which may be qualitatively viewed in a common phase diagram \cite{thom06c}.

\paragraph{{\nppdal}}

We next turn to the question of further actinide superconductors that are neither based on uranium nor plutonium. An important element in this respect is neptunium which is adjacent to plutonium. The Wigner-Seitz radius thereby suggests that the f-electrons in Np are in an itinerant state. Examination of spectroscopy and physical properties shows that the 5f states of Np are beginning to show the effects of localization, however, the metal is still fairly delocalized \cite{moor09}.

Recently heavy fermion superconductivity has also been discovered in {\nppdal} \cite{aoki07,griv08}. This represents the first Np-based superconducting system. It is interesting to compare the properties of this system with the Pu-based heavy fermion superconductors. The crystal structure of {\nppdal} is ZrNi$_2$Al$_5$ type body-centered tetragonal, space group I4/mmm with atomic positions Np $(0,0,0)$, Pd(1) $(1,1/2,0.1467)$, Pd(2) $(0,0,1/2)$ and Al $(0,0,0.255)$. The lattice constants establish a particularly anisotropic material, c=14.716{\AA} and a=4.148\,{\AA}. Electronic structure calculations suggest itinerant 5f electrons \cite{yama08}.

The normal state is characterized by a Fermi liquid specific heat with $\gamma=0.2\,{\rm J/mol\,K^2}$. In contrast, the magnetic susceptibility is temperature independent for the c-axis, but diverges all the way until superconductivity sets in. This and the linear temperature dependence of the electrical resistivity for the a-axis clearly signal NFL properties. 

The normal state susceptibility shows a Curie-Weiss temperature dependence with a fluctuating moment $\mu_{eff}^{ab}=3.22\,{\rm \mu_B/Np}$,  $\mu_{eff}^{c}=3.06\,{\rm \mu_B/Np}$ for the ab-plane and c-axis, respectively, that is intermediate to the free Np $5f^3$ free ion value of $3.62\,{\rm/Np}$ and the Np $5f^4$ configuration with $2.68\,{\rm/Np}$. The Curie-Weiss susceptibility extends all the way down to the onset of superconductivity at $T_s=4.9\,{\rm K}$. The tetragonal c-axis is magnetically hard. The electrical resistivity is characteristic of a good metal and decreases monotonically from its room temperature value of $\sim65\,{\rm \mu\Omega cm}$ down to $T_s$. Just above $T_s$ the resistivity is linear in temperature, characteristic of charge carrier scattering by critical fluctuations consistent with the Curie Weiss susceptibility. The extrapolated residual resistivity is $\rho_0\approx5\,{\rm \mu\Omega cm}$.  Despite the evidence for strongly temperature dependent fluctuations in the normal state specific heat show the behavior of a Fermi liquid with an enhanced $\gamma=0.2\,{\rm J/mol\,K^2}$.

The superconducting transition is accompanied by a pronounced $\lambda$ anomaly in the specific heat, where $\Delta C/\gamma T_s=2.33$. This is characteristic of strong-coupling superconductivity. The temperature dependence of the specific heat in the superconducting state is highly unconventional, following initially a $T^2$ dependence that settles into a $T^3$ dependence below $\sim1.8\,{\rm K}$.  The $T^3$ dependence of the low temperature specific heat is consistent with point nodes in the superconducting gap. In combination with the antiferromagnetic fluctuations inferred from the normal state susceptibility this suggests a $d$-wave state with point nodes. 

The initial slopes of $H_{c2}$ near $T_s$ are anomalously large with $dH_{c2}^{ab}/dT=-6.4\,{\rm T/K}$ and $dH_{c2}^{c}/dT=-31\,{\rm T/K}$, as for the Pu based superconductors. However, $H_{c2}$ is highly anisotropic and in comparison the Pu-based systems reduced, where $H_{c2}^{ab}(T\to0)=3.7\,{\rm T}$ and  $H_{c2}^{c}(T\to0)=14.3\,{\rm T}$. This suggests considerable paramagnetic limiting of $H_{c2}$. The d.c. magnetization shows that the lower critical field $H_{c1}=0.008\,{\rm T}$, coherence length $\xi=94\,{\rm \AA}$, penetration depth $\lambda =2600\,{\rm \AA}$ and Ginzburg-Landau parameter $\kappa=28$. For the c-axis the magnetization suggests first order behavior at low temperatures, akin {\cecoin} (for the a-axis $H_{c2}$ is too large). This implies also the possibility for an FFLO state.

$^{27}$Al NMR in single crystal {\nppdal}  \cite{chud08} shows a broadening of the NMR spectra when entering the superconducting state, consistent with a flux line lattice. Further, there is no coherence peak and the spin-lattice relaxation rate, $1/T_1$ shows a cubic temperature dependence. Both, the spin-lattice relaxation rate as well as the Knight shift point at line-nodes and strong coupling d-wave superconductivity.

Changes of the temperature dependence of the resistivity under magnetic field suggests the vicinity to a quantum critical point; the $T^2$ coefficient decrease as if it is singular at $H_{c2}$. Interestingly pressure suppresses $T_s$ above 57\,kbar, reminiscent of a superconducting dome \cite{hond08b}. This is also consistent with a vicinity to quantum criticality.

As for the Pu-based superconductors it is not clear, where the entropy of the magnetic fluctuations is dumped in the superconducting state. Based on the striking similarity of the Pu and Np superconductor it interesting to speculate on the possible implications of the paramagnetic limiting as the only difference. Since Pu is closer to the localization of the f electron, this may suggest an important role of charge fluctuations \cite{schl89}. In fact, similar considerations as for the vicinity of a valence instability discussed above may also apply here and charge density fluctuations may promote the superconductive pairing \cite{mont04,onis00}.

\subsection{Border of polar order}
\label{supe-bord-pola-orde}

For systems where the quasiparticle dressing cloud is dominated by excitations of the crystal electric fields an interesting question concerns, whether the quasiparticle interactions also include attractive components that may stabilize superconductivity. A scenario of this kind has been proposed for {\updal} as discussed in section \ref{updal}. However, for {\updal} the superconductivity coexists with large-moment antiferromagnetism where $T_s\ll T_N$.  In turn the interplay of the crystal field excitations with the antiferromagnetic order is of considerable complexity and essentially not accessible directly experimentally due to the strong hybridization of the 5f electrons with the conduction electrons. 

In comparison to U-based compounds, Pr-based compounds generally show distinct crystal electric field excitations. The quasiparticle dressing clouds in the Fermi liquid regime in pure Pr were, for instance, identified as being excitonic \cite{lonz88}. In recent years heavy-fermion superconductivity has been discovered in {\prossb} and related compounds (cf table \ref{table-skutt}). There is now growing consensus that the superconductivity in {\prossb} may be mediated by the exchange of quadrupolar fluctuations. In the following we first review the properties of {\prossb}. For more detailed reviews we refer to \cite{aoki07b,mapl05,mapl08,hass08}. The section concludes with a paragraph on {\prrup}, in which superconductivity emerges, when an insulating state is suppressed at 110\,kbar.

\begin{table}
\centering
\caption{Key properties of Pr-based heavy-fermion superconductors and siblings exhibiting conventional superconductivity. Missing table entries may reflect more complex behavior discussed in the text. References are given in the text.}
\label{table-skutt}
\begin{tabular}{lllllllll}
\hline\noalign{\smallskip}
& {\prossb}  & {\prrusb} & {\prrup}  \\
\noalign{\smallskip}\hline\noalign{\smallskip}
structure & cubic & cubic & cubic \\
space group & $Im\bar{3}$ & $Im\bar{3}$ & $Im\bar{3}$ \\
$a$({\AA}) & 9.302 & & - \\
$\Delta_{CEF}$(meV) & 7 & 64& \\
%$c$({\AA}) & - & - \\
%$c/a$ & - & - \\
\noalign{\smallskip}\hline
state & SC, AFQ & SC & IN, SC \\
$T_{c}$(K) & 1.3 (at 9\,T) & & 62  \\
$\vec{Q}$ & $(0,0,1)$ & & - \\
$\mu_{ord}¥$($\mu_{B}$) & 0.085 & & - \\
$\gamma {\rm (J/mol K^2)}$ & 0.5 & 0.059 & -   \\
%$T_K {\rm (K)}$ & - & - \\
\noalign{\smallskip}\hline
$T_{s}$(K) & 1.85 & 1.3 & 1.8 \\
& & & ($p>110\,{\rm kbar}$) \\
$\Delta÷C/\gamma_{n}T_{s}$ & $>5$ & & -  \\
$H_{c2}$(T) & 2.3 & 0.2 & 2 \\
$dH_{c2}/dT$(T/K) & -1.9 &  & -  \\
%$H^{c}_{c2}$(T) & - & - \\
%$\partial÷H^{c}_{c2}/\partial÷T$(T/K) & - & - \\
\noalign{\smallskip}\hline
$\xi_{0}$({\AA}) & 120 & 400 & -  \\
$\lambda_{0}$({\AA}) & 3440 & 3650 & -  \\
$\kappa_{GL}$({\AA}) & 28 & 9 &  \\
%$\kappa_{GL,c}$({\AA}) & - & - \\
%\noalign{\smallskip}\hline
%$\partial÷T_{s}/\partial÷p_{[100]}$(mK/GPa)& - & -  \\
%$\partial÷T_{s}/\partial÷p_{[001]}$(mK/GPa)& - & -  \\
%$\partial÷T_{s}/\partial÷p_{V}$(mK/GPa)& - &  -  \\
%$\partial÷T_{s}/\partial÷p$(mK/GPa)& - &  -  \\
\noalign{\smallskip}\hline
%$C/T$ & - & - \\
%$\kappa$ & - & - \\
%$1/T_{1}$ & - & -  \\
%$\lambda$ & - & - \\
%\noalign{\smallskip}\hline
year of discovery & 2002 & 2005 & 2004 \\
\noalign{\smallskip}\hline
\end{tabular}
\end{table}
 
\paragraph{{\prossb}}
\label{para-prossb}

{\prossb} belongs to the rare-earth-filled skutterudites, a class of systems with an exceptionally rich spectrum of vastly different ground states. Examples include insulating and metallic behavior, long range magnetic and polar order as well as conventional and unconventional superconductivity \cite{sale03,aoki05,mapl05,sato07,mapl08}. 

The large variety of electronic behaviors may be traced to the unusual crystal structure, which for the case of {\prossb} consists of a stiff icosahedron Sb cage typical of binary skutterudites, filled with a loosely bound Pr ion. The Pr ion is presumably in an off-center position \cite{goto04}. The space group of the crystal structure is $Im\bar{3}$, where the local point symmetry of the rare-earth ion is tetrahedral, $T_h(m3)$, which does not include a 4-fold rotation axis. Consequently the crystal electric fields split the $J=4$ multiplet of the Pr$^{3+}$ ions into a $\Gamma_1$ singlet, a non-Kramers nonmagnetic doublet $\Gamma_{23}$ and two triplets $\Gamma_4^{1,2}$ \cite{take01}. As for all rare-earth filled skutterudites the Pr-ion exhibits 'rattling' modes, leading to almost dispersionless low-energy phonons as seen in Raman scattering \cite{ogit08}.  In neutron scattering the rattling modes result in large Debye-Waller factors and in Raman scattering a second order phonon peak has been observed \cite{kane06,goto04}. Even though the Pr ion is only loosely bound, the p-f hybridization is expected to be large because of the cage of Sb atoms surrounding it.

The resistivity of {\prossb} decreases monotonically with temperature and displays a roll-off around 10\,K followed by a superconducting transition at $T_s=1.85\,{\rm K}$ \cite{baue02a}.  The susceptibility displays a broad maximum around 3\,K and the specific heat exhibits a pronounced Schottky anomaly. The features in the resistivity, susceptibility and specific heat are due to thermally populated CEF-split Pr$^{3+}$ energy levels. Two differing crystal field schemes were initially proposed, a $\Gamma_1$ singlet ground state and $\Gamma_4$ triplet first excited state \cite{aoki02} and vice versa \cite{mapl02,voll03}. Inelastic neutron scattering \cite{gore04} and detailed measurements of the magnetic field dependence \cite{aoki02} have settled this issue and it is now accepted \cite{baue06c}, that the ground state is a $\Gamma_1$ singlet, followed by a $\Gamma_4$ triplet first excited state. 

Quantum oscillatory studies show Fermi surface sheets consistent with localized 4f electrons \cite{suga02,suga08}. In comparison with other systems in this series the Fermi surface lacks nesting and compares well with that of {\laossb}. The similarity of the Fermi surface topology is underscored by Hall effect and thermopower measurements, which are similar for both compounds \cite{suga05}. 

It was immediately recognized that {\prossb} represents the first example of a Pr-based heavy-fermion superconductor \cite{baue02a}. Although the low temperature specific heat is dominated by a Schottky anomaly around 2\,K, it is possible to infer a strongly enhanced linear term in the normal state specific heat $C/T\approx0.2$ to $0.75\,{\rm J/mol\,K^2}$ (for a comprehensive discussion of the analysis of $C(T)$ see \cite{grub06} and references therein). A related large anomaly is observed in the specific heat at the superconducting transition, $\Delta C/T_s\approx0.5\,{\rm J/mol\,K^2}$, which, depending on the strength of the coupling, also points to a large value of $\gamma$. Finally,$H_{c2}\sim2.2\,{\rm T}$ is close to the orbital limit $H^{orb}_{c2}=2.4\,{\rm T}$, inferred from the experimentally observed variation $dH_{c2}/dT\approx-1.9\,{\rm T/K}$ near $T_s$. The large value of $dH_{c2}/dT$ also supports heavy fermion superconductivity. 

An increasing number of experimental data suggest that the superconductivity in {\prossb} is unconventional. $\mu$-SR shows that the superconductivity is accompanied by time reversal symmetry breaking \cite{aoki03}. The penetration depth measurements show a temperature dependence of the penetration depth $\lambda\propto T^2$ and superfluid density $\rho_s\propto T^2$ down to $0.3T_s$ \cite{chia03}. The zero temperature penetration depth $\lambda=3440\,{\rm \AA}$ is comparatively short. The data for $\lambda$ and $\rho_s$ are consistent with point nodes of strong-coupling superconductivity with $\Delta(0)/k_BT_s=2.6$. This is contrasted by Sb-NMR of the spin-lattice relaxation rate, which lacks a coherence peak and shows a temperature dependence consistent with an isotropic energy gap of a very strong-coupling state \cite{kote03b}. A well-developed superconducting gap, which is nearly isotropic is also observed in tunneling spectroscopy \cite{sude04}. Small angle neutron scattering in {\prossb} has revealed an asymmetry of the flux line lattice that suggests a p-wave superconducting state \cite{huxl04b}. 

\begin{figure}
\includegraphics[width=.35\textwidth,clip=]{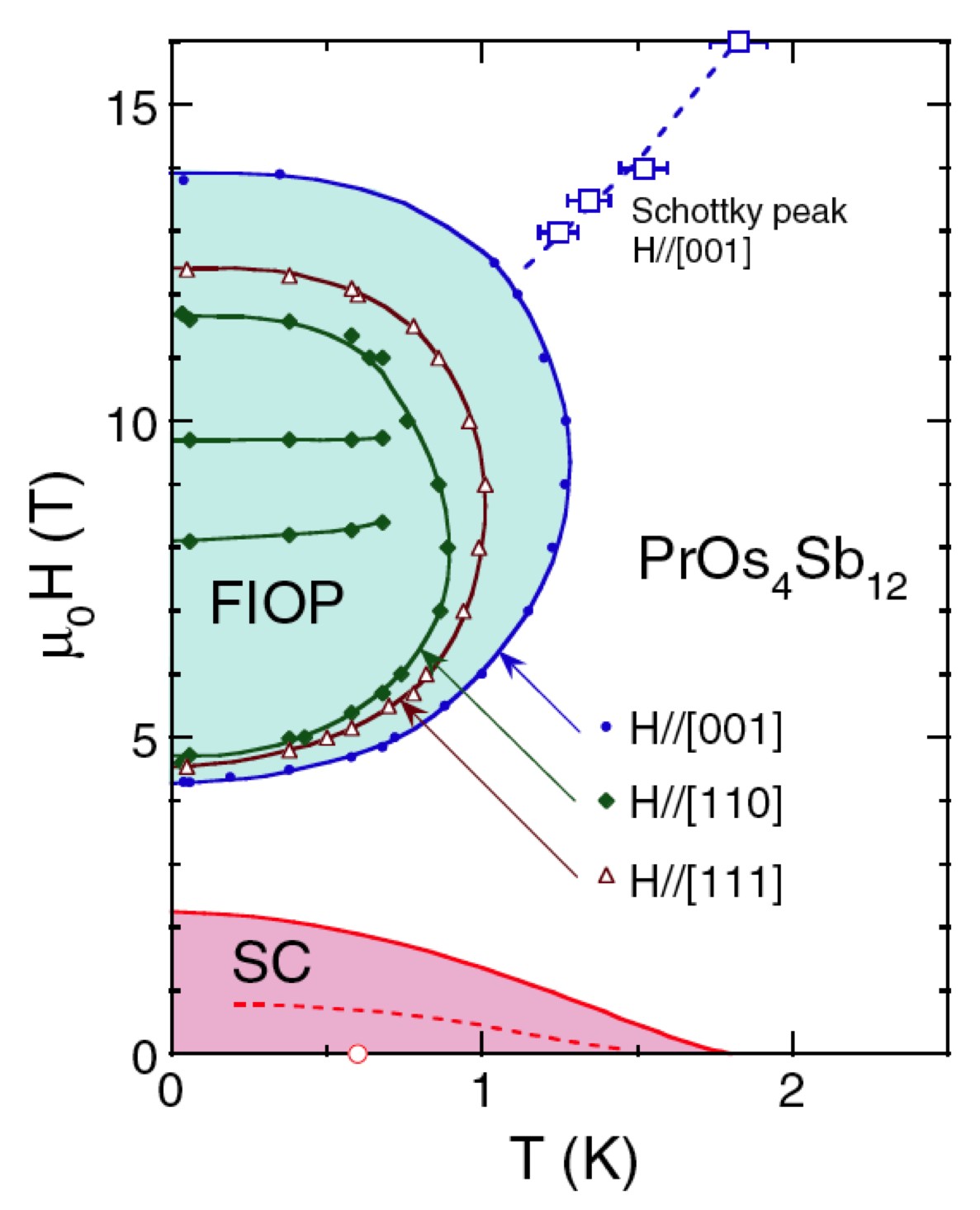}
\caption{Magnetic field versus temperature phase diagram of {\prossb}. In high magnetic field an ordered state is stabilized that is driven by the level crossing of the crystal electric fields under magnetic field. Plot taken from \cite{aoki07}.}
\label{pd-pros4sb12} 
\end{figure}

The case for unconventional superconductivity in {\prossb} is underscored by the observation of conventional superconductivity in the Pr-filled skutterudites {\prrusb} \cite{fred05} and PrRu$_4$As$_{12}$  ($T_s=2.4\,{\rm K}$) \cite{shir97}, as well as the La-filled skutterudites LaRu$_4$As$_{12}$ ($T_s=10.4\,{\rm K}$), LaFe$_4$P$_{12}$ ($T_s=4.1\,{\rm K}$), LaRu$_4$P$_{12}$ ($T_s=7.2\,{\rm K}$), LaRu$_4$Sb$_{12}$ ($T_s=3.4\,{\rm K}$) and LaOs$_4$Sb$_{12}$ ($T_s=.74\,{\rm K}$) \cite{mapl02,miya04,sato03}. Remarkably, upon doping {\prossb} by La on the Pr site and Ru on the Os site the heavy-fermion superconductivity gradually turns into conventional superconductivity. This suggests, that a certain stability of the heavy fermion superconductivity against defects exists.

A controversial question in {\prossb} concerns, whether the superconductivity consists of multiple superconducting phases and/or multiband superconductivity. The specific heat and thermal expansion display a double superconducting transition \cite{baue02a,voll03,meas04,aoki02,rotu04,oesc04}. The similarity of the observed behavior across a large number of different samples seems to suggest that the behavior is intrinsic. However, recent studies of a very high quality single crystal show only a single transition \cite{seyf06,meas08}.  A detailed study of samples with a double transition in the specific heat using micro-Hall probe and magneto-optical imaging reveal considerable inhomogeneities that question a bulk nature of the double transition \cite{kasa08b}. The double transition is also reflected in the susceptibility \cite{cich05,meas04,meas08,fred04,grub06} and resistivity \cite{meas04}, which points at an extrinsic origin. Multiband superconductivity has been suggested on the basis of thermal conductivity measurements, which readily return to the normal state behavior in small magnetic fields \cite{seyf05,seyf06}. Further, $H_{c2}$ shows positive curvature near $T_s$ \cite{meas04}.  Multi-band superconductivity has also been inferred from Sb-NQR studies \cite{yogi08}, which supports a fully gapped large Fermi surface that drives strong-coupling superconductivity accompanied by a small Fermi surface with line nodes.

Additional transitions to further superconducting states have been inferred from magnetothermal transport  \cite{izaw03}, the low field magnetization \cite{cich05} and Andreev reflections \cite{ture08}. As for the magnetothermal transport, a change of symmetry is observed at fairly high fields $\sim 1\,{\rm T}$, while the magnetization shows a pronounced enhancement of the lower critical field and critical current density below $\sim0.5\,{\rm K}$. The transitions in the magnetothermal transport and magnetization are unrelated. Both await further clarification in terms of other experimental quantities. 

Inelastic neutron scattering suggests that quadrupolar fluctuations are involved in the superconducting pairing \cite{kuwa05,raym08b}. A clear dispersion is found for the transition $\Gamma_1$ to $\Gamma_4^{(2)}$ for $\vec{Q}=(\zeta,0,0)$ in zero magnetic field. Both the excitation energy and scattering intensity exhibit a minimum at $\vec{q}=(1,0,0)$, the ordering wave vector of the field induced antiferroquadrupolar order described below. The excitations hence are quadrupolar and not magnetic. When entering the superconducting state, the excitation energy and its width decrease, signaling an interplay that may either be due to a freezing out of the damping by particle-hole excitations or an indication that the quadrupolar excitations are directly involved in the pairing. In particular, for low temperatures and magnetic fields the energy of this excitation compares with the superconducting gap. This suggests that superconducting pairing may be mediated by this excitation.

We finally turn to the remarkable vicinity of long range polar order and superconductivity in {\prossb}. A pronounced phase transition emerges above $\sim4\,{\rm T}$ that reaches 1.3\,K at 9\,T followed by a decreases and suppression above 13\,T as summarized in Fig..\,\ref{pd-pros4sb12} \cite{aoki02,voll03,suga05,taya03,rotu04,oesc04}. The large entropy released at this phase transition clearly shows that the 4f electrons are involved in the ordering process. Neutron diffraction reveals a small antiferromagnetic modulation in the high field phase \cite{kohg03,kane07}. For field $\vec{H}\parallel[0,0,1]$ and $\vec{H}\parallel[1,1,0]$ the superlattice has wave vector $\vec{q}=(1,0,0)$, where the corresponding ordered moment of $\mu_{ord}=0.025\,{\rm \mu_B/Pr}$ represents only a few \% of the uniform magnetization.

It is possible to show that this modulation results from $\Gamma_5$-type antiferroquadrupolar interactions  \cite{shii04a,shii04b}. Within this scenario the anisotropy of the field induced ordered phase is due to the tetrahedral point symmetry $T_h$ of the Pr ion. The antiferroquadrupolar order is driven by the Zeeman splitting and the crossing of the lower triplet with the singlet level at 9\,T. It is interesting to note that the ordering wave-vector corresponds to the nesting wave-vector in {\prrup} \cite{lee01,hao04} and {\prfep} \cite{iwas02}, which display anomalous ordering transitions.

%Theoretically, the nature of the superconductivity, the possible double transition and the high-field phase have attracted considerable interest.  The proposed pairing symmetries include chiral p-wave states (albeit for the incorrect crystal field scheme \cite{miya03b}), anisotropic s and s- plus d-wave pairing \cite{gory03}. The stability of the superconductivity under doping has stimulated a mixed parity formalism for centro-symmetric systems \cite{serg04b}. In a weak coupling analysis the double transition could only be accounted for with two accidentally degenerate singlet representations \cite{mukh06}. The high field phase finally, was attributed to an interplay of the quadrupolar moments of the Pr 4f electrons with the Zeeman splitting of the crystal electric fields \cite{shii04b,shii04c,otsu05}.

\paragraph{{\prrup}}

The Pr filled skutterudite compounds {\prrup} exhibits a metal insulator transition at $T_{MI}=62\,{\rm K}$, that defies an explanation in terms of magnetic or charge ordering \cite{seki97}. Under hydrostatic pressure $T_{MI}$ varies only weakly, but additional anomalies emerge below $T_{MI}$ that suggest further ordering transitions. Above 110\,kbar {\prrup} turns metallic with a superconductivity below $T_s\sim1.8\,{\rm K}$ \cite{miya04}. The upper critical field of this superconducting state is rather high  $H_{c2}\approx2\,{\rm T}$. Whether or not the superconductivity is unconventional awaits further clarification, where the similarity of $T_s$ and $H_{c2}$ with {\prossb} is interesting to note.

%%%%%%%%%%%%%%%%%%%%%%%%%%%%%%%%%%%%%%%%%%%%%%%%%
\section{MULTIPLE PHASES
\label{mult-supe-phas}}

\subsection{Order parameter transitions}

Many of the superconducting phases of intermetallic compounds reviewed in this paper are candidates for unconventional superconductivity with complex superconducting order parameters. They may in turn display various symmetry broken superconducting phases. In the following we summarize the evidence for such multiple superconducting phases. At present the only stoichiometric superconductor, where multiple superconducting phases are observed beyond doubts is the archetypical heavy-fermion system {\upt}, which will be addressed first. This is followed by short summaries on further candidates for such phases, where prominent examples are {\prossb} and {\uthbe}.

\subsubsection{Superconducting phases of {\upt}
\label{symm-brea-fiel}}

The normal state properties of {\upt} have been introduced in section \ref{upt}. At low temperature {\upt} displays a peculiar form of commensurate antiferromagnetic order below $T_N=5\,{\rm K}$ with tiny magnetic moments, that appears to be related to a highly dynamic magnetic ground state. The antiferromagnetic order is only observed in neutron scattering and the metallic state shares the properties of a strongly renormalized Fermi liquid. In this heavy fermion ground state superconductivity appears below $T_s=0.54\,{\rm K}$. While heavy-fermion superconductivity in its own right would already be quite remarkable, it is the observation of three superconducting phases that has attracted tremendous scientific interest. In the following we briefly review the superconducting phase diagram in {\upt}. A detailed account may be found in \cite{joyn02}.

The first indication for multiple superconducting phases was observed in the ultrasound attenuation in applied magnetic fields and in $H_{c2}$. The bulk property that exhibits the most distinct evidence of multiple superconducting phases is the specific heat, where two transitions are seen. The transition temperatures are $T_s^+=0.530\,{\rm K}$ and $T_s^-=0.480\,{\rm K}$ \cite{fish89}. Thus the splitting is of the order $\sim$10\% of $T_s$ and rather small. With respect to the linear term of the normal state specific heat, $\gamma_n=0.44\,{\rm J/mol\,K^2}$, the anomalies at $T_s$ are given by $\Delta C^+/\gamma T_s^+=0.545$ and $\Delta C^-/\gamma T_s^-=0.272$. The ratio of the height of the upper to the lower anomalies is hence about 2:1. Even though the specific heat anomalies are substantial, they are small by comparison to the BCS value. This indicates nodes in the superconducting gap. Another signature of nodes in the gap is a linear decrease of $C/T$ below $T_s$ down to 0.1\,K, below which a pronounced upturn is observed \cite{bris94}. 

\begin{figure}
\includegraphics[width=.35\textwidth,clip=]{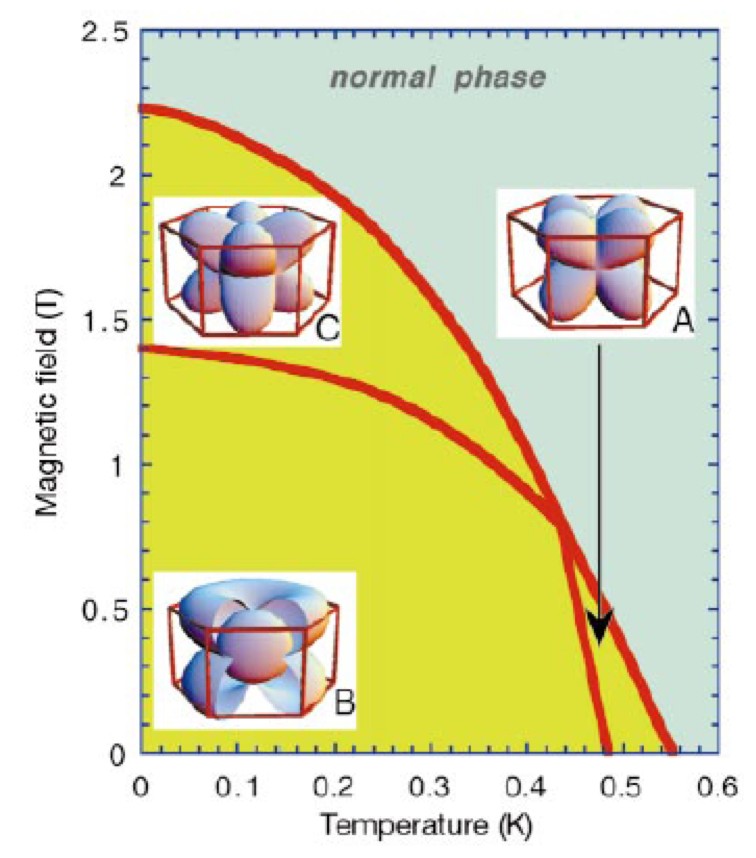}
\caption{Superconducting phases of {\upt} as a function of magnetic field. The insets show the nodal structures of the E$_{1u}$ representations, proposed on the basis of small angle neutron scattering of the flux line lattice. Plot taken from \cite{huxl00}.}
\label{pd-upt3-b} 
\end{figure}

Applied magnetic field has been found to reduce $T_s^+$ and $T_s^-$ at different initial rates without significant broadening for field parallel and perpendicular to the c-axis  as shown in Fig.\,\ref{pd-upt3-b} \cite{hass89,hass90}. The transition merges at a tetracritical point $(H^*,T_H^*)$, where for $H\parallel \hat{c}$: $H^*=0.4\,{\rm T}$, $T^*_H=T_s^+-0.1\,{\rm K}$. For $H\perp \hat{c}$: $H^*=0.8\,{\rm T}$, $T^*_H=T_s^+-0.15\,{\rm K}$. Tetracriticality has been confirmed by ultrasound attenuation \cite{aden90,brul90}, dilatometry \cite{vand93b} and the magnetocaloric effect \cite{boge93}. The general consensus has become, that {\upt} exhibits three superconducting phases referred to as A, B and C. Phases A and B support a Meissner and a Shubnikov phase below and above $H_{c1}$. As a function of temperature $H_{c1}$ shows a sudden increase in slope at $T_s^-$ \cite{vinc91}. Qualitatively the three component phase diagram contrasts an extrinsic origin, where the phase transition lines may be expected to have similar field dependences. 

In general $H_{c2}$ in {\upt} exceeds Pauli limiting. The anisotropy of $H\parallel\hat{c}$ and $H\perp\hat{c}$ changes at around 0.2\,K with $H_{c2}^{\parallel}<H_{c2}^{\perp}$ at low temperatures and $H_{c2}^{\parallel}>H_{c2}^{\perp}$ near $T_s$ \cite{shiv86}. The presence of the three superconducting phases requires to distinguish coherence lengths and penetration depths according to these phases. On the one hand, the zero temperature value of $H_{c2}$ is characteristic of the C phase, where $H_{c2}^{\parallel}(T\to0)=2.1\,{\rm T}$ and $H_{c2}^{\perp}(T\to0)=2.8\,{\rm T}$. The anisotropy of $H_{c2}$ may be accounted for by an anisotropic mass enhancement. The coherence length inferred from $H_{c2}$ then is $\xi\approx120\,{\rm \AA}$. On the other hand, the initial slope of $H_{c2}$ with temperature near $T_s^+$ is characteristic of the A phase, where $dH^{\parallel}_{c2}/dT\vert_{T_s^+}=-7.2\pm0.6\,{\rm T/K}$ and $dH^{\perp}_{c2}/dT\vert_{T_s^+}=-4.4\pm0.3\,{\rm T/K}$. When accounting for this anisotropy also in terms of the effective mass enhancement, it is possible to obtain an estimate of the Ginzburg-Landau parameter $\kappa_{\rm GL}=44$. In other words {\upt} is a strong type 2 superconductor. Some simple estimates arrive at values of the penetration depth of the order $\lambda_{\parallel}(T\to0)=4500\,{\rm \AA}$ and $\lambda_{\perp}(T\to0)=7400\,{\rm \AA}$, consistent with the short coherence length estimated for the C phase. It can finally be shown that weak coupling theory yields the same value of $\kappa_{\rm GL}$. This implies that {\upt} is still fairly well described in a weak coupling approximation.

The effect of hydrostatic pressure on the superconducting transitions and the antiferromagnetic order strongly suggests, that the antiferromagnetic order is instrumental for the symmetry breaking between the different superconducting phases. In the specific heat the two superconducting transitions are found to decrease at different rates, eventually merging into a single transition above $\sim$3\,kbar \cite{trap91}. At the same time neutron scattering establishes that the ordered moment decreases under pressure and vanishes above $\sim$3\,kbar, while $T_N$ is essentially not affected by pressure \cite{hayd92}.

Numerous other experimental probes suggest unconventional pairing and provide important hints as to the precise nature of the gap symmetry. For instance, in a recent small angle neutron scattering study the magnetic field dependence of the flux line lattice has been established. The upshot of this study is that the three superconducting phases belong to the $E_{2u}$ symmetry \cite{huxl00} (see also \cite{cham01} for theoretical considerations on the flux line lattice). For an extended review and critical discussion of the various theoretical scenarios we refer to \cite{joyn02}. Despite the large body of studies the search for the correct order parameter symmetry has not been entirely conclusive so far.

\subsubsection{Further candidates}

Nearly all of the systems covered in this review in one way or the other may be candidates for multiple superconducting phases. The nature of these phases may be quite different, representing either different order parameter symmetries or real space modulations with different ordered state. In the following we draw attention to candidates, which await further clarification.

\paragraph{\cecusi}

As reviewed in sections \ref{cemt} and \ref{supe-vale-inst} recent high pressure studies in pure and Ge doped {\cecusi} reveal the presence of two superconducting domes (Fig.\,\ref{PD-CeCuSiGe}). At low pressures this material is a candidate for magnetically mediated pairing driven by the vicinity to an antiferromagnetic quantum critical point. At high pressures a second dome emerges and it has been argued that this superconducting phase is related to fluctuations in the charge density of a valence transition \cite{yuan04,holm04}.

\paragraph{\cenige}

At ambient pressure {\cenige} displays an incipient form of superconductivity. It has been argued that the ambient pressure behavior is reminiscent of {\cepdsi} in the vicinity of the critical pressure. Under pressure the signatures of superconductivity vanish. At high pressures an additional superconducting transition emerges as shown in Fig.\,\ref{PD-CePdeSi+CeNiGe} \cite{gros97}. In principle this second superconducting dome may hint at an additional superconducting phase, but little is known about this state.

\paragraph{\ceirin}

Pure single crystals of {\ceirin} display a difference of the temperature of a zero resistance transition, $T_{s1}=0.75\,{\rm K}$ and the bulk superconducting transition in the specific heat, $T_{s2}=0.4\,{\rm K}$. It is tempting to attribute the resistive transition to sample inhomogeneities. However, if the two transitions are intrinsic, it may signal the presence of two superconducting instabilities, where the first transition corresponds to incipient superconductivity.

\paragraph{\updal}
In {\updal} single crystals grown with an Al-rich starting composition showed particularly sharp superconducting transitions at $T_s$ in the resistivity \cite{sako93}. This suggested an improved sample quality. Remarkably the specific heat, thermal expansion and elastic constants in these samples revealed an additional anomaly around 0.6\,K well below $T_s$ \cite{sato94,mats94,sako94}. The nature of this transition has so far not been settled. Either it signals an additional superconducting transition akin the double transition observed in {\upt}, or it corresponds to another ordering transition. For the first case, it is conceivable that the antiferromagnetic order of {\updal} represents the symmetry breaking field. In the latter case, it is possible that the emerging order leads to an additional symmetry breaking of the superconducting order that may stabilize additional superconducting phases. 

\paragraph{\urusi
\label{urusi-mult-phase}}
Early studies of the specific heat of the superconducting transition in {\urusi} showed features reminiscent of the double transition in {\upt} \cite{hass91}. Detailed studies in high quality single crystals did not confirm the first findings.  Keeping in mind that the tiny-moment antiferromagnetism in {\upt} represents the symmetry breaking field, that stabilizes the different superconducting phases, it seems plausible that the same might occur in {\urusi}. However, the antiferromagnetism in {\urusi} seems to be related to an impurity phase. Moreover, under pressure the superconductivity vanishes, when large moment antiferromagnetism appears. The observed change of curvature in $H_{c2}$ of {\urusi} has motivated considerations of the possible formation of a FFLO state. However, as discussed below {\urusi} does not develop a FFLO state. In turn it is currently accepted that {\urusi} does not support additional superconducting phases.

\paragraph{\ube
\label{uthbe-phases}}

Doping {\ube} with Th results in the phase diagram shown in Fig.\,\ref{pd-uthbe13} \cite{ott86}. For $x_1=0.02<x<x_2=0.042$ two successive transitions at $T_{s1}>T_{s2}$ are observed in the specific heat. The onset of superconductivity is thereby at $T_{s1}$. The pressure dependence of Th-doped samples also suggest the existence of two superconducting phases \cite{lamb86}, where an investigation of the lower critical field suggests that $T_{s2}$ indeed marks the onset of another superconducting phase \cite{rauc87}. A group theoretical analysis of these properties has been reported in
\cite{luky89,makh92}.  However, it still seems unsettled whether the lower transition at $T_{s2}$ indeed represents another superconducting transition \cite{kuma87,sigr89,mart00} or the onset of a defect induced form of magnetic order as suggested by $\mu$-SR \cite{heff86}. 

\paragraph{\uge}
Pressure and magnetic field studies suggest that the superconductivity in {\uge} is driven by the first order transition between the FM1 and FM2 ferromagnetic phases (Fig.\,\ref{pd-uge2}). The superconductivity hence exists in the presence of two different forms of ferromagnetic order. Theoretical considerations have shown, that the order parameter symmetry in ferromagnetic superconductors depends on the orientation of the ferromagnetic moment. Experimental evidence that tentatively supports different superconducting phases in the FM1 and FM2 state may be seen in the discontinuity of $T_s$ at $p_x$ and the reentrance of $H_{c2}$ for pressures just above $p_x$ and field applied along the crystallographic a-axis. However, as discussed in section \ref{uge}, the magnetic anisotropy of {\uge} remains unchanged under pressure. It therefore appears unlikely that the superconducting phases in the FM1 and FM2 state are fundamentally different. Further studies will have clarify this issue.

\paragraph{\urhge}
One of the most unusual phase diagrams amongst all f-electron superconductors is observed in {\urhge}. As a function of magnetic field superconductivity is at first suppressed, but reappears at high magnetic fields, when the ordered moment is forced to rotate from the c-axis to the b-axis. The phase diagram yields up to three different superconducting phases: the zero field state, the high field state below $H_R$, where the moment is almost rotated into the b-axis and finally above $H_R$, where the moment is essentially aligned with the b-axis. As for {\uge} the allowed order parameter symmetries have been worked out for the orthorhombic crystal structure.

\paragraph{{\ceptsi} {\&} {CeMX$_3$}}

The pressure versus temperature phase diagram of the four non-centrosymmetric heavy-fermion superconductors is dominated by a decreases of the Neel temperature. The transiton line crosses the superconducting dome in the middle, so that the phase diagram is comprised of a regime where $T_N>T_s$ and a regime where $T_N$ has vanished. These two regimes are in principle candidates for differences in the order parameter.

\begin{figure}
\includegraphics[width=.35\textwidth,clip=]{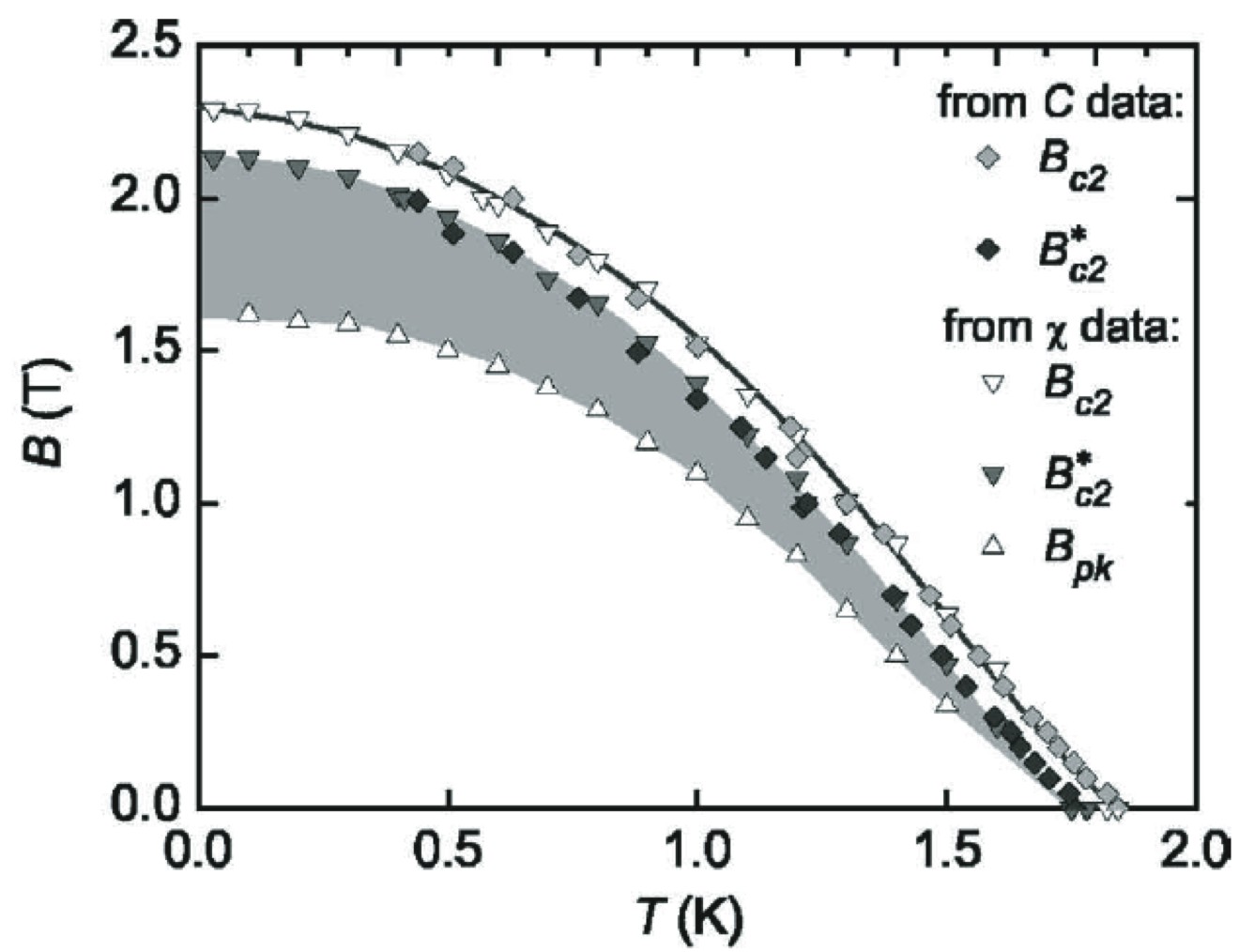}
\caption{Superconducting phase diagram of {\prossb}. Plot taken from \cite{grub06}.}
\label{pd-sc-pros4sb12}
\end{figure}

\paragraph{\prossb}
The superconducting state of {\prossb} exhibits several features that have been interpreted as tentative evidence for multiple superconducting phases. The specific heat of {\prossb} displays two superconducting transitions \cite{meas04,voll03,huxl04b}, where doping by Ru and La stabilizes the upper transition while mechanical thinning stabilizes the lower transition. However, the origin of the double transition is a controversial issue, where recent studies suggest that it may of extrinsic origin \cite{seyf06,meas08,kasa08b} (for details see section \ref{supe-bord-pola-orde}). As shown in Fig.\,\ref{pd-sc-pros4sb12} tentative transition lines in the susceptibility and a variety of other quantities may be traced all the way to zero temperature. Studies of the thermal conductivity \cite{izaw03} also suggest multiple superconducting phases, but with a different phase diagram that is not matched by any other property. Finally, high precision measurements of the magnetization suggest the possible existence of yet another transition line at very low fields \cite{cich05}. A comprehensive discussion along with detailed measurements of the specific heat and AC susceptibility have been given by \cite{grub06}.

%%%%%%%%%%%%%%%%%%%%%%%%

%\newpage
\subsection{Textures}
\label{textures}

An important fundamental and technological question in condensed matter systems are weak interactions that cause the formation of intermediate- and long-scale textures. The f-electron superconductors reviewed here exhibit several forms of electronic order and thus posses different types of characteristic rigidities. As far as the superconducting state is concerned these are the coherence length and penetration depth, while the magnetic state is characterized by the spin-wave stiffness, spin-orbit coupling, CEF pinning potential and dipolar interactions. As a first example the competition of exchange splitting with superconducting pairing is addressed. This competition may result in real-space modulations of the superconductivity and spin polarization as reviewed in section [\ref{FFLO-phas}]. The possible interplay of ferromagnetic domain structures and superconductivity is briefly addressed in section [\ref{doma-stru-supe-ferr}]. 

\label{text-supe}

\subsubsection{Fulde-Ferrell-Larkin-Ovchinnikov states
\label{FFLO-phas}}

The novel forms of superconductivity of interest in the present review are characterized by real space modulations and anisotropies of the superconducting gap function that are caused by a loss of symmetries beyond those of the underlying crystal structure. In turn the phase rigidity of the superconducting condensate in these superconductors yields changes of sign in momentum space. An entirely different class of novel superconducting states was predicted by Fulde, Ferrel, Larkin and Ovchinnikov (FFLO) \cite{fuld64,lark65}. As opposed to changes in momentum space in the FFLO state the order parameter changes sign in real space. In its original version the FFLO state considered superconductivity in the presence of a strong uniform exchange field. The Cooper pairs thereby form between Zeeman split parts of the Fermi surface, so that pairing with a finite momentum $\vec{q}$ is stabilized, where $({\vec{k}\uparrow}, {-\vec{k}+{\vec{q}\downarrow}})$. In the following we will briefly review the current status of FFLO states in the f-electron superconductors addressed in this paper. Detailed reviews may be found, e.g, in \cite{bris97,mats07,casa04}; for recent theory see \cite{houz06,houz07}.

Despite intense efforts, only a small number of candidate materials could be identified that may support an FFLO state, notably heavy fermion superconductors and quasi-two-dimensional organic superconductors for fields parallel to the layers \cite{gloo93,yin93,burk94,shim94,dupu95,tach96,buzd97,grue66}.  This may be traced back to the rather severe conditions under which the FFLO state is expected to form. As a first precondition, pair breaking in applied magnetic fields must be dominated by paramagnetic limiting and not orbital limiting \cite{grue66}. Second, impurities are detrimental to the FFLO state, making high-purity samples a key requirement \cite{asla69,taka69}. Third, large anisotropies of the Fermi surface are favorable to the FFLO state.

FFLO considered the effects of a uniform exchange field on s-wave superconductors. In the presence of pure orbital limiting the superconducting transition is second order at all magnetic fields and the superconductivity is unchanged by the exchange field. In contrast, in the presence of pure Pauli limiting the superconducting transition changes in finite fields from second to first order for temperatures below $T^{\dagger}=0.56\,T_s$ \cite{sain69,kett99}. Below $T^{\dagger}$ an inhomogeneous form of superconductivity stabilizes, in which the Cooper pairs support a finite momentum $({\vec{k}\uparrow}, {-\vec{k}+{\vec{q}\downarrow}})$. 

In the bulk properties the signature of the FFLO state is an increase of $H_{c2}$ below $T^{\dagger}$, that may be accompanied by a change of curvature. The size of this increase depends sensitively on the anisotropy of the Fermi surface ranging from 7\% of the Pauli limit for three dimensions \cite{fuld64,lark65,taka69,sain69}, over 42\% for two dimensions \cite{burk94,shim94,aoi74,bula74} to a divergence for one dimension \cite{suzu83,mach84}. Microscopically the FFLO state consists in spatial modulations of the superconductivity in real space, for which the order parameter may be given in general as $\Delta(\vec{r})=\sum_{m=1}^{M}\Delta_m\exp^{i\vec{q}_m\cdot\vec{r}}$ \cite{fuld64,lark65,shim98,bowe02,mora04,comb05,mora05,wang06b}. The superposition of degenerate components then yields a rich variety of symmetries of the real space modulations, e.g., hexagonal, square, triangular and one-dimensional modulations. 

It has long been appreciated that the stringent requirements for an FFLO state may be satisfied in superconductors with short coherence length, because the orbital limiting field diverges as $H^{orb}_{c2}\propto 1/\xi^2$ so that Pauli limiting may dominate. Prime examples are the heavy fermion superconductors reviewed here. The situation for an FFLO state then involves (i) a finite admixture of orbital limiting, (ii) the coexistence of antiferro- or ferromagnetic order, and (iii) anisotropic (unconventional) order parameter symmetries. The exploration of these issues has stimulated a large number of theoretical studies \cite{grue66,tach96,shim96,sugi06,adac03,houz06,shim97a,buzd96a,buzd96b,klei00,yang04}. For a recent review of these studies we refer to \cite{mats07}.

The question, whether FFLO states exist in heavy fermion superconductors has been explored in a number of systems. For instance, the AC susceptibility, magnetization, ultrasound velocity and thermal expansion near $H_{c2}$ in CeRu$_2$ and {\updal} exhibit the characteristics of a peak effect \cite{haga96,steg96,tach96,taka96,gege96}. It is now broadly accepted that these features do not yield microscopic characteristics related to a FFLO state, but instead may be due to subtle forms of defect related pinning. Further candidates for an FFLO state are {\urusi} and {\ube}, which display additional contributions in $H_{c2}$ \cite{buzd96a,buzd96b,glem99,bris95}. For {\urusi} this contribution is seen for the c-axis and rather small. In contrast {\ube} displays a change of curvature in $H_{c2}(T)$. It has been shown that these features are consistent with a vicinity to the FFLO formation, but the FFLO state does not form. Possible explanations include the sample purity, which is very good but does not reach the exceptionally clean limit required. Candidates for a FFLO state that have been identified recently in specific heat studies under magnetic field are {\pucoga}, {\purhga} \cite{javo07} and {\nppdal} \cite{aoki07}.

\begin{figure}
\includegraphics[width=.35\textwidth,clip=]{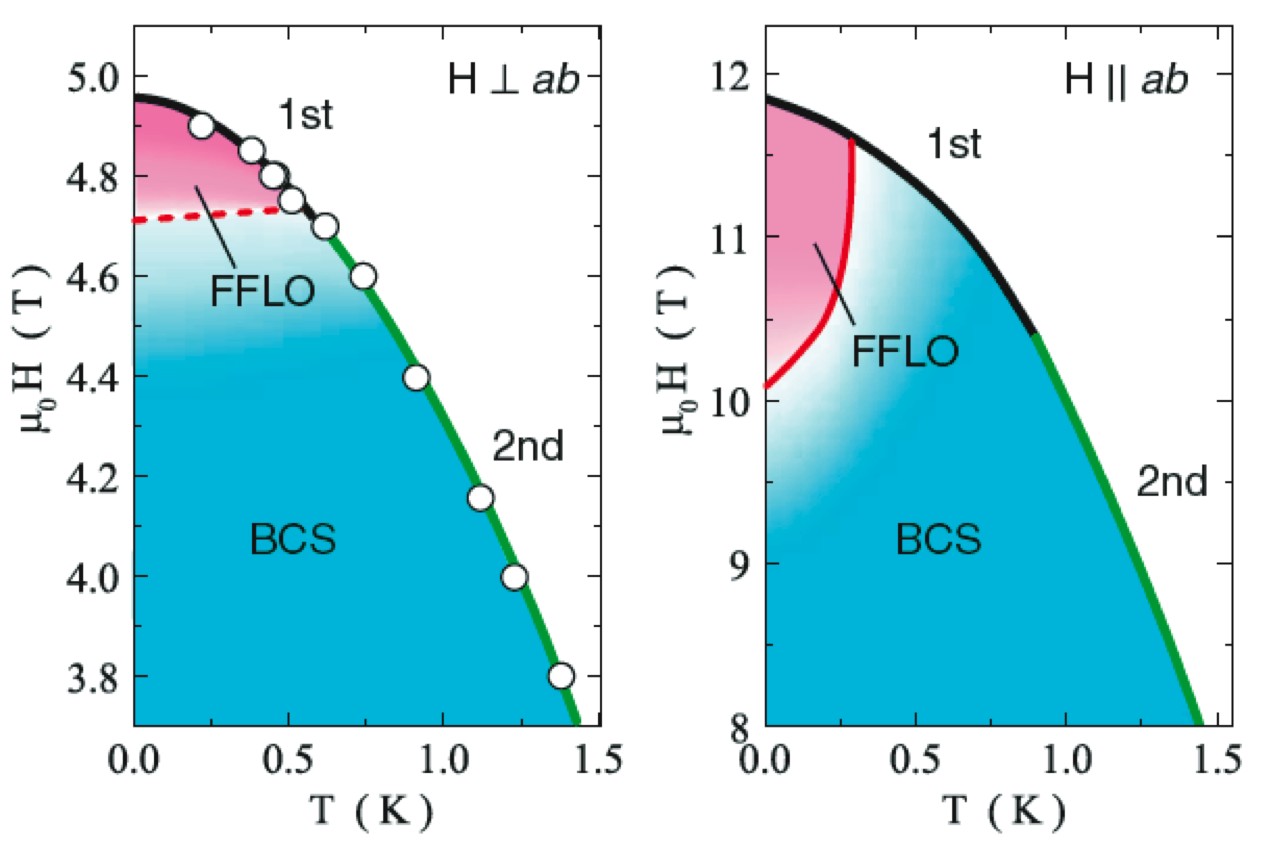}
\caption{Superconducting phase diagram of {\cecoin}. In the low temperature limit a body of evidence suggests the formation of a FFLO state (pink shading). Plot taken from \cite{mats07}.}
\label{pd-cecoin5-fflo}
\end{figure}

The perhaps best candidate of an FFLO state known to date has been identified in {\cecoin} (Fig.\,\ref{pd-cecoin5-fflo}). Several features in the superconducting phase diagram have been observed uniquely in {\cecoin}. The specific heat \cite{rado03,bian03b}, magnetization, \cite{grat06} magnetostriction \cite{corr07}, thermal conductivity \cite{capa04b}, penetration depth \cite{mart05}, ultrasound velocity \cite{wata04} and NMR Knight shift \cite{kaku05,kuma06,mitr06} show that the transition at $H_{c2}$ is first order for $T<0.3\,T_s$ and $T<0.4\,T_s$ for field parallel and perpendicular to the c-axis, respectively. This is the behavior expected for paramagnetic limiting of $H_{c2}$, where the samples studied were readily in the ultra-pure limit, i.e., the coherence length is only a small fraction of the charge carrier mean free path. It is thereby helpful to note that the orbital limit $H_{c2,ab}^{orb}=38.6\,{\rm T}$ and $H_{c2,c}^{orb}=11.7\,{\rm T}$ inferred from the initial slope of $H_{c2}$ near $T_s$ substantially exceeds the experimentally observed values of $H_{c2}$. The corresponding values of the Maki parameter near 5 exceed by a large margin the threshold of 1.8, above which a FFLO state may be expected. 

Specific heat and torque magnetization first identified a second order phase transition line in the superconducting state that branches off from $H_{c2} (T)$ at a temperature well below that of the change from second to first order and decreases with decreasing temperature (see Fig.\,\ref{pd-cecoin5-fflo}) \cite{micl06,bian02}. The presence of this line was confirmed in subsequent measurements of the penetration depth \cite{mart05}, thermal conductivity \cite{capa04b}, ultrasound velocity \cite{wata04}, magnetization \cite{grat06}, magnetostriction \cite{corr07} and NMR \cite{kaku05,mitr06}. The resulting phase pocket is a strong contender for a FFLO state.

The size of the novel phase pocket is anisotropic and considerably smaller in a field perpendicular to the ab-plane. The transition field is weakly temperature dependent for field direction perpendicular to the ab-plane and strongly field dependent for field direction parallel to the ab-plane. The anisotropy suggests that the FFLO state is more stable for field parallel to the ab-plane. This may be related to the two-dimensional character of the Fermi surface and the anisotropy of the spin fluctuation spectra, where the latter appear to be involved in the pairing interactions as discussed in section \ref{cemin-system}.

Key characteristics observed for the novel phase pocket may be summarized as follows. The penetration depth in the ab-plane increases, consistent with a decrease of the superfluid density \cite{mart05}. The thermal conductivity, providing a directional probe of the quasiparticle spectrum, is highly anisotropic. For heat current parallel to the applied field the thermal conductivity is enhanced, while it has not been possible to clarify changes of the thermal conductivity for heat current transverse to the applied field. As this behavior is somewhat counterintuitive, it has been proposed that the interplay of vortex lattice, spatial order parameter modulation and nodal gap structure results in an effective increase of vortex cores in the nodal plane \cite{capa04b}.

The flux line lattice in {\cecoin} has been studied, e.g., by the ultrasound velocity \cite{wata04,iked06}, which provides information on the pinning of the vortex cores by defects. Notably it is possible to extract information on the $c_{44}$ dispersive flux line tilt mode. A careful analysis of the decrease observed in $c_{44}$ implies a decrease of the superconducting volume fraction. Small angle neutron scattering reported so far did not meet the scattering condition necessary to probe the FFLO state \cite{bian08}. Microscopic information on the FFLO regime is also provided by NMR spectra of the In(1) and In(2) sites in the {\cein}- and CoIn$_2$ layers, respectively \cite{kaku05,mitr06,youn07,kuma06, sing01}. In the FFLO regime a key feature for both field directions is the appearance of a second resonance line in the superconducting state, where the lines are close to the values of the normal and superconducting state, respectively. It is currently unresolved if the NMR spectra for field parallel to the ab-plane also reveal antiferromagnetic components of the vortex cores \cite{kaku05,micl06,youn07,sing01}. Moreover, Cd doping of {\cecoin} leads to a rapid suppression of the first order behavior of $H_{c2}$, but Hg doping only smears out the phase pocket without change of characteristic temperatures \cite{toki08}. These studies support a nonmagnetic origin of the phase pocket, in the spirit of the original FFLO proposal.

\subsubsection{Magnetic domains versus flux lines
\label{doma-stru-supe-ferr}}

An issue that has not yet been explored experimentally concerns the interplay of the length scales characteristic of superconductivity with those characteristic of competing or coexisting forms of order. For the case of the superconducting ferromagnets several papers have explored this question from a theoretical point of view, e.g., \cite{soni98,soni02,buzd03}. 

%%%%%%%%%%%%%%%%%%%%%%%%%%%%%%%%%%%%%%%%%%%%%%%%%
\section{PERSPECTIVES}

Even though the first example of a heavy-fermion superconductor, {\cecusi}, was discovered nearly 30 years ago, an impressive series of new systems with surprising combinations of properties have come to light only recently This has resulted in two developments. First, more systems are different and we are only beginning to distinguish new classes of systems that are outside these general patterns. Second, the more general experimental ingredients controlling unconventional superconductivity are finally becoming apparent. In the following we briefly summarize these new developments.

Dominant interactions that control the properties of f-electron compounds may be summarized as follows: (i) the degree of f-electron localization,  (ii) crystal electric fields,  (iii) spin and orbital degrees of freedom and their coupling, and (iv) electron-phonon interactions. Amongst the large variety of f-electron superconductors that have been discovered in recent years, there are candidates where any one of the first three interaction channels appears to dominate the pairing interactions. For instance most of the members of the series {\cemt} and {\cemin} are candidates for antiferromagnetically mediated pairing. The U-compounds {\uge}, URhGe and UCoGe are candidates for ferromagnetically mediated pairing. Systems like {\prossb} are candidates for pairing by quadrupolar fluctuations, while {\cecusi} at high pressure or the Pu-based superconductors are candidates for valence fluctuations of the f-electrons and thus electron density. For instance, DMFT calculations reveal the fluctuating valence of Pu between 4, 5 and 6, ending in an average f-occupancy of 5.2. Despite their great microscopic differences all of these systems may be combined in a single graph shown in Fig.\,\ref{tc-tsf}, where the superconducting transition temperature (here denoted as $T_c$) is compared on logarithmic scales with characteristic temperature scale $T_0$ of the correlations \cite{sarr07}. Note that because $T_0$ represents essentially an effective band width, this does not capture just spin fluctuation mediated pairing. 

Regarding the bulk properties of the f-electrons superconductors reviewed here a host of characteristics suggests unconventional superconductivity with a complex nodal structure of the superconducting gap. A particularly remarkable property concerns the large upper critical field. In the immediate vicinity of a quantum critical point these upper critical fields become additionally enhanced. Examples include URhGe, {\cerhsiot} and {\ceirsiot}. It will be every interesting to learn more about the mechanism underlying this exceptional enhancement.

A common theme for many of the systems covered in this review is the vicinity of the superconductivity to inherent Fermi surface instabilities. In the bulk properties this may be seen in the observed deviations from Fermi liquid behavior. As a rather remarkable microscopic piece of information quantum oscillatory studies under pressure reveal changes of the Fermi surface topology precisely where the superconductivity is most pronounced. Examples include {\cerhsi}, {\cein}, {\cerhin} and {\uge}.  This contrasts the traditional Ansatz to treat superconductivity as a property of stable Fermi liquids. It may therefore be highly instructive to investigate both theoretically and experimentally scenarios of superconductivity in the vicinity of Fermi surface reconstructions. For the case of the high-$T_c$ cuprates this question has been explored extensively in a variety of scenarios, such as Pomeranchuk instabilities, preformed pairs, orbital currents and stripes.  In this context it is interesting to consider, whether the recent discovery of electron-pockets of the Fermi surface in a hole doped cuprate actually hits on yet another analogy of the superconducting phases of f-electron compounds \cite{pfle07b}.

Many compounds discovered so far exhibit superconductivity in the vicinity of zero temperature instabilities. Examples are the systems like {\cemt}, {\cemin}, {\uge}, URhGe, UCoGe, UIr, {\ceptsi} and CeMX$_3$. It has thereby been noticed that moderate anisotropies of the electronic and crystal structure promote the occurrence of superconductivity, while full inversion symmetry of the crystal structure does not seem to be a precondition. These studies suggest as a requirement for superconductive pairing the need to balance stronger interactions that would otherwise lead to other forms of order such as magnetism. Although this is an important theme, it is also important to keep in mind that the recent discoveries were made by following this approach experimentally to start with. It is then interesting to note that a number of compounds are also quite insensitive to pressure. Examples are {\cecuge} above $p_c$, {\updal}, {\unial}, {\ube}. This implies either, that we do not have an appropriate control parameter to change the particular balance in these compounds, or it suggests that unconventional forms of superconductivity exist, that are much more robust and do not require the vicinity to a zero temperature instability.

Experimentally the types of f-electron superconductors observed so far enforce the question why heavy-fermion superconductivity has only been observed in systems containing Ce, Pr, U, Pu and Np? There is a priori no reason, why compounds based on other f-electron elements should not also exhibit unconventional forms of superconductivity. Clearly, as concerns the electronic properties of these compounds the understanding must be far from complete. For instance, superconductivity has recently been reported in the Yb-boride $\beta$-YbAlB$_4$ \cite{naka08b} and pure Eu metal \cite{debe09}.  

Last but not least, the importance of high sample quality cannot be emphasized enough. It is not just that the unconventional superconductivity tends to be extraordinarily sensitive to defects. Well characterized high quality single crystals are also essential to unravel the precise nature of the superconductivity alongside any other electronic properties in these compounds. Once high quality samples are available, controlled experimental techniques to systematically screen the evolution of these materials as a function of a non-thermal control parameter have become the outstanding tool. 

We conclude this review with the remark, that it is generally very difficult to assign unambiguously the possible pairing interactions to a single interaction channel in a number of the f-electron compounds. For example in {\updal} both an antiferromagnetically mediated and excitonic pairing mechanism have been proposed. This underscores quite generally the need for a description based on a coupling of two or more correlated subsystems. From a purely esthetic point of view complex coupled systems tend to appear less beautiful, because they are generally over-parametrized and less tractable. However, the very need to consider these complexities also emphasizes the enormous potential for new and entirely unexpected phenomena, many of which are yet to be discovered.

%%%%%%%%%%%%%%%%%%%%%%%%%%%%%%%%%%%%%%%%%%%%%%%%%
\section*{Acknowledgments}
I wish to thank P. B\"oni. R. Hackl and M. Sigrist for carefully reading the manuscript. Comments by S. Fujimoto, T. Kobayashi, V. P. Mineev, K. Moore, J. D. Thompson and F. Steglich are gratefully acknowledged. 

%\bibliographystyle{apsrmp}
%\bibliography{RMP-references}
%\end{document} 
%%%%%%%%%%%%%%%%%%%%%%%%%%%%%%%%%%%%%%%%%%%%%%%%%
%%%%%%%%%%%%%%%%%%%%%%%%%%%%%%%%%%%%%%%%%%%%%%%%%

\end{document}